\documentclass[11pt,preprint]{article}
\pdfoutput=1
\usepackage{jheppub}
\usepackage[usenames,dvipsnames]{xcolor}
\usepackage{mathrsfs}

\usepackage{tikz}
\usetikzlibrary{decorations.pathmorphing,decorations.pathreplacing,calligraphy,calc,snakes,cd,external,arrows.meta,patterns}

\usepackage[export]{adjustbox}
\usepackage{hyperref}
\usepackage{bbm}
\usepackage{mathtools}
\usepackage{empheq}
\usepackage{nccmath}
\usepackage{blkarray}
\usepackage{esint}
\usepackage{microtype}
\usepackage{slashed}

\let\oldFootnote\footnote
\newcommand\nextToken\relax

\renewcommand\footnote[1]{%
    \oldFootnote{#1}\futurelet\nextToken\isFootnote}

\newcommand\isFootnote{%
    \ifx\footnote\nextToken\textsuperscript{,}\fi}
    
\hypersetup{
    pdfencoding=unicode,
	colorlinks=true,
	urlcolor=Maroon,
	linkcolor=RoyalBlue,
	citecolor=Maroon,
	pdftitle={Crossing beyond scattering amplitudes},
	pdfauthor={Simon Caron-Huot, Mathieu Giroux, Holmfridur S. Hannesdottir, Sebastian Mizera},
	pdfdisplaydoctitle=true,
	pdfstartview=FitH,
	linktocpage=true
}

\def\>{\rangle}
\def\<{\langle}

\def\vac#1{\<0|#1|0\>}
\def\adag{a^\dagger}
\def\bdag{b^\dagger}
\def\cM{\mathcal{M}}

\def\le{\preceq}
\def\.{{\cdot}}

\def\d{\mathrm{d}}

\def\be{\begin{equation}}
\def\ee{\end{equation}}
\def\zb{\bar{z}}
\def\Li{\mathrm{Li}}
\def\Cut{\mathrm{Cut}}

\DeclareMathOperator{\disc}{Disc}
\def\M{\mathcal{M}}
\def\e{\mathrm{e}}

\def\eps{\varepsilon}
\def\tri{\mathrm{tri}}
\def\ot{\leftarrow}
\def\gammaE{\gamma_{\text{E}}}
\renewcommand{\leq}{\leqslant}
\renewcommand{\le}{\leqslant}
\renewcommand{\geq}{\geqslant}
\renewcommand{\ge}{\geqslant}

\usetikzlibrary{decorations.markings}
\usetikzlibrary{decorations.pathmorphing}
\tikzset{->-/.style={decoration={
  markings,
  mark=at position #1 with {\arrow{>}}},postaction={decorate}}}
\tikzset{-latex-/.style={decoration={
  markings,
  mark=at position #1 with {\arrow{latex}}},postaction={decorate}}}
\tikzset{-latex-reverse-/.style={decoration={
  markings,
  mark=at position #1 with {\arrow{Latex[reversed, scale=0.8]}}},postaction={decorate}}}

\newcommand\doublelineGRAY[3]{%
    \draw[color=gray!10] ($(#1)+(#3:0.035)$) -- ($(#2)+(#3:0.035)$);
    \draw[color=gray!10] ($(#1)+(#3:-0.035)$) -- ($(#2)+(#3:-0.035)$);
}
\newcommand\doubleline[3]{%
    \draw[Maroon,dash pattern=on 1pt off 0.6pt] ($(#1)+(#3:0.035)$) -- ($(#2)+(#3:0.035)$);
    \draw[RoyalBlue] ($(#1)+(#3:-0.035)$) -- ($(#2)+(#3:-0.035)$);
}
\newcommand\doublelineR[3]{%
    \draw[Maroon,dash pattern=on 1pt off 0.6pt] ($(#1)$) -- ($(#2)$);
}
\newcommand\doublelineB[3]{%
    \draw[RoyalBlue] ($(#1)$) -- ($(#2)$);
}
\usetikzlibrary{shapes.misc}
\tikzset{cross/.style={cross out, draw=black, very thick, minimum size=2*(#1-\pgflinewidth), inner sep=0pt, outer sep=0pt}, 
cross/.default={1pt}}

\def\centerarc[#1](#2)(#3:#4:#5)%
    { \draw[#1] ($(#2)+({#5*cos(#3)},{#5*sin(#3)})$) arc (#3:#4:#5); }
\tikzset{
    vector/.style={
        decoration={snake, aspect=0.75, mirror, segment length=2mm},
        decorate
    },
	photon/.style={decorate, decoration={snake, amplitude=1pt, segment length=6pt}
	}
}

\definecolor{charcoal}{HTML}{343837}

\def\AA{\mathcal{A}}
\def\DD{\mathcal{D}}
\def\D{\mathrm{D}}
\def\rd{\mathrm{d}}

\def\GL{\mathrm{GL}}
\def\rotateup{\scalebox{1.05}{$\curvearrowleft$}}
\def\rotatedown{\rotatebox[origin=c]{180}{$\curvearrowleft$}}
\def\onembox{\mathrm{box}}
\def\boxx{\mathrm{box}}
\def\leq{\leqslant}
\def\geq{\geqslant}
\def\Im{\mathrm{Im}\,}
\def\Re{\mathrm{Re}\,}
\def\Z{\mathcal{Z}}
\def\N{\mathcal{N}}
\def\R{\mathcal{R}}
\def\B{\mathcal{B}}
\def\cut{\mathrm{Cut}}

\def\Exp{\mathrm{Exp}}

\DeclareMathOperator*{\sumint}{%
\mathchoice%
  {\ooalign{$\displaystyle\sum$\cr\hidewidth$\displaystyle\int$\hidewidth\cr}}
  {\ooalign{\raisebox{.14\height}{\scalebox{.7}{$\textstyle\sum$}}\cr\hidewidth$\textstyle\int$\hidewidth\cr}}
  {\ooalign{\raisebox{.2\height}{\scalebox{.6}{$\scriptstyle\sum$}}\cr$\scriptstyle\int$\cr}}
  {\ooalign{\raisebox{.2\height}{\scalebox{.6}{$\scriptstyle\sum$}}\cr$\scriptstyle\int$\cr}}
}

\definecolor{darkgreen}{rgb}{0.0, 0.4, 0.0}

\usepackage[titles]{tocloft}
\usepackage{dirtytalk}
\usepackage{changepage}
\usepackage{subcaption}
\usepackage{float}
\usepackage{enumitem}  
\usepackage{stackengine}
\usepackage{bm}

 \DeclarePairedDelimiterXPP\EV[1]{E}(){}{
 
 #1}
 \DeclarePairedDelimiterXPP\Var[1]{V}(){}{
 
 #1}

\usetikzlibrary{fadings}
\tikzfading[name=fade right,
right color=transparent!0,
left color=transparent!100]

\colorlet{choral}{RoyalBlue}
\colorlet{darkred}{Maroon}
\definecolor{neworange}{HTML}{E99F0C}
\colorlet{Orange}{neworange}
\colorlet{orange}{neworange}
\colorlet{darkorange}{neworange}
\tikzset{
    partial ellipse/.style args={#1:#2:#3}{
        insert path={+ (#1:#3) arc (#1:#2:#3)}
    }
}

\tikzset{gradRtoB/.style={
    postaction={
        decorate,
        decoration={
            markings,
            mark=at position \pgfdecoratedpathlength-0.5pt with {\arrow[blue,line width=#1] {}; },
            mark=between positions 0 and \pgfdecoratedpathlength-0pt step 0.5pt with {
                \pgfmathsetmacro\myval{multiply(divide(
                    \pgfkeysvalueof{/pgf/decoration/mark info/distance from start}, \pgfdecoratedpathlength),100)};
                \pgfsetfillcolor{RoyalBlue!\myval!Maroon!};
                \pgfpathcircle{\pgfpointorigin}{#1};
                \pgfusepath{fill};}
}}}}

\newcommand\shadetext[2][]{%
  \setbox0=\hbox{{#2}}%
  \tikz[baseline=0]\path [#1] \pgfextra{\rlap{\copy0}} (0,-\dp0) rectangle (\wd0,\ht0);%
}

\title{Crossing beyond scattering amplitudes}
\usepackage{orcidlink}
\author[\!a,b,\orcidlink{0000-0002-7005-9652}]{Simon Caron-Huot,}\emailAdd{schuot@physics.mcgill.ca}
\author[\!a,\orcidlink{0000-0002-2672-634X}]{Mathieu Giroux,}\emailAdd{mathieu.giroux2@mail.mcgill.ca}
\author[\!b,\orcidlink{0000-0002-5440-2086}]{Holmfridur S. Hannesdottir,}\emailAdd{hofie@ias.edu}
\author[\,b,\orcidlink{0000-0002-8066-5891}]{\\ Sebastian Mizera}\emailAdd{smizera@ias.edu}

\affiliation{$^a$Department of Physics, McGill University, 3600 Rue University, Montr\'eal, H3A 2T8, QC Canada}
\affiliation{$^b$Institute for Advanced Study, Einstein Drive, Princeton, NJ 08540, USA}

\abstract{
We find that different asymptotic measurements in quantum field theory can be related to one another through new versions of \emph{crossing symmetry}. Assuming analyticity, we conjecture generalized crossing relations for multi-particle processes and the corresponding paths of analytic continuation. We prove them to all multiplicity at tree-level in quantum field theory and string theory. We illustrate how to practically perform analytic continuations on loop-level examples using different methods, including unitarity cuts and differential equations. We study the extent to which anomalous thresholds away from the usual physical region can cause an analytic obstruction to crossing when massless particles are involved. In an appendix, we review and streamline historical proofs of four-particle crossing symmetry in gapped theories.
}

\begin{document} 

\addtocontents{toc}{\protect\thispagestyle{empty}}

\maketitle

\thispagestyle{empty}

\setcounter{page}{3}
\allowdisplaybreaks
\newpage
\section{Introduction}

Crossing symmetry states that particles can be interpreted as anti-particles traveling backward in time. More precisely, it means that scattering amplitudes in different kinematic channels are analytic continuations of one and the same function. This proposal equips us with a powerful computational tool: if one, for example, knows the scattering amplitude for electron-positron annihilation, $e^-e^+ \to \gamma \gamma$, then the amplitude for Compton scattering $e^- \gamma \to e^- \gamma$ can simply be obtained by analytic continuation, as illustrated below in Fig.~\ref{fig:introcross1}.
Since its proposal in 1954~\cite{Gell-Mann:1954ttj}, crossing symmetry has grown to be widely accepted as a fact of life with many applications, including to dispersion relations, though its theoretical foundations have remained arguably on shaky grounds beyond the simplest cases.
For example, even though Feynman diagrams for crossed processes \emph{look} the same, since they are obviously built out of the same propagators and vertices with simply some energy signs changed, this by no means guarantees that the resulting functions are \emph{continuously} related by a path of analytic continuation.

In the case of $2\to2$ scattering of stable particles in mass-gapped theories, crossing symmetry was rigorously proven by Bros, Epstein, and Glaser in the 1960s~\cite{Bros:1964iho,Bros:1965kbd} within the framework of axiomatic field theory. The proof involves two main parts: first,  showing that there exists a region in the \textit{off-shell} kinematic space for which Green's functions corresponding to two physical channels are the same, and second---vastly more difficult---that there exists an analytic continuation between these Green's functions purely within the \textit{on-shell} kinematics. Since then, there has been intermittent progress on crossing symmetry, including an extension of the proof to the $2\to3$ case in \cite{Bros:1972jh,Bros:1985gy}, and more recently a perturbative proof of crossing in the planar limit for any multiplicities and masses in \cite{Mizera:2021fap}.

In this paper, we take a step towards understanding crossing for scattering at higher points by asking: assuming that an analytic path interchanging some incoming particles to outgoing anti-particles exists, what is the result of the continuation? For instance, what could we reasonably expect to be the result of crossing the photon and the positron in the $e^- \gamma \to e^- e^- e^+$ process? Naively, based on the physical arguments of a symmetry between particles and anti-particles, one could expect that the answer is the amplitude in the crossed channel, or perhaps its complex conjugate.

\begin{figure}[t]
    \centering
    \begin{tikzpicture}[line width=1,draw=charcoal,yshift=-1]
    \begin{scope}[xshift=-175,yshift=25]
        \draw[] (0.1,0) -- (0.3,0);
    \end{scope}
  \draw[->] (-6.5, 0) -- (6.5, 0);
  \draw[->] (0, -0.3) -- (0, 4.5);
  \node[] at (6,4.0) {$s$};
  \draw[shift={(6,4)},scale=0.5] (-0.5,0.5) -- (-0.5,-0.5) -- (0.5,-0.5);
  \centerarc[<->,Maroon,thick](0,-1)(35:145:5);
  \begin{scope}[xshift=-165,yshift=25,scale=0.7]
  \draw[photon,RoyalBlue] (2,0) -- ++(-155:1.45) node[left] {\small$\gamma$};
  \draw[->,photon,RoyalBlue] (2,0) -- ++(-155:1.15);
\draw[Maroon] (2,0) -- ++(25:1.35) node[right] {\small$e^+$};
\draw[Maroon,-<] (2,0) -- ++(25:1.15);
\draw[RoyalBlue] (2,0) -- ++(-25:1.35) node[right] {\small$e^-$};
\draw[RoyalBlue,-<] (2,0) -- ++(-25:1.15);
  \draw[photon,Maroon] (2,0) -- ++(155:1.45) node[left] {\small$\gamma$};
  \draw[->,photon,Maroon] (2,0) -- ++(155:1.15);
\filldraw[fill=gray!5, line width=1.3pt](2,0) circle (0.6) node[yshift=1] {$S^\dag$};
  \end{scope}
    \begin{scope}[xshift=90,yshift=25,scale=0.7]
\draw[Maroon] (2,0) -- ++(-155:1.35) node[left] {\small$e^-$};
\draw[Maroon,->] (2,0) -- ++(-155:1.15);
  \draw[photon,Maroon] (2,0) -- ++(25:1.45) node[right] {\small$\gamma$};
  \draw[-<,photon,Maroon] (2,0) -- ++(25:1.15);
\draw[RoyalBlue] (2,0) -- ++(-25:1.35) node[right] {\small$e^-$};
\draw[RoyalBlue,-<] (2,0) -- ++(-25:1.15);
  \draw[photon,RoyalBlue] (2,0) -- ++(155:1.45) node[left] {\small$\gamma$};
  \draw[->,photon,RoyalBlue] (2,0) -- ++(155:1.15);
\filldraw[fill=gray!5, line width=1.3pt](2,0) circle (0.6) node {$S$};
  \end{scope}
\end{tikzpicture}
    \caption{Crossing symmetry illustrated on a $2\to2$ scattering processes. If we cross an electron and photon in Compton scattering, $e^- \gamma \to e^- \gamma$ on the right, we end up with electron-photon annihilation, $e^+ e^- \to \gamma \gamma$ on the left. The particles that cross are colored in red. The Mandelstam invariant $s$ crosses from being positive in one process to negative in the other. Note that time flows from right to left in these diagrams.}
    \label{fig:introcross1}
\end{figure}
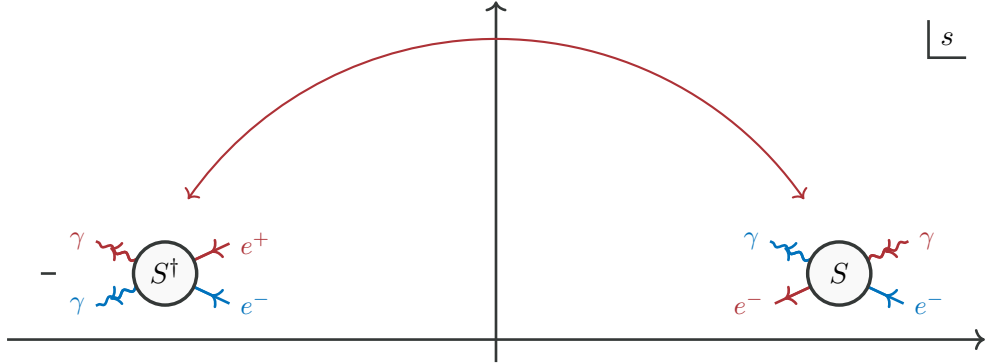

Remarkably, it turns out that crossing is much richer: it does not only relate scattering amplitudes to each other, but rather, it relates different types of observables!  A simple way to motivate this idea is to recall why amplitudes are analytic in the first place.  The traditional explanation involves microcausality: vanishing of spacelike commutators makes momentum-space correlators analytic in some domains, as reviewed below.
The appearance of commutators suggests that analytic continuation can modify the operator ordering, and thus lead to observables in which the time ordering of operators is relaxed.

There are many contexts in which such observables occur \cite{Caron-Huot:2023vxl}. For example, when the scattered objects have a large degeneracy of internal states, exclusive S-matrix elements between individual initial and final states are typically entropically suppressed $\sim \e^{-S/2}$ and can exhibit a high sensitivity to the details of initial and final states.
In such situations, ``in-in'' expectation values, where one inclusively sums over unobserved final states, are typically closer to experiment and simultaneously easier to calculate using either many-body tools or semi-classical approximations.  As another example, according to the Kosower--Maybee--O'Connell (KMOC) formalism~\cite{Kosower:2018adc,Cristofoli:2021vyo}, on-shell in-in observables can describe gravitational waveforms as detected by the LIGO--Virgo--KAGRA detectors \cite{LIGOScientific:2016aoc,LIGOScientific:2017vwq,LIGOScientific:2017ync}, whether or not the participating black holes or compact objects have a large entropy. The waveform in a related scattering problem is then computed as the expectation value of a graviton in the background of black-hole scattering (which can be related to the emission from bound orbits using effective field theory logic). 
The KMOC formalism has recently been used to compute gravitational waveforms in perturbation theory by modeling the black holes as heavy scalars~\cite{Brandhuber:2023hhy,Herderschee:2023fxh,Elkhidir:2023dco,Georgoudis:2023lgf,Caron-Huot:2023vxl}.

The general non-time-ordered observables that will appear in this paper can be characterized by arbitrary strings of creation/annihilation operators, $\adag$/$a$ and
$\bdag$/$b$, that, respectively, act in the far past and future.
Besides in-in expectation values and inclusive cross sections, these include more exotic out-of-time-order amplitudes that have been classified in a recent paper~\cite{Caron-Huot:2023vxl}, which also gave various methods to compute them.  Thus, we will find that
analyticity of the S-matrix unites families of asymptotic observables. This statement is powerful, since knowing one asymptotic observable in a family, others can be computed simply by analytic continuation. 
It also means that aspects of one observable, such as symbol properties, automatically translate to the others.

The crossing paths we will consider involve a set of particles that are highly boosted along a specific lightcone direction, say the \say{$p^+$} direction.  We will use a single complex variable $z$ to apply a complex boost to these particles while staying entirely on-shell, thus rotating some momenta from incoming to outgoing and vice versa. This complex boost coincides with a standard CPT transformation \cite{Streater:1989vi}, however one which acts on only a \emph{subset} of the momenta.
We stress that, because this complex boost is not applied to all momenta, crossing symmetry does not merely follow from CPT invariance of amplitudes. 

\begin{figure}[t]
    \centering
    \begin{tikzpicture}[line width=1,draw=charcoal]
    \begin{scope}[xshift=-195,yshift=25]
        \draw[] (0.1,0) -- (0.3,0);
    \end{scope}
  \draw[->] (-6.5, 0) -- (6.5, 0);
  \draw[->] (0, -0.3) -- (0, 4.5);
  \node[] at (6,4.0) {$z$};
  \draw[shift={(6,4)},scale=0.5] (-0.5,0.5) -- (-0.5,-0.5) -- (0.5,-0.5);
  \centerarc[<->,Maroon,thick](0,-1)(35:145:5);
  \begin{scope}[xshift=-140,yshift=25,scale=0.7]
\draw[RoyalBlue] (0,0.3) -- (-1.2,0.3) node[left] {\small$e^-$};
\draw[RoyalBlue,->] (0,0.3) -- (-1,0.3);
\draw[RoyalBlue] (0,-0.3) -- (-1.2,-0.3) node[left] {\small$e^-$};
\draw[RoyalBlue,->] (0,-0.3) -- (-1,-0.3);
\draw[Maroon] (2,0.3) -- (3.2,0.3) node[right] {\small$e^-$};
\draw[Maroon,-<] (2,0.3) -- (3,0.3);
\draw[RoyalBlue] (2,-0.3) -- (3.2,-0.3) node[right] {\small$e^-$};
\draw[RoyalBlue,-<] (2,-0.3) -- (3,-0.3);
\draw[photon,Maroon] (2,0) -- ++(135:1.40) node[left] {\small$\gamma$};
\draw[->,photon,Maroon] (2,0) -- ++(135:1.15);
\filldraw[fill=gray!30](0,-0.3) rectangle (2,0.3);
\draw (2,0.3) -- (0.95,0.3);
\draw (2,-0.3) -- (0.95,-0.3);
\draw[] (1,0) node {$\medmath{X}$};
\filldraw[fill=gray!5, line width=1.3pt](0,0) circle (0.6) node[] {$S$};
\filldraw[fill=gray!5, line width=1.3pt](2,0) circle (0.6) node[yshift=1] {$S^\dag$};
\draw[dashed,orange] (1,1.2) -- (1,-0.8);
  \end{scope}
    \begin{scope}[xshift=105,yshift=25,scale=0.7]
\draw[RoyalBlue] (2,0) -- ++(-145:1.35) node[left] {\small$e^-$};
\draw[RoyalBlue,->] (2,0) -- ++(-145:1.15);
\draw[RoyalBlue] (2,0) -- ++(180:1.35) node[left] {\small$e^-$};
\draw[RoyalBlue,->] (2,0) -- ++(180:1.15);
\draw[photon,Maroon] (3.2,0.3) node[right] {\small$\gamma$} -- (2,0.3);
\draw[->,photon,Maroon] (3.2,0.3) -- (2.8,0.3);
\draw[RoyalBlue] (2,-0.3) -- (3.2,-0.3) node[right] {\small$e^-$};
\draw[RoyalBlue,-<] (2,-0.3) -- (3,-0.3);
\draw[Maroon] (2,0) -- ++(145:1.35) node[left] {\small$e^+$};
\draw[->,Maroon] (2,0) -- ++(145:1.15);
\filldraw[fill=gray!5, line width=1.3pt](2,0) circle (0.6) node {$S$};
  \end{scope}
\end{tikzpicture}
    \caption{Crossing symmetry illustrated on a $2\to3$ processes. We cross an electron (photon) in the scattering process $e^- \gamma \to e^- e^- e^+ $ from outgoing (incoming), using a path parameterized by $z$. The particles that cross are colored in red. The original scattering amplitude is obtained for $z>0$, but after the analytic continuation to $z<0$, we land on the inclusive observable for measuring a photon in the background of $e^-e^-$ scattering. The observable is obtained as the conjugated S-matrix for the process $e^- e^- \to \gamma X$ times the S-matrix for $X \to e^- e^-$, where $X$ contains all possible states which must be summed and integrated over.}
    \label{fig:introcross2}
\end{figure}
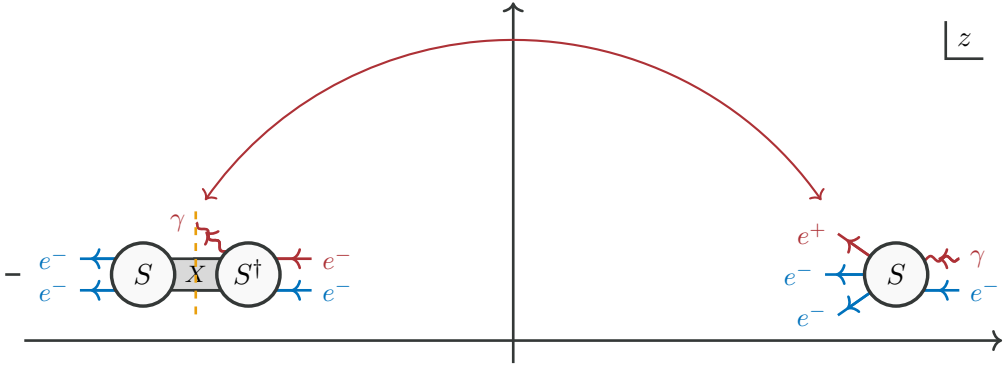

The result of applying this continuation to the $2\to 3$ process $e^- \gamma \to e^- e^- e^+$ is shown in Fig.~\ref{fig:introcross2}: for $z<0$ it yields an inclusive measurement of the photon field in the background of an $e^- e^-$.  Replacing the electrons with black holes (or other compact objects), and the photon by a graviton, yields the gravitational waveform mentioned above. This is a striking result: inclusive waveforms can be computed by analytically continuing the \textit{exclusive} amplitude for a graviton and black hole to scatter into two black holes and an anti-black hole!%
\footnote{
Analytic continuations relating inclusive cross-sections to exclusive scattering
amplitudes have been discussed in the old Regge theory literature, see \cite[Ch.~6]{Brower:1974yv}.
}

Starting from a scattering amplitude $AB \to CD$, our main conjectured
\textit{crossing equation} predicts what happens when we cross an incoming single-particle state $B$ with an outgoing $C$:
\begin{equation}
\begin{gathered}
\begin{tikzpicture}[line width=1]
\def\Ang{30};
\def\CAng{150};
\def\CosVal{0.8660};
\def\SinVal{0.5};
\def\CstVal{0.1};
\begin{scope}[xshift=0]
\coordinate (c) at (0,0);
\draw[->-=.35,Maroon] (c)++(50:1) node[right] {\footnotesize$2$} -- (c);
\draw[->-=.80,Maroon] (c) -- ++(130:1) node[left] {\footnotesize$3$};
\draw[RoyalBlue] ($(c) + (\SinVal*\CstVal,\CosVal*\CstVal)$)++(-\Ang:1.15) -- ($(c)+(\SinVal*\CstVal,\CosVal*\CstVal)$);
\draw[RoyalBlue] ($(c) + (-\SinVal*\CstVal,-\CosVal*\CstVal)$)++(-\Ang:1.15) -- ($(c)+(-\SinVal*\CstVal,-\CosVal*\CstVal)$);
\draw[RoyalBlue] ($(c)+(0,0)$)++(-\Ang:1.15) -- ($(c)+(0,0)$);
\draw[RoyalBlue] ($(c) + (\SinVal*\CstVal,-\CosVal*\CstVal)$)++(-\CAng:1.15) -- ($(c)+(\SinVal*\CstVal,-\CosVal*\CstVal)$);
\draw[RoyalBlue] ($(c) + (-\SinVal*\CstVal,\CosVal*\CstVal)$)++(-\CAng:1.15) -- ($(c)+(-\SinVal*\CstVal,\CosVal*\CstVal)$);
\draw[RoyalBlue] ($(c)+(0,0)$)++(-\CAng:1.15) -- ($(c)+(0,0)$);
\filldraw[fill=gray!5](0,0) circle (0.4) node {$S$};
\draw [pen colour={gray},
    decorate, 
    decoration = {calligraphic brace,
        raise=5pt,
        amplitude=2pt}] (1.0,0.4+0.65) --  (1.0,0.4)
node[pos=0.5,right=10pt,Maroon]{$B$};
\draw [pen colour={gray},
    decorate, 
    decoration = {calligraphic brace,
        raise=5pt,
        amplitude=2pt}] (1.0,-0.2) --  (1.0,-0.2-0.65)
node[pos=0.5,right=10pt,RoyalBlue]{$A$};
\draw [pen colour={gray},
    decorate, 
    decoration = {calligraphic brace,
        raise=5pt,
        amplitude=2pt}] (-1.0,0.4) -- (-1.0,0.4+0.65) 
node[pos=0.5,left=10pt,Maroon]{$C$};
\draw [pen colour={gray},
    decorate, 
    decoration = {calligraphic brace,
        raise=5pt,
        amplitude=2pt}] (-1.0,-0.2-0.65) -- (-1.0,-0.2)
node[pos=0.5,left=10pt,RoyalBlue]{$D$};
\end{scope}
\node[align=center] at (3,0.5) {\footnotesize cross ${\textcolor{Maroon}{2} \leftrightarrow \textcolor{Maroon}{3}}$};
\draw[-latex] (2.2,0) -- (3.8,0);
\draw[line width=0.8] (4.4,0) -- (4.65,0);
\begin{scope}[xshift=200]
\coordinate (c) at (0,0);
\draw[->-=.35,Maroon] (c)++(50:1) node[right] {\footnotesize$\bar{3}$} -- (c);
\draw[->-=.80,Maroon] (c) -- ++(130:1) node[left] {\footnotesize$\bar{2}$};
\draw[RoyalBlue] ($(c) + (\SinVal*\CstVal,\CosVal*\CstVal)$)++(-\Ang:1.85) -- ($(c)+(\SinVal*\CstVal,\CosVal*\CstVal)$);
\draw[RoyalBlue] ($(c) + (-\SinVal*\CstVal,-\CosVal*\CstVal)$)++(-\Ang:1.85) -- ($(c)+(-\SinVal*\CstVal,-\CosVal*\CstVal)$);
\draw[RoyalBlue] ($(c)+(0,0)$)++(-\Ang:1.85) -- ($(c)+(0,0)$);
\draw[RoyalBlue] ($(c) + (\SinVal*\CstVal,-\CosVal*\CstVal)$)++(-\CAng:1.85) -- ($(c)+(\SinVal*\CstVal,-\CosVal*\CstVal)$);
\draw[RoyalBlue] ($(c) + (-\SinVal*\CstVal,\CosVal*\CstVal)$)++(-\CAng:1.85) -- ($(c)+(-\SinVal*\CstVal,\CosVal*\CstVal)$);
\draw[RoyalBlue] ($(c)+(0,0)$)++(-\CAng:1.85) -- ($(c)+(0,0)$);
\filldraw[fill=gray!30,rotate=-30](0,-0.2) rectangle (1,0.2);
\filldraw[fill=gray!30,rotate=-150](0,-0.2) rectangle (1,0.2);
\filldraw[fill=gray!5](0,0) circle (0.4) node[yshift=1] {$S^\dag$};
\filldraw[fill=gray!5]($(0,0)+(-30:1.2)$) circle (0.4) node {$S$};
\filldraw[fill=gray!5]($(0,0)+(-150:1.2)$) circle (0.4) node {$S$};
\node[] at (-0.54,-0.28) {\tiny $Y$};
\node[] at (0.48,-0.28) {\tiny $X$};
\draw [pen colour={gray},
    decorate, 
    decoration = {calligraphic brace,
        raise=5pt,
        amplitude=2pt}] (1.0,0.4+0.65) --  (1.0,0.4)
node[pos=0.5,right=10pt,Maroon]{$\bar{C}$};
\draw [pen colour={gray},
    decorate, 
    decoration = {calligraphic brace,
        raise=5pt,
        amplitude=2pt}] (1.6,-0.55) --  (1.6,-1.2)
node[pos=0.5,right=10pt,RoyalBlue]{$A$};
\draw [pen colour={gray},
    decorate, 
    decoration = {calligraphic brace,
        raise=5pt,
        amplitude=2pt}] (-1.0,0.4) -- (-1.0,0.4+0.65) 
node[pos=0.5,left=10pt,Maroon]{$\bar{B}$};
\draw [pen colour={gray},
    decorate, 
    decoration = {calligraphic brace,
        raise=5pt,
        amplitude=2pt}] (-1.6,-1.2) --  (-1.6,-0.55)
node[pos=0.5,left=10pt,RoyalBlue]{$D$};
\draw[dashed,orange] (0.5,-1) -- (0.5,1);
\draw[dashed,orange] (-0.5,-1) -- (-0.5,1);
\end{scope}
\end{tikzpicture}
\end{gathered}
\label{eq:crossingintro}
\end{equation}
Following \cite{Caron-Huot:2023vxl}, we use conventions in which operator
products are taken from right to left, as in bra-ket notation (so time flows to the left in the $S$-blobs). 
The dashed lines represent cuts that sum over the intermediate on-shell states $X$ and $Y$ and integrate over their phase space with positive energies.

The above conjecture will be motivated by rigorous manipulations of off-shell correlators. We thus expect it to hold when the corresponding analyticity properties extend to the mass shell, which we will physically argue for in some situations.  Rigorously proving that last part, similarly to \cite{Bros:1964iho,Bros:1965kbd,Bros:1971ghu,Bros:1985gy}, will require efforts that are beyond the scope of this paper.
Our main goal is to elucidate what statements of crossing and microcausality can actually be valid beyond $2 \to 2$, and to test them in examples. Our results are not restricted to axiomatic field theory, and we consider various processes order-by-order in perturbation theory, as well as individual diagrams with massless and massive internal particles in dimensional regularization.

In Sec.~\ref{sec:Smatrix} we recall the main axioms of S-matrix theory and how they naturally allow for generalized (non-time-ordered) amplitudes;
we also review the classical arguments linking microcausality and analyticity, highlighting subtleties with the on-shell limits.
In Sec.~\ref{sec:crossing} we combine these ideas to formulate the crossing equation \eqref{eq:crossingintro}, illustrating in simple tree-level and one-loop cases why all factors are necessary.
We speculate in Sec.~\ref{sec:conjectureII} on a minimal extension to general multi-particle clusters $A,B,C$ and $D$. Intriguingly, this extension makes diagrammatic sense, but cannot be written as a string of operators acting on a Hilbert space.

We will stress that crossing is already non-trivial at \textit{tree level} if one pays close attention to the $i\varepsilon$ prescriptions in the propagators of (unstable) heavy intermediate particles. In Sec.~\ref{sec:treelevelProof} we give a combinatorial proof that our conjecture (including its speculative multi-particle extension) holds at tree level for any multiplicity.

In Sec.~\ref{sec:fivepoint} we continue our investigations to one loop
and consider all master integrals for massless five-particle amplitudes. The crossing equation relates various amplitudes and cuts in different channels, which we test using a variety of techniques for analytic continuation of (i) closed-form expressions, (ii) loop-momentum Feynman integrals, (iii) Schwinger-parametrized integrals, and (iv) differential equations. The crossing equation is verified by comparing these results to direct computations of unitarity cuts. 
We also highlight the existence of moves which relate (ordinary) amplitudes to amplitudes, leading to non-trivial relations between integration constants in different channels for the differential equation method.

The prerequisite for crossing to work, i.e., that amplitudes are actually analytic on the mass shell along the considered path, is not merely a mathematical detail waiting to be eventually ironed out.
In Sec.~\ref{sec:anomalous} we give a non-trivial counterexample,
involving massive external legs and a massless internal particle,
where our crossing conjecture fails as a result of landing
us on the ``wrong side'' of an anomalous threshold branch cut. We introduce an extension of the Coleman--Norton picture of anomalous thresholds to illustrate how this phenomenon is caused by internal particles moving faster than the crossed external ones, thus highlighting
the importance of a mass gap in the arguments from the last century.
However, we will also observe that even in gapless theories,
certain crossing moves, including the one in Fig.~\ref{fig:introcross2}, appear to be immune from this phenomenon,
suggesting that a mass gap is not always necessary.

In Sec.~\ref{sec:strings} we analyze the predictions of crossing for tree-level string amplitudes. Analytic continuation can be achieved directly by contour manipulations on the moduli space of Riemann surfaces. After reviewing how to see unitarity cuts from this perspective, we give an example of an in-in observable in string theory by crossing from a time-ordered amplitude and numerically plot the corresponding space-time geometry.
In Sec.~\ref{sec:conclusions} we present our conclusions.

In App.~\ref{app:axiomatic} we offer simplified physicist-level versions of the axiomatic field theory proofs of crossing symmetry and of analyticity near the mass shell.  We highlight powerful simplifications and theorems that have been developed in the 1970's, a decade after the most-cited results were initially obtained, in the hope that these will help answer the questions raised in this work. The remaining appendices give details of technical computations: embedding-space formulae for cuts in App.~\ref{sec:app_embeddingspace}, differential equations for massless pentagon diagram in App.~\ref{app:details}, and numerical approach to solving differential equations in App.~\ref{app:numerics}.

\paragraph{Notations and conventions}

We work in $\D$ spacetime dimensions, and take momentum components of $p^\mu$ to be $p^\mu = (p^0, p^1, \ldots, p^{\D-1})$. The spatial part of $p^\mu$ is labeled with $\vec{p} = (p^1, \ldots, p^{\D-1})$. We work in mostly-plus signature where $p^2=-(p^0)^2+(p^1)^2 + \ldots + (p^{\D-1})^2$ and all-outgoing conventions in which $p_i^0 < 0$ for incoming particles and $p_i^0 > 0$ for the outgoing ones. When using lightcone coordinates, we define
\begin{equation}
    p^{\pm} = p^0\pm p^{\D-1}\,, \quad p^\perp = (p^1, \ldots, p^{\D-2}) \,,
\end{equation}
and hence $p^2 = - p^+ p^- + (p^{\perp})^2$.
To summarize, we will use the following terminology:
\begin{alignat}{3}
    \text{Future timelike:} \quad & p^0 >0 \,, & \quad p^2<0 \,, \\
    \text{Past timelike:} \quad & p^0 <0 \,, & \quad p^2<0 \,, \\
    \text{Spacelike:} \quad & \bullet & \quad p^2>0 \,.
\end{alignat}
We use the notation $p_{ij \cdots k}^\mu = p_i^\mu + p_j^\mu + \ldots + p_k^\mu$, and $s_{ij \cdots k} = - p_{ij \cdots k}^2$ for Mandelstam invariants.

The connected interacting part $T$ of the S-matrix is obtained as usual with
\be
    S=\mathbbm{1}+ i T+\ldots\,,
\ee 
where the dots include disconnected products of $T$'s (that are only present at high multiplicity).
The interacting part of the scattering amplitude, $\mathcal{M}$, is obtained as the matrix element of $T$ with the overall momentum-conserving delta function factored out: $\mathcal{M}_{f \ot i} = \langle f | T | i \rangle (2\pi)^\D \delta^\D (\sum_i p_i^\mu)$. 

We draw diagrams to follow the ordering of operators in bra-ket notation, so that time flows towards the \emph{left} in scattering amplitudes $S$ or $\cM$, and towards the right in conjugated amplitudes $S^\dag$ or $\cM^\dag$ factors.

\section{Background}
\label{sec:Smatrix}
In this section, we review some background material on the S-matrix, ranging from the algebra of asymptotic creation/annihilation operators to causality and crossing symmetry.
\subsection{Review of S-matrix axioms}
In this paper, we use manipulations in axiomatic field theory to motivate our formulae and conjectures for scattering amplitudes and other asymptotic observables. In a recent paper~\cite{Caron-Huot:2023vxl}, we detailed the necessary axioms of S-matrix theory, explained the blob notation and how to efficiently compute asymptotic observables. Here, we provide a brief review of these points.

We assume the existence of an \textit{asymptotic algebra} separately in the far past and far future. In the far past, we have asymptotic creation (annihilation) operators, $a_i^\dag$ ($a_i$), for each particle $i$, which satisfy the usual commutation relations 
\be
[a_i,a^\dagger_j] =
\delta_{i,j}\ 2 |p_i^{0}|\ (2\pi)^{\D-1} \delta^{\D-1}(\vec{p}_i-\vec{p}_j)\,.
\ee
The Kronecker delta function $\delta_{i,j}$ ensures that the flavor and spin indices, which are included in the labels $i$ and $j$, are the same and $p_i^{0}$ is the energy of particle $i$. We assume an analogous algebra in the far future, where the operators are labeled as $b_i$ and $b_i^\dag$.\footnote{Note that the literature often uses $a_i^{\text{in}}$ and $a_i^{\text{out}}$ for the operators $a_i$ and $b_i$, respectively. Here we use $a$'s and $b$'s instead for clarity.} These operators act on equivalent Hilbert spaces. In particular, we assume the existence of a time-invariant vacuum, $|0\rangle$, which is annihilated by all the $a_i$ and all the $b_i$. The operators in the past and future are related by a unitary evolution operator, $S$, according to
\be
b_i=S^\dag a_i S\,,
\ee
and its conjugate $b_i^\dag=S^\dag a_i^\dag S$. The evolution operator acts trivially on the vacuum,
\be
S |0\rangle = S^\dag |0\rangle = |0 \rangle.
\ee
We also assume that one-particle states are stable, which means that
\be
S a_i^\dag |0 \rangle = S^\dag a_i^\dag |0 \rangle = a_i^\dag |0 \rangle,
\ee
along with the analogous equations obtained by replacing $a_i^\dag$ with $b_i^\dag$. We frequently use the shorthand notation $ | i \cdots j \rangle \equiv a_i^\dag \cdots a_j^\dag | 0 \rangle$ and $ \langle i \cdots j | \equiv \langle 0 | a_i \cdots a_j$. Notice, in particular, that states such as $ | i \cdots j \rangle$ are taken by default to live in the far past. 

This setup can be applied to any quantum field theory in which the conventional S-matrix makes sense, such as pion scattering in four dimensions or perturbative QCD in dimensional regularization.

Equipped with these rules, we can build various objects that we call \textit{asymptotic observables} by stringing together creation and annihilation operators. For example, we can string together three $b_i$'s and two $a_i^\dag$'s to get
\begin{equation}
    \langle 0 | b_5 b_4 b_3 a_2^\dag a_1^\dag | 0 \rangle = \langle 0 | S^\dag a_5 S S^\dag a_4 S S^\dag a_3 S a_2^\dag a_1^\dag | 0 \rangle
    = \langle 0 | a_5 a_4 a_3 S a_2^\dag a_1^\dag | 0 \rangle 
    =
    \langle 3 4 5 | S | 1 2 \rangle \,,
\end{equation}
which is the usual (time-ordered) scattering amplitude for $1 2 \to 345$. To match with the time-ordering in these expressions, we will henceforth write such an amplitude using the mirrored convention $345 \ot 12$. An analogous derivation shows that the $m \ot n$ scattering amplitude $\langle n+1 \cdots n+m | S | 1 2 \cdots n\rangle$ is obtained by stringing $m$ operators of $b_i$ type with $n$ operators of $a_i^\dag$ type.

Scattering amplitudes comprise only a small number of possible strings of creation and annihilation operators. We can, for example,  form the following observable,
\begin{equation}
\Exp_3 \equiv 
    \langle 0 | a_5 a_4 b_3 a_2^\dag a_1^\dag | 0 \rangle
    =
    \langle 4 5 | S^\dag a_3 S | 1 2 \rangle = \sumint_X \<4 5|S^\dagger |X\> \<X3|S|1 2\>\,,
\end{equation}
where we have inserted a complete set of states $\mathbbm{1} = \sumint_X |X\rangle \langle X|$ in the rightmost equality. This represents the inclusive amplitude for measuring the particle labeled by $3$ in the background of $4 5 \ot 1 2$ scattering, and it is important enough so that we give it a name: $\Exp_3$. As discussed in detail in \cite{Caron-Huot:2023vxl}, this type of measurement is relevant when the scattered particles have many internal states, in which the exclusive scattering amplitude for producing particle $3$ is suppressed. The relevant object is rather an inclusive observable, where the in-states $|1 2 \rangle$ and $| 4 5\rangle$ are integrated against the same wavefunction, and we measure $3$ after the collision, while being inclusive over anything that might have happened after the scattering. It has been used, for example, in the KMOC formalism~\cite{Kosower:2018adc} to compute the gravitational waveform emitted by scattering of two black holes.

At higher points we can form even more asymptotic observables. For example, at six points, we can build an out-of-time-ordered correlator (OTOC), $\vac{a_6 b_5a_4\bdag_3\adag_2\adag_1}$, or inclusive cross sections, such as $\lim_{4 \to 3} \vac{a_6a_5 \bdag_4 b_3 \adag_2\adag_1} = \sumint_X \< 5 6|S^\dagger|3X\> \<X3|S|1 2\>$. As the number of external particles $n$ grows larger, scattering amplitudes account for ever-smaller fraction of the total possibilities for forming asymptotic observables to ultimately become of measure zero.

We represent asymptotic observables diagrammatically by rewriting all operators in the future using the relation $b_i = S^\dag a_i S$, and inserting a complete set of states. As an example, for the OTOC above, we write
\begin{equation}
    \vac{b_6 b_5 a_4\bdag_3\adag_2\adag_1} = \sumint_{X,Y} \langle 5 6 | S |X \rangle \langle X 4 |  S^\dag |Y 3\rangle \langle Y | S | 1 2 \rangle \,,
\end{equation}
which we then represent diagrammatically as
\begin{equation}
    \adjustbox{valign=c}{
\begin{tikzpicture}[line width=1,scale=0.9]
\draw[<-, color=gray!55] (2,1) -- (1,1) node[above,midway]{\footnotesize{time}};
\draw[->, color=gray!55] (0.4,1) -- (-0.7,1) node[above, midway]{\footnotesize{time}};
\draw[->, color=gray!55] (3.5,1) -- (2.5,1) node[above,midway]{\footnotesize{time}};
\draw[] (0,0.3) -- (-1.2,0.3) node[left] {\small$5$};
\draw[->] (0,0.3) -- (-1,0.3);
\draw[] (0,-0.3) -- (-1.2,-0.3) node[left] {\small$6$};
\draw[->] (0,-0.3) -- (-1,-0.3);
\draw[] (3,0.3) -- (4.2,0.3) node[right] {\small$1$};
\draw[-<] (3,0.3) -- (4,0.3);
\draw[] (3,-0.3) -- (4.2,-0.3) node[right] {\small$2$};
\draw[-<] (3,-0.3) -- (4,-0.3);
\draw[] (1.5,0) -- (2.5,-1) node[right] {\small$3$};
\draw[-<] (1.5,0) -- (2.3,-0.8);
\draw[] (0.5,1) node[above] {\small$4$} -- (1.5,0);
\draw[<-] (0.7,0.8) -- (1.5,0);
\filldraw[fill=gray!30](0,-0.3) rectangle (1.5,0.3);
\filldraw[fill=gray!30](1.5,-0.3) rectangle (3,0.3);
\filldraw[fill=gray!5, line width=1.3pt](0,0) circle (0.6) node {$S$};
\filldraw[fill=gray!5, line width=1.3pt](1.5,0) circle (0.6) node[yshift=1] {$S^\dag$};
\filldraw[fill=gray!5, line width=1.3pt](3,0) circle (0.6) node {$S$};
\draw (2.25,0.3) -- (2.26,0.3);
\draw (2.25,-0.3) -- (2.26,-0.3);
\draw (0.75,0.3) -- (0.76,0.3);
\draw (0.75,-0.3) -- (0.76,-0.3);
\draw[dashed,orange] (0.75,1.2) -- (0.75,-1.2);
\draw[dashed,orange] (2.25,1.2) -- (2.25,-1.2);
\draw[] (0.75,0) node {\small $\medmath{X}$};
\draw[] (2.25,0) node {\small $\medmath{Y}$};
\end{tikzpicture}}
\end{equation}
The diagram portrays how we broke the OTOC into the following contributions: the scattering amplitude for $Y \ot 1 2$, the conjugated amplitude for $X4 \ot Y 3$, and the amplitude for $5 6 \ot X$. Here, both $X$ and $Y$ are inclusively summed and integrated over, that is, we add a phase-space integral of the form $\sumint_X = \sum_{X_j \in X} \prod_{i \in X_j} \int \frac{\rd^{\D}q_i }{(2\pi)^{\D-1}} \delta^+ (q_i^2+m_i^2)$ for any set of states $X_j$ that can take part in the process.
Note that to mirror the convention for the ordering of asymptotic creation/annihilation operators, our convention is that time flows from right to left in the blob diagrams. 

\subsection{Causality and crossing: Naive, but not too naive}

Crossing symmetry of $2\ot 2$ scattering amplitudes is well-established in the context of axiomatic quantum field theory for theories with a mass gap, i.e., where all particles in the spectrum have positive masses \cite{Bros:1965kbd}.  In this subsection, we briefly review the argument in a way that will highlight the new complications at higher multiplicities ($n>4$ particles). To avoid clutter, we discuss below a real scalar theory with a single field $\phi$, normalized such that its two-point function features a canonically normalized pole. We comment on spin later in the next section.

The conventional axiomatic approach involves considering \emph{retarded commutators} of currents in scattering states \cite{Bros:1965kbd,Sommer:1970mr,Bogolyubov:1990kw},
\begin{equation} \label{R amplitude}
    \mathcal{R}(p_3,p_2) = \int \text{d}^\D x_3~\text{d}^\D x_2\, \e^{-ip_2{\cdot}x_2-ip_3{\cdot}x_3}\, \langle 0|\DD\, [j(x_3),j(x_2)]_R\, \AA |0\rangle\,.
\end{equation}
Here, the \emph{current} $j(x)=i(-\partial_x^2+m^2)\phi(x)$
is the derivative of a local field $\phi(x)$ and the \emph{retarded product} of fields is defined as\footnote{The $R$-product \eqref{R product} is to be understood as an operation on fields, similar to the time-ordering symbol, such that derivatives in the current
are allowed to act on the $\theta$-function: $[\partial^\mu \phi(x),\phi(y)]_R\equiv \partial_x^\mu\left([\phi(x),\phi(y)]_R\right)$.
In this setup, \eqref{R amplitude} is analogous to the standard time-ordered LSZ formula, where derivatives acting on $\theta$ are also necessary to correctly account for vertices where two or more external legs join.}
\be
 [\phi(x),\phi(y)]_R = [\phi(x),\phi(y)] \theta(x^0-y^0)\,.
 \label{R product}
\ee
Note that the effect of each $(-\partial_x^2+m^2)$ in the correlation function is simply to amputate external-leg poles at $p^2+m^2=0$ (after Fourier transform) as in the standard LSZ reduction formula. The retarded product is sandwiched between two strings of asymptotic creation and annihilation operators $\AA$ and $\DD$, for example $\AA = a_1^\dag$ and $\DD = b_4$.

As we will see, for real positive energy $p_3^0$, \eqref{R amplitude} evaluates to the amplitude for a certain process, and for real negative $p_3^0$, it evaluates to that of another process when the momenta are taken to be on shell.
The key idea is then to use analyticity to relate these two processes.
It turns out that analyticity will come about because the retarded product \eqref{R product} vanishes unless $x^\mu$ lies inside the \emph{future} lightcone $V^+$ of $y^\mu$ thanks to the $\theta$-function. Consequently, its Fourier transform \eqref{R amplitude} is analytic in some domain. It includes, off-shell, the upper-half energy plane. Let us now make all of this precise.

\paragraph{Retarded product for real momenta}

The first step is to evaluate \eqref{R amplitude} for real and on-shell momenta.  On-shell, the current is a total derivative, namely
\be \label{total derivative}
j(p)\equiv\int \text{d}^\D x\, \e^{-ip{\cdot} x} j(x)
\xrightarrow{\stackrel{\text{on-shell:}}{p^2\to -m^2}}  \int \text{d}^\D x \frac{\partial}{\partial x^\mu}
\left[ \e^{-ip{\cdot} x}\ (-i\partial^\mu_x+p^\mu) \phi(x)\right]\,.
\ee

Observe that this is solely an identity about the Fourier transform of a distribution, and we have not yet made use of the equations of motion. Instead, the dynamics comes into play through the presumption that the product $\e^{-ip{\cdot}x}\phi(x)$ undergoes rapid oscillations as $x$ approaches infinity, \emph{except} potentially along the paths $x^\mu \propto \pm p^\mu$ of particles heading towards infinity. As a result, after integrating against a smooth function of the (on-shell) momentum $p$, on-shell currents are reduced to surface terms along the direction of particles, thereby naturally giving rise to asymptotic creation and annihilation operators:

\be \label{j from a and b}
\lim_{p^2\to -m^2} j(p) \equiv \left\{\begin{array}{ll}
 \adag_{-p} - \bdag_{-p} \qquad & p^0<0 \quad(\text{incoming})\,,\\
 b_p - a_p \qquad & p^0>0 \quad(\text{outgoing})\,.
 \end{array}\right.
\ee
The argument presented above is recognizable from textbook derivations of the LSZ reduction formula (see, for instance, \cite{Srednicki:2007qs}). That said, the $a_p$ and $\bdag_{-p}$ terms are usually omitted in textbooks, given that they vanish for vacuum time-ordered correlators, wherein fields only carry negative frequency components in the past (and positive frequency in the future). Nevertheless, when different operator orderings are taken into consideration, all terms matter. Many examples of reduction formulae obtained from \eqref{j from a and b} were recently discussed in \cite{Caron-Huot:2023vxl}.

When applying the total-derivative argument \eqref{total derivative} to the Fourier transform of \eqref{R product} with respect to $x_2$, one finds surface terms exclusively at past infinity because of the step function:
\be \label{j from a and b 2}
\lim_{p_2^2\to -m^2} [j(x_3),j(p_2)]_R = 
\begin{cases}
    ~[j(x_3),\adag_2-\bdag_2] & p_2^0<0\,,\\
  -[j(x_3),a_2-b_2] \qquad & p_2^0>0\,,
\end{cases}
\ee
where we used the shorthand notation $\adag_2=\adag_{-p_2}$ and $a_{2}=a_{p_2}$, and similarly for the $b$'s/$\bdag$'s. 

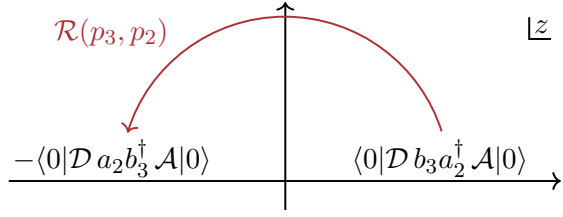
\begin{figure}
     \centering
     \adjustbox{valign=c}{
\tikzset{every picture/.style={line width=0.75pt}}  
\begin{tikzpicture}[x=0.75pt,y=0.75pt,yscale=-1,xscale=1]
\draw[<-]    (178,73.5) -- (178,178.5) ;
\draw[->]    (38.5,163.86) -- (317,163.86) ;
\draw[yshift=-13]   (312,109.5) -- (301,109.5) -- (301,98.5);
\draw  [draw opacity=0] (98.43,139.55) .. controls (108.82,105.6) and (140.4,80.92) .. (177.75,80.92) .. controls (214.83,80.92) and (246.22,105.25) .. (256.84,138.81) -- (177.75,163.86) -- cycle ; \draw[<-]  [color=Maroon  ,draw opacity=1 ] (98.43,139.55) .. controls (108.82,105.6) and (140.4,80.92) .. (177.75,80.92) .. controls (214.83,80.92) and (246.22,105.25) .. (256.84,138.81) ;  
\draw[yshift=-13] (302,98.9) node [anchor=north west][inner sep=0.75pt]  [font=\normalsize]  {$z$};
\draw (210,141.4) node [anchor=north west][inner sep=0.75pt]  [font=\normalsize]  {$\langle 0| \DD\, b_{3} a_{2}^{\dagger}\, \AA |0\rangle$};
\draw (39,141.4) node [anchor=north west][inner sep=0.75pt]  [font=\normalsize]  {$-\langle 0| \DD\, a_{2} b_{3}^{\dagger} \,\AA |0\rangle$};
\draw (60,80) node [anchor=north west][inner sep=0.75pt]  [font=\normalsize,color=Maroon  ,opacity=1 ]  {$\mathcal{R}( p_{3} ,p_{2})$};
\end{tikzpicture}
    }
     \caption{The retarded product provides a path of analytic continuation between the two observables shown on the right and on the left.}
     \label{fig:rprodAC}
 \end{figure}

Below, the regime relevant to crossing is when $p_3$ and $p_2$ have the largest energies in the problem. By momentum conservation, their energies necessarily have opposite signs, and we also assume that $p_2^\mu+p_3^\mu\neq 0$, i.e., the two particles are not forward with each other.
The Fourier transform with respect to $x_3$ of the first line of \eqref{j from a and b 2}
then reduces to $[b_3-a_3,\adag_2-\bdag_2]=[b_3,\adag_2]$.
Furthermore, we can drop one term in the commutator since $b_3 \AA|0\>=0$ due to energy considerations (if $|p_3^0|$ is larger than the energy of the states in $\AA$). Analogous comments apply to the second line in \eqref{j from a and b 2}.
Thus, on shell, we have
\be \label{R reduction}
\mathcal{R}(p_3,p_2) = \left\{\begin{array}{ll}
\< 0|\DD\, b_3 \adag_2\, \AA |0\>, \qquad&
p_3^0\sim {-}p_2^0>0,\quad p_3^0\gg \mbox{all other energies}\,,\\
-\<0|\DD\, a_2 \bdag_3\, \AA |0\>, &
p_2^0\sim{-}p_3^0>0,\quad p_2^0\gg \mbox{all other energies}\,.
\end{array}\right.
\ee
This completes our evaluation of $\mathcal{R}$ for real momenta.
Note that when either $\AA$ or $\DD$ creates a multi-particle state, the second line of \eqref{R reduction} is not a conventional scattering amplitude.
Fig.~\ref{fig:rprodAC} sketches how analytic continuation in energy relates these two objects. The deformation parameter $z$ will be defined more carefully later in Sec.~\ref{sec:conjectureI}.

\paragraph{Analyticity of the retarded product}
Analyticity of $\mathcal{R}$ at complex momenta
finds its root in the support of \eqref{R product}, which implies analyticity of the (generally off-shell) correlation function \eqref{R amplitude} at fixed real $(p_2+p_3)^\mu$ when
a positive-timelike imaginary part is added to $p_3^\mu$:
\be
\label{R analyticity}
\mathcal{R}(p_3, p_2) \mbox{ is analytic in }
D \equiv \{ (p_2,p_3) ~|~ (p_2+p_3)^\mu\in\mathbb{R}^{1,3} \mbox{ and }
{\rm Im}\,p_3^\mu \in V^+\}\,,
\ee
where $V^+\equiv \{ p^\mu \in \mathbb{R}^{1,3}~|~ p^0>|\vec{p}|\}$ is the set of
positive-timelike vectors.

Equations~\eqref{R reduction} and \eqref{R analyticity}
are the building blocks from which one would like to establish crossing symmetry. Before moving on, it is instructive to highlight two important points:
\begin{enumerate}[labelwidth=!,leftmargin=*,label=(\roman*)]
    \item \label{point:1}
For $2\to 2$ scattering, the amplitude \eqref{R amplitude} is a function of the two independent Mandelstam invariants
\begin{equation}
    s=-(p_1+p_2)^2 \quad \text{and} \quad t=-(p_2+p_3)^2\,.
\end{equation}
Using $\AA = a_1^\dagger$, $\DD = b_4$, and the stability condition mentioned earlier,
\eqref{R reduction} simplifies to
\be\label{eq:Rst}
\mathcal{R}(s,t) = \left\{\begin{array}{ll}
{}_{\rm out}\< 34|\mathbbm{1}|12\>_{\rm in} \quad &\mbox{$s$-channel kinematics}\,, \\
-{}_{\rm in}\< 24|\mathbbm{1}|13\>_{\rm out} \quad &\mbox{$u$-channel kinematics}\,.\end{array}\right.
\ee
The opposite arrangements of \say{in} and \say{out} states in \eqref{eq:Rst} explains why fixed-$t$ analyticity
connects the $s$-channel amplitude
to the \emph{complex conjugate} of the $u$-channel amplitude.
    \item \label{point:2}
At face value, the statements \eqref{R reduction} and \eqref{R analyticity} have non-overlapping validity.
This is the well-known difficulty in establishing crossing: the axiomatic domain $D$ does not intersect the mass shell.
In fact, it is simply not possible to add a small timelike imaginary part to a real momentum
without leaving the mass shell:
\be
{\rm Im}\,p_3^\mu\in V^+
\quad\implies\quad 
{\rm Re}\,p_3 \cdot {\rm Im}\,p_3 \neq 0 
\quad\implies\quad 
p_3^2\neq-m_3^2\in\mathbb{R}\,.
\ee
The traditional \say{work-around} involves two steps \cite{Bros:1965kbd}.
First, we consider off-shell correlators as a function of $s$ and complexified masses
$-p_i^2\mapsto m_i^2+\xi$, where $\xi\in\mathbb{C}$. For $2\ot 2$ scattering, by writing
$s=m_3^2+m_4^2-2p_3{\cdot}p_4+2\xi$, one sees that \eqref{R analyticity} contains an off-shell domain of the form
\begin{equation}
    D \supset D_1 \equiv \{s,m_3^2\in \mathbb{C}~|~
   {\rm Im}\,s>0,\, {\rm Re}\,\xi < \xi_0<0\}\,,
\end{equation}
where $\xi_0$ is fixed by the masses and the kinematics in the transverse plane (see App.~\ref{app:axiomatic}). Next, we need to show that the on-shell limit is sufficiently mild for amputated correlators such as $\mathcal{R}$, and
specifically that they enjoy \emph{local analyticity} in a neighborhood of the mass shell
for real physical momenta:
\begin{equation}
    D_2 \equiv \{s\in \mathbb{R}~|~|s|> s_0 \} \times \{ \xi \in \mathbb{D}_r(\xi_0)\}\,.
\end{equation}
Here, $\mathbb{D}_r(\xi_0)$ denotes an open disk centered at $\xi_0<0$ with radius $r$ large enough for the disk to contain the mass shell (the origin in the $\xi$-plane: $\xi=0$).
    Then, general properties of complex functions of two variables imply that the analyticity domain $D_\mathcal{R}$ of $\mathcal{R}$ contains the mass shell as well as complex $s$, at least for sufficiently large $s$. To put it more clearly, we have
    \begin{equation}
         D_\mathcal{R}\supset D_1 \cup D_2 ~\implies~
        D_\mathcal{R}\supset
         \{ s\in \mathbb{C}~|~{\rm Im}\,s>0
         ,\; \mbox{$|s|$ ``large enough''  and}~ \xi=0\}\,.
    \end{equation}
For the reader's benefit, this mathematical step is reviewed at a physicist's level of rigor in App.~\ref{app:axiomatic}.  The upshot is that crossing paths like those in Fig.~\ref{fig:rprodAC} exist
and remain on shell at all times.
To our knowledge, such arguments have been carried out rigorously for $2\ot 2$ scattering processes but have never been completed for generic $n\ot m$ processes (excluding the $3\ot 2$ case \cite{Bros:1985gy}).
\end{enumerate}

The central objective of this paper is to address and resolve point \ref{point:1} for $n$-point processes and to formulate a reasonable statement of crossing in the general case.  We propose to take seriously the idea that the two objects in \eqref{eq:Rst} (or more generally in \eqref{R reduction}) are related by an on-shell analytic continuation, and leave a rigorous proof addressing point \ref{point:2} to future work.

\section{Crossing symmetry} \label{sec:crossing}

In this section, we introduce the \emph{crossing equation} \eqref{eq:crossing23}, which relates asymptotic observables of the type discussed above.

\subsection{Main proposal: Crossing two particles in an \texorpdfstring{$n$}{n}-point amplitude}
\label{sec:conjectureI}

To state our proposal precisely, it is useful to introduce a complex variable $z$, which naturally generalizes the $s$-plane to $n$-point kinematics. We first go to the Lorentz frame in which the lightcone components of the momenta $p_2$ and $p_3$ are real and directly opposite, i.e., $p_3^\pm = - p_2^\pm$. Next, we apply the deformation parameter $z$, such that
\begin{equation}\begin{aligned}\label{eq:crossingpath0}
p_2^\mu(z) = \big(\phantom{+}z p_2^+,\, \phantom{+}\tfrac{1}{z} p_2^-,\, p_2^\perp\big)\,,
\\
p_3^\mu(z) = \big({-}z p_2^+,\, {-}\tfrac{1}{z} p_2^-,\, p_3^\perp\big)\,,
\end{aligned}\end{equation}
where $p_2^\pm < 0$.
It is straightforward to verify that momentum conservation $\sum_{i=1}^{n} p_i(z) = 0$ and on-shell conditions $p_i(z)^2 = p_i^2$ are preserved for any complex value of $z$.

Owing to point \ref{point:2} mentioned earlier, we only expect crossing to work for ``large enough'' $z$.  Indeed, to avoid extraneous terms
in \eqref{R reduction}, $z$ needs to be sufficiently large
that $|z p_i^+|>|p_j^+|$ for all $i\in \{2,3\}$ and $j\in A\cup D$, where $A$ and $D$ denote the sets of particles created and annihilated within the $\AA$ and $\DD$ strings respectively.
We will consider paths that take the form of large arcs across the $z$ half-plane:
\begin{equation}
\adjustbox{valign=c}{
\begin{tikzpicture}
  \draw[->,thick] (-1.5, 0) -- (1.5, 0);
  \draw[->,thick] (0, -0.5) -- (0, 1.25);
  \node[] at (1.1,1.05) {$z$};
  \draw[] (1.35,0.85) -- (0.9,0.85) -- (0.9,1.25);
  \draw[Maroon,fill=Maroon,thick] (0.5,0.0) circle (0.05);
  \draw[Maroon,fill=Maroon,thick] (1,0.1) circle (0.05);
  \draw[Maroon,fill=Maroon,thick] (-1,0.1) circle (0.05);
  \draw[Maroon,fill=Maroon,thick] (-0.5,0.0) circle (0.05);
  \node[below] at (-0.5,0) {\footnotesize{$-1$}};
  \node[below] at (0.5,0) {\footnotesize{$1$}};
  \draw[Maroon,thick] (1,0.1) arc (0:180:1);
  \draw[->,Maroon,thick] (1,0.1) arc (0:135:1);
\end{tikzpicture}
}
\label{eq:crossingpath3}
\end{equation}
The Mandelstam invariants that involve both $2$ and $3$ simultaneously, or neither of them, remain fixed. On the other hand, the invariants that get deformed are of the form $s_{ijk\ldots}$ with $i\in \{2,3\}$ and $j,k,\ldots \in A\cup D$,
\be
s_{ijk\ldots}(z) = -(p_i(z)+p_j + p_k + \ldots)^2 = z\, p^+_i (p^-_j + p^-_k + \ldots) + \mathcal{O}(z^0)\,,
\ee
and hence for large enough $z$, they rotate counter-clockwise and remain within one half-plane, from being large positive to large negative or vice versa. Our proposal is that such paths
indeed connect the two observables in \eqref{R reduction}, as was shown in Fig.~\ref{fig:rprodAC}.
We summarize it in the following relation:
\be\label{eq:crossing23a}
\left[\langle 0 | \DD\, b_3 a_2^\dag\,  \AA | 0 \rangle \right]_{{\curvearrowleft}z} = - \langle 0 | \DD\, a_2 b_3^\dag\, \AA | 0 \rangle \,.
\ee
We call \eqref{eq:crossing23a} the \emph{crossing equation}. Its simplest instance is when $\AA = \prod_{i \in A} a_i^\dag$ gives a multi-particle incoming state and $\DD = \prod_{i \in D} b_i$ creates a multi-particle outgoing state. Before the continuation, this choice results in the ordinary time-ordered scattering amplitude $S_{D3 \ot 2A}$. The crossing equation in this case can be written as
\be
\left[ S_{D 3 \ot 2 A} \right]_{{\curvearrowleft}z} = \left[ \langle D | a_3 S a_2^\dag  | A \rangle \right]_{{\curvearrowleft}z} = - \langle D | S a_2 S^\dag a_3^\dag S | A \rangle = - \sumint_{X,Y} S_{D \ot Y} S^\dag_{Y2 \ot 3X} S_{X \ot A}\, ,
\ee
where on the right-hand side we inserted two resolutions of identity. In the diagrammatic language, we have 
\begin{equation}
\adjustbox{valign=c}{
\begin{tikzpicture}[line width=1]
\def\Ang{30};
\def\CAng{150};
\def\CosVal{0.8660};
\def\SinVal{0.5};
\def\CstVal{0.1};
\begin{scope}[xshift=0]
\coordinate (c) at (0,0);
\draw[->-=.35,Maroon] (c)++(50:1) node[right] {\footnotesize$2$} -- (c);
\draw[->-=.80,Maroon] (c) -- ++(130:1) node[left] {\footnotesize$3$};
\draw[RoyalBlue] ($(c) + (\SinVal*\CstVal,\CosVal*\CstVal)$)++(-\Ang:1.15) -- ($(c)+(\SinVal*\CstVal,\CosVal*\CstVal)$);
\draw[RoyalBlue] ($(c) + (-\SinVal*\CstVal,-\CosVal*\CstVal)$)++(-\Ang:1.15) -- ($(c)+(-\SinVal*\CstVal,-\CosVal*\CstVal)$);
\draw[RoyalBlue] ($(c)+(0,0)$)++(-\Ang:1.15) -- ($(c)+(0,0)$);
\draw[RoyalBlue] ($(c) + (\SinVal*\CstVal,-\CosVal*\CstVal)$)++(-\CAng:1.15) -- ($(c)+(\SinVal*\CstVal,-\CosVal*\CstVal)$);
\draw[RoyalBlue] ($(c) + (-\SinVal*\CstVal,\CosVal*\CstVal)$)++(-\CAng:1.15) -- ($(c)+(-\SinVal*\CstVal,\CosVal*\CstVal)$);
\draw[RoyalBlue] ($(c)+(0,0)$)++(-\CAng:1.15) -- ($(c)+(0,0)$);
\filldraw[fill=gray!5, very thick](0,0) circle (0.4) node {$S$};
\draw [pen colour={gray},
    decorate, 
    decoration = {calligraphic brace,
        raise=5pt,
        amplitude=2pt}] (1.0,-0.2) --  (1.0,-0.2-0.65)
node[pos=0.5,right=10pt,RoyalBlue]{$A$};
\draw [pen colour={gray},
    decorate, 
    decoration = {calligraphic brace,
        raise=5pt,
        amplitude=2pt}] (-1.0,-0.2-0.65) -- (-1.0,-0.2)
node[pos=0.5,left=10pt,RoyalBlue]{$D$};
\end{scope}
\node[align=center] at (3,0.5) {\footnotesize cross ${\textcolor{Maroon}{2} \leftrightarrow \textcolor{Maroon}{3}}$};
\draw[-latex] (2.2,0) -- (3.8,0);
\draw[line width=0.8] (4.4,0) -- (4.65,0);
\begin{scope}[xshift=200]
\coordinate (c) at (0,0);
\draw[->-=.35,Maroon] (c)++(50:1) node[right] {\footnotesize$\bar{3}$} -- (c);
\draw[->-=.80,Maroon] (c) -- ++(130:1) node[left] {\footnotesize$\bar{2}$};
\draw[RoyalBlue] ($(c) + (\SinVal*\CstVal,\CosVal*\CstVal)$)++(-\Ang:1.85) -- ($(c)+(\SinVal*\CstVal,\CosVal*\CstVal)$);
\draw[RoyalBlue] ($(c) + (-\SinVal*\CstVal,-\CosVal*\CstVal)$)++(-\Ang:1.85) -- ($(c)+(-\SinVal*\CstVal,-\CosVal*\CstVal)$);
\draw[RoyalBlue] ($(c)+(0,0)$)++(-\Ang:1.85) -- ($(c)+(0,0)$);
\draw[RoyalBlue] ($(c) + (\SinVal*\CstVal,-\CosVal*\CstVal)$)++(-\CAng:1.85) -- ($(c)+(\SinVal*\CstVal,-\CosVal*\CstVal)$);
\draw[RoyalBlue] ($(c) + (-\SinVal*\CstVal,\CosVal*\CstVal)$)++(-\CAng:1.85) -- ($(c)+(-\SinVal*\CstVal,\CosVal*\CstVal)$);
\draw[RoyalBlue] ($(c)+(0,0)$)++(-\CAng:1.85) -- ($(c)+(0,0)$);
\filldraw[fill=gray!30,rotate=-30](0,-0.2) rectangle (1,0.2);
\filldraw[fill=gray!30,rotate=-150](0,-0.2) rectangle (1,0.2);
\filldraw[fill=gray!5, very thick](0,0) circle (0.4) node[yshift=1] {$S^\dag$};
\filldraw[fill=gray!5, very thick]($(0,0)+(-30:1.2)$) circle (0.4) node {$S$};
\filldraw[fill=gray!5, very thick]($(0,0)+(-150:1.2)$) circle (0.4) node {$S$};
\node[] at (-0.54,-0.29) {\tiny $Y$};
\node[] at (0.48,-0.29) {\tiny $X$};
\draw [pen colour={gray},
    decorate, 
    decoration = {calligraphic brace,
        raise=5pt,
        amplitude=2pt}] (1.6,-0.55) --  (1.6,-1.2)
node[pos=0.5,right=10pt,RoyalBlue]{$A$};
\draw [pen colour={gray},
    decorate, 
    decoration = {calligraphic brace,
        raise=5pt,
        amplitude=2pt}] (-1.6,-1.2) --  (-1.6,-0.55)
node[pos=0.5,left=10pt,RoyalBlue]{$D$};
\draw[dashed,orange] (0.5,-1) -- (0.5,1);
\draw[dashed,orange] (-0.5,-1) -- (-0.5,1);
\end{scope}
\end{tikzpicture}
}
\label{eq:crossing23}
\end{equation}
Here, red color-coding marks which particles are being crossed. If they carried charges, after crossing $2$ and $3$ become their own anti-particles, $\bar{2}$ and $\bar{3}$.
The more general case \eqref{eq:crossing23a} can also be written in the blob picture if we simply embed the above relation in a bigger diagram. We remind the reader that $S$ and $S^\dag$ contain both connected and disconnected terms.

This is the main proposal which we will explore in this paper, for both massive and massless particles.
To be fully precise, our conjecture is that this holds whenever the retarded product is analytic in a neighborhood of the mass shell, meaning that there is no obstruction from point \ref{point:2} above. This condition is non-trivial and a tentative criterion is proposed in Sec.~\ref{sec:anomalous}.

\paragraph{Spinning particles} Let us emphasize that the above proposal can also be straightforwardly stated for amplitudes of spinning particles, even though all our examples will involve only scalars. All that is needed is to deform the external polarization spinors and vectors in such a way that the transverse conditions ($p_j{\cdot}\varepsilon_j(z)=0$) and/or the Dirac equation ($(\slashed{p}_j(z) - m_j)u_j(z)=0$) remain satisfied at all times.
Since the deformations in \eqref{eq:crossingpath0} coincide with a boost along the 3-direction, a simple way to achieve this is to apply the same boost to the polarizations of all particles $j\in B\cup C$, which gives, respectively, for vectors and spinors:\footnote{
 Alternatively, one could view spinning amplitudes as collections of scalar amplitudes multiplying suitable polarization structures, but note that this approach can introduce extra branch cuts \cite{Trueman:1964zza,williams1967construction,Cohen-Tannoudji:1968lnm,Hara:1971kj}.
}
\begin{equation}\begin{aligned}
\varepsilon_j^\mu(z)&= \big(z \varepsilon_j^+,\, \tfrac{1}{z} \varepsilon_j^-,\, \varepsilon_j^\perp\big)\,, \\
u_j(z) &= z^{-iS^{03}} u_j \simeq \mbox{diag}\left(\tfrac{1}{\sqrt{z}},\sqrt{z},\sqrt{z},\tfrac{1}{\sqrt{z}}\right)\cdot u_j \qquad\mbox{(D=4)}\,.
\end{aligned}\label{continuations spinor}\end{equation}
Here, $S^{03}=\frac{i}{2} \gamma^0 \gamma^3$ is the generator
of a Lorentz boost (assuming that the Clifford algebra is written as
$\{\gamma^\mu,\gamma^\nu\}=-2\eta^{\mu\nu}\mathbbm{1}$, recall that we work in mostly-plus signature) and \say{$\simeq$} gives the special case for a four-dimensional Dirac spinor in the standard chiral basis \cite{Peskin:1995ev}.
Note that for fermions, the spinors thus obtained at the $z<0$ endpoint of \eqref{continuations spinor} may differ by overall ``little group'' phases
compared with some predetermined standard choice for the corresponding momentum. In addition, for fermions, the minus sign in \eqref{R reduction} would be absent, since the retarded product involves an anticommutator.

\subsection{A simple tree-level example}
\label{sec:treelevel_intro}

The relation \eqref{eq:crossing23} can be interpreted in two ways:
as a linear relation between out-of-time-ordered amplitudes, \emph{or} as a non-linear relation between conventional time-ordered amplitudes.

This latter perspective gives predictions that can be immediately tested.  Consider, for example, a theory where the following tree-level process is possible, involving a heavy intermediate particle of mass $M$:
\begin{equation}\label{eq:M_12_345}
i\cM_{345 \ot 12}=
\begin{gathered}
\adjustbox{valign=c}{
\begin{tikzpicture}[baseline= {($(current bounding box.base)+(10pt,10pt)$)},line width=1, scale=0.7]
\coordinate (a) at (0,0) ;
\coordinate (b) at (1,0) ;
\coordinate (c) at ($(b)+(-40:1)$);
\draw[] (a) -- (b);
\draw[] (b) -- (c);
\draw[Maroon] (b) -- ++ (30:1) node[right] {\footnotesize$2$};
\draw[Maroon] (c) -- ++ (-150:1) node[left]{\footnotesize$3$};
\draw[RoyalBlue] (c) -- ++ (-30:1) node[right] {\footnotesize$1$};
\draw[RoyalBlue] (a) -- (-150:1) node[left] {\footnotesize$4$};
\draw[RoyalBlue] (a) -- (150:1) node[left] {\footnotesize$5$};
\fill[black,thick] (a) circle (0.07);
\fill[black,thick] (b) circle (0.07);
\fill[black,thick] (c) circle (0.07);
\end{tikzpicture}
}
\end{gathered}
=
\frac{-ig^3}{(-s_{45}+M^2-i\eps)(-s_{13}+M^2)}\,.
\end{equation}
It might sound strange that crossing has something to say about tree-level amplitudes, which are rational functions.
The point is that amplitudes are really boundary values of (sums of) analytic functions and so, under crossing, they can contain
$\delta$-function contributions. Below, we wish to check whether these are also correctly reproduced. 
For this reason, it is important to carefully keep track of the $i\eps$ factors.

In \eqref{eq:M_12_345}, we can omit the $i\eps$ prescription for $s_{13}$ since this invariant is spacelike in the considered kinematics and, consequently, its propagator cannot vanish. However, we need to keep $i\eps$ for the timelike Mandelstam invariant $s_{45}$.
We now analytically continue the kinematics so as to 
swap the energies of particles 2 and 3.
The invariant $s_{13}$ starts negative and rotates in the counter-clockwise direction, which means it ends up below the positive axis, while the invariant $s_{45}$ stays fixed:
\begin{equation}
\adjustbox{valign=c}{
\begin{tikzpicture}
  \draw[->,thick] (-1.5, 0) -- (1.5, 0);
  \draw[->,thick,white] (0, -1.6) -- (0, 1.6);
  \draw[->,thick] (0, -1.25) -- (0, 1.25);
  \node[] at (1.1,1.05) {$s_{ij}$};
  \draw[RoyalBlue,fill=RoyalBlue,thick] (0.5,0.1) circle (0.05);
  \node[] at (0.35,0.4) {$\textcolor{RoyalBlue}{s_{45}}$};
  \draw[] (1.35,0.85) -- (0.8,0.85) -- (0.8,1.25);
  \draw[Maroon,fill=Maroon,thick] (-1.1,0.0) circle (0.05);
  \draw[->,Maroon,thick] (-1.1,0.0) arc (0:178:-1.1);
  \node[] at (-0.35,-0.6) {$\textcolor{Maroon}{s_{13}}$};
\end{tikzpicture}
}
\end{equation}
The result of analytic continuation can be written as
\begin{equation}
 \left[i\cM_{345 \ot 12}\right]_{ 
 \raisebox{\depth}{\scalebox{1}[-1]{$\curvearrowright$}}s_{13}} =
 \frac{-ig^3}{(-s_{45}+M^2-i\eps)(-s_{13}+M^2+i\eps)}\,, \label{5pt tree crossing}
\end{equation}
where now $s_{13} > 0$ and its $-i\eps$ prescription matters.
Notice how the second propagator ended up with the ``wrong'' $i\eps$ prescription. This fact is significant because this pole is accessible in $245 \ot 13$ kinematics.\footnote{
Strictly speaking, the pole cannot be exactly on the real axis since $M$ must be a resonance --- obviously, if $M$ can be kinematically produced, it can decay.
However, in the narrow-width approximation this does not affect the present calculations:
the two-particle cut near resonance is proportional to
\begin{equation}
    2{\rm Re} \left(\frac{-i}{-s_{45}+M^2-iM\Gamma}\right) \approx 2\pi\delta(-s_{45}+M^2) + \mathcal{O}(\Gamma)\,.   
\end{equation}
A more detailed discussion can be found in \cite{Caron-Huot:2023vxl}.}
This is a similar situation to how the crossing transformation for $34\ot 12$ scattering ends up on the ``wrong side'' of the $u$-channel branch cut for a general non-perturbative amplitude (see Fig.~\ref{fig:introcross1}).
Here, however, the $s_{45}$ pole is still on the correct side, so we cannot identify the right-hand side with a complex conjugated amplitude. \emph{It is a different object}.

The crossing equation in~\eqref{eq:crossing23} gives a definite prediction for the result of analytic continuation. It should be equal to the complex-conjugated five-particle amplitude, plus a non-linear term where the heavy particle $M$ is produced, namely 
\begin{subequations}
\begin{align}
 \left[i\cM_{345 \ot 12}\right]_{\raisebox{\depth}{\scalebox{1}[-1]{$\curvearrowright$}}s_{13}}
& \stackrel{?}{=} i\cM^{\dag}_{245\ot 13} +
 [i\cM_{45\ot M}]\, 2\pi\delta(-s_{45}+M^2)\,[i\cM^{\dag}_{M2\ot 13}]
\label{5pt crossing tree test}
\\ &=
\frac{-ig^3}{(-s_{45}+M^2+i\eps)(-s_{13}+M^2+i\eps)}
+\frac{2\pi \delta(-s_{45}+M^2) g^3}{-s_{13}+M^2+i\eps}
\\ &= \left( \frac{1}{-s_{45}+M^2+i\eps}+ 2\pi i \delta(-s_{45}+M^2)\right) \frac{-ig^3}{-s_{13}+M^2+i\eps}\,.
\label{eq:5pttree_c}
\end{align}
\end{subequations}
Using the familiar identity $\frac{1}{x\pm i\eps}=\text{PV}\frac{1}{x}\mp i\pi\delta(x)$, this agrees precisely with \eqref{5pt tree crossing}! Diagrammatically:
\begin{equation}
\left[
\begin{gathered}
\adjustbox{valign=c}{
\begin{tikzpicture}[baseline= {($(current bounding box.base)+(10pt,10pt)$)},line width=1, scale=0.7]
\coordinate (a) at (0,0) ;
\coordinate (b) at (1,0) ;
\coordinate (c) at ($(b)+(-40:1)$);
\draw[] (a) -- (b);
\draw[] (b) -- (c);
\draw[Maroon] (b) -- ++ (30:1) node[right] {\footnotesize$2$};
\draw[Maroon] (c) -- ++ (-150:1) node[left]{\footnotesize$3$};
\draw[RoyalBlue] (c) -- ++ (-30:1) node[right] {\footnotesize$1$};
\draw[RoyalBlue] (a) -- (-150:1) node[left] {\footnotesize$4$};
\draw[RoyalBlue] (a) -- (150:1) node[left] {\footnotesize$5$};
\fill[black,thick] (a) circle (0.07);
\fill[black,thick] (b) circle (0.07);
\fill[black,thick] (c) circle (0.07);
\end{tikzpicture}
}
\end{gathered}
\right]_{\raisebox{\depth}{\scalebox{1}[-1]{$\curvearrowright$}}s_{13}}
=
\adjustbox{valign=c}{
\begin{tikzpicture}[baseline= {($(current bounding box.base)+(10pt,10pt)$)},line width=1, scale=0.7]
\coordinate (a) at (0,0) ;
\coordinate (b) at (1,0) ;
\coordinate (c) at ($(b)+(-40:1)$);
\draw[] (a) -- (b);
\draw[] (b) -- (c);
\draw[Maroon] (b) -- ++ (150:1) node[yshift=3,left] {\footnotesize$\bar{2}$};
\draw[Maroon] (c) -- ++ (30:1) node[right]{\footnotesize$\bar{3}$};
\draw[RoyalBlue] (c) -- ++ (-30:1) node[right] {\footnotesize$1$};
\draw[RoyalBlue] (a) -- (-150:1) node[left] {\footnotesize$4$};
\draw[RoyalBlue] (a) -- (150:1) node[left] {\footnotesize$5$};
\fill[black,thick] (a) circle (0.07);
\fill[black,thick] (b) circle (0.07);
\fill[black,thick] (c) circle (0.07);
\draw[dashed,orange] (-0.5,-1.2) -- (-0.5,1);
\draw[dashed,orange] (2.3,-1.2) -- (2.3,1);
\end{tikzpicture}
}
+
\adjustbox{valign=c}{
\begin{tikzpicture}[baseline= {($(current bounding box.base)+(10pt,10pt)$)},line width=1, scale=0.7]
\coordinate (a) at (0,0) ;
\coordinate (b) at (1,0) ;
\coordinate (c) at ($(b)+(-40:1)$);
\draw[] (a) -- (b);
\draw[] (b) -- (c);
\draw[Maroon] (b) -- ++ (150:1) node[yshift=3,left] {\footnotesize$\bar{2}$};
\draw[Maroon] (c) -- ++ (30:1) node[right]{\footnotesize$\bar{3}$};
\draw[RoyalBlue] (c) -- ++ (-30:1) node[right] {\footnotesize$1$};
\draw[RoyalBlue] (a) -- (-150:1) node[left] {\footnotesize$4$};
\draw[RoyalBlue] (a) -- (150:1) node[left] {\footnotesize$5$};
\fill[black,thick] (a) circle (0.07);
\fill[black,thick] (b) circle (0.07);
\fill[black,thick] (c) circle (0.07);
\draw[dashed,orange] (0.5,-1.2) -- (0.5,1);
\draw[dashed,orange] (2.3,-1.2) -- (2.3,1);

\end{tikzpicture}
}
\end{equation}
Note that a cut through the $s_{13}$ propagator is not allowed, because it does not fit in the pattern \eqref{eq:crossing23}.

It is easy to verify that the proposal \eqref{eq:crossing23} also works for any other $3\ot 2$ tree diagram: $\cM^{\dag}$ gives the correct continuation of the amplitude except for propagators involving $s_{45}$, which stays fixed during the rotation.
Indeed, after the rotation and with $z$ still large, the only timelike invariants are $s_{13}$, $s_{24}$, $s_{25}$ and $s_{45}$.  The first three rotate counter-clockwise and end up below the axis, as appropriate for $\cM^{\dag}$.  Propagators involving $s_{45}$, however, do not get complex conjugated, and this is fixed by the $S$ blob (which is only different from identity where $\delta(-s_{45}+M^2)$ has support).

We conclude that the proposed crossing relation in \eqref{eq:crossing23} is the simplest option that is not obviously disproved by tree-level considerations.  
A proposal which would not get the correct $i\varepsilon$'s at tree level would stand no chance of landing on the correct branch for the more intricate functions that appear at loop level.

\subsection{First loop-level examples}\label{sec:triloop}

As a glimpse into how the crossing relations work at loop level, let us verify it for the massless scalar triangle amplitude $\cM_{3456 \ot 12}$ in four spacetime dimensions,
\begin{equation}
\begin{gathered}
\begin{tikzpicture}[line width=1]
\draw[] (0,0) -- (1,0.5) -- (1,-0.5) -- (0,0);
\draw[Maroon] (1.5,0.75) -- (1,0.5);
\draw[RoyalBlue] (1,0.5) -- (0.5,0.75);
\draw[RoyalBlue] (1.5,-0.75) -- (1,-0.5);
\draw[Maroon] (1,-0.5) -- (0.5,-0.75);
\draw[RoyalBlue] (-0.5,0.25) -- (0,0) -- (-0.5,-0.25);
\node[] at (-1.5,0) {$s_{45}$};
\node[] at (1,1.2) {$s_{26}$};
\node[] at (1,-1.2) {$s_{13}$};
\node[scale=0.75,RoyalBlue] at (0.3,0.8) {$6$};
\node[scale=0.75,Maroon] at (1.7,0.8) {$2$};
\node[scale=0.75,RoyalBlue] at (1.7,-0.8) {$1$};
\node[scale=0.75,Maroon] at (0.3,-0.8) {$3$};
\node[scale=0.75,RoyalBlue] at (-0.8,0.2) {$4$};
\node[scale=0.75,RoyalBlue] at (-0.8,-0.2) {$5$};
\end{tikzpicture}
\end{gathered}
\hspace{1cm}
\begin{gathered}
\begin{tikzpicture}[line width=1]
\draw[] (0,0) -- (1,0.5) -- (1,-0.5) -- (0,0);
\draw[->,-latex reversed] (0,0) -- (0.5,0.25);
\draw[->,-latex reversed] (1,0.5) -- (1,-0.05);
\draw[->,-latex reversed] (1,-0.5) -- (0.4,-0.2);
\draw[] (1.5,0.75) -- (1,0.5) -- (0.5,0.75);
\draw[] (1.5,-0.75) -- (1,-0.5) -- (0.5,-0.75);
\draw[] (-0.5,0.25) -- (0,0) -- (-0.5,-0.25);
\node[] at (-1.5,0) {$p_{45}$};
\node[] at (1,1.3) {$p_{26}$};
\node[] at (1,-1.2) {$p_{13}$};
\node[] at (1.7,0) {$\ell-p_{13}$};
\node[] at (0.25,-0.5) {$\ell$};
\node[] at (-0.15,0.5) {$\ell+p_{45}$};
\draw[->,gray,line width=0.7] (-0.5,0) -- (-0.8,0);
\draw[->,gray,line width=0.7] (1,0.8) --++(90:0.3);
\draw[->,gray,line width=0.7] (1,-0.7) --++(-90:0.3);
\end{tikzpicture}
\end{gathered}
\label{eq:tridiagram}
\end{equation}
Here (as always in this paper) time flows from right to left and we use an all-outgoing convention for the momenta.
The amplitude is given by
\begin{equation}
        i \cM^\text{tri} = \int \frac{\d^4 \ell}{\pi^2} \frac{1}{[\ell^2-i\varepsilon][(\ell-p_{13})^2-i\varepsilon] [(\ell+p_{45})^2-i\varepsilon]} \,,
        \label{eq:Mtriloopmom}
\end{equation}
where we use the conventions that vertices in the Feynman diagram corresponding to $i{\cal M}$ come with a factor of $(-i)$, the propagator of an edge with momentum $q_e$ is taken to be $\frac{-i}{q_e^2 + m_e^2 - i\varepsilon}$, and we multiply by an overall factor of $(-1)^{\text{V}}$ where $\text{V}$ is the total number of vertices. These factors of $i$ contribute $i^{-\text{L}}$ to ${\cal M}$.
The expression in~\eqref{eq:Mtriloopmom} can easily be evaluated to be
\begin{equation}
    \cM^\text{tri} = \frac{-2}{s_{13}(z-\zb)} \left[ \Li_2(z)-\Li_2 (\zb) + \frac{1}{2} \left( \log z + \log \zb\right) \left( \log(1-z) - \log(1-\zb) \right) \right], \label{eq:Mtri}
\end{equation}
where we have introduced two independent variables $z$ and $\zb$, which relate to kinematic variables as follows
\begin{equation}
    z \zb = \frac{s_{26}}{s_{13}} \quad \text{and} \quad (1-z)(1-\zb) = \frac{s_{45}}{s_{13}} \,.
    \label{eq:zzbar}
\end{equation}
The formula \eqref{eq:Mtri} is valid in the Euclidean region $R^\text{E}$ described below.
Different kinematic regions can be obtained by continuations in $z$ and $\zb$.
Due to the symmetry in $z \leftrightarrow \zb$, we can take $\zb<z$ without loss of generality. The branch points of the amplitude are at $z=0$, $z=1$ and $z=\infty$, as well as the corresponding $\zb=0$, $\zb=1$ and $\zb=\infty$. Moreover, there is a branch point at $z=\zb$.
We can separate the different kinematic regions in the space of $z$ and $\zb$ according to the signs of the invariants $s_{26}$, $s_{45}$ and $s_{13}$, as follows:
\begin{equation}
\begin{gathered}
\begin{tikzpicture}
  \fill [darkgreen!20, domain=-2:2, variable=\x]
  (1,0) -- (1,1) -- (0,0) -- cycle;
  \fill [darkgreen!20, domain=-2:2, variable=\x]
  (1, 1) -- (2,1) -- (2,2) -- cycle;
  \fill [darkgreen!20, domain=-2:2, variable=\x]
  (-1,-1) -- (0,-1) -- (0,0) -- cycle;
  \fill [Maroon!10, domain=-2:2, variable=\x]
  (0,-1) -- (1,-1) -- (1,0) -- (0,0) -- cycle;
  \fill [RoyalBlue!10, domain=-2:2, variable=\x]
  (1,-1) -- (2,-1) -- (2,0) -- (1,0) -- cycle;
  \fill [brown!10, domain=-2:2, variable=\x]
  (1,0) -- (2,0) -- (2,1) -- (1,1) -- cycle;
  \draw[domain=-1:2, smooth, variable=\x, darkgreen] plot ({\x}, {\x});
  \draw[domain=0:2, smooth, variable=\x, black] plot ({\x}, {0});
  \draw[domain=1:2, smooth, variable=\x, black] plot ({\x}, {1});
  \draw[domain=-1:1,variable=\y,black] plot({1},\y);
  \draw[domain=-1:0,variable=\y,black] plot({0},\y);
  \draw[->,thick] (-1, 0) -- (2.2, 0) node[right] {$z$};
  \draw[->,thick] (0, -1) -- (0, 2.2) node[above] {$\zb$};
  \node[] at (0.5,-0.5) {$R^{(26)}$};
  \node[] at (1.5,-0.5) {$R^{(13)}$};
  \node[] at (1.5,0.5) {$R^{(45)}$};
  \node[] at (0.5,1.25) {$R^\text{E}$};
  \draw[->,line width=0.07] (0.95,1.25) to[out=30,in=180] (1.75,1.5);
  \draw[->,line width=0.07] (0.5,0.9) to[out=-90,in=180] (0.75,0.5);
  \draw[->,line width=0.07] (0.35,0.9) to[out=-130,in=90] (-0.15,-0.35);
\end{tikzpicture}
\end{gathered}
\end{equation}
where the regions are labeled with a superscript of the Mandelstam invariants that are positive. For example, $R^{(26)}$ is the region in $(z,\zb)$ where $s_{26}>0$, $s_{45}<0$ and $s_{13}<0$, which corresponds to $0<z<1$ and $\zb<0$. We have labeled the Euclidean region, where the invariants are either all positive or all negative, as $R^\text{E}$. When two invariants are positive, we have the relation $R^{(13)(45)} = R^{(26)}$, and its permutations.

When computing the amplitude in any other kinematic region than the one where $0<z,\zb<1$, we must specify the imaginary parts of $z$ and $\zb$ to land on the correct branches of the polylogarithms in \eqref{eq:Mtri}. Using that the signs of the imaginary parts of the invariants can be implemented as $s_{ij} \to s_{ij} + i\varepsilon$, we get the following assignments of infinitesimal imaginary parts of $z$ and $\zb$ for the scattering amplitude $\cM$:
\begin{subequations}
    \begin{align}
    z & \to z + i\varepsilon\quad \text{and} \quad \zb \to \zb - i \varepsilon \qquad \text{in } \quad R^{(26)} \,, \quad R^{(45)} \,, \quad R^{(26)(45)}\,,
    \label{eq:zzb1}
    \\ 
    z & \to z - i\varepsilon\quad \text{and} \quad \zb \to \zb + i \varepsilon \qquad \text{in } \quad R^{(13)} \,,\quad R^{(13)(45)} \,, \quad R^{(13)(26)}\,.
    \label{eq:zzb2}
\end{align}
\end{subequations}

The crossing of particles $2$ and $3$ starting from $\cM^\text{tri}_{3456\ot 12}$
will provide a first non-trivial example of the crossing equation at loop level. Let us start by working out the result of the rotation via \eqref{eq:crossingpath3}. During the crossing step, both $s_{26}$ and $s_{13}$ rotate in a large semicircle in the lower half-plane to become timelike. The corresponding rotation of $z$ and $\zb$ is achieved with $z$ rotating clockwise around $1$ and $\zb$ fixed,
\begin{equation}
\begin{gathered}
\begin{tikzpicture}
  \draw[->,thick] (-1.5, 0) -- (1.5, 0);
  \draw[->,thick] (0, -1.25) -- (0, 1.25);
  \node[] at (1.1,1.05) {$s_{ij}$};
  \draw[] (1.35,0.85) -- (0.8,0.85) -- (0.8,1.25);
  \draw[RoyalBlue,fill=RoyalBlue,thick] (0.24,0.12) circle (0.05);
  \node[] at (0.24,0.32) {\color{RoyalBlue}$s_{45}$};
  \draw[Maroon,fill=Maroon,thick] (-1,0) circle (0.05);
  \draw[->,Maroon,thick] (-1,0) arc (0:175:-1);
  \draw[Maroon,fill=Maroon,thick] (-0.8,0) circle (0.05);
  \draw[->,Maroon,thick] (-0.8,0) arc (0:175:-0.8);
  \node[] at (-0.8,-1) {\color{Maroon} $s_{13}$};
  \node[] at (-0.3,-0.4) {\color{Maroon} $s_{26}$};
  \draw[black!80,fill=black!80,thick] (0,0) circle (0.05);
  \draw[decorate, decoration={zigzag, segment length=6, amplitude=2}, black!80] (0,0) -- (1.5,0);
\end{tikzpicture}
\hspace{1.5cm}
\begin{tikzpicture}
  \draw[->,thick] (-1, 0) -- (2.2, 0);
  \draw[->,thick] (0, -1) -- (0, 1.5);
  \node[] at (1.8,1.3) {$z$};
  \draw[] (2.0,1.1) -- (1.6,1.1) -- (1.6,1.5);
  \draw[decorate, decoration={zigzag, segment length=6, amplitude=2}, black!80] (1,0) -- (2.17,0);
  \draw[black!80,fill=black!80,thick] (1,0) circle (0.05);
  \draw[decorate, decoration={zigzag, segment length=6, amplitude=2}, black!80] (0,0) -- (-1,0);
  \draw[black!80,fill=black!80,thick] (0,0) circle (0.05);
  \draw[Maroon,fill=Maroon,thick] (1.2,0.1) circle (0.05);
  \draw[->,Maroon,thick] (1.2,0.1) arc (0:-160:0.35);
\end{tikzpicture}
\end{gathered}
\hspace{1.5cm}
\begin{gathered}
\begin{tikzpicture}
  \draw[->,thick] (-1, 0) -- (2.2, 0);
  \draw[->,thick] (0, -1) -- (0, 1.5);
  \node[] at (1.8,1.3) {$\zb$};
  \draw[] (2.0,1.1) -- (1.6,1.1) -- (1.6,1.5);
  \draw[decorate, decoration={zigzag, segment length=6, amplitude=2}, black!80] (1,0) -- (2.17,0);
  \draw[black!80,fill=black!80,thick] (1,0) circle (0.05);
  \draw[decorate, decoration={zigzag, segment length=6, amplitude=2}, black!80] (0,0) -- (-1,0);
  \draw[black!80,fill=black!80,thick] (0,0) circle (0.05);
  \draw[Maroon,fill=Maroon,thick] (0.2,0) circle (0.05);
  \tikzset{ma/.style={decoration={markings,mark=at position 0.7 with {\arrow[scale=0.8]{>}}},postaction={decorate}}}
  \draw[ma][Maroon,thick] (0.2,0) arc (0:360:-0.2);
\end{tikzpicture}
\end{gathered}
\end{equation}
After the rotation, we end up with the amplitude in the region $0<\zb<z<1$, in addition to the discontinuity across the branch cut that starts at $z=1$. Since the amplitude and the complex amplitude are the same in this region, we can write the result of the crossing as
\begin{equation}
 \left[\cM^\text{tri}_{3456 \ot 12 }\right]_{\substack{{\rotatedown
 s_{13}}\\ {\rotatedown s_{26}}}}
 =
 \cM^\text{tri\dag}_{2456 \ot 13} + \text{Disc}_{z>1} \cM^\text{tri}_{3456 \ot 12}\,,
 \label{eq:rotDisc}
\end{equation}
where $\cM^\text{tri\dag}_{2456 \ot 13}$ is given by the expression in \eqref{eq:Mtri}, and the discontinuity across the branch cut at $z>1$ is defined as $\text{Disc}_{z>1} \cM^\text{tri}_{3456 \ot 12}  =  \left. \cM^\text{tri}_{3456 \ot 12} \right\vert_{z+i\varepsilon}-\left. \cM^\text{tri}_{3456 \ot 12} \right\vert_{z-i\varepsilon}$, which we compute to be
\begin{equation}
    \text{Disc}_{z>1} \cM^\text{tri}_{3456 \ot 12}  = \frac{-2\pi i}{s_{13}(z-\zb)} \log \frac{z}{\zb}\,,
    \label{eq:discM}
\end{equation}
for $z>1$ and $s_{13}<0$.

Next, we compare this result with the prediction from the crossing equation in \eqref{eq:crossing23}. After crossing, the physical channel will be the one in which $1$ and $3$ are incoming, with two separate contributions that fit the blob pattern, namely
\begin{align}
    \left.
    \begin{gathered}
    \begin{tikzpicture}[line width=1, scale=1]
    \begin{scope}[xshift=4,yshift=-2]
    \coordinate (a) at (0,0);
    \coordinate (b) at (-1,0.5);
    \coordinate (c) at (-1,-0.5);
    \coordinate (d) at (-1,0);
    \draw[] (a) -- (b) -- (c) -- (a);
    \draw[RoyalBlue] (b)-- ++(170:0.5) node[left,scale=0.75] {$6$};
    \draw[Maroon] (b)-- ++(30:0.5) node[right,scale=0.75,yshift=5] {$2$};
    \draw[RoyalBlue] (c)-- ++(-170:0.5) node[left,scale=0.75,xshift=-3] {$4$};
    \draw[RoyalBlue] (c)-- ++(-140:0.5) node[left,scale=0.75,xshift=0,yshift=-4] {$5$};
    \draw[RoyalBlue] (a)-- ++(30:0.5) node[right,scale=0.75,xshift=-3] {$1$};
    \draw[Maroon] (a)-- ++(-130:0.5) node[left,below,scale=0.75,xshift=-3] {$3$};
    \end{scope}
    \begin{scope}[xshift=-50pt,yshift=31.9pt]  
    \draw[line width=1, line cap=round,yshift=-5pt] (0,-2.2) -- (-0.135,-2.2) (0,0) -- (-0.135,0) (-0.135,-2.2) -- (-0.135,0);
    \end{scope}
    \begin{scope}[xshift=22pt,yshift=31.9pt,xscale=-1]  
    \draw[line width=1, line cap=round,yshift=-5pt] (0,-2.2) -- (-0.135,-2.2) (0,0) -- (-0.135,0) (-0.135,-2.2) -- (-0.135,0);
    \end{scope}
    \begin{scope}[xshift=0pt,yshift=-35pt]
    \draw[line width=0.4, line cap=round] (-1.66,0) -- (-1.66,-0.135) (0.7,0) -- (0.7,-0.135) (0.7,-0.135)  -- (-1.66,-0.135) node[below, midway,yshift=2pt]{$\subset \cM$};
    \node[] at (1.6,0) {$_{\substack{\rotatedown
 s_{13} \\  \rotatedown s_{26}}}$};
    \end{scope}
    \end{tikzpicture}
    \end{gathered}
    \right.
\stackrel{?}{=}
\qquad
    \begin{gathered}
    \begin{tikzpicture}[baseline= {($(current bounding box.base)+(10pt,10pt)$)},line width=1, scale=1]
    \coordinate (a) at (0,0) ;
    \coordinate (b) at (-1,0.5) ;
    \coordinate (c) at (-1,-0.5);
    \coordinate (d) at (-1,0);
    \draw[] (a) -- (b) -- (c) -- (a);
    \draw[RoyalBlue] (b)-- ++(170:0.5) node[left,scale=0.75] {$6$};
    \draw[Maroon] (b)-- ++(140:0.5) node[left,scale=0.75,yshift=5] {$\bar{2}$};
    \draw[RoyalBlue] (c)-- ++(-170:0.5) node[left,scale=0.75,xshift=-3] {$4$};
    \draw[RoyalBlue] (c)-- ++(-140:0.5) node[left,scale=0.75,xshift=0,yshift=-4] {$5$};
    \draw[RoyalBlue] (a)-- ++(30:0.5) node[right,scale=0.75,xshift=-3] {$1$};
    \draw[Maroon] (a)-- ++(-30:0.5) node[right,scale=0.75,xshift=-3] {$\bar{3}$};
    \draw[dashed,Orange] ($(d)+(180:0.25)+(0,1)$) -- ($(d)+(180:0.25)+(0,-1)$);
    \draw[dashed,Orange] ($(a)+(0:0.25)+(0,1)$) -- ($(a)+(0:0.25)+(0,-1)$);
    \begin{scope}[xshift=0pt,yshift=-25pt]
    \draw[line width=0.4, line cap=round,yshift=-7] (-1.2,0) -- (-1.2,-0.135) (0.18,0)  -- (0.18,-0.135) (-1.2,-0.135) -- (0.18,-0.135) node[below, midway,yshift=2pt]{$\subset \cM^\dag$};
    \draw[line width=0.4, line cap=round,yshift=-7] (-2.2,0) -- (-2.2,-0.135) (-1.28,0)  -- (-1.28,-0.135) (-2.2,-0.135) -- (-1.28,-0.135) node[below, midway,yshift=0pt]{$\subset S$};
    \draw[line width=0.4, line cap=round,yshift=-7] (0.28,0) -- (0.28,-0.135) (0.9,0)  -- (0.9,-0.135) (0.28,-0.135) -- (0.9,-0.135) node[below, midway,yshift=0pt]{$\subset S$};
    \end{scope}
    \end{tikzpicture}
    \end{gathered}
    \quad+\quad
    \begin{gathered}
    \begin{tikzpicture}[baseline= {($(current bounding box.base)+(10pt,10pt)$)},line width=1, scale=1]
    \coordinate (a) at (0,0) ;
    \coordinate (b) at (-1,0.5) ;
    \coordinate (c) at (-2,-0.5);
    \coordinate (d) at (-1,0);
    \draw[] (a) -- (b) -- (c) -- (a);
    \draw[RoyalBlue] (b)-- ++(170:0.5) node[left,scale=0.75] {$6$};
    \draw[Maroon] (b)-- ++(140:0.5) node[left,scale=0.75,yshift=5] {$\bar{2}$};
    \draw[RoyalBlue] (c)-- ++(-170:0.5) node[left,scale=0.75,xshift=-3] {$4$};
    \draw[RoyalBlue] (c)-- ++(-140:0.5) node[left,scale=0.75,xshift=0,yshift=-4] {$5$};
    \draw[RoyalBlue] (a)-- ++(30:0.5) node[right,scale=0.75,xshift=-3] {$1$};
    \draw[Maroon] (a)-- ++(-30:0.5) node[right,scale=0.75,xshift=-3] {$\bar{3}$};
    \draw[dashed,Orange] ($(d)+(180:0.25)+(0,1)$) -- ($(d)+(180:0.25)+(0,-1)$);
    \draw[dashed,Orange] ($(a)+(0:0.25)+(0,1)$) -- ($(a)+(0:0.25)+(0,-1)$);
    \begin{scope}[xshift=0pt,yshift=-25pt]
    \draw[line width=0.4, line cap=round,yshift=-7] (-1.2,0) -- (-1.2,-0.135) (0.18,0)  -- (0.18,-0.135) (-1.2,-0.135) -- (0.18,-0.135) node[below, midway,yshift=2pt]{$\subset \cM^\dag$};
    \draw[line width=0.4, line cap=round,yshift=-7] (-2.8,0) -- (-2.8,-0.135) (-1.28,0)  -- (-1.28,-0.135) (-2.8,-0.135) -- (-1.28,-0.135) node[below, midway,yshift=0pt]{$\subset S$};
    \draw[line width=0.4, line cap=round,yshift=-7] (0.28,0) -- (0.28,-0.135) (0.9,0)  -- (0.9,-0.135) (0.28,-0.135) -- (0.9,-0.135) node[below, midway,yshift=0pt]{$\subset S$};
    \end{scope}
    \end{tikzpicture}
    \end{gathered}
\label{eq:cuts45Mtri}
\end{align}
The prediction is therefore that the amplitude after the rotation should be equal to the conjugated amplitude plus a cut in the $s_{45}$ channel, both of which are computed in the region $2456 \ot 13$:
\begin{equation}
   \left[\cM^\text{tri}_{3456 \ot 12 }\right]_{ \substack{\rotatedown
 s_{13}\\ \rotatedown s_{26}}}
 \stackrel{?}{=}
 \cM^\text{tri\dag}_{2456 \ot 13} +\cut_{s_{45}} \cM^\text{tri}_{2456 \ot 13}\,,\label{eq:predictionCrossing0}
\end{equation}
where cut operation $\cut_{s_{45}} \cM^\text{tri}_{2456 \ot 13}$ is defined through the rightmost picture in~\eqref{eq:cuts45Mtri}.\footnote{In general, by $\Cut(i{\cal M})$ we denote the result of replacing all cut propagators according to $\frac{-i}{q_e^2 + m_e^2 - i\varepsilon} \to 2\pi \delta^{\pm}(q_e^2 + m_e^2)$ and complex-conjugating all the vertices and propagators in the $S^\dag$ blob, where the $\pm$ signs are selected according to the energy flow. In the conventions explained below \eqref{eq:Mtriloopmom} and after absorbing the overall minus sign on the right-hand side of the crossing equation \eqref{eq:crossing23}, the overall factors of $i$ in $\Cut\, \M$ work out to be $-i^{-\text{L}_L + \text{L}_M - \text{L}_R - 1}$, where $\text{L}_L$, $\text{L}_M$, and $\text{L}_R$ are the number of loops in the left, middle, and right blobs respectively.}
To verify this prediction, we compute the cut as
\begin{equation}
        \cut_{s_{45}} \cM^\text{tri}_{2456 \ot 13}  = - \left(2\pi \right)^2  \int \frac{\d^4 \ell}{i\pi^2} \frac{\delta^-[\ell^2] \delta^+\big[(\ell+p_{45})^2\big]}{(\ell-p_{13})^2+i\varepsilon}\,,
\end{equation}
where the momentum labelings were given in~\eqref{eq:tridiagram}.
We work in the center-of-mass frame of $s_{45}$, where momentum conservation and the delta functions impose 
\begin{equation}
    \ell^0 = \ell^0_\ast = -\frac{[s_{13}(1-z)(1-\zb)]^{1/2}}{2} \,,
\end{equation}
as well as
\begin{equation}
    p_{13}^0 = \frac{s_{13}(z+\zb-2)}{2 [s_{13} (1-z)(1-\zb)]^{1/2}} \quad \text{and} \quad
    \vert \vec{p}_{13} \vert = \frac{z-\zb}{2} \left[ \frac{s_{13}}{(1-z)(1-\zb)} \right]^{1/2} \,.
\end{equation}
Putting all this together, we get
\begin{equation}\label{eq:cut2Particles}
    \Cut_{s_{45}} \cM^\text{tri}_{2456 \ot 13}= - i\pi \int_{-1}^1  \frac{\d \cos \theta}{s_{13}-2 \ell^0_\ast (p_{13}^0-\vert \vec{p}_{13}\vert \cos \theta)} = \frac{-2 \pi i}{s_{13} (z-\zb)} \log \frac{z}{\zb}\,.
\end{equation}
Comparing the last expression with \eqref{eq:rotDisc} validates \eqref{eq:predictionCrossing0}, in accordance with the prediction of the crossing equation in \eqref{eq:crossing23}.

\subsection{\label{sec:individual-cuts}Crossing predicts individual cuts}

\begin{figure}
    \centering
        \begin{subfigure}[c]{0.24\textwidth}
        \centering
\begin{tikzpicture}
  \draw[->,thick] (-1, 0) -- (2.2, 0);
  \draw[->,thick] (0, -1) -- (0, 1.5);
  \node[] at (1.8,1.3) {$z$};
  \draw[] (2.0,1.1) -- (1.6,1.1) -- (1.6,1.5);
  \draw[decorate, decoration={zigzag, segment length=6, amplitude=2}, black!80] (1,0) -- (2.17,0);
  \draw[black!80,fill=black!80,thick] (1,0) circle (0.05);
  \draw[decorate, decoration={zigzag, segment length=6, amplitude=2}, black!80] (0,0) -- (-1,0);
  \draw[black!80,fill=black!80,thick] (0,0) circle (0.05);
  \draw[->,Maroon,thick] (1.2,0.1) arc (0:-160:0.35);
  \draw[->,darkgreen,dashed,thick] (1.2,0.1) arc (0:280:0.45);
  \draw[-,darkgreen,dashed,thick] (1.2,0.1) arc (0:330:0.45);
  \draw[darkgreen,fill=Maroon,thick] (1.2,0.1) circle (0.05);
  \draw[darkgreen,fill=darkgreen,thick] (1.2,-0.1) circle (0.05);
  \node[scale=0.75] at (1,0.7) {$\text{\color{darkgreen} unitarity path}$};
  \node[scale=0.75] at (1.2,-0.65) {$\text{\color{Maroon} crossing path}$};
\end{tikzpicture}
\end{subfigure}
\hfill
\begin{subfigure}[c]{0.24\textwidth}
    \centering
\begin{tikzpicture}
\tikzset{ma/.style={decoration={markings,mark=at position 0.5 with {\arrow[scale=0.8]{>}}},postaction={decorate}}}
\tikzset{mar/.style={decoration={markings,mark=at position 0.5 with {\arrowreversed[scale=0.8]{>}}},postaction={decorate}}}
  \draw[->,thick] (-1, 0) -- (2.2, 0);
  \draw[->,thick] (0, -1) -- (0, 1.5);
  \node[] at (1.8,1.3) {$\zb$};
  \draw[] (2.0,1.1) -- (1.6,1.1) -- (1.6,1.5);
  \draw[decorate, decoration={zigzag, segment length=6, amplitude=2}, black!80] (1,0) -- (2.17,0);
  \draw[black!80,fill=black!80,thick] (1,0) circle (0.05);
  \draw[decorate, decoration={zigzag, segment length=6, amplitude=2}, black!80] (0,0) -- (-1,0);
  \draw[black!80,fill=black!80,thick] (0,0) circle (0.05);
  \draw[darkgreen,fill=Maroon,thick] (0.2,0) circle (0.05);
  \node[scale=1] at (-0.6,0.75) {$\substack{\text{\color{white} crossing} \\ \text{\color{white} path}}$};
  \node[scale=1] at (-0.6,0.75) {$\substack{\text{\color{Maroon} crossing} \\ \text{\color{Maroon} path}}$};
  \node[scale=1] at (1,0.7) {$\substack{\text{\color{darkgreen} unitarity} \\ \text{\color{darkgreen} path}}$};
  \draw[ma][darkgreen,dashed,thick] (0.2,0) arc (0:-360:-0.30);
  \draw[ma][Maroon,thick] (0.2,0) arc (0:-360:-0.20);
\end{tikzpicture}
\end{subfigure}
\hfill
\begin{subfigure}[c]{0.5\textwidth}
    \centering
\begin{equation*}
    \left[
    \adjustbox{valign=c, scale={0.8}{0.8}}{\begin{tikzpicture}[baseline= {($(current bounding box.base)+(10pt,10pt)$)},line width=1, scale=0.7]
    \coordinate (a) at (0,0) ;
    \coordinate (b) at (-1,0.5) ;
    \coordinate (c) at (-1,-0.5);
    \coordinate (d) at (-1,0);
    \draw[] (a) -- (b) -- (c) -- (a);
    \draw[RoyalBlue] (b)-- ++(170:0.5) node[left,scale=0.75] {$6$};
    \draw[Maroon] (b)-- ++(30:0.5) node[left,scale=0.75,yshift=5] {$2$};
    \draw[RoyalBlue] (c)-- ++(-170:0.5) node[left,scale=0.75,xshift=-3] {$4$};
    \draw[RoyalBlue] (c)-- ++(-140:0.5) node[left,scale=0.75,xshift=0,yshift=-4] {$5$};
    \draw[RoyalBlue] (a)-- ++(30:0.5) node[right,scale=0.75,xshift=-3] {$1$};
    \draw[Maroon] (a)-- ++(-130:0.5) node[left,below,scale=0.75,xshift=-3] {$3$};
    \end{tikzpicture}}
    \right]_{\substack{\rotatedown
 s_{13} \\ \rotatedown s_{26}}}=
    \adjustbox{valign=c, scale={0.8}{0.8}}{\begin{tikzpicture}[baseline= {($(current bounding box.base)+(10pt,10pt)$)},line width=1, scale=0.7]
    \coordinate (a) at (0,0) ;
    \coordinate (b) at (-1,0.5) ;
    \coordinate (c) at (-1,-0.5);
    \coordinate (d) at (-1,0);
    \draw[] (a) -- (b) -- (c) -- (a);
    \draw[RoyalBlue] (b)-- ++(170:0.5) node[left,scale=0.75] {$6$};
    \draw[Maroon] (b)-- ++(140:0.5) node[left,scale=0.75,yshift=5] {$\bar{2}$};
    \draw[RoyalBlue] (c)-- ++(-170:0.5) node[left,scale=0.75,xshift=-3] {$4$};
    \draw[RoyalBlue] (c)-- ++(-140:0.5) node[left,scale=0.75,xshift=0,yshift=-4] {$5$};
    \draw[RoyalBlue] (a)-- ++(30:0.5) node[right,scale=0.75,xshift=-3] {$1$};
    \draw[Maroon] (a)-- ++(-30:0.5) node[right,scale=0.75,xshift=-3] {$\bar{3}$};
    \draw[dashed,Orange] ($(d)+(180:0.25)+(0,1)$) -- ($(d)+(180:0.25)+(0,-1)$);
    \draw[dashed,Orange] ($(a)+(0:0.25)+(0,1)$) -- ($(a)+(0:0.25)+(0,-1)$);
    \end{tikzpicture}}+\adjustbox{valign=c, scale={0.8}{0.8}}{
    \begin{tikzpicture}[baseline= {($(current bounding box.base)+(10pt,10pt)$)},line width=1, scale=0.7]
    \coordinate (a) at (0,0);
    \coordinate (b) at (-1,0.5);
    \coordinate (c) at (-2,-0.5);
    \coordinate (d) at (-1,0);
    \draw[] (a) -- (b) -- (c) -- (a);
    \draw[RoyalBlue] (b)-- ++(170:0.5) node[left,scale=0.75] {$6$};
    \draw[Maroon] (b)-- ++(140:0.5) node[left,scale=0.75,yshift=5] {$\bar{2}$};
    \draw[RoyalBlue] (c)-- ++(-170:0.5) node[left,scale=0.75,xshift=-3] {$4$};
    \draw[RoyalBlue] (c)-- ++(-140:0.5) node[left,scale=0.75,xshift=0,yshift=-4] {$5$};
    \draw[RoyalBlue] (a)-- ++(30:0.5) node[right,scale=0.75,xshift=-3] {$1$};
    \draw[Maroon] (a)-- ++(-30:0.5) node[right,scale=0.75,xshift=-3] {$\bar{3}$};
    \draw[dashed,Orange] ($(d)+(180:0.25)+(0,1)$) -- ($(d)+(180:0.25)+(0,-1)$);
    \draw[dashed,Orange] ($(a)+(0:0.25)+(0,1)$) -- ($(a)+(0:0.25)+(0,-1)$);
    \end{tikzpicture}}
    \end{equation*}
    \end{subfigure}
\vspace{0.5cm}
\par
\begin{subfigure}[c]{0.24\textwidth}
    \centering
\begin{tikzpicture}
\tikzset{ma/.style={decoration={markings,mark=at position 0.5 with {\arrow[scale=0.8]{>}}},postaction={decorate}}}
\tikzset{mar/.style={decoration={markings,mark=at position 0.5 with {\arrowreversed[scale=0.8]{>}}},postaction={decorate}}}
  \draw[->,thick] (-1, 0) -- (2.2, 0);
  \draw[->,thick] (0, -1) -- (0, 1.5);
  \node[] at (1.8,1.3) {$z$};
  \draw[] (2.0,1.1) -- (1.6,1.1) -- (1.6,1.5);
  \draw[decorate, decoration={zigzag, segment length=6, amplitude=2}, black!80] (1,0) -- (2.17,0);
  \draw[black!80,fill=black!80,thick] (1,0) circle (0.05);
  \draw[decorate, decoration={zigzag, segment length=6, amplitude=2}, black!80] (0,0) -- (-1,0);
  \draw[black!80,fill=black!80,thick] (0,0) circle (0.05);
  \draw[->,Maroon,thick] (0.8,0) arc (0:-160:-0.35);
  \draw[-,darkgreen,dashed,thick] (0.8,0) arc (0:360:0.20);
  \draw[darkgreen,fill=Maroon,thick] (0.8,0.0) circle (0.05);
  \node[scale=0.75] at (1.2,-0.65) {$\text{\color{darkgreen} unitarity path}$};
  \node[scale=0.75] at (1,0.7) {$\text{\color{Maroon} crossing path}$};
\end{tikzpicture}
\end{subfigure}
\hfill
\begin{subfigure}[c]{0.24\textwidth}
    \centering
\begin{tikzpicture}
\tikzset{ma/.style={decoration={markings,mark=at position 0.55 with {\arrow[scale=0.8]{>}}},postaction={decorate}}}
\tikzset{mar/.style={decoration={markings,mark=at position 0.5 with {\arrowreversed[scale=0.8]{>}}},postaction={decorate}}}
  \draw[->,thick] (-1, 0) -- (2.2, 0);
  \draw[->,thick] (0, -1) -- (0, 1.5);
  \node[] at (1.8,1.3) {$\zb$};
  \draw[] (2.0,1.1) -- (1.6,1.1) -- (1.6,1.5);
  \draw[decorate, decoration={zigzag, segment length=6, amplitude=2}, black!80] (1,0) -- (2.17,0);
  \draw[black!80,fill=black!80,thick] (1,0) circle (0.05);
  \draw[decorate, decoration={zigzag, segment length=6, amplitude=2}, black!80] (0,0) -- (-1,0);
  \draw[black!80,fill=black!80,thick] (0,0) circle (0.05);
  \draw[darkgreen,fill=Maroon,thick] (-0.5,0.2) circle (0.05);
  \node[scale=1] at (-0.6,0.75) {$\substack{\text{\color{Maroon} crossing} \\ \text{\color{Maroon} path}}$};
  \node[scale=1] at (1,0.7) {$\substack{\text{\color{darkgreen} unitarity} \\ \text{\color{darkgreen} path}}$};
  \draw[->,darkgreen,dashed,thick] (-0.5,0.2) arc (-21.8:-300:-0.539);
  \draw[-,darkgreen,dashed,thick] (-0.5,0.2) arc (-21.8:-360:-0.539);
  \draw[ma][Maroon,thick] (-0.5,0.2) arc (0:-360:0.15);
\draw[darkgreen,fill=darkgreen,thick] (-0.51,-0.1) circle (0.05);
\end{tikzpicture}
\end{subfigure}
\hfill
\begin{subfigure}[c]{0.5\textwidth}
    \centering
\begin{equation*}
    \left[
    \adjustbox{valign=c, scale={0.8}{0.8}}{\begin{tikzpicture}[baseline= {($(current bounding box.base)+(10pt,10pt)$)},line width=1, scale=0.7]
    \coordinate (a) at (0,0) ;
    \coordinate (b) at (-1,0.5) ;
    \coordinate (c) at (-1,-0.5);
    \coordinate (d) at (-1,0);
    \draw[] (a) -- (b) -- (c) -- (a);
    \draw[Maroon] (b)-- ++(170:0.5) node[left,scale=0.75] {$2$};
    \draw[RoyalBlue] (b)-- ++(30:0.5) node[left,scale=0.75,yshift=5] {$6$};
    \draw[RoyalBlue] (c)-- ++(-170:0.5) node[left,scale=0.75,xshift=-3] {$4$};
    \draw[RoyalBlue] (c)-- ++(-140:0.5) node[left,scale=0.75,xshift=0,yshift=-4] {$5$};
    \draw[RoyalBlue] (a)-- ++(30:0.5) node[right,scale=0.75,xshift=-3] {$1$};
    \draw[Maroon] (a)-- ++(-30:0.5) node[right,scale=0.75,xshift=-3] {$3$};
    \end{tikzpicture}}
    \right]_{ \substack{\rotatedown s_{26} \\ \rotateup s_{13}}}=
    \adjustbox{valign=c, scale={0.8}{0.8}}{\begin{tikzpicture}[baseline= {($(current bounding box.base)+(10pt,10pt)$)},line width=1, scale=0.7]
    \coordinate (a) at (0,0) ;
    \coordinate (b) at (-1,0.5) ;
    \coordinate (c) at (-1,-0.5);
    \coordinate (d) at (-1,0);
    \draw[] (a) -- (b) -- (c) -- (a);
    \draw[RoyalBlue] (b)-- ++(10:0.5) node[right,scale=0.75] {$6$};
    \draw[Maroon] (b)-- ++(40:0.5) node[right,scale=0.75,yshift=5] {$\bar{2}$};
    \draw[RoyalBlue] (c)-- ++(-170:0.5) node[left,scale=0.75,xshift=-3] {$4$};
    \draw[RoyalBlue] (c)-- ++(-140:0.5) node[left,scale=0.75,xshift=0,yshift=-4] {$5$};
    \draw[RoyalBlue] (a)-- ++(30:0.5) node[right,scale=0.75,xshift=-3] {$1$};
    \draw[Maroon] (a)-- ++(-120:0.5) node[below,scale=0.75,xshift=-3] {$\bar{3}$};
    \draw[dashed,Orange] ($(d)+(180:0.25)+(0,1)$) -- ($(d)+(180:0.25)+(0,-1)$);
    \draw[dashed,Orange] ($(a)+(0:0.25)+(0,1)$) -- ($(a)+(0:0.25)+(0,-1)$);
    \end{tikzpicture}}
    +
    \adjustbox{valign=c, scale={0.8}{0.8}}{\begin{tikzpicture}[baseline= {($(current bounding box.base)+(10pt,10pt)$)},line width=1, scale=0.7]
    \coordinate (a) at (0,0);
    \coordinate (b) at (-1,0.5);
    \coordinate (c) at (-2,-0.5);
    \coordinate (d) at (-1,0);
    \draw[] (a) -- (b) -- (c) -- (a);
    \draw[RoyalBlue] (b)-- ++(10:0.5) node[right,scale=0.75] {$6$};
    \draw[Maroon] (b)-- ++(40:0.5) node[right,scale=0.75,yshift=5] {$\bar{2}$};
    \draw[RoyalBlue] (c)-- ++(-170:0.5) node[left,scale=0.75,xshift=-3] {$4$};
    \draw[RoyalBlue] (c)-- ++(-140:0.5) node[left,scale=0.75,xshift=0,yshift=-4] {$5$};
    \draw[RoyalBlue] (a)-- ++(30:0.5) node[right,scale=0.75,xshift=-3] {$1$};
    \draw[Maroon] (a)-- ++(-120:0.5) node[below,scale=0.75,xshift=-3] {$\bar{3}$};
    \draw[dashed,Orange] ($(d)+(180:0.25)+(0,1)$) -- ($(d)+(180:0.25)+(0,-1)$);
    \draw[dashed,Orange] ($(a)+(0:0.25)+(0,1)$) -- ($(a)+(0:0.25)+(0,-1)$);
    \end{tikzpicture}}
\end{equation*}
    \end{subfigure}
    \caption{Paths of analytic continuation prescribed by unitarity and crossing. \textbf{Top:} The dashed green path is the one prescribed by unitarity. Note that $\zb$ stays fixed both when evaluating the unitarity equation and during the crossing path. The red path is the one prescribed by crossing, and it immediately enters the second sheet upon the start of the analytic continuation. \textbf{Bottom:} Unitarity path and crossing path for the process $456 \ot 123$ in the region $R^{(13)(45)}$. In this region, $\zb<0$, $0<z<1$ and we take $\zb \to \zb+i\varepsilon$. The crossing path prescribes a rotation in $\zb$ to have a positive imaginary part on the second sheet, and $z$ rotates clockwise around $1$, into a new kinematic channel $R^{(26)(45)}$.}
    \label{fig:unitarity_crossing}
\end{figure}
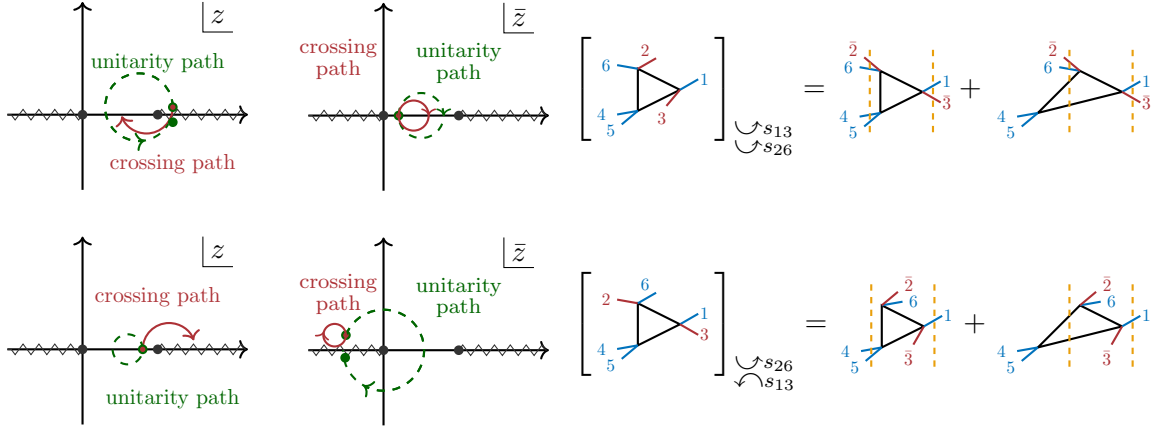

In the triangle example above, we can compare the result with what we could have predicted using unitarity. Unitarity would tell us that the imaginary part of the amplitude in a given kinematic channel is equal to the sum over all cuts in that channel. For the triangle diagram above, unitarity gives
\be 
    2 i \, \Im \cM^\text{tri}_{3456 \ot 12} = \cM^\text{tri}_{3456 \ot 12} - \cM^\text{tri \dag}_{3456 \ot 12}
    = \sum_{\substack{\text{all allowed}\\i,j}} \cut^{\text{U}}_{s_{ij}} \cM^{\tri}_{3456 \ot 12}  \,.
    \label{eq:unitarity}
\ee
We warn the reader that the definition of ``Cut'' here is subtly but consequentially different from the one used before. The unitarity cuts (or ``Cutkosky-rules'' cuts) in \eqref{eq:unitarity} instruct us to complex conjugate the amplitude to the right of the cut~\cite{Cutkosky:1960sp}. We denote this fact with a superscript ``$\text{U}$'' standing for ``unitarity cut''. As can be deduced from~\eqref{eq:zzb1}, the difference between the amplitude and its conjugate in this case is simply given by the difference between $\cM$ evaluated at $z+i\varepsilon$ and $z-i\varepsilon$. Furthermore, the only allowed cut in the $3456 \ot 12$ kinematics is the one in the $s_{45}$ channel. That is, the left-hand side of~\eqref{eq:unitarity} is simply given by the discontinuity across the branch cut at $z>1$:
\be 
    \disc_{z>1} \cM^\text{tri}_{3456 \ot 12}
    = \cut^{\text{U}}_{s_{45}} \cM^{\tri}_{3456 \ot 12} \qquad \text{(unitarity prediction)}\,.
    \label{eq:unitaritypred}
\ee 
Since the unitarity cut happens to agree with $\Cut_{s_{45}} \cM^{\tri}_{2456 \ot 13}$ up to a sign, this is the same prediction as we obtained using the crossing equation in \eqref{eq:rotDisc}. However, the paths that led to this are different, as we have shown in Fig.~\ref{fig:unitarity_crossing} (top). The unitarity path takes us between the value of $\cM^{\mathrm{tri}}$ above and below the branch cuts, while staying entirely on the first sheet in the $z$-plane. Note that $\zb$ remains fixed along both the unitarity and crossing paths. On the other hand, the crossing path immediately enters the second sheet of the $z$-plane at the beginning of the analytic continuation. In this example, it is easy to see that despite the different paths, the analytic continuation of the cut is straightforward and the crossing path can simply be extended to a full circle. As a result,~\eqref{eq:unitaritypred} simply becomes a consistency condition for the crossing prediction. We emphasize that, despite arriving at this conclusion, the two paths are conceptually different.

The story changes if we instead compare crossing and unitarity in a kinematic channel where more than one cut is allowed, as shown in Fig.~\ref{fig:unitarity_crossing} (bottom). Unitarity tells us to compare $\cM^{\mathrm{tri}}$ on the first sheet, above and below the branch cut in $\zb$, which leads to
\be 
    2 i \Im \cM^\text{tri}_{456 \ot 123}
    = \cut^{\text{U}}_{s_{45}} \cM^{\tri}_{456 \ot 123} + \cut^{\text{U}}_{s_{13}} \cM^{\tri}_{456 \ot 123} \qquad \text{(unitarity prediction)}\,,
\ee
where, as usual, the right-hand side of the cut is given by conjugated Feynman rules. The crossing prediction, however, is 
\be 
  \left[\cM^\text{tri}_{245 \ot 136 }\right]_{ \substack{\rotatedown s_{26}\\ \rotateup s_{13}}}
  =
  \cM^\text{tri \dag}_{345 \ot 126} + \cut_{s_{45}} \cM^{\text{tri}}_{345 \ot 126} \qquad \text{(crossing prediction)} \,. \label{eq:crpred0}
\ee 
To avoid clutters, we leave implicit in this formula that $\Cut_{s_{45}} \cM^{\mathrm{tri}}$ should be evaluated with the middle part between the orange cuts in Fig.~\ref{fig:unitarity_crossing} conjugated. Note that the endpoint of the crossing path corresponds to an $i \varepsilon$ prescription in $z$ and $\zb$ that cannot be represented by any of the choices in~\eqref{eq:zzb1} and~\eqref{eq:zzb2}. This reflects rather explicitly the fact that the object obtained through crossing is \emph{not} a conventional time-ordered amplitude.

There are two things to note about the result in \eqref{eq:crpred0}. First, the crossing prediction is not equivalent to the one obtained by unitarity, and it is not obvious how to connect the two. Second, if we can compute the amplitude in different kinematic regions, the crossing equation gives a new way to predict the \emph{individual} cut contribution $\Cut_{s_{45}}$ in this channel: it corresponds to the difference between $\cM^\text{tri \dag}_{345 \ot 126}$ (i.e., $\cM^{\text{tri}}$ with $\zb \to \zb + i\varepsilon$ and $z \to z-i\varepsilon$) and $\cM^\text{tri}$ with $\zb \to \zb + i\varepsilon$ and $z \to z+i\varepsilon$. In this case, this cut is not equal to the total discontinuity in the kinematic region $R^{(26) (45)}$, since the discontinuity would also include contributions from cuts in the $s_{26}$ channel. The crossing equation, therefore, yields a prediction distinct from that of total discontinuities in a kinematic channel, underpinning some of the recent developments regarding the relation between cuts and discontinuities~\cite{Abreu:2014cla,Bourjaily:2020wvq}. In this example, it is perhaps not surprising that the $\Cut_{s_{45}}$ operation is equivalent to taking the discontinuity in $z>1$, as we found in the region where $0<\zb<1$. The non-trivial information we get from the crossing equation, however, is which branches we land on, since the cut $\Cut_{s_{45}}\cM^{\text{tri}}_{345 \ot 126}$ generally has non-vanishing real and imaginary parts.

\paragraph{All-loop predictions}
We can extend our previous discussion to a prediction of the $\Cut_{s_{45}}$ operation to the $L$-loop triangle ladder diagram, whose full expression is given by~\cite{Usyukina:1993ch}
\be 
    \mathcal{M}^{\text{tri ladder}}=
    \frac{(-1)^L}{L! (z{-}\zb)} \frac{1}{s_{13} s_{45}^{L{-}1}}
    \sum_{j=L}^{2L}
    \frac{\left(-1\right)^j j! \left[ \log z{+}\log\zb \right]^{2L{-}j}}{\left(j{-}L\right)!\left(2L{-}j\right)!} \left[ \text{Li}_{j}(z)-\text{Li}_{j}(\zb)\right]\,,
    \label{eq:triladder}
\ee 
and the analytic continuations in $z$ and $\zb$ given in~\eqref{eq:zzb1} and~\eqref{eq:zzb2}.

Starting in the region $R^{(13)(45)}$, we cross $2\leftrightarrow3$ and end in the region $R^{(26)(45)}$ with $0<z<1$ and $\zb<0$.
The prediction of the crossing equation is exactly analogous to the one for the one-loop triangle: the analytically continued amplitude should be equal to the conjugated amplitude plus the cuts in the $s_{45}$ channel, namely
\begin{align}
    \left[
    \begin{gathered}
    \begin{tikzpicture}[baseline= {($(current bounding box.base)+(10pt,10pt)$)},line width=1, scale=1]
    \coordinate (a) at (0,0) ;
    \coordinate (b) at (-1,0.5) ;
    \coordinate (c) at (-1,-0.5);
    \coordinate (d) at (-1,0);
    \draw[] (a) -- (b) -- (c) -- (a);
    \draw[] (-0.8,-0.4) -- (-1,-0.3);
    \draw[] (-1,-0.1) -- (-0.6,-0.3);
    \draw[] (-0.2,-0.1) -- (-1,0.3);
    \begin{scope}[shift={(-0.7,-0.1)}, rotate=60]
    \foreach \i in {0, 0.15, 0.075} {
        \fill (\i,0) circle (0.5pt);
    }
    \end{scope}
    \draw[Maroon] (b)-- ++(170:0.5) node[left,scale=0.75] {$2$};
    \draw[RoyalBlue] (b)-- ++(30:0.5) node[left,scale=0.75,yshift=5] {$6$};
    \draw[RoyalBlue] (c)-- ++(-170:0.5) node[left,scale=0.75,xshift=-3] {$4$};
    \draw[RoyalBlue] (c)-- ++(-140:0.5) node[left,scale=0.75,xshift=0,yshift=-4] {$5$};
    \draw[RoyalBlue] (a)-- ++(30:0.5) node[right,scale=0.75,xshift=-3] {$1$};
    \draw[Maroon] (a)-- ++(-30:0.5) node[right,scale=0.75,xshift=-3] {$3$};
    \end{tikzpicture}
    \end{gathered}
    \right]_{\substack{\rotatedown
 s_{13} \\ \rotatedown s_{26}}}
 \hspace{-0.4cm}
    \stackrel{\substack{\text{crossing} \\ \text{prediction}}}{=}
    \quad
    \begin{gathered}
    \begin{tikzpicture}[baseline= {($(current bounding box.base)+(10pt,10pt)$)},line width=1, scale=1]
    \coordinate (a) at (0,0) ;
    \coordinate (b) at (-1,0.5) ;
    \coordinate (c) at (-1,-0.5);
    \coordinate (d) at (-1,0);
    \draw[] (a) -- (b) -- (c) -- (a);
    \draw[] (-0.8,-0.4) -- (-1,-0.3);
    \draw[] (-1,-0.1) -- (-0.6,-0.3);
    \draw[] (-0.2,-0.1) -- (-1,0.3);
    \begin{scope}[shift={(-0.7,-0.1)}, rotate=60]
    \foreach \i in {0, 0.15, 0.075} {
        \fill (\i,0) circle (0.5pt);
    }
    \end{scope}
    \draw[RoyalBlue] (b)-- ++(10:0.5) node[right,scale=0.75] {$6$};
    \draw[Maroon] (b)-- ++(40:0.5) node[right,scale=0.75,yshift=5] {$\bar{2}$};
    \draw[RoyalBlue] (c)-- ++(-170:0.5) node[left,scale=0.75,xshift=-3] {$4$};
    \draw[RoyalBlue] (c)-- ++(-140:0.5) node[left,scale=0.75,xshift=0,yshift=-4] {$5$};
    \draw[RoyalBlue] (a)-- ++(30:0.5) node[right,scale=0.75,xshift=-3] {$1$};
    \draw[Maroon] (a)-- ++(-120:0.5) node[left,scale=0.75,yshift=-3] {$\bar{3}$};
    \draw[dashed,Orange] ($(d)+(180:0.25)+(0,1)$) -- ($(d)+(180:0.25)+(0,-1)$);
    \draw[dashed,Orange] ($(a)+(0:0.25)+(0,1)$) -- ($(a)+(0:0.25)+(0,-1)$);
    \end{tikzpicture}
    \end{gathered}
    \qquad
    + \sum_{\substack{\text{all cuts} \\ \text{in } s_{45}}}
    \begin{gathered}
    \begin{tikzpicture}[baseline= {($(current bounding box.base)+(10pt,10pt)$)},line width=1, scale=1]
    \coordinate (a) at (0,0) ;
    \coordinate (b) at (-1,0.5) ;
    \coordinate (c) at (-2,-0.5);
    \coordinate (d) at (-1,0);
    \draw[] (a) -- (b) -- (c) -- (a);
    \draw (-1.6,-0.4) -- (-1.8, -0.3); 
    \draw (-1.8,-0.45) -- (-1.9, -0.4); 
    \draw (-1.0,-0.25) -- (-1.5, 0); 
    \draw (-0.2,-0.05) -- (-1.1, 0.4); 
    \begin{scope}[shift={(-0.9,0.0)}, rotate=40]
    \foreach \i in {0, 0.15, 0.075} {
        \fill (\i,0) circle (0.5pt);
    }
    \end{scope}
    \draw[RoyalBlue] (b)-- ++(10:0.5) node[right,scale=0.75] {$6$};
    \draw[Maroon] (b)-- ++(40:0.5) node[right,scale=0.75,yshift=5] {$\bar{2}$};
    \draw[RoyalBlue] (c)-- ++(-170:0.5) node[left,scale=0.75,xshift=-3] {$4$};
    \draw[RoyalBlue] (c)-- ++(-140:0.5) node[left,scale=0.75,xshift=0,yshift=-4] {$5$};
    \draw[RoyalBlue] (a)-- ++(30:0.5) node[right,scale=0.75,xshift=-3] {$1$};
    \draw[Maroon] (a)-- ++(-120:0.5) node[left,scale=0.75,yshift=-3] {$\bar{3}$};
    \draw[dashed,Orange] ($(d)+(180:0.25)+(0,1)$) -- ($(d)+(180:0.25)+(0,-1)$);
    \draw[dashed,Orange] ($(a)+(0:0.25)+(0,1)$) -- ($(a)+(0:0.25)+(0,-1)$);
    \end{tikzpicture}
    \end{gathered}
\label{eq:Cutladder}
\end{align}
where the middle part between the orange cuts is conjugated as before.

In this case, individual cuts in the $s_{45}$ channel are not straightforward to compute. Instead of checking the crossing equation, we can use it to make a non-trivial prediction: namely that the sum of cuts on the right-hand side of~\eqref{eq:Cutladder} is equal to the discontinuity across the $z>1$ branch cut of~\eqref{eq:triladder}, while keeping $\zb<0$ fixed above the branch cut (i.e., taking $\zb \to \zb + i \varepsilon$):
\be  \label{cuts45 pred}
    \sum_{\substack{\text{all cuts} \\ \text{in } s_{45}}} \Cut_{s_{45}} \mathcal{M}^{\text{tri ladder}}
    \stackrel{?}{=}
    \frac{(-1)^L\left(s_{45}\right)^{1-L}}{L! s_{13} (z-\zb)}
    \sum_{j=L}^{2L}
    \frac{\left(-1\right)^j j! \left( \log(-z \zb) + i \pi \right)^{2L-j}}{\left(j-L\right)!\left(2L-j\right)!} \frac{2\pi i \log^{j-1} z  }{(j-1)!}\,.
\ee
The $i \pi$'s in this expression represent the non-trivial prediction of the crossing equation. More generally, the crossing equation computes $\Cut_{s_{45}}$, including taking into account which branches we land on. Of course, the prediction \eqref{cuts45 pred} is admittedly not too surprising either since the calculation amounted to taking an $s_{45}$ discontinuity. The interesting aspect of the method is that it explains clearly \emph{how} to take the discontinuity even in situations that are not controlled by unitarity (this cut is not equal to the imaginary part of the amplitude).

In Sec.~\ref{sec:fivepoint}, we will further test the crossing equation in numerous additional loop-level examples. 

\subsection{\label{sec:conjectureII}Conjectural extension: Crossing multiple particles?}

Let us now discuss our conjecture for a generalization of the crossing equation to crossing multiple particles at the same time. For concreteness, we start with the process $CD \ot AB$, where each of $A$, $B$, $C$, and $D$ is a non-empty set of particles. We then cross the clusters $B$ and $C$.
In light-cone coordinates, the analytic continuation is described by the following deformation of the momenta for each particle in the $B$ and $C$ clusters:
\begin{equation}\begin{aligned}
p_b^\mu(z) = \big(z p_b^+,\, \tfrac{1}{z} p_b^-,\, p_b^\perp\big) \quad\text{for all } b \in B\, ,
\label{eq:crossingpath}
\\
p_c^\mu(z) = \big(z p_c^+,\, \tfrac{1}{z} p_c^-,\, p_c^\perp\big) \quad\text{for all } c \in C\, ,
\end{aligned}\end{equation}
where $p^{\pm}_b < 0 < p^\pm_c$ and $-p_i^2 = p_i^+ p_i^- - (p_i^\perp)^2$.
The on-shell conditions are preserved.
Moreover, in order for momentum conservation to remain satisfied along the deformation, we work in any Lorentz frame in which the condition
\be
\sum_{b\in B} p^{\pm}_b + \sum_{c \in C} p^{\pm}_c = \sum_{a\in A} p^{\pm}_a + \sum_{d \in D} p^{\pm}_d = 0\, ,
\ee
is satisfied. Note that, by Lorentz invariance, we could equally have
deformed the momenta for the clusters $A$ and $D$, i.e., the
$B \leftrightarrow C$ and $A \leftrightarrow D$ crossing paths are equivalent. We perform the same deformation along a large arc in the $z$ plane as illustrated previously in \eqref{eq:crossingpath3}. Only the Mandelstam invariants $s_I$ for which $I$ and its complement $\bar{I}$ both contain at least one label from $B \cup C$ and one from $A \cup D$ get $z$-deformed. They behave as $s_I(z) \sim z$ at large $z$. For example, if $s_I$ starts off timelike ($s_I > 0$), it rotates along a large arc in the upper half-plane to being spacelike ($s_I < 0$).

Equation \eqref{eq:crossing23} gives our main conjecture for the crossing of \emph{two} elementary (stable) particles. However, as noted above that equation, the path which crosses two stable particles $b\leftrightarrow c$ is equivalent to one which crosses the conjugate clusters $A\leftrightarrow D$, for any number of legs.
We can use this observation to motivate a simple candidate crossing equation valid for arbitrary clusters of particles $A,B,C$ and $D$. We simply tack an $S$ blob onto each cluster:
\begin{equation}
\adjustbox{valign=c}{
\begin{tikzpicture}[line width=1]
\def\Ang{40};
\def\CAng{140};
\def\CosVal{0.766044};
\def\SinVal{0.642788};
\def\CstVal{0.1};
\begin{scope}[xshift=0]
\coordinate (c) at (0,0);
\coordinate (d) at (0,0.4);
\draw[RoyalBlue] ($(c) + (\SinVal*\CstVal,\CosVal*\CstVal)$)++(-\Ang:1.15) -- ($(c)+(\SinVal*\CstVal,\CosVal*\CstVal)$);
\draw[RoyalBlue] ($(c) + (-\SinVal*\CstVal,-\CosVal*\CstVal)$)++(-\Ang:1.15) -- ($(c)+(-\SinVal*\CstVal,-\CosVal*\CstVal)$);
\draw[RoyalBlue] ($(c)+(0,0)$)++(-\Ang:1.15) -- ($(c)+(0,0)$);
\draw[RoyalBlue] ($(c) + (\SinVal*\CstVal,-\CosVal*\CstVal)$)++(-\CAng:1.15) -- ($(c)+(\SinVal*\CstVal,-\CosVal*\CstVal)$);
\draw[RoyalBlue] ($(c) + (-\SinVal*\CstVal,\CosVal*\CstVal)$)++(-\CAng:1.15) -- ($(c)+(-\SinVal*\CstVal,\CosVal*\CstVal)$);
\draw[RoyalBlue] ($(c)+(0,0)$)++(-\CAng:1.15) -- ($(c)+(0,0)$);
\draw[Maroon] ($(c) + (\SinVal*\CstVal,\CosVal*\CstVal)$)++(\CAng:1.15) -- ($(c)+(\SinVal*\CstVal,\CosVal*\CstVal)$);
\draw[Maroon] ($(c) + (-\SinVal*\CstVal,-\CosVal*\CstVal)$)++(\CAng:1.15) -- ($(c)+(-\SinVal*\CstVal,-\CosVal*\CstVal)$);
\draw[Maroon] ($(c)+(0,0)$)++(\CAng:1.15) -- ($(c)+(0,0)$);
\draw[Maroon] ($(c) + (\SinVal*\CstVal,-\CosVal*\CstVal)$)++(\Ang:1.15) -- ($(c)+(\SinVal*\CstVal,-\CosVal*\CstVal)$);
\draw[Maroon] ($(c) + (-\SinVal*\CstVal,\CosVal*\CstVal)$)++(\Ang:1.15) -- ($(c)+(-\SinVal*\CstVal,\CosVal*\CstVal)$);
\draw[Maroon] ($(c)+(0,0)$)++(\Ang:1.15) -- ($(c)+(0,0)$);
\filldraw[fill=gray!5, very thick] (0,0) circle (0.4) node {$S$};
\draw [pen colour={gray},
    decorate, 
    decoration = {calligraphic brace,
        raise=5pt,
        amplitude=2pt}] (1.0,0.4+0.65) --  (1.0,0.4)
node[pos=0.5,right=10pt,Maroon]{$B$};
\draw [pen colour={gray},
    decorate, 
    decoration = {calligraphic brace,
        raise=5pt,
        amplitude=2pt}] (1.0,-0.4) --  (1.0,-0.4-0.65)
node[pos=0.5,right=10pt,RoyalBlue]{$A$};
\draw [pen colour={gray},
    decorate, 
    decoration = {calligraphic brace,
        raise=5pt,
        amplitude=2pt}] (-1.0,0.4) -- (-1.0,0.4+0.65) 
node[pos=0.5,left=10pt,Maroon]{$C$};
\draw [pen colour={gray},
    decorate, 
    decoration = {calligraphic brace,
        raise=5pt,
        amplitude=2pt}] (-1.0,-0.4-0.65) -- (-1.0,-0.4)
node[pos=0.5,left=10pt,RoyalBlue]{$D$};
\end{scope}
\node[align=center] at (3,0.5) {\footnotesize cross ${\textcolor{Maroon}{B} \leftrightarrow \textcolor{Maroon}{C}}$};
\draw[-latex] (2.2,0) -- (3.8,0) node[right, xshift=7pt]{$-$};
\begin{scope}[xshift=200]
\coordinate (c) at (0,0);
\coordinate (d) at (0,0.4);
\draw[RoyalBlue] ($(c) + (\SinVal*\CstVal,\CosVal*\CstVal)$)++(-\Ang:1.85) -- ($(c)+(\SinVal*\CstVal,\CosVal*\CstVal)$);
\draw[RoyalBlue] ($(c) + (-\SinVal*\CstVal,-\CosVal*\CstVal)$)++(-\Ang:1.85) -- ($(c)+(-\SinVal*\CstVal,-\CosVal*\CstVal)$);
\draw[RoyalBlue] ($(c)+(0,0)$)++(-\Ang:1.85) -- ($(c)+(0,0)$);
\draw[RoyalBlue] ($(c) + (\SinVal*\CstVal,-\CosVal*\CstVal)$)++(-\CAng:1.85) -- ($(c)+(\SinVal*\CstVal,-\CosVal*\CstVal)$);
\draw[RoyalBlue] ($(c) + (-\SinVal*\CstVal,\CosVal*\CstVal)$)++(-\CAng:1.85) -- ($(c)+(-\SinVal*\CstVal,\CosVal*\CstVal)$);
\draw[RoyalBlue] ($(c)+(0,0)$)++(-\CAng:1.85) -- ($(c)+(0,0)$);
\draw[Maroon] ($(c) + (\SinVal*\CstVal,\CosVal*\CstVal)$)++(\CAng:1.85) -- ($(c)+(\SinVal*\CstVal,\CosVal*\CstVal)$);
\draw[Maroon] ($(c) + (-\SinVal*\CstVal,-\CosVal*\CstVal)$)++(\CAng:1.85) -- ($(c)+(-\SinVal*\CstVal,-\CosVal*\CstVal)$);
\draw[Maroon] ($(c)+(0,0)$)++(\CAng:1.85) -- ($(c)+(0,0)$);
\draw[Maroon] ($(c) + (\SinVal*\CstVal,-\CosVal*\CstVal)$)++(\Ang:1.85) -- ($(c)+(\SinVal*\CstVal,-\CosVal*\CstVal)$);
\draw[Maroon] ($(c) + (-\SinVal*\CstVal,\CosVal*\CstVal)$)++(\Ang:1.85) -- ($(c)+(-\SinVal*\CstVal,\CosVal*\CstVal)$);
\draw[Maroon] ($(c)+(0,0)$)++(\Ang:1.85) -- ($(c)+(0,0)$);
\filldraw[fill=gray!30,rotate=-\Ang](0,-0.2) rectangle (1,0.2);
\filldraw[fill=gray!30,rotate=-\CAng](0,-0.2) rectangle (1,0.2);
\filldraw[fill=gray!30,rotate=\Ang](0,-0.2) rectangle (1,0.2);
\filldraw[fill=gray!30,rotate=\CAng](0,-0.2) rectangle (1,0.2);
\filldraw[fill=gray!5, very thick](0,0) circle (0.4) node[yshift=1] {$S^\dag$};
\filldraw[fill=gray!5, very thick]($(0,0)+(-\Ang:1.2)$) circle (0.4) node {$S$};
\filldraw[fill=gray!5, very thick]($(0,0)+(-\CAng:1.2)$) circle (0.4) node {$S$};
\filldraw[fill=gray!5, very thick]($(0,0)+(\Ang:1.2)$) circle (0.4) node {$S$};
\filldraw[fill=gray!5, very thick]($(0,0)+(\CAng:1.2)$) circle (0.4) node {$S$};
\draw [pen colour={gray},
    decorate, 
    decoration = {calligraphic brace,
        raise=5pt,
        amplitude=2pt}] (1.6,0.6+0.65) --  (1.6,0.6)
node[pos=0.5,right=10pt,Maroon]{$\bar{C}$};
\draw [pen colour={gray},
    decorate, 
    decoration = {calligraphic brace,
        raise=5pt,
        amplitude=2pt}] (1.6,-0.75) --  (1.6,-0.75-0.65)
node[pos=0.5,right=10pt,RoyalBlue]{$A$};
\draw [pen colour={gray},
    decorate, 
    decoration = {calligraphic brace,
        raise=5pt,
        amplitude=2pt}] (-1.6,0.6) -- (-1.6,0.6+0.65) 
node[pos=0.5,left=10pt,Maroon]{$\bar{B}$};
\draw [pen colour={gray},
    decorate, 
    decoration = {calligraphic brace,
        raise=5pt,
        amplitude=2pt}] (-1.6,-0.75-0.65) --  (-1.6,-0.75)
node[pos=0.5,left=10pt,RoyalBlue]{$D$};
\draw[dashed,orange] (0.45,-1) -- (0.45,1);
\draw[dashed,orange] (-0.45,-1) -- (-0.45,1);
\node[] at (-0.47,-0.39) {\tiny $Y$};
\node[] at (0.47,-0.39) {\tiny $X$};
\node[] at (-0.43,0.34) {\tiny $Z$};
\node[] at (0.47,0.34) {\tiny $W$};
\end{scope}
\end{tikzpicture}
}
\label{eq:crossing_speculation}
\end{equation}
In terms of an equation, this translates to
\begin{equation}
    \left[S_{CD \ot AB}\right]_{{\curvearrowleft}z} =  - \sumint_{X,Y,Z,W} S_{B \ot Z} S_{D \ot Y} S^\dag_{Y Z \ot X W} S_{W \ot C} S_{X \ot A} \,.
\label{eq:crossing_speculation_eq}
\end{equation}
Once again, a more general version of this crossing conjecture can be obtained by embedding \eqref{eq:crossing_speculation} in a larger blob diagram.
The two-particle crossing equation in \eqref{eq:crossing23} is obtained as special case
where $B$ and $C$ are single-particle states, so that we can ignore the $S$ blocks acting on them, thanks to the stability condition.

From the viewpoint of local field theory, the multi-particle extension of the crossing equation~\eqref{eq:crossing_speculation}
is somewhat mysterious, since the disconnected products of $S$ blobs cannot be obtained from
the LSZ reduction of any correlation function (which would necessarily involve a chain of $S$'s as in Sec.~\ref{sec:Smatrix}).  
Nonetheless, we will find in Sec.~\ref{sec:treelevelProof} that it passes very non-trivial checks at tree level. However, in Sec.~\ref{sec:analytic-obstruction}, we are going to discuss a one-loop example with massless kinematics in which the conjecture \eqref{eq:crossing_speculation} meets an analytic obstruction due to an anomalous threshold.

\subsection{Webs of observables}
\label{sec:webs}

\begin{figure}
    \centering
    \begin{subfigure}[c]{0.1\textwidth}
        \centering
           \adjustbox{valign=c}{\tikzset{every picture/.style={line width=0.75pt}}   
\begin{tikzpicture}[x=0.75pt,y=0.75pt,yscale=-1,xscale=1]
\tikzset{ma/.style={decoration={markings,mark=at position 0.5 with {\arrow[scale=0.5]{>}}},postaction={decorate}}}
\tikzset{mar/.style={decoration={markings,mark=at position 0.5 with {\arrowreversed[scale=0.5]{>}}},postaction={decorate}}}
\draw  [white, fill=gray!3  ,fill opacity=1 ] (177.4,116.33) .. controls (171.1,116.33) and (166,111.23) .. (166,104.93) -- (166,27.54) .. controls (166,21.24) and (171.1,16.14) .. (177.4,16.14) -- (211.6,16.14) .. controls (217.9,16.14) and (223,21.24) .. (223,27.54) -- (223,104.93) .. controls (223,111.23) and (217.9,116.33) .. (211.6,116.33) -- cycle ;
\draw  [fill=gray!5  ,fill opacity=1 ][very thick]  (181.91,97.26) .. controls (181.91,90.17) and (187.66,84.42) .. (194.75,84.42) .. controls (201.84,84.42) and (207.59,90.17) .. (207.59,97.26) .. controls (207.59,104.36) and (201.84,110.1) .. (194.75,110.1) .. controls (187.66,110.1) and (181.91,104.36) .. (181.91,97.26) -- cycle ;
\draw[ma][line cap=round]    (188.59,85.96) -- (171.1,86.14) ;
\draw[ma][line cap=round]    (189.1,108.56) -- (171.61,108.73) ;
\draw[ma][line cap=round]    (219.43,86.48) -- (201.94,86.65) ;
\draw[ma][line cap=round]    (219.43,108.05) -- (201.94,108.22) ;
\draw  [fill=gray!5  ,fill opacity=1 ][very thick]  (183.24,34.6) .. controls (183.24,27.51) and (188.99,21.76) .. (196.08,21.76) .. controls (203.17,21.76) and (208.92,27.51) .. (208.92,34.6) .. controls (208.92,41.69) and (203.17,47.44) .. (196.08,47.44) .. controls (188.99,47.44) and (183.24,41.69) .. (183.24,34.6) -- cycle ;
\draw[mar][line cap=round]    (201.7,45.54) -- (220.19,45.76) ;
\draw[mar][line cap=round]    (201.67,22.94) -- (220.16,23.16) ;
\draw[ma][line cap=round]    (188.76,23.81) -- (170.27,23.98) ;
\draw[ma][line cap=round]    (188.76,45.38) -- (170.27,45.55) ;
\draw[<->,line width=0.5]   (195,75) -- (195,57.33) node[below,yshift=-50]{$(n=4)$};

\draw (189.1,92.36) node [anchor=north west][inner sep=0.75pt]  [font=\scriptsize]  {$S$};
\draw (189.76,28.21) node [anchor=north west][inner sep=0.75pt]  [font=\scriptsize]  {$S^{\dagger }$};
\end{tikzpicture}}
    \end{subfigure}\qquad
    \begin{subfigure}[c]{0.4\textwidth}
        \centering
         \adjustbox{valign=c}{\tikzset{every picture/.style={line width=0.75pt}}
\begin{tikzpicture}[x=0.75pt,y=0.75pt,yscale=-1,xscale=1]
\tikzset{ma/.style={decoration={markings,mark=at position 0.5 with {\arrow[scale=0.5]{>}}},postaction={decorate}}}
\tikzset{mar/.style={decoration={markings,mark=at position 0.5 with {\arrowreversed[scale=0.5]{>}}},postaction={decorate}}}
\draw[white, fill=gray!3  ,fill opacity=1] (84.3,145.68) .. controls (84.3,134.3) and (93.52,125.08) .. (104.9,125.08) -- (305.7,125.08) .. controls (317.08,125.08) and (326.3,134.3) .. (326.3,145.68) -- (326.3,207.48) .. controls (326.3,218.86) and (317.08,228.08) .. (305.7,228.08) -- (104.9,228.08) .. controls (93.52,228.08) and (84.3,218.86) .. (84.3,207.48) -- cycle ;
\draw  [fill=gray!30  ,fill opacity=1 ][thick]  (255.63,138.86) -- (291.01,138.86) -- (291.01,151.24) -- (255.63,151.24) -- cycle ;
\draw  [fill=gray!5  ,fill opacity=1 ][very thick]  (234.4,145.68) .. controls (234.4,138.7) and (240.06,133.04) .. (247.03,133.04) .. controls (254.01,133.04) and (259.67,138.7) .. (259.67,145.68) .. controls (259.67,152.66) and (254.01,158.31) .. (247.03,158.31) .. controls (240.06,158.31) and (234.4,152.66) .. (234.4,145.68) -- cycle ;
\draw[ma][line cap=round]    (239.2,136.11) -- (221,136.28) ;
\draw[ma][line cap=round]    (241.47,156.8) -- (223.28,156.97) ;
\draw  [fill=gray!5  ,fill opacity=1 ][very thick]  (309.71,145.7) .. controls (309.69,152.68) and (304.01,158.32) .. (297.04,158.3) .. controls (290.06,158.28) and (284.42,152.61) .. (284.44,145.63) .. controls (284.46,138.65) and (290.13,133.01) .. (297.11,133.03) .. controls (304.08,133.05) and (309.73,138.72) .. (309.71,145.7) -- cycle ;
\begin{scope}[xshift=-4,yshift=-1]
\draw[ma][line cap=round]    (295.11,156.8) -- (277.47,161.39) ;
\end{scope}
\draw[mar][line cap=round]    (302.66,134.56) -- (320.86,134.44) ;
\draw[mar][line cap=round]    (303.11,156.8) -- (321.3,156.68) ;
\draw  [fill= gray!30 ,fill opacity=1 ][thick]  (154.54,154.16) -- (119.16,154.43) -- (119.07,142.04) -- (154.45,141.78) -- cycle ;
\draw  [fill=gray!5  ,fill opacity=1 ][very thick]  (175.72,147.18) .. controls (175.77,154.16) and (170.16,159.85) .. (163.18,159.91) .. controls (156.2,159.96) and (150.5,154.35) .. (150.45,147.37) .. controls (150.4,140.39) and (156.01,134.69) .. (162.99,134.64) .. controls (169.97,134.58) and (175.67,140.2) .. (175.72,147.18) -- cycle ;
\draw[mar][line cap=round]    (170.99,156.78) -- (189.19,156.48) ;
\draw[mar][line cap=round]    (168.56,136.11) -- (186.75,135.81) ;
\draw  [fill=gray!5  ,fill opacity=1 ][very thick]  (100.41,147.72) .. controls (100.38,140.74) and (106.01,135.06) .. (112.99,135.03) .. controls (119.97,135) and (125.65,140.63) .. (125.68,147.6) .. controls (125.72,154.58) and (120.09,160.27) .. (113.11,160.3) .. controls (106.13,160.33) and (100.45,154.7) .. (100.41,147.72) -- cycle ;
\begin{scope}[xshift=4,yshift=2.5]
\draw[mar][line cap=round]    (115.11,154.3) -- (132.28,160.14) ;
\end{scope}
\draw[ma][line cap=round]    (107.54,158.81) -- (89.35,159.06) ;
\draw[ma][line cap=round]    (106.93,136.57) -- (88.74,136.83) ;
\draw  [fill=gray!5  ,fill opacity=1 ][very thick]  (125.48,210.51) .. controls (125.48,203.42) and (131.23,197.67) .. (138.32,197.67) .. controls (145.41,197.67) and (151.16,203.42) .. (151.16,210.51) .. controls (151.16,217.6) and (145.41,223.35) .. (138.32,223.35) .. controls (131.23,223.35) and (125.48,217.6) .. (125.48,210.51) -- cycle ;
\draw[ma][line cap=round]    (162.97,210.34) -- (151.16,210.51) ;
\draw[ma][line cap=round]    (132.16,199.21) -- (113.67,199.38) ;
\draw[ma][line cap=round]    (132.67,221.81) -- (114.18,221.98) ;
\draw[ma][line cap=round]    (164,199.72) -- (145.51,199.89) ;
\draw[ma][line cap=round]    (164,221.29) -- (145.51,221.46) ;
\draw  [fill=gray!5  ,fill opacity=1 ][very thick]  (258.81,209.84) .. controls (258.81,202.75) and (264.56,197) .. (271.65,197) .. controls (278.74,197) and (284.49,202.75) .. (284.49,209.84) .. controls (284.49,216.93) and (278.74,222.68) .. (271.65,222.68) .. controls (264.56,222.68) and (258.81,216.93) .. (258.81,209.84) -- cycle ;
\draw[mar][line cap=round]    (277.27,220.78) -- (295.76,221.01) ;
\draw[mar][line cap=round]    (277.24,198.18) -- (295.73,198.4) ;
\draw[ma][line cap=round]    (263.33,199.05) -- (244.84,199.23) ;
\draw[ma][line cap=round]    (263.33,220.63) -- (244.84,220.8) ;
\draw[mar][line cap=round]    (245,209.76) -- (258.81,209.84) ;
\draw[<->, line width=0.5]    (140,174) -- (140,191) ;
\draw[<->, line width=0.5]    (240,201) -- (183,163) node[below,midway,yshift=-40]{$(n=5)$};
\draw[<->, line width=0.5]    (271.5,174.5) -- (271.5,191.5) ;

\draw (241.71,140.8) node [anchor=north west][inner sep=0.75pt]  [font=\scriptsize]  {$S$};
\draw (265.8,139.84) node [anchor=north west][inner sep=0.75pt]  [font=\scriptsize]  {$X$};
\draw (291.09,139.0) node [anchor=north west][inner sep=0.75pt]  [font=\scriptsize]  {$S^{\dagger }$};
\draw (157.95,142.18) node [anchor=north west][inner sep=0.75pt]  [font=\scriptsize]  {$S$};
\draw (131.33,143.1) node [anchor=north west][inner sep=0.75pt]  [font=\scriptsize]  {$X$};
\draw (106.1,140.89) node [anchor=north west][inner sep=0.75pt]  [font=\scriptsize]  {$S^{\dagger }$};
\draw (133.16,205.61) node [anchor=north west][inner sep=0.75pt]  [font=\scriptsize]  {$S$};
\draw (266.49,204.94) node [anchor=north west][inner sep=0.75pt]  [font=\scriptsize]  {$S$};

\draw [color=Orange  ,draw opacity=1 ] [dashed]  (271.97,129.35) -- (271.5,165.85) ;
\draw [color=Orange  ,draw opacity=1 ] [dashed]  (137.74,165.85) -- (137.88,126.35) ;  

\end{tikzpicture}}
    \end{subfigure}
\caption{Crossing relations obtained from \eqref{eq:crossing23} for $n=4$ and $n=5$ external legs. For $n=5$ there is a disjoint family involving the complex conjugates of the shown objects.}
\label{fig:web45} 
\end{figure}
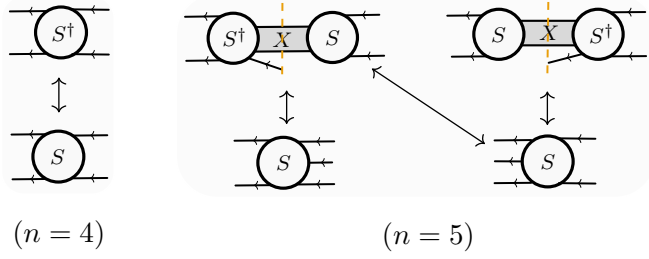

At this stage, we can ask whether every observable can be related to any other using the above crossing rules.
Surprisingly, we find that the answer turns out to be no already for $n = 5$. This is illustrated in the diagrams in Fig.~\ref{fig:web45}, where a pair of observables is joined if and only if it participates in the same crossing relation.

We recall from \cite[Tab.~1]{Caron-Huot:2023vxl} that there are $2$ and $8$ measurement types for $n=4$ and $5$ respectively. Crossing moves for $n=4$ only relate amplitudes to complex-conjugated amplitudes. For $n=5$, there are two separate families with $4$ measurement types each. For example, $S_{2 \ot 3}$ turns out to be connected to $S_{3 \ot 2}$ by a composition of two crossing moves, but $S_{2 \ot 3}$ and $S^\dag_{2 \ot 3}$ are not.

\begin{figure}
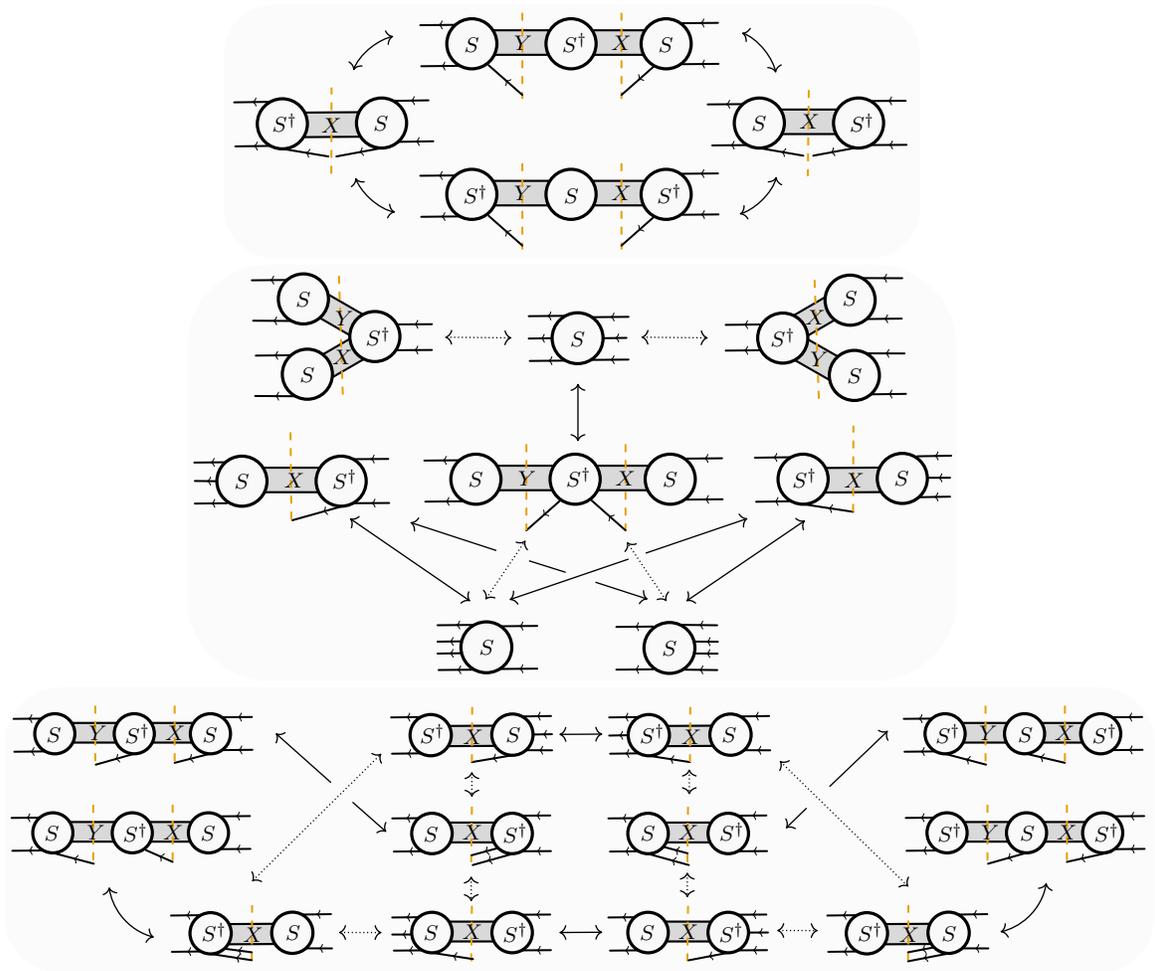

    \centering
    \begin{subfigure}[c]{0.7\textwidth}
        \centering
            \adjustbox{valign=c}{\input{tikz/web4}}
    \end{subfigure}
    \begin{subfigure}[c]{0.76\textwidth}
        \centering
              \adjustbox{valign=c}{\input{tikz/web3}}
    \end{subfigure}
    \begin{subfigure}[c]{0.99\textwidth}
        \centering
            \adjustbox{valign=c}{\input{tikz/web5}}
    \end{subfigure}
    \caption{\label{fig:web6} Webs of relations between asymptotic measurements with $n=6$.
    The solid lines indicate the moves that involve the two-particle relation \eqref{eq:crossing23}, whereas the dashed lines display the three-particle crossing using \eqref{eq:crossing_speculation}.
    The middle family is disjoint from its complex conjugate, while the other two are self-conjugate.}
\end{figure}

For $n=6$ particles, there are 28 measurement types according to
the classification in \cite{Caron-Huot:2023vxl}, but
the more speculative relation \eqref{eq:crossing_speculation} adds four new objects
(the two at the ends of the middle panel top line in Fig.~\ref{fig:web6} and their complex conjugates),
for a total of 32. They split into four families: with $4$, $8$, $8$, and $12$ entries each, as shown in Fig.~\ref{fig:web6}.

Notably, we find that $S_{4 \ot 2}$, $S_{2 \ot 4}$, and $S_{3 \ot 3}$ all lie in the same family, but their complex conjugates do not.  Also, our crossing relations do not seem to relate the
inclusive cross sections (leftmost diagram in the first family) to the amplitude.  We find this somewhat surprising since some relations of this kind have been discussed in the 1970s in the context of Regge theory (see \cite[Ch. 6]{Brower:1974yv}).
We stress that we are solely using the specific analytic continuation encoded by the crossing equations \eqref{eq:crossing23} and its possible extension \eqref{eq:crossing_speculation}, and that
there might exist other ones reconnecting the families.

Finally, let us note that it is also interesting to consider how crossing relates conventional time-ordered amplitudes with different labels
(not just measurement types). For example, one may be interested in relating different channels of the same diagram. This will be discussed and exemplified further near the end of Sec.~\ref{sec:fivepoint}, and we anticipate that such relations between kinematics could be useful in the context of differential equation approaches to Feynman integrals, alleviating the need to separately compute initial conditions in each channel \cite{chicherin:2020oor}.

\paragraph{\bf Crossing a single particle} Note that a single particle cannot be crossed to become its own anti-particle with one deformation because such a deformation would violate momentum conservation. However, we can achieve it by a composition of multiple crossing moves described above. First, notice that because of the symmetry of \eqref{eq:crossing_speculation}, its right-hand side can be alternatively obtained by starting with the process $AD \ot BC$ and crossing $A \leftrightarrow B$ instead. In equations,
\begin{equation}\label{eq:crossing-BC-AB}
    \left[S_{AD \ot BC}\right]_{\rotateup z'} =  - \sumint_{X,Y,Z,W} S_{B \ot Z} S_{D \ot Y} S^\dag_{Y Z \ot X W} S_{W \ot C} S_{X \ot A} \,,
\end{equation}
where the analytic continuation is obtained by deforming the momenta of particles $A$ and $B$ instead of $B$ and $C$ with a complex parameter $z'$, just as in \eqref{eq:crossingpath}. This is a different continuation from that used in \eqref{eq:crossing_speculation_eq}. But by combining these two crossing moves, we obtain a two-leg path of continuation between $S_{CD \ot AB}$ and $S_{AD \ot BC}$. In the $n=4$ case, it is the path described in the original work of BEG \cite{Bros:1965kbd}: it starts from the $s$-channel, passes through the $u$-channel (from the wrong side of branch cuts) in the intermediate step, and then lands on the $t$-channel. Let us introduce the following shorthand notation:
\be
S_{CD \ot AB} \;\xrightarrow[(B,C);\; (A,\bar{B})]{\text{cross}}\; S_{\bar{A}D \ot B\bar{C}}\,,
\ee
for this analytic continuation. For later convenience, we will additionally keep track of which labels correspond to anti-particles by adding bars. For example, $B$ is a particle on the right-hand side because it was crossed an even number of times, but $\bar{A}$ and $\bar{C}$ are anti-particles because they got crossed an odd number of times.

Let us now isolate a single particle $n$ and divide the remaining ones into four non-empty sets $I$, $J$, $K$, $L$, such that we are dealing with the amplitude $S_{KL n \ot IJ}$. We then use the following chain of the above moves \cite{Mizera:2021fap}:
\be\label{eq:crossing-single}
S_{KL n\ot IJ} \;\xrightarrow[(J,KL);\; (I,\bar{J})]{\text{cross}}\;
S_{n\bar{I}\ot J\bar{K}\bar{L}} \;\xrightarrow[(\bar{L},n);\; (J\bar{K},L)]{\text{cross}}\;
S_{\bar{I}\bar{J}K\ot \bar{L}\bar{n}} \;\xrightarrow[(\bar{n},\bar{I}\bar{J});\; (\bar{L},n)]{\text{cross}}\;
S_{KL\ot \bar{n}IJ}\,.
\ee
The scattering amplitude on the right-hand side is $S_{KL \ot \bar{n}IJ}$, which differs from the starting point of the continuation only by the fact that the outgoing particle $n$ is now an incoming anti-particle $\bar{n}$.

For amplitudes with planar ordering $(IJKLn)$, it is known perturbatively that there are no anomalous thresholds in any of the six upper half-planes involved in this deformation, which means that \eqref{eq:crossing-single} can be established rigorously \cite{Mizera:2021fap}. In other cases, \eqref{eq:crossing-single} depends on the validity of the crossing conjecture \eqref{eq:crossing_speculation_eq}.

\section{Tree-level proof of the cluster crossing equation}
\label{sec:treelevelProof}

In this section, we prove the crossing conjecture \eqref{eq:crossing_speculation} for an arbitrary scalar $n$-point tree-level diagram. In our conventions, the amplitude for such a diagram in the channel $CD\ot AB$ is given by
\begin{equation}\label{eq:generalTL}
    i\cM_{CD \ot AB} = (-i g)^{|\mathcal{E}|+1}\prod_{I\in\mathcal{E}} \frac{-i}{-s_{I}+m_I^2-i \varepsilon} \,,
\end{equation}
where the product runs over the set $\mathcal{E}$ of all internal edges in the diagram, and $I$ denotes the set of external momenta which enter that edge of the tree. To avoid clutter, we have taken the coupling constant for every vertex to be the same (labeled with $g$); the generalization to arbitrary coupling constants is trivial. Paralleling the discussion in Sec.~\ref{sec:conjectureII}, a similar proof can be constructed diagram-by-diagram for non-scalar scattering amplitudes after gauge fixing.

Of course, since tree amplitudes are rational functions of momenta made by multiplying simple ingredients (propagators and vertices), it is obvious that the expressions in different kinematics all admit closely related expressions. The non-trivial thing we will check is that the $i\varepsilon$'s work out precisely as predicted by \eqref{eq:crossing_speculation}.

Applying the kinematic deformation from Sec.~\ref{sec:conjectureI}, the set of Mandelstam invariants $s_I = -p_I^2$ appearing in a given diagram partitions into those that do depend on the deformation parameter $z$ and those that do not. Let us call the two sets $\Z$ and $\N$ respectively. The set $\Z$ consists of those $s_I$ such that $I$ and its complement $\bar{I}$ both contain at least one label from the crossed sets $B,C$ and at least one from the fixed set $A,D$. In those situations, $p_I^+ \sim z$ and $p_I^- \sim 1$, such that the corresponding Mandelstam invariant is deformed as $s_I \sim z$. 
The set $\N$ contains all remaining invariants, which are $z$-independent since they have either $p_I^\pm \sim z^{\pm 1}$ or $p_I^\pm \sim 1$. 

The analytic continuation illustrated in \eqref{eq:crossingpath3} has the effect that all $s_{I \in \Z}$ either start spacelike and rotate to become timelike with wrong-sign $-i\varepsilon$, or start timelike and rotate to become spacelike in which case their $i\varepsilon$ is no longer needed. The tree-level amplitude \eqref{eq:generalTL} therefore crosses to
\begin{equation}
    \big[ i\cM_{CD \ot AB} \big]_{\rotatedown \Z} = (- i g)^{|\mathcal{N}|+1} (i g)^{|\mathcal{Z}|} \prod_{I \in \N} \frac{-i}{-s_{I}+m_I^2-i \varepsilon}
    \prod_{J \in \Z} \frac{i}{-s_{J}+m_J^2+i \varepsilon}\,,
    \label{eq:treerotation}
\end{equation}
where we have denoted with $|\mathcal{N}|$ and $|\mathcal{Z}|$ the total numbers of invariants in the sets $\mathcal{N}$ and $\mathcal{Z}$, respectively. 
To condense this expression, we have not distinguished between spacelike and timelike edges, since $\pm i\varepsilon$ can be freely added to the spacelike ones anyway with no effect.

The goal is now to prove the crossing conjecture~\eqref{eq:crossing_speculation}.
To this end, we have to sum over all possible ways of placing the vertices of the tree into the form~\eqref{eq:crossing_speculation}, and compare the answer with the explicit result obtained in \eqref{eq:treerotation}. Before diving into the full proof, we start with a simple example that illustrates the intuition.

\subsection{Simple example revisited}
Let us circle back to the example from Sec.~\ref{sec:treelevel_intro}, and show how the crossing relation works out for the diagram
\begin{equation}\label{eq:M_12_345v2}
i\cM_{345 \ot 12} \;=\quad
\begin{gathered}
\begin{tikzpicture}[baseline= {($(current bounding box.base)+(10pt,10pt)$)},line width=1, scale=0.8]
\coordinate (a) at (0,0) ;
\coordinate (b) at (1,0) ;
\coordinate (c) at ($(b)+(-40:1)$);
\draw[] (a) -- (b);
\draw[] (b) -- (c);
\draw[Maroon,dash pattern=on 1pt off 0.6pt] (b) -- ++ (30:1) node[right] {\footnotesize$2$};
\draw[Maroon,dash pattern=on 1pt off 0.6pt] (c) -- ++ (-150:1) node[left]{\footnotesize$3$};
\draw[RoyalBlue] (c) -- ++ (-30:1) node[right] {\footnotesize$1$};
\draw[RoyalBlue] (a) -- (-150:1) node[left] {\footnotesize$4$};
\draw[RoyalBlue] (a) -- (150:1) node[left] {\footnotesize$5$};
\fill[black,thick] (a) circle (0.07);
\fill[black,thick] (b) circle (0.07);
\fill[black,thick] (c) circle (0.07);
\end{tikzpicture}
\end{gathered}
=\quad
\frac{-ig^3}{(-s_{45}+m_{45}^2-i\eps)(-s_{13}+m_{13}^2)}\,,
\end{equation}
where we start in the channel $345\ot 12$, and aim to cross particles 2 and 3.

In the notation of this section, $I=\{1,3\}$ and $\bar{I}=\{2,4,5\}$ for $s_{13}$, and the intersections of each with both $A\cup D$ and $B\cup C$ are non-empty. Therefore, the $s_{13}$-channel index belongs to $\Z$. Conversely, for $s_{45}$, we have $I=\{4,5\}$ and $\bar{I}=\{1,2,3\}$, and, in particular, $I\cap(B\cup C)=\varnothing$. Hence, the $s_{45}$-channel index belongs to $\N$.
As we found earlier in \eqref{5pt tree crossing}, the amplitude rotates to the expression
\begin{equation}
 \left[i\cM_{345 \ot 12}\right]_{ \rotatedown s_{13}} =
 \frac{-ig^3}{(-s_{45}+m_{45}^2-i\eps)(-s_{13}+m_{13}^2+i\eps)}\,,\label{treecrossing_proofsec}
\end{equation}
where we have used the rotation from \eqref{eq:crossingpath} and~\eqref{eq:crossingpath3}. In the crossed channel where 1 and 3 are incoming, this tree diagram can be drawn as
\begin{equation}
    \begin{gathered}
    \begin{tikzpicture}[baseline= {($(current bounding box.base)+(10pt,10pt)$)},line width=1.2, scale=0.8]
    \coordinate (a) at (0,0) ;
    \coordinate (b) at (1,0) ;
    \coordinate (c) at ($(b)+(0:1)$);
    \draw[RoyalBlue] (a) -- (b);
    \draw[Maroon,dash pattern=on 1pt off 0.6pt] ($(b)+(90:0.035)$) -- ($(c)+(90:0.035)$);
    \draw[RoyalBlue] ($(b)+(90:-0.035)$) -- ($(c)+(90:-0.035)$);
    \draw[Maroon,dash pattern=on 1pt off 0.6pt] (b) -- ++ (150:1) node[left,yshift=5pt] {\footnotesize$\bar{2}$};
    \draw[Maroon,dash pattern=on 1pt off 0.6pt] (c) -- ++ (30:1) node[right]{\footnotesize$\bar{3}$};
    \draw[RoyalBlue] (c) -- ++ (-30:1) node[right] {\footnotesize$1$};
    \draw[RoyalBlue] (a) -- (-150:1) node[left] {\footnotesize$4$};
    \draw[RoyalBlue] (a) -- (150:1) node[left] {\footnotesize$5$};
    \fill[black,thick] (a) circle (0.07);
    \fill[black,thick] (b) circle (0.07);
    \fill[black,thick] (c) circle (0.07);
\end{tikzpicture}
\end{gathered}
\end{equation}
Here, we have the marked the internal edges that, as $z\to \infty$, have the lightcone components $p_I^+ \sim z$ in red (as well as dashed), and those with $p_I^- \sim 1$ in blue (as well as solid). More precisely, we have
\begin{equation}
p_{13}^+ \sim z, \quad p_{13}^- \sim 1 \qquad\mathrm{and}\qquad p_{45}^+ \sim 1, \quad p_{45}^- \sim 1\, .
\end{equation}
Notice that the edge with both red and blue marking is the only one that rotates since $s_{13} \sim z$.
As we will see later, this color-coding will prove useful in the classification needed for proving the crossing equation for any tree-level diagram.

Before we proceed, let us gain some more intuition by exploring all the ways this diagram can fit into the blob pattern of the crossing equation in \eqref{eq:crossing_speculation}. One placement that is clearly allowed is when all of the vertices are within the $S^\dagger$ blob, which gives the conjugated amplitude in this channel (with an overall minus sign out front):
\begin{equation}
    i\cM_1 = 
    \begin{gathered}
    \begin{tikzpicture}[baseline= {($(current bounding box.base)+(10pt,10pt)$)},line width=1.2, scale=0.8]
    \coordinate (a) at (0,0) ;
    \coordinate (b) at (1,0) ;
    \coordinate (c) at ($(b)+(0:1)$);
    \draw[RoyalBlue] (a) -- (b);
    \draw[Maroon,dash pattern=on 1pt off 0.6pt] ($(b)+(90:0.035)$) -- ($(c)+(90:0.035)$);
    \draw[RoyalBlue] ($(b)+(90:-0.035)$) -- ($(c)+(90:-0.035)$);
    \draw[Maroon,dash pattern=on 1pt off 0.6pt] (b) -- ++ (150:1) node[left,yshift=5pt] {\footnotesize$\bar{2}$};
    \draw[Maroon,dash pattern=on 1pt off 0.6pt] (c) -- ++ (30:1) node[right]{\footnotesize$\bar{3}$};
    \draw[RoyalBlue] (c) -- ++ (-30:1) node[right] {\footnotesize$1$};
    \draw[RoyalBlue] (a) -- (-150:1) node[left] {\footnotesize$4$};
    \draw[RoyalBlue] (a) -- (150:1) node[left] {\footnotesize$5$};
    \fill[black,thick] (a) circle (0.07);
    \fill[black,thick] (b) circle (0.07);
    \fill[black,thick] (c) circle (0.07);
    \draw[dashed,Orange] ($(a)+(180:0.5)+(0,0.7)$) -- ($(a)+(180:0.5)+(0,-0.7)$);
    \draw[dashed,Orange] ($(c)+(0:0.5)+(0,0.7)$) -- ($(c)+(0:0.5)+(0,-0.7)$);
    \end{tikzpicture}
    \end{gathered}
    = \frac{-ig^3}{(-s_{45}+m_{45}^2+i\eps)(-s_{13}+m_{13}^2+i\eps)} \,.
    \label{eq:tree_term1}
\end{equation}
Another obvious one is when the blue propagator is cut, i.e., when particles 4 and 5 combine in one of the $S$-blobs on the left,
\begin{equation}
    i\cM_2 = 
    \begin{gathered}
    \begin{tikzpicture}[baseline= {($(current bounding box.base)+(10pt,10pt)$)},line width=1.2, scale=0.8]
    \coordinate (a) at (0,0) ;
    \coordinate (b) at (1,0) ;
    \coordinate (c) at ($(b)+(0:1)$);
    \draw[RoyalBlue] (a) -- (b);
    \draw[Maroon,dash pattern=on 1pt off 0.6pt] ($(b)+(90:0.035)$) -- ($(c)+(90:0.035)$);
    \draw[RoyalBlue] ($(b)+(90:-0.035)$) -- ($(c)+(90:-0.035)$);
    \draw[Maroon,dash pattern=on 1pt off 0.6pt] (b) -- ++ (150:1) node[left,yshift=5pt] {\footnotesize$\bar{2}$};
    \draw[Maroon,dash pattern=on 1pt off 0.6pt] (c) -- ++ (30:1) node[right]{\footnotesize$\bar{3}$};
    \draw[RoyalBlue] (c) -- ++ (-30:1) node[right] {\footnotesize$1$};
    \draw[RoyalBlue] (a) -- ++ (-150:1) node[left] {\footnotesize$4$};
    \draw[RoyalBlue] (a) -- ++ (150:1) node[left] {\footnotesize$5$};
    \fill[black,thick] (a) circle (0.07);
    \fill[black,thick] (b) circle (0.07);
    \fill[black,thick] (c) circle (0.07);
    \draw[dashed,Orange] ($(c)+(0:0.5)+(0,0.7)$) -- ($(c)+(0:0.5)+(0,-0.7)$);
    \draw[dashed,Orange] ($(a)+(0:0.5)+(0,0.7)$) -- ($(a)+(0:0.5)+(0,-0.7)$);
    \end{tikzpicture}
    \end{gathered}
    = -2 \pi i \delta(-s_{45}+m_{45}^2)\frac{ig^3}{-s_{13}+m_{13}^2+i\eps} \,.
    \label{eq:tree_term2}
\end{equation}
In fact, these two contributions add up to give the correct answer after rotation,
\begin{equation}
    i\cM_1 + i\cM_2 = \frac{-ig^3}{(-s_{45}+m_{45}^2-i\eps)(-s_{13}+m_{13}^2+i\eps)} = \left[i\cM_{345 \ot 12}\right]_{\substack{\rotatedown s_{13}\\ \rotateup s_{12}}} \,.
\end{equation}

Here and below, we use the notation that the factor between the two vertical dashed lines is $S^\dagger$, while the factors outside the dashed lines represent the $S$ factors acting on $A$, $B$, $C$ and $D$ in \eqref{eq:crossing_speculation}.   A cut going through an external line does not affect it, since it is already on-shell.

One can think of other ways of sharing the vertices among the $S$ and $S^\dag$ factors,
for example as
\begin{equation}
    \begin{gathered}
    \begin{tikzpicture}[baseline= {($(current bounding box.base)+(10pt,10pt)$)},line width=1.2, scale=0.8]
    \coordinate (a) at (0,0) ;
    \coordinate (b) at (1,0) ;
    \coordinate (c) at ($(b)+(0:1)$);
    \draw[RoyalBlue] (a) -- (b);
    \draw[Maroon,dash pattern=on 1pt off 0.6pt] ($(b)+(90:0.035)$) -- ($(c)+(90:0.035)$);
    \draw[RoyalBlue] ($(b)+(90:-0.035)$) -- ($(c)+(90:-0.035)$);
    \draw[Maroon,dash pattern=on 1pt off 0.6pt] (b) -- ++ (150:1) node[left,yshift=5pt] {\footnotesize$\bar{2}$};
    \draw[Maroon,dash pattern=on 1pt off 0.6pt] (c) -- ++ (30:1) node[right]{\footnotesize$\bar{3}$};
    \draw[RoyalBlue] (c) -- ++ (-30:1) node[right] {\footnotesize$1$};
    \draw[RoyalBlue] (a) -- (-150:1) node[left] {\footnotesize$4$};
    \draw[RoyalBlue] (a) -- (150:1) node[left] {\footnotesize$5$};
    \fill[black,thick] (a) circle (0.07);
    \fill[black,thick] (b) circle (0.07);
    \fill[black,thick] (c) circle (0.07);
    \draw[dashed,Orange] ($(c)+(180:0.5)+(0,0.7)$) -- ($(c)+(180:0.5)+(0,-0.7)$);
    \draw[dashed,Orange] ($(a)+(0:0.5)+(0,0.7)$) -- ($(a)+(0:0.5)+(0,-0.7)$);
    \end{tikzpicture}
    \end{gathered}
    \hspace{0.7cm}
    \begin{gathered}
    \begin{tikzpicture}[baseline= {($(current bounding box.base)+(10pt,10pt)$)},line width=1.2, scale=0.8]
    \coordinate (a) at (1.5,-0.5);
    \coordinate (b) at (1,0) ;
    \coordinate (c) at ($(b)+(0:1)$);
    \draw[RoyalBlue] (a) -- (b);
    \draw[Maroon,dash pattern=on 1pt off 0.6pt] ($(b)+(90:0.035)$) -- ($(c)+(90:0.035)$);
    \draw[RoyalBlue] ($(b)+(90:-0.035)$) -- ($(c)+(90:-0.035)$);
    \draw[Maroon,dash pattern=on 1pt off 0.6pt] (b) -- ++ (150:1) node[left,yshift=5pt] {\footnotesize$\bar{2}$};
    \draw[Maroon,dash pattern=on 1pt off 0.6pt] (c) -- ++ (30:1) node[right]{\footnotesize$\bar{3}$};
    \draw[RoyalBlue] (c) -- ++ (-30:1) node[right] {\footnotesize$1$};
    \draw[RoyalBlue] (a) -- ++ (-170:1.4) node[left] {\footnotesize$4$};
    \draw[RoyalBlue] (a) -- ++ (170:1.4) node[left] {\footnotesize$5$};
    \fill[black,thick] (a) circle (0.07);
    \fill[black,thick] (b) circle (0.07);
    \fill[black,thick] (c) circle (0.07);
    \draw[dashed,Orange] ($(c)+(0:0.5)+(0,0.7)$) -- ($(c)+(0:0.5)+(0,-0.7)$);
    \draw[dashed,Orange] ($(b)+(0:0.25)+(0,0.7)$) -- ($(b)+(0:0.25)+(0,-0.7)$);
    \end{tikzpicture}
    \end{gathered}
    \label{eq:tree_terms_notallowed}
\end{equation}
However, the left diagram is not allowed by the factorization of the blobs in \eqref{eq:crossing_speculation} (the $S$ factors must separately act within $A$, $B$, $C$ and $D$, and consequently, the rightmost vertex must be on the left of the first cut). The right diagram vanishes by conservation of momentum flow in the $p^-$ components. 

In fact, there is yet another reason why neither of them is allowed: the red-blue line can actually never be cut, as we will show in Sec.~\ref{sec:noRBcuts}. One can check that no placements of vertices are allowed apart from the ones spelled out in \eqref{eq:tree_term1} and~\eqref{eq:tree_term2}. We have therefore proven that the crossing equation~\eqref{eq:crossing_speculation_eq}  holds for this diagram.

\subsection{Classification into red and blue edges}

Armed with the insight from this simple example, we now turn to proving crossing for any tree diagram. As we just saw, it is useful to look at momentum conservation for the $p^+$ and $p^-$ components separately, both to determine which edges can potentially go on-shell, and also to exclude diagrams that are not allowed by momentum conservation in each component separately. We therefore use the same red and blue labeling as in the previous example, which gives us a simple way to exclude many of the cut diagrams that naively seem to fit into \eqref{eq:crossing_speculation}.

As before, for any tree diagram, let us label the edges that have $p_I^+ \sim z$ in red, $p_I^- \sim 1$ in blue, as $z \to \infty$. Any given edge might be red (dashed), blue (solid), red-blue, or remain uncolored (black/dotted) if its momentum does not satisfy the aforementioned scaling. In the absence of a black edge, an example diagram looks like
\begin{equation}
    \begin{gathered}
    \begin{tikzpicture}[line width=1.2, scale=0.8]
    \coordinate (a) at (0,0);
    \coordinate (b) at (1,0);
    \coordinate (c) at (2,0);
    \coordinate (d) at ($(c)+(-150:1)$);
    \coordinate (e) at ($(a)+(-150:1)$);
    \coordinate (k) at ($(a)+(180:1)$);
    \coordinate (f) at ($(k)+(150:1)$);
    \coordinate (g) at ($(f)+(150:1)$);
    \coordinate (h) at ($(f)+(30:1)$);
    \coordinate (i) at ($(f)+(-150:2)$);
    \coordinate (j) at ($(f)+(-150:1)$);
    \coordinate (l) at ($(b)+(50:1)$);
    \coordinate (m) at ($(l)+(30:1)$);
    \coordinate (o) at ($(j)+(150:1)$);
    \doubleline{a}{b}{90};
    \doubleline{b}{c}{90};
    \draw[Maroon,dash pattern=on 1pt off 0.6pt] (c) -- (d);
    \draw[RoyalBlue] (a) -- (e);
    \doubleline{a}{k}{90};
    \doubleline{k}{f}{30};
    \draw[RoyalBlue] (k) -- ++ (30:1);
    \draw[Maroon,dash pattern=on 1pt off 0.6pt] (f) -- (g);
    \draw[Maroon,dash pattern=on 1pt off 0.6pt] (f) -- (h);
    \doubleline{f}{i}{150};
    \draw[Maroon,dash pattern=on 1pt off 0.6pt] (i) -- ++ (-160:1);
    \draw[RoyalBlue] (i) -- ++ (160:1);
    \draw[Maroon,dash pattern=on 1pt off 0.6pt] (i) -- ++ (-30:1);
    \doubleline{j}{o}{30};
    \draw[RoyalBlue] (o) -- ++ (160:1);
    \draw[Maroon,dash pattern=on 1pt off 0.6pt] (o) -- ++ (-160:1);
    \draw[RoyalBlue] (c) -- ++ (30:1);
    \draw[RoyalBlue] (c) -- ++ (0:1);
    \draw[Maroon,dash pattern=on 1pt off 0.6pt] (c) -- ++ (-30:1);
    \draw[Maroon,dash pattern=on 1pt off 0.6pt] (d) -- ++ (-30:1);
    \draw[Maroon,dash pattern=on 1pt off 0.6pt] (d) -- ++ (-150:1);
    \draw[RoyalBlue] (e) -- ++ (160:1);
    \draw[RoyalBlue] (e) -- ++ (-160:1);
    \doubleline{l}{m}{150};
    \doubleline{b}{l}{130};
    \draw[Maroon,dash pattern=on 1pt off 0.6pt] (l) -- ++ (150:1);
    \draw[RoyalBlue] (m) -- ++ (20:1);
    \draw[Maroon,dash pattern=on 1pt off 0.6pt] (m) -- ++ (-20:1);
    \draw[Maroon,dash pattern=on 1pt off 0.6pt] (h) -- ++ (20:1);
    \draw[Maroon,dash pattern=on 1pt off 0.6pt] (h) -- ++ (-20:1);
    \foreach \n in {a,b,c,d,e,f,h,i,j,k,l,m,o}
    {\fill[black,thick] (\n) circle (0.07);}
    \end{tikzpicture}
    \end{gathered}
    \label{eq:tree_generic}
\end{equation}
We notice some additional features of this classification. First, the set of red edges $\R$ must be connected by itself thanks to momentum conservation. It also needs to be connected to all external particles in the set $B\cup C$.
Likewise, the set of blue edges $\B$ is a tree connected to $A\cup D$. We have therefore established that the original tree diagram contains a connected red tree $\R$, and a connected blue tree $\B$:
\begin{equation}
    \begin{gathered}
    \R = 
    \begin{tikzpicture}[line width=1.2, scale=0.6,baseline= {($(current bounding box.base)-(2pt,2pt)$)}]
    \coordinate (a) at (0,0);
    \coordinate (b) at (1,0);
    \coordinate (c) at (2,0);
    \coordinate (d) at ($(c)+(-150:1)$);
    \coordinate (e) at ($(a)+(-150:1)$);
    \coordinate (k) at ($(a)+(180:1)$);
    \coordinate (f) at ($(k)+(150:1)$);
    \coordinate (g) at ($(f)+(150:1)$);
    \coordinate (h) at ($(f)+(30:1)$);
    \coordinate (i) at ($(f)+(-150:2)$);
    \coordinate (j) at ($(f)+(-150:1)$);
    \coordinate (l) at ($(b)+(50:1)$);
    \coordinate (m) at ($(l)+(30:1)$);
    \coordinate (o) at ($(j)+(150:1)$);
    \doublelineR{a}{b}{90};
    \doublelineR{b}{c}{90};
    \draw[Maroon,dash pattern=on 1pt off 0.6pt] (c) -- (d);
    \doublelineR{a}{k}{90};
    \doublelineR{k}{f}{30};
    \draw[Maroon,dash pattern=on 1pt off 0.6pt] (f) -- (g);
    \draw[Maroon,dash pattern=on 1pt off 0.6pt] (f) -- (h);
    \doublelineR{f}{i}{150};
    \draw[Maroon,dash pattern=on 1pt off 0.6pt] (i) -- ++ (-160:1);
    \draw[Maroon,dash pattern=on 1pt off 0.6pt] (i) -- ++ (-30:1);
    \doublelineR{j}{o}{30};
    \draw[Maroon,dash pattern=on 1pt off 0.6pt] (o) -- ++ (-160:1);
    \draw[Maroon,dash pattern=on 1pt off 0.6pt] (c) -- ++ (-30:1);
    \draw[Maroon,dash pattern=on 1pt off 0.6pt] (d) -- ++ (-30:1);
    \draw[Maroon,dash pattern=on 1pt off 0.6pt] (d) -- ++ (-150:1);
    \doublelineR{l}{m}{150};
    \doublelineR{b}{l}{130};
    \draw[Maroon,dash pattern=on 1pt off 0.6pt] (l) -- ++ (150:1);
    \draw[Maroon,dash pattern=on 1pt off 0.6pt] (m) -- ++ (-20:1);
    \draw[Maroon,dash pattern=on 1pt off 0.6pt] (h) -- ++ (20:1);
    \draw[Maroon,dash pattern=on 1pt off 0.6pt] (h) -- ++ (-20:1);
    \foreach \n in {a,b,c,d,f,h,i,j,k,l,m,o}
    {\fill[Maroon] (\n) circle (0.04);}
    \end{tikzpicture}
    \end{gathered}
    \hspace{0.9cm}
    \begin{gathered}
    \B = 
    \begin{tikzpicture}[line width=1.2, scale=0.6,baseline= {($(current bounding box.base)-(2pt,2pt)$)}]
    \coordinate (a) at (0,0);
    \coordinate (b) at (1,0);
    \coordinate (c) at (2,0);
    \coordinate (d) at ($(c)+(-150:1)$);
    \coordinate (e) at ($(a)+(-150:1)$);
    \coordinate (k) at ($(a)+(180:1)$);
    \coordinate (f) at ($(k)+(150:1)$);
    \coordinate (g) at ($(f)+(150:1)$);
    \coordinate (h) at ($(f)+(30:1)$);
    \coordinate (i) at ($(f)+(-150:2)$);
    \coordinate (j) at ($(f)+(-150:1)$);
    \coordinate (l) at ($(b)+(50:1)$);
    \coordinate (m) at ($(l)+(30:1)$);
    \coordinate (o) at ($(j)+(150:1)$);
    \doublelineB{a}{b}{90};
    \doublelineB{b}{c}{90};
    \draw[RoyalBlue] (a) -- (e);
    \doublelineB{a}{k}{90};
    \doublelineB{k}{f}{30};
    \draw[RoyalBlue] (k) -- ++ (30:1);
    \doublelineB{f}{i}{150};
    \draw[RoyalBlue] (i) -- ++ (160:1);
    \doublelineB{j}{o}{30};
    \draw[RoyalBlue] (o) -- ++ (160:1);
    \draw[RoyalBlue] (c) -- ++ (30:1);
    \draw[RoyalBlue] (c) -- ++ (0:1);
    \draw[RoyalBlue] (e) -- ++ (160:1);
    \draw[RoyalBlue] (e) -- ++ (-160:1);
    \doublelineB{l}{m}{150};
    \doublelineB{b}{l}{130};
    \draw[RoyalBlue] (m) -- ++ (20:1);
    \foreach \n in {a,b,c,f,i,j,k,l,m,o}
    {\fill[RoyalBlue] (\n) circle (0.04);}
    \end{tikzpicture}
    \end{gathered}
    \label{eq:tree_generic_2}
\end{equation}

We can further ask if the red and blue trees are edge-disjoint or not. Let us consider the first case, $\R \cap \B = \varnothing$. It means that there might have been some black edges, but by momentum conservation, they could not be connected to any of the external legs. We conclude that there could be at most \emph{one} such black edge. Let us call it $e$. Moreover, it needs to be spacelike at large $z$, since the momentum flowing into it has $p_e^+ \sim 1$ and $p_e^- \sim z^{-1}$.

The remaining possibility is that $\R \cap \B \neq \varnothing$ with the intersection $\R \cap \B$ consisting of all the red-blue edges. The set of red-blue edges needs to be a tree by itself. If it was not, it would have at least two disconnected components. Such components could then be connected by a combination of a purely-red path and a purely-blue path, which would necessarily form a loop:
\begin{equation}
    \begin{gathered}
    \begin{tikzpicture}[line width=1.2, scale=0.8]
    \coordinate (a) at (0,0);
    \coordinate (b) at (1,0);
    \coordinate (c) at (2,0);
    \coordinate (d) at ($(c)+(-150:1)$);
    \coordinate (e) at ($(a)+(-150:1)$);
    \coordinate (k) at ($(a)+(180:1)$);
    \coordinate (f) at ($(k)+(150:1)$);
    \coordinate (g) at ($(f)+(150:1)$);
    \coordinate (h) at ($(f)+(30:1)$);
    \coordinate (i) at ($(f)+(-150:2)$);
    \coordinate (j) at ($(f)+(-150:1)$);
    \coordinate (l) at ($(b)+(50:1)$);
    \coordinate (m) at ($(l)+(30:1)$);
    \coordinate (o) at ($(j)+(150:1)$);
    \doubleline{a}{b}{90};
    \doubleline{b}{c}{90};
    \draw[Maroon,dash pattern=on 1pt off 0.6pt] (a) to [in=60,out=120] (k);
    \draw[RoyalBlue] (a) to [in=-60,out=-120] (k);
    \doubleline{k}{f}{30};
    \doubleline{f}{i}{150};
    \doubleline{j}{o}{30};
    \doubleline{l}{m}{150};
    \doubleline{b}{l}{130};
    \foreach \n in {a,b,c,f,i,j,k,l,m,o}
    {\fill[black,thick] (\n) circle (0.07);}
    \end{tikzpicture}
    \end{gathered}
    \label{eq:tree_loop}
\end{equation}
Therefore, since the diagram we started with is a tree, the number of connected components of $\R \cap \B$ needs to be one, i.e., it has to be a tree by itself.

The two possibilities of how the red and blue trees can be overlapping are therefore:
\begin{equation}
    \begin{gathered}
    \R \cap \B = \varnothing:
    \adjustbox{valign=c}{
    \begin{tikzpicture}[line width=1.2, scale=0.7,baseline= {($(current bounding box.base)-(12pt,12pt)$)}]
    \begin{scope}[xshift=0,yshift=0pt,scale=0.4]
    \coordinate (a) at (0,0);
    \coordinate (b) at (1,0);
    \coordinate (c) at (2,0);
    \coordinate (d) at ($(c)+(-150:1)$);
    \coordinate (e) at ($(a)+(-150:1)$);
    \coordinate (k) at ($(a)+(180:1)$);
    \coordinate (f) at ($(k)+(150:1)$);
    \coordinate (g) at ($(f)+(150:1)$);
    \coordinate (h) at ($(f)+(30:1)$);
    \coordinate (i) at ($(f)+(-150:2)$);
    \coordinate (j) at ($(f)+(-150:1)$);
    \coordinate (l) at ($(b)+(50:1)$);
    \coordinate (m) at ($(l)+(30:1)$);
    \coordinate (o) at ($(j)+(150:1)$);
    \doublelineR{a}{b}{90};
    \doublelineR{b}{c}{90};
    \draw[Maroon,dash pattern=on 1pt off 0.6pt] (c) -- (d);
    \doublelineR{a}{k}{90};
    \doublelineR{k}{f}{30};
    \draw[Maroon,dash pattern=on 1pt off 0.6pt] (f) -- (g);
    \draw[Maroon,dash pattern=on 1pt off 0.6pt] (f) -- (h);
    \doublelineR{f}{i}{150};
    \draw[Maroon,dash pattern=on 1pt off 0.6pt] (i) -- ++ (-160:1);
    \draw[Maroon,dash pattern=on 1pt off 0.6pt] (i) -- ++ (-30:1);
    \doublelineR{j}{o}{30};
    \draw[Maroon,dash pattern=on 1pt off 0.6pt] (o) -- ++ (-160:1);
    \draw[Maroon,dash pattern=on 1pt off 0.6pt] (c) -- ++ (-30:1);
    \draw[Maroon,dash pattern=on 1pt off 0.6pt] (d) -- ++ (-30:1);
    \draw[Maroon,dash pattern=on 1pt off 0.6pt] (d) -- ++ (-150:1);
    \doublelineR{l}{m}{150};
    \doublelineR{b}{l}{130};
    \draw[Maroon,dash pattern=on 1pt off 0.6pt] (l) -- ++ (150:1);
    \draw[Maroon,dash pattern=on 1pt off 0.6pt] (m) -- ++ (-20:1);
    \draw[Maroon,dash pattern=on 1pt off 0.6pt] (h) -- ++ (20:1);
    \draw[Maroon,dash pattern=on 1pt off 0.6pt] (h) -- ++ (-20:1);
    \foreach \n in {a,b,c,d,f,h,i,j,k,l,m,o}
    {\fill[Maroon] (\n) circle (0.08);}
    \end{scope}
    \begin{scope}[shift={(0,-1)},scale=0.4]
    \coordinate (a) at (0,0);
    \coordinate (b) at (1,0);
    \coordinate (c) at (2,0);
    \coordinate (d) at ($(c)+(-150:1)$);
    \coordinate (e) at ($(a)+(-150:1)$);
    \coordinate (k) at ($(a)+(180:1)$);
    \coordinate (f) at ($(k)+(150:1)$);
    \coordinate (g) at ($(f)+(150:1)$);
    \coordinate (h) at ($(f)+(30:1)$);
    \coordinate (i) at ($(f)+(-150:2)$);
    \coordinate (j) at ($(f)+(-150:1)$);
    \coordinate (l) at ($(b)+(50:1)$);
    \coordinate (m) at ($(l)+(30:1)$);
    \coordinate (o) at ($(j)+(150:1)$);
    \doublelineB{a}{b}{90};
    \doublelineB{b}{c}{90};
    \draw[RoyalBlue] (a) -- (e);
    \doublelineB{a}{k}{90};
    \doublelineB{k}{f}{30};
    \draw[RoyalBlue] (k) -- ++ (30:1);
    \doublelineB{f}{i}{150};
    \draw[RoyalBlue] (i) -- ++ (160:1);
    \doublelineB{j}{o}{30};
    \draw[RoyalBlue] (o) -- ++ (160:1);
    \draw[RoyalBlue] (c) -- ++ (30:1);
    \draw[RoyalBlue] (c) -- ++ (0:1);
    \draw[RoyalBlue] (e) -- ++ (160:1);
    \draw[RoyalBlue] (e) -- ++ (-160:1);
    \doublelineB{l}{m}{150};
    \doublelineB{b}{l}{130};
    \draw[RoyalBlue] (m) -- ++ (20:1);
    \foreach \n in {a,b,c,f,i,j,k,l,m,o}
    {\fill[RoyalBlue] (\n) circle (0.08);}
    \end{scope}
    \draw[dotted] (0,0) -- (0,-1) node[left,midway]{$e$};
    \node[color=gray!100] at (0,-1.7) {\footnotesize $(-p_e^2<0)$} ;
    \end{tikzpicture}
    }
    \end{gathered}
    \hspace{1.5cm}
    \R \cap \B \neq \varnothing:
    \begin{gathered}
    \adjustbox{valign=c}{
    \begin{tikzpicture}[line width=1.2, scale=0.7,baseline= {($(current bounding box.base)-(2pt,2pt)$)}]
    \begin{scope}[shift={(0,0.06)},scale=0.4]
    \coordinate (a) at (0,0);
    \coordinate (b) at (1,0);
    \coordinate (c) at (2,0);
    \coordinate (d) at ($(c)+(-150:1)$);
    \coordinate (e) at ($(a)+(-150:1)$);
    \coordinate (k) at ($(a)+(180:1)$);
    \coordinate (f) at ($(k)+(150:1)$);
    \coordinate (g) at ($(f)+(150:1)$);
    \coordinate (h) at ($(f)+(30:1)$);
    \coordinate (i) at ($(f)+(-150:2)$);
    \coordinate (j) at ($(f)+(-150:1)$);
    \coordinate (l) at ($(b)+(50:1)$);
    \coordinate (m) at ($(l)+(30:1)$);
    \coordinate (o) at ($(j)+(150:1)$);
    \doublelineR{a}{b}{90};
    \doublelineR{b}{c}{90};
    \draw[Maroon,dash pattern=on 1pt off 0.6pt] (c) -- (d);
    \doublelineR{a}{k}{90};
    \doublelineR{k}{f}{30};
    \draw[Maroon,dash pattern=on 1pt off 0.6pt] (f) -- (g);
    \draw[Maroon,dash pattern=on 1pt off 0.6pt] (f) -- (h);
    \doublelineR{f}{i}{150};
    \draw[Maroon,dash pattern=on 1pt off 0.6pt] (i) -- ++ (-160:1);
    \draw[Maroon,dash pattern=on 1pt off 0.6pt] (i) -- ++ (-30:1);
    \doublelineR{j}{o}{30};
    \draw[Maroon,dash pattern=on 1pt off 0.6pt] (o) -- ++ (-160:1);
    \draw[Maroon,dash pattern=on 1pt off 0.6pt] (c) -- ++ (-30:1);
    \draw[Maroon,dash pattern=on 1pt off 0.6pt] (d) -- ++ (-30:1);
    \draw[Maroon,dash pattern=on 1pt off 0.6pt] (d) -- ++ (-150:1);
    \doublelineR{l}{m}{150};
    \doublelineR{b}{l}{130};
    \draw[Maroon,dash pattern=on 1pt off 0.6pt] (l) -- ++ (150:1);
    \draw[Maroon,dash pattern=on 1pt off 0.6pt] (m) -- ++ (-20:1);
    \draw[Maroon,dash pattern=on 1pt off 0.6pt] (h) -- ++ (20:1);
    \draw[Maroon,dash pattern=on 1pt off 0.6pt] (h) -- ++ (-20:1);
    \foreach \n in {a,b,c,d,f,h,i,j,k,l,m,o}
    {\fill[Maroon] (\n) circle (0.08);}
    \end{scope}
    \begin{scope}[shift={(0,0)},scale=0.4]
    \coordinate (a) at (0,0);
    \coordinate (b) at (1,0);
    \coordinate (c) at (2,0);
    \coordinate (d) at ($(c)+(-150:1)$);
    \coordinate (e) at ($(a)+(-150:1)$);
    \coordinate (k) at ($(a)+(180:1)$);
    \coordinate (f) at ($(k)+(150:1)$);
    \coordinate (g) at ($(f)+(150:1)$);
    \coordinate (h) at ($(f)+(30:1)$);
    \coordinate (i) at ($(f)+(-150:2)$);
    \coordinate (j) at ($(f)+(-150:1)$);
    \coordinate (l) at ($(b)+(50:1)$);
    \coordinate (m) at ($(l)+(30:1)$);
    \coordinate (o) at ($(j)+(150:1)$);
    \doublelineB{a}{b}{90};
    \doublelineB{b}{c}{90};
    \draw[RoyalBlue] (a) -- (e);
    \doublelineB{a}{k}{90};
    \doublelineB{k}{f}{30};
    \draw[RoyalBlue] (k) -- ++ (30:1);
    \doublelineB{f}{i}{150};
    \draw[RoyalBlue] (i) -- ++ (160:1);
    \doublelineB{j}{o}{30};
    \draw[RoyalBlue] (o) -- ++ (160:1);
    \draw[RoyalBlue] (c) -- ++ (30:1);
    \draw[RoyalBlue] (c) -- ++ (0:1);
    \draw[RoyalBlue] (e) -- ++ (160:1);
    \draw[RoyalBlue] (e) -- ++ (-160:1);
    \doublelineB{l}{m}{150};
    \doublelineB{b}{l}{130};
    \draw[RoyalBlue] (m) -- ++ (20:1);
    \foreach \n in {a,b,c,f,i,j,k,l,m,o}
    {\fill[RoyalBlue] (\n) circle (0.08);}
    \end{scope}
    \end{tikzpicture}
    }
    \end{gathered}
    \label{eq:tree_generic_3}
\end{equation}
In the rest of this section, we focus on the second case, since the first one will be entirely analogous.

\subsection{Absence of cuts through red-blue edges}
\label{sec:noRBcuts}
Next, we would like to show that cuts through red-blue edges are always disallowed by kinematic considerations. As the simplest non-trivial example, consider the following diagram:
\tikzset{->-/.style={decoration={
  markings,
  mark=at position #1 with {\arrow{>}}},postaction={decorate}}}
\tikzset{-<-/.style={decoration={
  markings,
  mark=at position #1 with {\arrow{<}}},postaction={decorate}}}
\begin{equation}
    \begin{tikzpicture}[baseline= {($(current bounding box.base)+(5pt,5pt)$)},line width=1.2, scale=0.8]
    \coordinate (a) at (0,0) ;
    \coordinate (b) at (3,0) ;
    \draw[RoyalBlue] (a) -- ++ (-150:1) node[left] {$d$};
    \draw[Maroon,->-=.75,dash pattern=on 1pt off 0.6pt] (a) -- ++ (150:1) node[left] {$c$};
    \draw[Maroon,->-=.75,dash pattern=on 1pt off 0.6pt] (a) -- ++ (180:1) node[left] {$c$};
    \draw[Maroon,-<-=.75,dash pattern=on 1pt off 0.6pt] (a) -- ++ (30:1) node[right] {$b$};
    \draw[Maroon,-<-=.75,dash pattern=on 1pt off 0.6pt] (b) -- ++ (30:1) node[right] {$b$};
    \draw[Maroon,-<-=.75,dash pattern=on 1pt off 0.6pt] (b) -- ++ (0:1) node[right] {$b$};
    \draw[Maroon,->-=.75,dash pattern=on 1pt off 0.6pt] (b) -- ++ (150:1) node[above] {$c$};
    \draw[RoyalBlue] (b) -- ++ (-30:1) node[right] {$a$};
    \doubleline{b}{a}{50};
    \fill[black,thick] (a) circle (0.05);
    \fill[black,thick] (b) circle (0.05);
    \end{tikzpicture}
\end{equation}
Here, $a \in A$, $b \in B$, etc.
Denoting the momentum running through the red-blue edge as $p$, and setting $s=-p^2$ as the kinematic invariant and $m$ as the mass of the edge, the amplitude is $i \mathcal{M} = \frac{-i (-i g)^2}{-s+m^2- i\varepsilon}$. Depending on the momenta, $s$ can be either positive or negative, and the energy flow in the red-blue edge can be either positive or negative.

The amplitude after crossing $s$ from spacelike to timelike becomes
\begin{equation}
    [i \mathcal{M}]_{\rotatedown s} = \frac{-i (-ig )^2}{-s+m^2+i\varepsilon}\,.
    \label{eq:simple_tree_fullres}
\end{equation}
To compare this result with the crossing equation, we draw and compute all diagrams that are consistent with~\eqref{eq:crossing_speculation}:
\begin{align}
\begin{gathered}
    \begin{tikzpicture}[line width=1,scale=0.7]
    \begin{scope}[xshift=0]
    \coordinate (a) at (0,0) ;
    \coordinate (b) at (3,0) ;
    \draw[RoyalBlue,-<-=.75] (a) -- ++ (-30:1) node[right] {$a$};
    \draw[Maroon,->-=.75,dash pattern=on 1pt off 0.6pt] (a) -- ++ (-150:1) node[below] {$\bar{b}$};
    \draw[Maroon,->-=.75,dash pattern=on 1pt off 0.6pt] (a) -- ++ (150:1) node[above] {$\bar{b}$};
    \draw[Maroon,-<-=.75,dash pattern=on 1pt off 0.6pt] (a) -- ++ (30:1) node[right] {$\bar{c}$};
    \draw[black!80,->] (1.8,0.2) -- (1.4,0.2);
    \node[black!80] at (1.6,0.6) {$p$};
    \draw[Maroon,-<-=.75,dash pattern=on 1pt off 0.6pt] (b) -- ++ (30:1) node[right] {$\bar{c}$};
    \draw[Maroon,-<-=.75,dash pattern=on 1pt off 0.6pt] (b) -- ++ (-30:1) node[right] {$\bar{c}$};
    \draw[Maroon,->-=.75,dash pattern=on 1pt off 0.6pt] (b) -- ++ (150:1) node[above] {$\bar{b}$};
    \draw[RoyalBlue,->-=.75] (b) -- ++ (-150:1) node[left] {$d$};
    \doubleline{b}{a}{50};
    \fill[black,thick] (a) circle (0.05);
    \fill[black,thick] (b) circle (0.05);
    \draw[orange,dashed] (-0.5,1)--(-0.5,-1);
    \draw[orange,dashed] (3.5,1)--(3.5,-1);
    \end{scope}
    \end{tikzpicture}
    \qquad
    \begin{tikzpicture}[line width=1,scale=0.7]
    \begin{scope}[xshift=200]
    \coordinate (a) at (0,0) ;
    \coordinate (b) at (3,0) ;
    \draw[RoyalBlue,-<-=.75] (a) -- ++ (-30:1) node[right] {$a$};
    \draw[Maroon,->-=.75,dash pattern=on 1pt off 0.6pt] (a) -- ++ (-150:1) node[below] {$\bar{b}$};
    \draw[Maroon,->-=.75,dash pattern=on 1pt off 0.6pt] (a) -- ++ (150:1) node[above] {$\bar{b}$};
    \draw[Maroon,->-=.75,dash pattern=on 1pt off 0.6pt] (a) -- ++ (30:1) node[right] {$\bar{c}$};
    \draw[black!80,->] (1.8,0.2) -- (1.4,0.2);
    \draw[Maroon,->-=.75,dash pattern=on 1pt off 0.6pt] (b) -- ++ (30:1) node[right] {$\bar{c}$};
    \draw[Maroon,->-=.75,dash pattern=on 1pt off 0.6pt] (b) -- ++ (-30:1) node[right] {$\bar{c}$};
    \draw[Maroon,->-=.75,dash pattern=on 1pt off 0.6pt] (b) -- ++ (150:1) node[above] {$\bar{b}$};
    \draw[RoyalBlue,->-=.75] (b) -- ++ (-150:1) node[left] {$d$};
    \doubleline{b}{a}{50};
    \fill[black,thick] (a) circle (0.05);
    \fill[black,thick] (b) circle (0.05);
    \draw[orange,dashed] (-0.5,1)--(-0.5,-1);
    \draw[orange,dashed] (1.5,1)--(1.5,-1);
    \end{scope}
    \end{tikzpicture}
    \end{gathered}
    \\
    \begin{gathered}
    \begin{tikzpicture}[line width=1,scale=0.7]
    \begin{scope}[xshift=400]
    \coordinate (a) at (0,0) ;
    \coordinate (b) at (3,0) ;
    \draw[RoyalBlue,-<-=.75] (a) -- ++ (-30:1) node[right] {$a$};
    \draw[Maroon,->-=.75,dash pattern=on 1pt off 0.6pt] (a) -- ++ (-150:1) node[below] {$\bar{b}$};
    \draw[Maroon,->-=.75,dash pattern=on 1pt off 0.6pt] (a) -- ++ (150:1) node[above] {$\bar{b}$};
    \draw[Maroon,->-=.75,dash pattern=on 1pt off 0.6pt] (a) -- ++ (30:1) node[right] {$\bar{c}$};
    \draw[black!80,->] (1.8,0.2) -- (1.4,0.2);
    \draw[Maroon,->-=.75,dash pattern=on 1pt off 0.6pt] (b) -- ++ (30:1) node[right] {$\bar{c}$};
    \draw[Maroon,->-=.75,dash pattern=on 1pt off 0.6pt] (b) -- ++ (-30:1) node[right] {$\bar{c}$};
    \draw[Maroon,->-=.75,dash pattern=on 1pt off 0.6pt] (b) -- ++ (150:1) node[above] {$\bar{b}$};
    \draw[RoyalBlue,->-=.75] (b) -- ++ (-150:1) node[left] {$d$};
    \doubleline{b}{a}{50};
    \fill[black,thick] (a) circle (0.05);
    \fill[black,thick] (b) circle (0.05);
    \draw[orange,dashed] (1.3,1)--(1.3,-1);
    \draw[orange,dashed] (1.7,1)--(1.7,-1);
    \end{scope}
    \end{tikzpicture}
    \qquad
    \begin{tikzpicture}[line width=1,scale=0.7]
    \begin{scope}[xshift=600]
    \coordinate (a) at (0,0) ;
    \coordinate (b) at (3,0) ;
    \draw[RoyalBlue,-<-=.75] (a) -- ++ (-30:1) node[right] {$a$};
    \draw[Maroon,->-=.75,dash pattern=on 1pt off 0.6pt] (a) -- ++ (-150:1) node[below] {$\bar{b}$};
    \draw[Maroon,->-=.75,dash pattern=on 1pt off 0.6pt] (a) -- ++ (150:1) node[above] {$\bar{b}$};
    \draw[Maroon,->-=.75,dash pattern=on 1pt off 0.6pt] (a) -- ++ (30:1) node[right] {$\bar{c}$};
    \draw[black!80,->] (1.8,0.2) -- (1.4,0.2);
    \draw[Maroon,->-=.75,dash pattern=on 1pt off 0.6pt] (b) -- ++ (30:1) node[right] {$\bar{c}$};
    \draw[Maroon,->-=.75,dash pattern=on 1pt off 0.6pt] (b) -- ++ (-30:1) node[right] {$\bar{c}$};
    \draw[Maroon,->-=.75,dash pattern=on 1pt off 0.6pt] (b) -- ++ (150:1) node[above] {$\bar{b}$};
    \draw[RoyalBlue,->-=.75] (b) -- ++ (-150:1) node[left] {$d$};
    \doubleline{b}{a}{50};
    \fill[black,thick] (a) circle (0.05);
    \fill[black,thick] (b) circle (0.05);
    \draw[orange,dashed] (1.5,1)--(1.5,-1);
    \draw[orange,dashed] (3.5,1)--(3.5,-1);
    \end{scope}
    \end{tikzpicture}
    \end{gathered}
\end{align}
Naively, it looks like the sum of all of these contributions to the crossing conjecture is
\begin{equation}
    [i \mathcal{M}]_{\rotatedown s} \stackrel{?}{=} i (-ig )^2 \left[ \frac{1}{-s+m^2+i\varepsilon} + 2\pi i \delta(s) - 2\pi i \delta(s) + 2\pi i \delta(s)  \right] = \frac{-i (-ig )^2}{-s+m^2-i\varepsilon} \,,
    \label{eq:crossing_wrong}
\end{equation}
which would not agree with the answer in \eqref{eq:simple_tree_fullres}. When writing~\eqref{eq:crossing_wrong}, we have, however, ignored one crucial restriction: the positive-energy flow across the orange cuts. In all of the diagrams above except the first one, the red-blue edge is forced by momentum conservation to have a negative and large $p^-$ component. If $p^+$ is positive, this means that $s$ must be spacelike, and if $p^+$ is negative, then positive energy flow across the cuts cannot be satisfied, since the energy of the edge is proportional to $p^+{+}p^-$. In either case, we conclude that the red-blue edge cannot be cut.

This argument generalizes to bigger diagrams: a spectator from the set $A$ or $C$ cannot attach in any of the $S$ blobs on the left-hand side of \eqref{eq:crossing_speculation}, and vice versa for spectators from $B$ and $D$. Say $\R \cap \B$ was contained in one of the $S$ blobs, for example the top-left one shown in the diagram below:
\begin{align}
    \R \cap \B \neq \varnothing: \hspace{1cm} & 
    \begin{gathered}
    \begin{tikzpicture}[line width=1.2, scale=0.7]
    \begin{scope}[shift={(0,0)},scale=0.3]
    \draw[gray!20,fill=none] (-3.8,0) circle (3.3);
    \node[] at (-3.8,2) {$S$};
    \draw[gray!20,fill=none] (-3.8,-8) circle (3.3);
    \node[] at (-3.8,-6) {$S$};
    \draw[gray!20,fill=none] (11.8,0) circle (3.3);
    \node[] at (11.8,2) {$S$};
    \draw[gray!20,fill=none] (11.8,-8) circle (3.3);
    \node[] at (11.8,-6) {$S$};
    \fill[fill=gray!10,pattern=north east lines,pattern color=black!20] ($(4,-4)+(4,8)$) rectangle ($(4,-4)+(-4,-8)$);
    \coordinate (aa) at (-3.8,-8);
    \draw[RoyalBlue] (aa) -- ++ (150:2);
    \draw[RoyalBlue] (aa) -- ++ (-150:2);
    \draw[RoyalBlue] (aa) to [out=-20,in=-70] (4,-4);
    \coordinate (bb) at (11.8,-8);
    \draw[RoyalBlue] (bb) -- ++ (45:2);
    \draw[RoyalBlue] (bb) -- ++ (-45:2);
    \draw[RoyalBlue] ($(bb) + (0.5,0.5)$) -- ++ (0:2);
    \draw[RoyalBlue] (bb) to [out=160,in=-30] (4.5,-4);
\coordinate (cc) at (11.8,0);
    \draw[Maroon, dash pattern=on 1pt off 0.6pt] (cc) -- ++ (60:2);
    \draw[Maroon, dash pattern=on 1pt off 0.6pt] (cc) -- ++ (0:2);
    \draw[Maroon, dash pattern=on 1pt off 0.6pt] ($(cc)+(1,0)$) -- ++ (-60:2);
    \draw[Maroon, dash pattern=on 1pt off 0.6pt] (cc) to [out=180,in=40] (4,-4);
    \coordinate (a) at (0,0);
    \coordinate (b) at (1,0);
    \coordinate (c) at (2,0);
    \coordinate (d) at ($(c)+(-150:1)$);
    \coordinate (e) at ($(a)+(-150:1)$);
    \coordinate (k) at ($(a)+(180:1)$);
    \coordinate (f) at ($(k)+(150:1)-(0.5,0)$);
    \coordinate (g) at ($(f)+(150:1)$);
    \coordinate (h) at ($(f)+(30:1)$);
    \coordinate (hp) at ($(f)+(30:0.9)$);
    \coordinate (i) at ($(f)+(-150:2)$);
    \coordinate (j) at ($(f)+(-150:1)$);
    \coordinate (l) at ($(b)+(50:1)$);
    \coordinate (m) at ($(l)+(30:1)$);
    \coordinate (o) at ($(j)+(150:1)$);
    \draw[Maroon, dash pattern=on 1pt off 0.6pt] (hp) -- ++ (150:1);
    \doublelineR{f}{i}{150};
    \draw[Maroon, dash pattern=on 1pt off 0.6pt] (o) -- ++ (160:1);
    \draw[Maroon, dash pattern=on 1pt off 0.6pt] (i) -- ++ (-160:1);
    \draw[Maroon, dash pattern=on 1pt off 0.6pt] (i) -- ++ (-30:1);
    \doublelineR{j}{o}{30};
    \draw[Maroon, dash pattern=on 1pt off 0.6pt] (o) -- ++ (-160:1);
    \doublelineR{$(h)-(0.3,0)$}{$(j)-(0.3,0)$}{150};
    \doubleline{h}{j}{150};
    \draw[RoyalBlue, line cap=round] ($(c)+(0,-3)$) to [out=150,in=-70] (j);
    \draw[RoyalBlue, line cap=round] ($(c)+(0.5,-1)$) to [out=130,in=10] (h);
    \draw[dashed,orange] (0,4) -- (0,-12);
    \draw[dashed,orange] (8,4) -- (8,-12);
    \begin{scope}[xshift=0pt,yshift=-350pt]
    \draw[line width=0.4, line cap=round] (-7,0) -- (-7,-0.501) (15,0) -- (15,-0.501) (-7,-0.501) -- (15,-0.501) node[below, midway]{Not allowed};
    \end{scope}    
    \draw[gray!20,fill=gray!20] (4,-4) circle (3.3);
    \node[] at (4,-3.8) {$S^\dagger$}; 
    \end{scope}
        \end{tikzpicture}
    \end{gathered}
    \hspace{2cm}
    \label{eq:tree_generic_4}
\end{align}
Then, there would need to be at least one blue line with energy flowing backwards across the unitarity cut, as shown in the figure, since factorization forbids the red-blue tree to contain only outgoing blue edges. 
This possibility is excluded by momentum conservation in the $p^-$ components. In conclusion, the full red-blue tree $\R \cap \B$ must be fully contained inside the $S^\dagger$ blob. By the same logic, we conclude that if the diagram instead has a black edge, it must be within $S^\dagger$. 

\subsection{Factorization into branches}

To summarize, we now have the following classification of all edges in a general tree diagram: they are either red, blue, red-blue or black. The red-blue ones can never be on-shell, and hence they cannot be cut in the crossing equation. In addition, there can at most be one black edge in the tree, and it must be spacelike by momentum conservation, so it also cannot be cut and must be within $S^\dagger$. Looking back at \eqref{eq:treerotation}, we see that the edges that always remain within the $S^\dagger$ blob in the crossing conjecture \eqref{eq:crossing_speculation}, are precisely the ones in the set $\Z$, and rotate to become past timelike, i.e., with a propagator $\frac{-i}{-s_J+m_J^2+i\varepsilon}$. However, the edges which are either red, blue, or black are in the set $\N$ and do not rotate under crossing. Hence, we arrive at the following classification for a generic tree-level diagram:
\begin{equation}
\mathrm{\textcolor{Maroon}{red}},\, \mathrm{\textcolor{RoyalBlue}{blue}}, \mathrm{black \ edges} \in \N \quad \text{and} \quad \text{\textcolor{Maroon}{red}-\textcolor{RoyalBlue}{blue} edges} \in \Z\, .
\label{eq:RBassignments}
\end{equation}
The sets $\R \cap \B$ and $\Z$ are equal. Thus, the second product in \eqref{eq:treerotation} is accounted for and we are left with explaining how to obtain the first product from the crossing conjecture.

What remains is to figure out how the red and blue edges can fit into the blob pattern \eqref{eq:crossing_speculation}. To this end, let us look at the original diagram after removing all the black and red-blue edges:
\begin{equation}
    \begin{gathered}
    \begin{tikzpicture}[line width=1.2, scale=0.8, circ/.style={line width=0.8, shape=circle, fill=white, inner sep=1pt, draw, node contents=}]
    \coordinate (a) at (0,0);
    \coordinate (b) at (1,0);
    \coordinate (c) at (2,0);
    \coordinate (d) at ($(c)+(-150:1)$);
    \coordinate (e) at ($(a)+(-150:1)$);
    \coordinate (k) at ($(a)+(180:1)$);
    \coordinate (f) at ($(k)+(150:1)$);
    \coordinate (g) at ($(f)+(150:1)$);
    \coordinate (h) at ($(f)+(30:1)$);
    \coordinate (i) at ($(f)+(-150:2)$);
    \coordinate (j) at ($(f)+(-150:1)$);
    \coordinate (l) at ($(b)+(50:1)$);
    \coordinate (m) at ($(l)+(30:1)$);
    \coordinate (o) at ($(j)+(150:1)$);
    \doublelineGRAY{a}{b}{90};
    \doublelineGRAY{b}{c}{90};
    \doublelineGRAY{a}{k}{90};
    \doublelineGRAY{k}{f}{30};
    \doublelineGRAY{f}{i}{150};
    \doublelineGRAY{j}{o}{30};
    \doublelineGRAY{l}{m}{150};
    \doublelineGRAY{b}{l}{130};
    \draw[Maroon,dash pattern=on 1pt off 0.6pt] ($(c)+(-90:0)$) -- ($(d)+(-90:0)$);
    \draw[RoyalBlue] (a) -- (e);
    \draw[RoyalBlue] (k) -- ++ (30:1);
    \draw[Maroon,dash pattern=on 1pt off 0.6pt] (f) -- (g);
    \draw[Maroon,dash pattern=on 1pt off 0.6pt] (f) -- (h);
    \draw[Maroon,dash pattern=on 1pt off 0.6pt] ($(i)+(-90:0)$) -- ++ (-160:1);
    \draw[RoyalBlue] ($(i)+(90:0)$) -- ++ (160:1);
    \draw[Maroon,dash pattern=on 1pt off 0.6pt] ($(i)+(-90:0)$) -- ++ (-30:1);
    \draw[RoyalBlue] ($(o)+(90:0)$) -- ++ (160:1);
    \draw[Maroon,dash pattern=on 1pt off 0.6pt] ($(o)+(90:0)$) -- ++ (-160:1);
    \draw[RoyalBlue] ($(c)+(90:0)$) -- ++ (30:1);
    \draw[RoyalBlue] ($(c)+(90:0)$) -- ++ (0:1);
    \draw[Maroon,dash pattern=on 1pt off 0.6pt] ($(c)+(-90:0)$) -- ++ (-30:1);
    \draw[Maroon,dash pattern=on 1pt off 0.6pt] ($(d)+(-90:0)$) -- ++ (-30:1);
    \draw[Maroon,dash pattern=on 1pt off 0.6pt] ($(d)+(-90:0)$) -- ++ (-150:1);
    \draw[RoyalBlue] (e) -- ++ (160:1);
    \draw[RoyalBlue] (e) -- ++ (-160:1);
    \draw[Maroon,dash pattern=on 1pt off 0.6pt] (l) -- ++ (150:1);
    \draw[RoyalBlue] ($(m)+(90:0)$) -- ++ (20:1);
    \draw[Maroon,dash pattern=on 1pt off 0.6pt] ($(m)+(-90:0)$) -- ++ (-20:1);
    \draw[Maroon,dash pattern=on 1pt off 0.6pt] (h) -- ++ (20:1);
    \draw[Maroon,dash pattern=on 1pt off 0.6pt] (h) -- ++ (-20:1);
    \foreach \n in {d,e,h}
    {\fill[black,thick] (\n) circle (0.07);}
    \foreach \n in {a,f,k,l,o,c,m,i}
    {\draw (\n) node[circ];}
    \end{tikzpicture}
    \end{gathered}
    \label{eq:tree_generic2}
\end{equation}
This procedure introduced a number of vertices (leaves of $\R \cap \B$ or $e$), which we denoted with white circles. Following the previous discussion, white vertices have to belong to $S^\dagger$, while the remaining vertices can be in either of the blobs.

We are going to view each of the disconnected components as a separate tree, for which a momentum $q$ is injected via the white vertex in $S^\dagger$. In fact, we can treat the red and blue parts separately. We call each such component a \textit{branch}. Since the tree diagram factorizes into a part that always stays within $S^\dagger$, as well as a number of branches, the final step of the proof involves proving crossing at the level of each branch separately.

\subsection{Proof of crossing for each branch}
In this last step of the proof of crossing at tree level, we consider a single branch of the tree, where momentum is injected through a single white vertex in $S^\dagger$. Each individual branch is independent of the deformation parameter $z$, and therefore we are left with proving an equality between two quantities, which we will show using unitarity. We can represent the contribution from each branch separately, i.e., the contribution where an off-shell momentum $q$ is injected into a part of the diagram, via a form factor $F(q)$ in this theory. Matrix elements of the form factor between some states $X$ and $Y$ are given by
\begin{equation}
    i F_{Y \ot X}(q) = {}_{\text{out}}\langle Y|i \mathcal{O}_F(q)|X\rangle_{\text{in}} = i \langle Y| S \mathcal{O}_F(q)|X\rangle \,,
    \label{eq:OF}
\end{equation}
where $\mathcal{O}_F$ is a local operator. We remind the reader that the discussion in this section applies diagram by diagram in gauge-fixed perturbation theory.

It is simple to work out how the Mandelstam invariants rotate for a branch: since all the propagators in the diagram belong to either the set $A \cup D$ or $B \cup C$, i.e., the set $\N$, meaning that all of the Mandelstam invariants remain fixed during crossing, and its contribution is equal to:
\begin{equation}
    i F^{\mathrm{branch}}_{D \ot A} = (-i g)^{|\mathcal{E}_{\text{branch}}|+1}\prod_{I \in \text{branch}} \frac{-i}{-s_{I} + m_I^2 - i\varepsilon} \,,
\label{eq:branch_crossing_result}
\end{equation}
where we have labeled the part of the tree amplitude originating from the branch with $F$. Note that since $B$ and $C$ do not appear in this equation, crossing them is trivial. So, to prove the crossing equation for a single branch, we have to show that the sum over all contributions that fit into the blob pattern of~\eqref{eq:crossing_speculation} simply results in $i F^{ \mathrm{branch}}_{D B \ot C A}$, as if the only contribution were that from a single $S$ blob.

To evaluate the prediction of crossing from~\eqref{eq:crossing_speculation}, we consider a blue branch without loss of generality, i.e., one whose external particles belong to either $A$ or $D$. When the sets $B$ and $C$ are crossed, the kinematics of the branch remain unchanged. The multi-particle crossing proposal predicts
\be 
    \left[i F^{\mathrm{branch}}_{D \ot A}\right]_{\rotatedown \Z} = i \sumint_{X,Y} \langle D | S | Y \rangle \langle Y | \mathcal{O}_F (q)  S^\dag | X \rangle \langle X | S | A \rangle \,,
\ee 
where we have used that $i F^\dag_{Y \ot X}(q) = {}_{\text{in}}\langle Y| i \mathcal{O}_F (q)|X\rangle_{\text{out}} = i \langle Y| \mathcal{O}_F(q) S^\dag |X\rangle$. Note that we have incorporated the minus sign from~\eqref{eq:crossing_speculation_eq} into the factors of $i$. This equation can be evaluated straightforwardly with unitarity, using that the sum is over complete sets of states $|X \rangle$ and $|Y \rangle$, to give 
\be 
    \left[i F^{\mathrm{branch}}_{D \ot A}\right]_{\rotatedown \Z} = i \langle D | S \mathcal{O}_F (q) | A \rangle = i F^{\mathrm{branch}}_{D  \ot A} \,.
\ee 
Diagrammatically, we can represent this equation as
\begin{equation}
\begin{gathered}
\begin{tikzpicture}[line width=1, circ/.style={line width=0.8, shape=circle, fill=white, inner sep=1pt, draw, node contents=}]
\def\Ang{30};
\def\CAng{150};
\def\CosVal{0.8660};
\def\SinVal{0.5};
\def\CstVal{0.1};
\def\Ang{30};
\def\CAng{150};
\def\CosVal{0.8660};
\def\SinVal{0.5};
\def\CstVal{0.1};
\begin{scope}[xshift=-150]
\draw[xshift=0] (-0.1,0.05) -- (0.1,0.05);
\draw[xshift=0] (-0.1,-0.05) -- (0.1,-0.05);
\begin{scope}[xshift=-100]
\coordinate (c) at (0,0);
\coordinate (d) at (0,0.4);
\coordinate (e) at (0,0.25);
\draw[RoyalBlue] ($(c) + (\SinVal*\CstVal,\CosVal*\CstVal)$)++(-\Ang:1.85) -- ($(c)+(\SinVal*\CstVal,\CosVal*\CstVal)$);
\draw[RoyalBlue] ($(c) + (-\SinVal*\CstVal,-\CosVal*\CstVal)$)++(-\Ang:1.85) -- ($(c)+(-\SinVal*\CstVal,-\CosVal*\CstVal)$);
\draw[RoyalBlue] ($(c)+(0,0)$)++(-\Ang:1.85) -- ($(c)+(0,0)$);
\draw[RoyalBlue] ($(c) + (\SinVal*\CstVal,-\CosVal*\CstVal)$)++(-\CAng:1.85) -- ($(c)+(\SinVal*\CstVal,-\CosVal*\CstVal)$);
\draw[RoyalBlue] ($(c) + (-\SinVal*\CstVal,\CosVal*\CstVal)$)++(-\CAng:1.85) -- ($(c)+(-\SinVal*\CstVal,\CosVal*\CstVal)$);
\draw[RoyalBlue] ($(c)+(0,0)$)++(-\CAng:1.85) -- ($(c)+(0,0)$);
\filldraw[fill=gray!30,rotate=-30](0,-0.2) rectangle (1,0.2);
\filldraw[fill=gray!30,rotate=-150](0,-0.2) rectangle (1,0.2);
\filldraw[fill=gray!5](0,0) circle (0.4) node {$i F^\dag$};
\filldraw[fill=gray!5]($(0,0)+(-30:1.2)$) circle (0.4) node {$S$};
\filldraw[fill=gray!5]($(0,0)+(-150:1.2)$) circle (0.4) node {$S$};
\draw[Fuchsia,decorate,decoration={snake,amplitude=.4mm,segment length=2mm,post length=0.01mm,pre length=0.01mm}] (e) -- ($(e)+(0,0.7)$) node[above] {\footnotesize$q$};
\draw (e) node[circ];
\node[] at (-0.54,-0.28) {\tiny $Y$};
\node[] at (0.48,-0.28) {\tiny $X$};
\draw [pen colour={gray},
    decorate, 
    decoration = {calligraphic brace,
        raise=5pt,
        amplitude=2pt}] (1.6,-0.55) --  (1.6,-1.2)
node[pos=0.5,right=10pt,RoyalBlue]{$A$};
\draw [pen colour={gray},
    decorate, 
    decoration = {calligraphic brace,
        raise=5pt,
        amplitude=2pt}] (-1.6,-1.2) --  (-1.6,-0.55)
node[pos=0.5,left=10pt,RoyalBlue]{$D$};
\draw[dashed,orange] (0.5,-1) -- (0.5,1);
\draw[dashed,orange] (-0.5,-1) -- (-0.5,1);
\end{scope}
\end{scope}
\def\Ang{30};
\def\CAng{150};
\def\CosVal{0.8660};
\def\SinVal{0.5};
\def\CstVal{0.1};
\begin{scope}[xshift=-80]
\coordinate (c) at (0,0);
\coordinate (d) at (0,0.4);
\coordinate (e) at (0,0.25);
\draw[RoyalBlue] ($(c) + (\SinVal*\CstVal,\CosVal*\CstVal)$)++(-\Ang:1.15) -- ($(c)+(\SinVal*\CstVal,\CosVal*\CstVal)$);
\draw[RoyalBlue] ($(c) + (-\SinVal*\CstVal,-\CosVal*\CstVal)$)++(-\Ang:1.15) -- ($(c)+(-\SinVal*\CstVal,-\CosVal*\CstVal)$);
\draw[RoyalBlue] ($(c)+(0,0)$)++(-\Ang:1.15) -- ($(c)+(0,0)$);
\draw[RoyalBlue] ($(c) + (\SinVal*\CstVal,-\CosVal*\CstVal)$)++(-\CAng:1.15) -- ($(c)+(\SinVal*\CstVal,-\CosVal*\CstVal)$);
\draw[RoyalBlue] ($(c) + (-\SinVal*\CstVal,\CosVal*\CstVal)$)++(-\CAng:1.15) -- ($(c)+(-\SinVal*\CstVal,\CosVal*\CstVal)$);
\draw[RoyalBlue] ($(c)+(0,0)$)++(-\CAng:1.15) -- ($(c)+(0,0)$);
\draw[Fuchsia,decorate,decoration={snake,amplitude=.4mm,segment length=2mm,post length=0.01mm,pre length=0.01mm}] (e) -- ($(d)+(0,0.5)$) node[above] {\footnotesize$q$};
\filldraw[Fuchsia] (e) circle (0.05);
\filldraw[fill=gray!5] (0,0) circle (0.4) node {$iF$};
\draw [pen colour={gray},
    decorate, 
    decoration = {calligraphic brace,
        raise=5pt,
        amplitude=2pt}] (1.0,-0.2) --  (1.0,-0.2-0.65)
node[pos=0.5,right=10pt,RoyalBlue]{$A$};
\draw [pen colour={gray},
    decorate, 
    decoration = {calligraphic brace,
        raise=5pt,
        amplitude=2pt}] (-1.0,-0.2-0.65) -- (-1.0,-0.2)
node[pos=0.5,left=10pt,RoyalBlue]{$D$};
\end{scope}
\begin{scope}[xshift=-40]
    \draw[dashed,orange] (-0.7,-1) -- (-0.7,1);
    \draw[dashed,orange] (-0.5,-1) -- (-0.5,1);
\end{scope}
\end{tikzpicture}
\label{eq:tree_unitarity2}
\end{gathered}
\end{equation}
Comparing this result with~\eqref{eq:branch_crossing_result}, we see that the crossing proposal and the explicit computation agree for the crossing if the tree diagram contains a single branch.

It only remains to make sure that the crossing proposal for multi-particle crossing~\eqref{eq:crossing_speculation} reproduces the correct overall sign in cases where the tree diagram contains multiple branches.
If we note that the crossing equation for branches,~\eqref{eq:tree_unitarity2} does not introduce an extra sign for the white vertex $v$, so we can combine the findings in this section to write the prediction of ~\eqref{eq:crossing_speculation} as
\begin{equation}
    \big[ i\cM_{CD \ot AB} \big]_{\rotatedown \Z} = (- i g)^{|\mathcal{N}|+|\mathcal{Z}|+1} \prod_{I \in \N} \frac{-i}{-s_{I}+m_I^2-i \varepsilon}
    \prod_{J \in \Z} \frac{-i}{-s_{J}+m_J^2+i \varepsilon}\,,
    \label{eq:treerotation2}
\end{equation}
where $\mathcal{N}$ is the set of invariants $s_I$ that appear in this diagram and stay fixed during crossing, while $\mathcal{Z}$ is the set of those that rotate. Note that $|\mathcal{N}|+|\mathcal{Z}|+1$ is the total number of vertices in the diagram. The minus sign in front is the one from the right-hand side of~\eqref{eq:crossing_speculation}. This expression agrees with~\eqref{eq:branch_crossing_result}, thus concluding the proof. 

Before concluding this section, let us make the comment that we could have proven~\eqref{eq:tree_unitarity2} at the level of individual diagrams using the largest-time equation~\cite{tHooft:1973wag,Veltman:1994wz}, which involves summing over all possibilities of the vertices being distributed between $S$ and $F^\dag$. As an example of how such an argument works out, we focus on the simplest diagram of a branch with only one propagator and show how~\eqref{eq:tree_unitarity2} works in perturbation theory. We have to show that there are non-trivial cancelations between the different ways of distributing the vertices of the branch into the crossing equation. We consider a branch with large minus components of all momenta (a blue branch), without loss of generality. To manifest the cancelation between different terms, we add and subtract diagrams for which $q$ attaches in the leftmost $S$-blob as follows:
\begin{subequations}
    \begin{align}
    & \quad\; \begin{tikzpicture}[baseline= {($(current bounding box.base)-(10pt,10pt)$)},line width=1, scale=0.6, circ/.style={line width=0.8, shape=circle, fill=white, inner sep=1pt, draw, node contents=}]
    \coordinate (a) at (0,0);
    \coordinate (e) at ($(a)+(-150:1)$);
    \coordinate (f) at ($(a)+(1,0.5)$);
    \coordinate (g) at ($(a)+(0,0.5)$);
    \draw[dashed,Orange] ($(a)+(0:0.5)+(0,0.7)$) -- ($(a)+(0:0.5)+(0,-1.2)$);
    \draw[dashed,Orange] ($(a)+(0:-1.5)+(0,0.7)$) -- ($(a)+(0:-1.5)+(0,-1.2)$);
    \draw[RoyalBlue] (e) -- ++ (160:1);
    \draw[RoyalBlue] (e) -- ++ (-160:1);
    \draw[RoyalBlue] (a) -- (e);
    \foreach \n in {e}
    {\fill[black,thick] (\n) circle (0.070001);}
    \draw[Fuchsia,decorate,decoration={snake,amplitude=.4mm,segment length=2mm,post length=0.01mm,pre length=0.01mm}] (a) -- (g) node[above] {\footnotesize$q$};
    \foreach \n in {a}
    {\draw (\n) node[circ];};
    \end{tikzpicture}
    \;+\;
    \begin{tikzpicture}[baseline= {($(current bounding box.base)-(10pt,10pt)$)},line width=1, scale=0.6, circ/.style={line width=0.8, shape=circle, fill=white, inner sep=1pt, draw, node contents=}]
    \coordinate (a) at (0,0);
    \coordinate (e) at ($(a)+(-150:2)+(0,0.5)$);
    \coordinate (f) at ($(a)+(1,0.5)$);
    \coordinate (g) at ($(a)+(0,0.5)$);
    \draw[dashed,Orange] ($(a)+(0:0.5)+(0,0.7)$) -- ($(a)+(0:0.5)+(0,-1.2)$);
    \draw[dashed,Orange] ($(a)+(0:-1.5)+(0,0.7)$) -- ($(a)+(0:-1.5)+(0,-1.2)$);
    \draw[RoyalBlue] (e) -- ++ (160:1);
    \draw[RoyalBlue] (e) -- ++ (-160:1);
    \draw[RoyalBlue] (a) -- (e);
    \foreach \n in {e}
    {\fill[black,thick] (\n) circle (0.070001);}
    \draw[Fuchsia,decorate,decoration={snake,amplitude=.4mm,segment length=2mm,post length=0.01mm,pre length=0.01mm}] (a) -- (g) node[above] {\footnotesize$q$};
    \foreach \n in {a}
    {\draw (\n) node[circ];};
    \end{tikzpicture}
    \;+\;
    \begin{tikzpicture}[baseline= {($(current bounding box.base)-(10pt,10pt)$)},line width=1, scale=0.6, circ/.style={line width=0.8, shape=circle, fill=white, inner sep=1pt, draw, node contents=}]
    \coordinate (a) at (0,0);
    \coordinate (a) at (0,0);
    \coordinate (f) at ($(a)+(1,0.5)$);
    \coordinate (e) at ($(f)-(0,1)$);
    \coordinate (g) at ($(a)+(0,0.5)$);
    \draw[dashed,Orange] ($(a)+(0:0.5)+(0,0.7)$) -- ($(a)+(0:0.5)+(0,-1.2)$);
    \draw[dashed,Orange] ($(a)+(0:-1.5)+(0,0.7)$) -- ($(a)+(0:-1.5)+(0,-1.2)$);
    \draw[RoyalBlue] (e) -- ($(b)-(0,0.8)$);
    \draw[RoyalBlue] (e) -- ($(b)-(0,1.1)$);
    \draw[RoyalBlue] (a) -- (e);
    \foreach \n in {e}
    {\fill[black,thick] (\n) circle (0.070001);}
    \draw[Fuchsia,decorate,decoration={snake,amplitude=.4mm,segment length=2mm,post length=0.01mm,pre length=0.01mm}] (a) -- (g) node[above] {\footnotesize$q$};
    \foreach \n in {a}
    {\draw (\n) node[circ];};
    \end{tikzpicture}\label{eq:anchored0}
    \\
    &
    =
    \underbracket[0.4pt]{\begin{tikzpicture}[baseline= {($(current bounding box.base)-(10pt,10pt)$)},line width=1, scale=0.6, circ/.style={line width=0.8, shape=circle, fill=white, inner sep=1pt, draw, node contents=}]
    \coordinate (a) at (0,0);
    \coordinate (e) at ($(a)+(-150:1)$);
    \coordinate (f) at ($(a)+(1,0.5)$);
    \coordinate (g) at ($(a)+(0,0.5)$);
    \fill[fill=RoyalBlue!10] ($(a)+(0.5,-1.2)$) rectangle ($(a)+(1.5,0.7)$);
    \draw[dashed,Orange] ($(a)+(0:0.5)+(0,0.7)$) -- ($(a)+(0:0.5)+(0,-1.2)$);
    \draw[dashed,Orange] ($(a)+(0:-1.5)+(0,0.7)$) -- ($(a)+(0:-1.5)+(0,-1.2)$);
    \draw[RoyalBlue] (e) -- ++ (160:1);
    \draw[RoyalBlue] (e) -- ++ (-160:1);
    \draw[RoyalBlue] (a) -- (e);
    \foreach \n in {e}
    {\fill[black,thick] (\n) circle (0.070001);}
    \draw[Fuchsia,decorate,decoration={snake,amplitude=.4mm,segment length=2mm,post length=0.01mm,pre length=0.01mm}] (a) -- (g) node[above] {\footnotesize$q$};
    \foreach \n in {a}
    {\draw (\n) node[circ];};
    \end{tikzpicture}}_{\textcolor{RoyalBlue}{1}}
    +
    \underbracket[0.4pt]{\begin{tikzpicture}[baseline= {($(current bounding box.base)-(10pt,10pt)$)},line width=1, scale=0.6, circ/.style={line width=0.8, shape=circle, fill=white, inner sep=1pt, draw, node contents=}]
    \coordinate (a) at (0,0);
    \coordinate (e) at ($(a)+(-150:2)+(0,0.5)$);
    \coordinate (f) at ($(a)+(1,0.5)$);
    \coordinate (g) at ($(a)+(0,0.5)$);
    \fill[fill=RoyalBlue!10] ($(a)+(0.5,-1.2)$) rectangle ($(a)+(1.5,0.7)$);
    \draw[dashed,Orange] ($(a)+(0:0.5)+(0,0.7)$) -- ($(a)+(0:0.5)+(0,-1.2)$);
    \draw[dashed,Orange] ($(a)+(0:-1.5)+(0,0.7)$) -- ($(a)+(0:-1.5)+(0,-1.2)$);
    \draw[RoyalBlue] (e) -- ++ (160:1);
    \draw[RoyalBlue] (e) -- ++ (-160:1);
    \draw[RoyalBlue] (a) -- (e);
    \foreach \n in {e}
    {\fill[black,thick] (\n) circle (0.070001);}
    \draw[Fuchsia,decorate,decoration={snake,amplitude=.4mm,segment length=2mm,post length=0.01mm,pre length=0.01mm}] (a) -- (g) node[above] {\footnotesize$q$};
    \foreach \n in {a}
    {\draw (\n) node[circ];};
    \end{tikzpicture}}_{\textcolor{RoyalBlue}{2}}
    +
    \underbracket[0.4pt]{\begin{tikzpicture}[baseline= {($(current bounding box.base)-(10pt,10pt)$)},line width=1, scale=0.6, circ/.style={line width=0.8, shape=circle, fill=white, inner sep=1pt, draw, node contents=}]
    \coordinate (a) at (0,0);
    \coordinate (f) at ($(a)+(1,0.5)$);
    \coordinate (e) at ($(f)-(0,1)$);
    \coordinate (g) at ($(a)+(0,0.5)$);
    \fill[fill=Maroon!10,pattern=north east lines,pattern color=Maroon!20] ($(a)+(0.5,-1.2)$) rectangle ($(a)+(1.5,0.7)$);
    \draw[dashed,Orange] ($(a)+(0:0.5)+(0,0.7)$) -- ($(a)+(0:0.5)+(0,-1.2)$);
    \draw[dashed,Orange] ($(a)+(0:-1.5)+(0,0.7)$) -- ($(a)+(0:-1.5)+(0,-1.2)$);
    \draw[RoyalBlue] (e) -- ($(b)-(0,0.8)$);
    \draw[RoyalBlue] (e) -- ($(b)-(0,1.1)$);
    \draw[RoyalBlue] (a) -- (e);
    \foreach \n in {e}
    {\fill[black,thick] (\n) circle (0.070001);}
    \draw[Fuchsia,decorate,decoration={snake,amplitude=.4mm,segment length=2mm,post length=0.01mm,pre length=0.01mm}] (a) -- (g) node[above] {\footnotesize$q$};
    \foreach \n in {a}
    {\draw (\n) node[circ];};
    \end{tikzpicture}}_{\textcolor{Maroon}{3}}
    +
    \underbracket[0.4pt]{\begin{tikzpicture}[baseline= {($(current bounding box.base)-(10pt,10pt)$)},line width=1, scale=0.6, circ/.style={line width=0.8, shape=circle, fill=white, inner sep=1pt, draw, node contents=}]
    \coordinate (a) at (0,0);
    \coordinate (b) at ($(a)-(2,0)$);
    \coordinate (e) at ($(a)+(-150:1)$);
    \coordinate (f) at ($(a)+(1,0.5)$);
    \coordinate (g) at ($(b)+(0,0.5)$);
    \fill[fill=RoyalBlue!10] ($(a)+(0.5,-1.2)$) rectangle ($(a)+(1.5,0.7)$);
    \draw[dashed,Orange] ($(a)+(0:0.5)+(0,0.7)$) -- ($(a)+(0:0.5)+(0,-1.2)$);
    \draw[dashed,Orange] ($(a)+(0:-1.5)+(0,0.7)$) -- ($(a)+(0:-1.5)+(0,-1.2)$);
    \draw[RoyalBlue] (e) -- ++ (180:1);
    \draw[RoyalBlue] (e) -- ++ (-160:1);
    \draw[RoyalBlue] (b) -- (e);
    \foreach \n in {e}
    {\fill[black,thick] (\n) circle (0.070001);}
    \draw[Fuchsia,decorate,decoration={snake,amplitude=.4mm,segment length=2mm,post length=0.01mm,pre length=0.01mm}] (b) -- (g) node[above] {\footnotesize$q$};
    \foreach \n in {b}
    {\draw (\n) node[circ];};
    \end{tikzpicture}}_{\textcolor{RoyalBlue}{4}}
    +
    \underbracket[0.4pt]{\begin{tikzpicture}[baseline= {($(current bounding box.base)-(10pt,10pt)$)},line width=1, scale=0.6, circ/.style={line width=0.8, shape=circle, fill=white, inner sep=1pt, draw, node contents=}]
    \coordinate (a) at (0,0);
    \coordinate (b) at ($(a)-(2,0)$);
    \coordinate (f) at ($(a)+(1,0.5)$);
    \coordinate (e) at ($(f)-(0,1)$);
    \coordinate (g) at ($(b)+(0,0.5)$);
    \fill[fill=Maroon!10,pattern=north east lines,pattern color=Maroon!20] ($(a)+(0.5,-1.2)$) rectangle ($(a)+(1.5,0.7)$);
    \draw[dashed,Orange] ($(a)+(0:0.5)+(0,0.7)$) -- ($(a)+(0:0.5)+(0,-1.2)$);
    \draw[dashed,Orange] ($(a)+(0:-1.5)+(0,0.7)$) -- ($(a)+(0:-1.5)+(0,-1.2)$);
    \draw[RoyalBlue] (e) -- ($(b)-(0,0.8)$);
    \draw[RoyalBlue] (e) -- ($(b)-(0,1.1)$);
    \draw[RoyalBlue] (b) -- (e);
    \foreach \n in {e}
    {\fill[black,thick] (\n) circle (0.070001);}
    \draw[Fuchsia,decorate,decoration={snake,amplitude=.4mm,segment length=2mm,post length=0.01mm,pre length=0.01mm}] (b) -- (g) node[above] {\footnotesize$q$};
    \foreach \n in {b}
    {\draw (\n) node[circ];};
    \end{tikzpicture}}_{\textcolor{Maroon}{5}}
    \label{eq:anchored1}
    \\
    &     \hspace{3cm} 
    +
    \underbracket[0.4pt]{\begin{tikzpicture}[baseline= {($(current bounding box.base)-(10pt,10pt)$)},line width=1, scale=0.6, circ/.style={line width=0.8, shape=circle, fill=white, inner sep=1pt, draw, node contents=}]
    \coordinate (a) at (0,0);
    \coordinate (b) at ($(a)-(2,0)$);
    \coordinate (e) at ($(a)+(-150:2)+(0,0.5)$);
    \coordinate (f) at ($(a)+(1,0.5)$);
    \coordinate (g) at ($(b)+(0,0.5)$);
    \fill[fill=RoyalBlue!10] ($(a)+(0.5,-1.2)$) rectangle ($(a)+(1.5,0.7)$);
    \draw[dashed,Orange] ($(a)+(0:0.5)+(0,0.7)$) -- ($(a)+(0:0.5)+(0,-1.2)$);
    \draw[dashed,Orange] ($(a)+(0:-1.5)+(0,0.7)$) -- ($(a)+(0:-1.5)+(0,-1.2)$);
    \draw[RoyalBlue] (e) -- ++ (160:1);
    \draw[RoyalBlue] (e) -- ++ (-160:1);
    \draw[RoyalBlue] (b) -- (e);
    \foreach \n in {e}
    {\fill[black,thick] (\n) circle (0.070001);}
    \draw[Fuchsia,decorate,decoration={snake,amplitude=.4mm,segment length=2mm,post length=0.01mm,pre length=0.01mm}] (b) -- (g) node[above] {\footnotesize$q$};
    \foreach \n in {b}
    {\draw (\n) node[circ];};
    \end{tikzpicture}}_{\textcolor{RoyalBlue}{6}}
    -
    \underbracket[0.4pt]{\begin{tikzpicture}[baseline= {($(current bounding box.base)-(10pt,10pt)$)},line width=1, scale=0.6, circ/.style={line width=0.8, shape=circle, fill=white, inner sep=1pt, draw, node contents=}]
    \coordinate (a) at (0,0);
    \coordinate (b) at ($(a)-(2,0)$);
    \coordinate (e) at ($(a)+(-150:1)$);
    \coordinate (f) at ($(a)+(1,0.5)$);
    \coordinate (g) at ($(b)+(0,0.5)$);
    \fill[fill=gray!10,pattern=dots,pattern color=black!80] ($(a)-(2.5,1.2)$) rectangle (-1.5,0.7);
    \draw[dashed,Orange] ($(a)+(0:0.5)+(0,0.7)$) -- ($(a)+(0:0.5)+(0,-1.2)$);
    \draw[dashed,Orange] ($(a)+(0:-1.5)+(0,0.7)$) -- ($(a)+(0:-1.5)+(0,-1.2)$);
    \draw[RoyalBlue] (e) -- ++ (180:1);
    \draw[RoyalBlue] (e) -- ++ (-160:1);
    \draw[RoyalBlue] (b) -- (e);
    \foreach \n in {e}
    {\fill[black,thick] (\n) circle (0.070001);}
    \draw[Fuchsia,decorate,decoration={snake,amplitude=.4mm,segment length=2mm,post length=0.01mm,pre length=0.01mm}] (b) -- (g) node[above] {\footnotesize$q$};
    \foreach \n in {b}
    {\draw (\n) node[circ];};
    \end{tikzpicture}}_{\textcolor{gray}{7}}
    -
    \underbracket[0.4pt]{\begin{tikzpicture}[baseline= {($(current bounding box.base)-(10pt,10pt)$)},line width=1, scale=0.6, circ/.style={line width=0.8, shape=circle, fill=white, inner sep=1pt, draw, node contents=}]
    \coordinate (a) at (0,0);
    \coordinate (b) at ($(a)-(2,0)$);
    \coordinate (e) at ($(a)+(-30:2)+(0,0.5)$);
    \coordinate (f) at ($(a)+(1,0.5)$);
    \coordinate (g) at ($(b)+(0,0.5)$);
    \fill[fill=gray!10,pattern=dots,pattern color=black!80] ($(a)-(2.5,1.2)$) rectangle (-1.5,0.7);
    \draw[dashed,Orange] ($(a)+(0:0.5)+(0,0.7)$) -- ($(a)+(0:0.5)+(0,-1.2)$);
    \draw[dashed,Orange] ($(a)+(0:-1.5)+(0,0.7)$) -- ($(a)+(0:-1.5)+(0,-1.2)$);
    \draw[RoyalBlue] (e) -- ++ (180:4);
    \draw[RoyalBlue] (e) -- ++ (-175:4);
    \draw[RoyalBlue] (b) -- (e);
    \foreach \n in {e}
    {\fill[black,thick] (\n) circle (0.070001);}
    \draw[Fuchsia,decorate,decoration={snake,amplitude=.4mm,segment length=2mm,post length=0.01mm,pre length=0.01mm}] (b) -- (g) node[above] {\footnotesize$q$};
    \foreach \n in {b}
    {\draw (\n) node[circ];};
    \end{tikzpicture}}_{\textcolor{gray}{8}}
    -
    \underbracket[0.4pt]{\begin{tikzpicture}[baseline= {($(current bounding box.base)-(10pt,10pt)$)},line width=1, scale=0.6, circ/.style={line width=0.8, shape=circle, fill=white, inner sep=1pt, draw, node contents=}]
    \coordinate (a) at (0,0);
    \coordinate (b) at ($(a)-(2,0)$);
    \coordinate (e) at ($(a)+(-150:2)+(0,0.5)$);
    \coordinate (f) at ($(a)+(1,0.5)$);
    \coordinate (g) at ($(b)+(0,0.5)$);
    \draw[dashed,Orange] ($(a)+(0:0.5)+(0,0.7)$) -- ($(a)+(0:0.5)+(0,-1.2)$);
    \draw[dashed,Orange] ($(a)+(0:-1.5)+(0,0.7)$) -- ($(a)+(0:-1.5)+(0,-1.2)$);
    \draw[RoyalBlue] (e) -- ++ (160:1);
    \draw[RoyalBlue] (e) -- ++ (-160:1);
    \draw[RoyalBlue] (b) -- (e);
    \foreach \n in {e}
    {\fill[black,thick] (\n) circle (0.070001);}
    \draw[Fuchsia,decorate,decoration={snake,amplitude=.4mm,segment length=2mm,post length=0.01mm,pre length=0.01mm}] (b) -- (g) node[above] {\footnotesize$q$};
    \foreach \n in {b}
    {\draw (\n) node[circ];};
    \end{tikzpicture}}_{9}
    \label{eq:anchored2}
\end{align}
\end{subequations}
In equations, the top line is given by
\begin{subequations}
\begin{align}
    i F^{\text{branch}} & =  \frac{- i}{-s_I+m_I^2-i\varepsilon} - 2\pi \Theta(-q^0) \delta(-s_I+m_I^2) - 2 \pi \Theta(q^0) \delta(-s_I+m_I^2)
    \\ & = \frac{- i}{-s_I+m_I^2+i\varepsilon} \,.
    \label{eq:iFbranch_pert}
\end{align}
\end{subequations}
Note that we only have modified the top line~\eqref{eq:anchored0} by a trivial zero: we added and then subtracted all the diagrams for which the vertex of the branch is in the $S$-blob on the left. This was not a random choice. Indeed, the diagrams are now organized in a way that makes manifest cancellations between them.
In particular, the diagrams \textcolor{RoyalBlue}{1}, \textcolor{RoyalBlue}{2}, \textcolor{RoyalBlue}{4} and \textcolor{RoyalBlue}{6}, which have the rightmost $S$ shaded in blue, cancel. This follows from unitarity: their sum precisely corresponds to all ways of distributing the two vertices between the $S$ on the left, and $F^\dag$ in the middle region; in equations,
\begin{subequations}
\begin{align}
    \textcolor{RoyalBlue}{1} +  \textcolor{RoyalBlue}{2}+  \textcolor{RoyalBlue}{4} + \textcolor{RoyalBlue}{6}
    & =
    \frac{- i}{-s_I+m_I^2-i\varepsilon} - 2\pi \Theta(-q^0) \delta(-s_I+m_I^2) \\ & \hspace{2cm} + \frac{ i}{-s_I+m_I^2+i\varepsilon} - 2 \pi \Theta(q^0) \delta(-s_I+m_I^2) = 0 \,.
\end{align}
\end{subequations}
This type of cancellation between placements of vertices between $S$ and $S^\dag$ is familiar from perturbative proofs of the cutting rules using the largest-time equation\footnote{These proofs often using a black-and-white-vertex notation (nothing to do with our black/white coding) to show how diagrams, their conjugates and cuts cancel among each other, see e.g.~\cite{Veltman:1994wz}. When mapping those proofs to our setup, the vertices in the leftmost $S$ blob would be white, and the vertices in the middle $S^\dag$ blob would be black. The sum over all configurations of black and white vertices cancel among each other.}.
Similarly, diagrams \textcolor{Maroon}{3} and \textcolor{Maroon}{5} cancel, since they correspond to all ways of distributing the white vertex between $S$ and $S^\dag$, keeping the black vertex fixed in the rightmost $S$ blob (shaded in red with diagonal lines):
\begin{equation}
    \textcolor{Maroon}{3} + \textcolor{Maroon}{5} = - 2\pi \Theta(q^0) \delta(-s_I+m_I^2) + 2\pi \Theta(q^0) \delta(-s_I+m_I^2) = 0 \,.
\end{equation}
Last, keeping the white vertex in the leftmost $S$ blob, represented with the grey dotted region, shows that diagrams \textcolor{gray}{7} and \textcolor{gray}{8} cancel:
\begin{equation}
    \textcolor{gray}{7} + \textcolor{gray}{8} = - 2\pi \Theta(q^0) \delta(-s_I+m_I^2) + 2\pi \Theta(q^0) \delta(-s_I+m_I^2) = 0 \,.
\end{equation}
The only remaining contribution is diagram 9, which corresponds to placing the whole diagram within the $S$ blob with an extra overall minus sign, which is precisely what we expect from \eqref{eq:branch_crossing_result}, and agrees with~\eqref{eq:iFbranch_pert}.

As a final remark, notice that since this proof of the crossing equation for branches presented in this subsection is based on unitarity, we never actually used that the diagram was a tree. Thus, we have also proven the crossing equation~\eqref{eq:crossing23}  in the case of diagrams in which two particles attach to the rest of the diagram at a single vertex, and those two particles are precisely the ones that are crossed from incoming to outgoing. 
\section{One-loop crossing for massless five-point amplitudes}
\label{sec:fivepoint}

In this section, we verify the crossing conjecture for all massless one-loop master integrals relevant for (up to) five-point functions in dimensional regularization around $\D=4$ dimensions. In doing so, we will employ various techniques to evaluate the master integrals and their cuts, including Schwinger parameters, the embedding space formalism, and differential equations.

For fully massless amplitudes, non-trivial crossings start at four points, where amplitudes simply analytically continue to complex-conjugated amplitudes using the crossing path~\eqref{eq:crossingpath}. Below, we will illustrate through a concrete example that this well-established conclusion is, indeed, in perfect harmony with our crossing equation \eqref{eq:crossing23}.

At five points, there is only one possibility for crossings of the process $S_{345\ot 12}$: we cross an incoming particle with an outgoing one.\footnote{At five point, the option of crossing an incoming particle (e.g., 2) with a cluster of two particles (e.g., 34) is not considered separately because it is equivalent to a single particle crossing (e.g., crossing 1 and 5).} Taking these two particles to be 2 and 3, this leads to
\begin{equation}
\begin{gathered}
\adjustbox{valign=t}{\begin{tikzpicture}[line width=1]
\draw[] (0,1.7) node {\small$\vac{b_5 b_4 b_3 a_2^\dag a_1^\dag}$};
\draw[RoyalBlue] (1.2,0.3) node[right] {\small$1$} -- (0,0.3);
\draw[Maroon] (0,0.5) -- (-1.2,0.5) node[left] {\small$3$};
\draw[Maroon,->] (0,0.5) -- (-1,0.5);
\draw[RoyalBlue,>-] (1,0.3) -- (0,0.3);
\draw[RoyalBlue] (0,0) node[right] {} -- (-1.2,0) node[left] {\small$4$};
\draw[Maroon] (1.2,-0.3) node[right] {\small$2$} -- (0,-0.3) node[left] {};
\draw[RoyalBlue,->] (0,0) -- (-1,0);
\draw[Maroon,>-] (1,-0.3) -- (0,-0.3);
\draw[RoyalBlue] (0,-0.5) node[right] {} -- (-1.2,-0.5) node[left] {\small$5$};
\draw[RoyalBlue,->] (0,-0.5) -- (-1,-0.5);
\filldraw[fill=gray!5,very thick](0,0) circle (0.6) node {$S$};
\draw[-latex] (2,0) -- (4,0) node[right, xshift=10pt]{$-$};
\node[align=center] at (3,0.5) {\footnotesize cross $\textcolor{Maroon}{2} \leftrightarrow \textcolor{Maroon}{3}$};
\end{tikzpicture}}
\quad
\adjustbox{valign=t}{\begin{tikzpicture}[line width=1]
\draw[] (1,1.7) node {\small$-\vac{b_5 b_4 a_2 \bdag_3\adag_1}$};
\draw[RoyalBlue] (0,0.3) -- (-1.2,0.3) node[left] {\small$4$};
\draw[RoyalBlue,->] (0,0.3) -- (-1,0.3);
\draw[RoyalBlue] (0,-0.3) -- (-1.2,-0.3) node[left] {\small$5$};
\draw[RoyalBlue,->] (0,-0.3) -- (-1,-0.3);
\draw[RoyalBlue] (2,0.3) -- (3.2,0.3) node[right] {\small$1$};
\draw[RoyalBlue,-<] (2,0.3) -- (3,0.3);
\draw[Maroon] (2,-0.3) -- (3.2,-0.3) node[right] {\small$\bar{3}$};
\draw[Maroon,-<] (2,-0.3) -- (3,-0.3);
\draw[Maroon] (2,0) -- (1,-1) node[left] {\small$\bar{2}$};
\draw[Maroon,->] (2,0) -- (1.2,-0.8);
\filldraw[fill=gray!30,very thick](0,-0.3) rectangle (2,0.3);
\draw[] (1,0) node {$X$};
\filldraw[fill=gray!5,very thick](0,0) circle (0.6) node {$S$};
\filldraw[fill=gray!5,very thick](2,0) circle (0.6) node[yshift=1] {$S^\dag$};
\draw[dashed,orange] (1,1.2) -- (1,-1.2);
\end{tikzpicture}}
\end{gathered}
\label{eq:5ptcross1_OL}
\end{equation}
In what follows, we will meticulously check the consistency of this statement against explicit examples. In particular, at the end of this section, we will have checked all possible crossings for up to five-point diagrams, and up to pentagon topology. 

Note that at one loop, non-trivial crossings begin with massless bubbles. Indeed, tadpoles either vanish (when massless) or are functions of the mass in the loop (when massive). Consequently, they remain unchanged under crossing and are not considered here.

Finally, towards the end of this section, we will explore further applications of crossing symmetry, such as how it can be used in practical calculations to relate time-ordered amplitudes with different labels. 

\subsection{Bubbles}

We start by establishing crossing for massless bubble integrals, which in $\D$ spacetime dimensions are given by
\begin{equation}
    \cM_{345\leftarrow12}^{\text{bub}} = \int \frac{\rd^\D \ell}{i\pi^{\D/2}} \frac{1}{[\ell^2-i\varepsilon][(\ell+p)^2-i\varepsilon]} \,.
\end{equation}
This integral is easy to perform; it evaluates to 
\begin{equation}\label{eq:bubles}
    \cM_{345\leftarrow12}^{\text{bub}} = \frac{\Gamma\big(\frac{\D}{2}-1\big)^2 \Gamma\big(2-\frac{\D}{2}\big)}{\Gamma(\D-2)}(p^2-i\varepsilon)^{\D/2-2} \,.
\end{equation}
Here, $p^2$ corresponds to the squared momentum entering the vertices, which at five points can be realized by four different configurations (up to permutations of indices):
\begin{equation}
\begin{gathered}
    \begin{tikzpicture}[scale=1]
        \draw[thick,white] (-0.5,0) to [in=150,out=75] (0.75,0.5) ;
        \draw[thick,white] (0.5,0) to [in=-30,out=75-180] (-0.75,-0.5) ;
        \draw[thick] (0.5,0) to [out=120,in=60] (-0.5,0);
        \draw[thick] (0.5,0) to [out=-120,in=-60] (-0.5,0);
        \draw[thick,RoyalBlue] (0.5,0) -- ++ (45:0.5);
        \draw[thick,Maroon] (0.5,0) -- ++ (-45:0.5);
        \draw[thick,RoyalBlue] (-0.5,0) -- ++ (180-45:0.5);
        \draw[thick,Maroon] (-0.5,0) -- ++ (180+45:0.5);
        \draw[thick,RoyalBlue] (-0.5,0) -- ++ (180:0.5);
        \node[scale=0.75] at (1,0.5) {\color{RoyalBlue}$1$};
        \node[scale=0.75] at (1,-0.5) {\color{Maroon}$2$};
        \node[scale=0.75] at (-1,-0.5) {\color{Maroon}$3$};
        \node[scale=0.75] at (-1.2,0) {\color{RoyalBlue}$4$};
        \node[scale=0.75] at (-1,0.5) {\color{RoyalBlue}$5$};
    \end{tikzpicture}
\end{gathered}
\hspace{0.5cm}
\begin{gathered}
    \begin{tikzpicture}[scale=1]
        \draw[thick] (0.5,0) to [out=120,in=60] (-0.5,0);
        \draw[thick] (0.5,0) to [out=-120,in=-60] (-0.5,0);
        \draw[thick,Maroon] (0.5,0) -- ++ (45:0.5);
        \draw[thick,RoyalBlue] (-0.5,0) -- ++ (180-45:0.5);
        \draw[thick,RoyalBlue] (-0.5,0) -- ++ (180+45:0.5);
        \draw[thick,RoyalBlue] (-0.5,0) to [in=150,out=75] (0.75,0.5) ;
        \draw[thick,Maroon] (0.5,0) to [in=-30,out=75-180] (-0.75,-0.5) ;
        \node[scale=0.75] at (0.9,0.75) {\color{RoyalBlue}$1$};
        \node[scale=0.75] at (1,0.45) {\color{Maroon}$2$};
        \node[scale=0.75] at (-0.9,-0.65) {\color{Maroon}$3$};
        \node[scale=0.75] at (-1,-0.4) {\color{RoyalBlue}$4$};
        \node[scale=0.75] at (-0.9,0.5) {\color{RoyalBlue}$5$};
    \end{tikzpicture}
\end{gathered}
\hspace{0.5cm}
\begin{gathered}
    \begin{tikzpicture}[scale=1]
        \draw[thick] (0.5,0) to [out=120,in=60] (-0.5,0);
        \draw[thick] (0.5,0) to [out=-120,in=-60] (-0.5,0);
        \draw[thick,RoyalBlue] (0.5,0) -- ++ (45:0.5);
        \draw[thick,RoyalBlue] (-0.5,0) -- ++ (180-45:0.5);
        \draw[thick,RoyalBlue] (-0.5,0) -- ++ (180+45:0.5);
        \draw[thick,Maroon] (-0.5,0) to [in=150,out=75] (0.75,0.5) ;
        \draw[thick,Maroon] (0.5,0) to [in=-30,out=75-180] (-0.75,-0.5) ;
        \node[scale=0.75] at (0.9,0.75) {\color{Maroon}$2$};
        \node[scale=0.75] at (1,0.45) {\color{RoyalBlue}$1$};
        \node[scale=0.75] at (-0.9,-0.65) {\color{Maroon}$3$};
        \node[scale=0.75] at (-1,-0.4) {\color{RoyalBlue}$4$};
        \node[scale=0.75] at (-0.9,0.5) {\color{RoyalBlue}$5$};
    \end{tikzpicture}
\end{gathered}
\hspace{0.5cm}
\begin{gathered}
    \begin{tikzpicture}[scale=1]
        \draw[thick,white] (-0.5,0) to [in=150,out=75] (0.75,0.5) ;
        \draw[thick,white] (0.5,0) to [in=-30,out=75-180] (-0.75,-0.5) ;
        \draw[thick] (0.5,0) to [out=120,in=60] (-0.5,0);
        \draw[thick] (0.5,0) to [out=-120,in=-60] (-0.5,0);
        \draw[thick,RoyalBlue] (0.5,0) -- ++ (45:0.5);
        \draw[thick,Maroon] (0.5,0) -- ++ (-45:0.5);
        \draw[thick,RoyalBlue] (-0.5,0) -- ++ (180-45:0.5);
        \draw[thick,RoyalBlue] (-0.5,0) -- ++ (180+45:0.5);
        \draw[thick,Maroon] (0.5,0) to [in=-30,out=75-180] (-0.75,-0.5) ;
        \node[scale=0.75] at (-0.9,-0.65) {\color{Maroon}$3$};
        \node[scale=0.75] at (1,0.5) {\color{RoyalBlue}$1$};
        \node[scale=0.75] at (1,-0.5) {\color{Maroon}$2$};
        \node[scale=0.75] at (-1,-0.4) {\color{RoyalBlue}$4$};
        \node[scale=0.75] at (-1,0.5) {\color{RoyalBlue}$5$};
    \end{tikzpicture}
\end{gathered}
\end{equation}

When crossing particles $2$ and $3$, the relevant rotations for each of the four diagrams above are summarized as
\begin{equation}
\adjustbox{valign=c}{\begin{tikzpicture}[scale=0.8]
  \draw[->,thick] (-1.5, 0) -- (1.5, 0);
  \draw[->,thick,white] (0, -1.6) -- (0, 1.6);
  \draw[->,thick] (0, -1.25) -- (0, 1.25);
  \node[] at (1.1,1.05) {\small$s_{ij}$};
  \node[] at (-0.35,0.6) {\small$\textcolor{Maroon}{s_{12}}$};
  \draw[] (1.35,0.85) -- (0.8,0.85) -- (0.8,1.25);
  \draw[Maroon,fill=Maroon,thick] (1,0.1) circle (0.05);
  \draw[->,Maroon,thick] (1,0.1) arc (0:180:1);
  \draw[decorate, decoration={zigzag, segment length=6, amplitude=2}, black!80] (0,0) -- (1.5,0);
  \draw[black!80,fill=black!80,thick] (0,0) circle (0.05);
\end{tikzpicture}}
\hspace{0.5cm}
\adjustbox{valign=c}{\begin{tikzpicture}[scale=0.8]
  \draw[->,thick] (-1.5, 0) -- (1.5, 0);
  \draw[->,thick,white] (0, -1.6) -- (0, 1.6);
  \draw[->,thick] (0, -1.25) -- (0, 1.25);
  \node[] at (1.1,1.05) {\small$s_{ij}$};
  \draw[RoyalBlue,fill=RoyalBlue,thick] (-0.4,0) circle (0.05);
  \draw[] (1.35,0.85) -- (0.8,0.85) -- (0.8,1.25);
  \node[] at (-0.35,0.3) {\small$\textcolor{RoyalBlue}{s_{23}}$};
  \draw[decorate, decoration={zigzag, segment length=6, amplitude=2}, black!80] (0,0) -- (1.5,0);
  \draw[black!80,fill=black!80,thick] (0,0) circle (0.05);
\end{tikzpicture}}
\hspace{0.5cm}
\adjustbox{valign=c}{\begin{tikzpicture}[scale=0.8]
  \draw[->,thick] (-1.5, 0) -- (1.5, 0);
  \draw[->,thick,white] (0, -1.6) -- (0, 1.6);
  \draw[->,thick] (0, -1.25) -- (0, 1.25);
  \node[] at (1.1,1.05) {\small$ s_{ij}$};
  \draw[] (1.35,0.85) -- (0.8,0.85) -- (0.8,1.25);
  \draw[decorate, decoration={zigzag, segment length=6, amplitude=2}, black!80] (0,0) -- (1.5,0);
  \draw[black!80,fill=black!80,thick] (0,0) circle (0.05);
  \draw[Maroon,fill=Maroon,thick] (-1.1,0.0) circle (0.05);
  \draw[->,Maroon,thick] (-1.1,0.0) arc (0:178:-1.1);
  \node[] at (-0.35,-0.6) {\small$\textcolor{Maroon}{s_{13}}$};
\end{tikzpicture}}
\hspace{0.5cm}
\adjustbox{valign=c}{\begin{tikzpicture}[scale=0.8]
  \draw[->,thick] (-1.5, 0) -- (1.5, 0);
  \draw[->,thick,white] (0, -1.6) -- (0, 1.6);
  \draw[->,thick] (0, -1.25) -- (0, 1.25);
  \node[] at (1.1,1.05) {\small$s_{ij}$};
  \draw[RoyalBlue,fill=RoyalBlue,thick] (0.5,0.1) circle (0.05);
  \node[] at (0.4,0.4) {\small$\textcolor{RoyalBlue}{s_{45}}$};
  \draw[] (1.35,0.85) -- (0.8,0.85) -- (0.8,1.25);
  \draw[decorate, decoration={zigzag, segment length=6, amplitude=2}, black!80] (0,0) -- (1.5,0);
  \draw[black!80,fill=black!80,thick] (0,0) circle (0.05);
\end{tikzpicture}}
\end{equation}
Since each bubble amplitude depends on a single invariant $s_{ij}$, it is clear that in the case of the first three diagrams, for which the relevant invariant $s_{ij}$ either ends up negative or on the wrong side of the cut (for $s_{13}$), the amplitudes analytically continue to complex conjugated amplitudes. There is no possible cut in the $s_{45}$ channel in these cases, so this prediction is consistent with the expectation from \eqref{eq:5ptcross1_OL}, since the $S$-blob is necessarily trivial.

In the case of the fourth diagram, the amplitude is unchanged under the rotation. The prediction from the crossing equation in \eqref{eq:5ptcross1_OL} is that we get the conjugated amplitude plus a cut in the $s_{45}$ channel, namely
\begin{equation}\label{eq:bubCross}
    \left[\adjustbox{valign=c}{\begin{tikzpicture}[line width=1,scale=1]
        \draw[thick,white] (-0.5,0) to [in=150,out=75] (0.75,0.5) ;
        \draw[thick,white] (0.5,0) to [in=-30,out=75-180] (-0.75,-0.5) ;
        \draw[thick] (0.5,0) to [out=120,in=60] (-0.5,0);
        \draw[thick] (0.5,0) to [out=-120,in=-60] (-0.5,0);
        \draw[thick,RoyalBlue] (0.5,0) -- ++ (45:0.5);
        \draw[thick,Maroon] (0.5,0) -- ++ (-45:0.5);
        \draw[thick,RoyalBlue] (-0.5,0) -- ++ (180-45:0.5);
        \draw[thick,RoyalBlue] (-0.5,0) -- ++ (180+45:0.5);
        \draw[thick,Maroon] (0.5,0) to [in=-30,out=75-180] (-0.75,-0.5) ;
        \node[Maroon,scale=0.75] at (-0.9,-0.6) {$\color{Maroon}3$};
        \node[RoyalBlue,scale=0.75] at (1,0.5) {$\color{RoyalBlue}1$};
        \node[Maroon,scale=0.75] at (1,-0.5) {$\color{Maroon}2$};
        \node[RoyalBlue,scale=0.75] at (-1,-0.4) {$\color{RoyalBlue}4$};
        \node[RoyalBlue,scale=0.75] at (-1,0.5) {$\color{RoyalBlue}5$};
\end{tikzpicture}}\right]_{\substack{\text{nothing}\\\text{rotates}}}\stackrel{?}{=}
\quad
\adjustbox{valign=c}{\begin{tikzpicture}[line width=1,scale=1]
        \draw[thick,white] (-0.5,0) to [in=150,out=75] (0.75,0.5) ;
        \draw[thick,white] (0.5,0) to [in=-30,out=75-180] (-0.75,-0.5) ;
        \draw[thick] (0.5,0) to [out=120,in=60] (-0.5,0);
        \draw[thick] (0.5,0) to [out=-120,in=-60] (-0.5,0);
        \draw[thick,RoyalBlue] (0.5,0) -- ++ (45:0.5);
        \draw[thick,Maroon] (0.5,0) -- ++ (-45:0.5);
        \draw[thick,RoyalBlue] (-0.5,0) -- ++ (180-45:0.5);
        \draw[thick,RoyalBlue] (-0.5,0) -- ++ (180+45:0.5);
        \draw[thick,Maroon] (0.5,0) to [in=-30,out=75-180] (-0.75,-0.5) ;
        \node[Maroon,scale=0.75] at (-0.87,-0.65) {$\color{Maroon}\bar{2}$};
        \node[RoyalBlue,scale=0.75] at (1,0.5) {$\color{RoyalBlue}1$};
        \node[Maroon,scale=0.75] at (1,-0.5) {$\color{Maroon}\bar{3}$};
        \node[RoyalBlue,scale=0.75] at (-1,-0.4) {$\color{RoyalBlue}4$};
        \node[RoyalBlue,scale=0.75] at (-1,0.5) {$\color{RoyalBlue}5$};
\end{tikzpicture}}^\dagger
+
\quad
\adjustbox{valign=c}
{\begin{tikzpicture}[line width=1,scale=1]
        \draw[thick,white] (-0.5,0) to [in=150,out=75] (0.75,0.5) ;
        \draw[thick,white] (0.5,0) to [in=-30,out=75-180] (-0.75,-0.5) ;
        \draw[thick] (0.5,0) to [out=120,in=60] (-0.5,0);
        \draw[thick] (0.5,0) to [out=-120,in=-60] (-0.5,0);
        \draw[thick,RoyalBlue] (0.5,0) -- ++ (45:0.5);
        \draw[thick,Maroon] (0.5,0) -- ++ (-45:0.5);
        \draw[thick,color=RoyalBlue] (-0.5,0) -- ++ (180-45:0.5);
        \draw[thick,color=RoyalBlue] (-0.5,0) -- ++ (180+45:0.5);
        \draw[thick,color=Maroon] (0.5,0) to [in=-30,out=75-180] (-0.75,-0.5) ;
        \node[scale=0.75,color=Maroon] at (-0.87,-0.65) {$\color{Maroon}\bar{2}$};
        \node[scale=0.75,color=RoyalBlue] at (1,0.5) {$\color{RoyalBlue}1$};
        \node[scale=0.75,color=Maroon] at (1,-0.5) {$\color{Maroon}\bar{3}$};
        \node[scale=0.75,color=RoyalBlue] at (-1,-0.4) {$\color{RoyalBlue}4$};
        \node[scale=0.75,color=RoyalBlue] at (-1,0.5) {$\color{RoyalBlue}5$};
        \draw[dashed,orange] (0,0.8) -- (0,-0.8);
\end{tikzpicture}}
\end{equation}
The right-hand side is a sum over the conjugated amplitude and the unitarity cut, so the two contributions sum up to the amplitude $\cM_{542\ot13}^{\text{bub}}$. The easiest way to see this explicitly is by noticing that the cut term in \eqref{eq:bubCross} is a unitarity cut (or Cutkosky cut), so it is equal to twice the imaginary part of the amplitude. Therefore, 
\begin{equation}
    \mbox{RHS of \eqref{eq:bubCross}}= \cM_{345\leftarrow12}^{\text{bub}\dagger}+2\text{Im} \cM_{345\leftarrow12}^{\text{bub}}= \cM_{345\leftarrow12}^{\text{bub}}\,,
\end{equation}
and the direct computation agrees with the crossing prediction from  \eqref{eq:5ptcross1_OL}.

In fact, an analogous argument also holds for crossing any clusters of particles for any two-point Feynman diagram by unitarity. Indeed, both evaluate to the same class of functions; by dimensional analysis, one has
\begin{equation}
\cM_{m+1 \ldots n \leftarrow 12\ldots m}^{\text{bub}}\equiv 
\adjustbox{valign=c}{\begin{tikzpicture}[line width=1,scale=0.6]
    \draw[] (-2,0) -- ++ (150:1);
    \draw[] (-2,0) -- ++ (-160:1);
    \draw[] (-2,0) -- ++ (-150:1);
    \draw[] (-2,0) -- ++ (-140:1);
    \draw[] (2,0) -- ++ (30:1);
    \draw[] (2,0) -- ++ (-20:1);
    \draw[] (2,0) -- ++ (-30:1);
    \draw[] (2,0) -- ++ (-40:1);
    \draw[] (-2,0) to [in=150,out=30] (2,0) ;
    \draw[] (-2,0) to [in=-150,out=-30] (2,0) ;
    \draw[] (-2,0) to [in=-160,out=-20] (2,0) ;
    \draw[] (-2,0) to [in=-170,out=-10] (2,0) ;
    \draw[] (-2,0) to [in=150,out=40] (3,1) ;
    \draw[] (-2,0) to [in=150,out=40] (3,1.5) ;
    \draw[] (2,0) to [in=-30,out=-140] (-3,-1) ;
    \draw[] (2,0) to [in=-30,out=-140] (-3,-1.5) ;
    \node[scale=0.5] at (-2.7,0.2) {$\vdots$};
    \node[scale=0.5] at (2.7,0.2) {$\vdots$};
    \node[scale=0.5] at (-1.0,0.25) {$\vdots$};
    \node[scale=0.5] at (1.0,0.25) {$\vdots$};
    \node[scale=0.5] at (-2.7,-1.3) {$\vdots$};
    \node[scale=0.5] at (2.7,1.5) {$\vdots$};
    \filldraw[fill=gray!5] (0,0) circle (0.8) node {$T$};
\end{tikzpicture}}
\propto (-s_{i,i+1,\ldots,j}-i\varepsilon)^{a}\,,
\end{equation}
 where both $a$ and the constant of proportionality are independent of $s_{i,i+1,\ldots,j}$.
 
\subsection{Triangles}
For the next loop integral at five points, we add one more propagator and consider the massless scalar triangle amplitude $\cM^\tri$,
\begin{equation}
\begin{gathered}
\begin{tikzpicture}[line width=1]
\draw[] (0,0) -- (1,0.5) -- (1,-0.5) -- (0,0);
\draw[] (1.4,0.9) -- (1,0.5);
\draw[] (1.5,-0.75) -- (1,-0.5) -- (0.5,-0.75);
\draw[] (-0.5,0.25) -- (0,0) -- (-0.5,-0.25);
\node[] at (-1.5,0) {$s_{45}$};
\node[] at (1,-1.2) {$s_{13}$};
\node[scale=0.75] at (1.6,1.0) {$2$};
\node[scale=0.75] at (1.7,-0.8) {$1$};
\node[scale=0.75] at (0.3,-0.8) {$3$};
\node[scale=0.75] at (-0.8,0.2) {$4$};
\node[scale=0.75] at (-0.8,-0.2) {$5$};
\end{tikzpicture}
\end{gathered}
\hspace{1cm}
\begin{gathered}
\begin{tikzpicture}[line width=1]
\draw[] (0,0) -- (1,0.5) -- (1,-0.5) -- (0,0);
\draw[->,-latex reversed] (0,0) -- (0.5,0.25);
\draw[->,-latex reversed] (1,0.5) -- (1,-0.05);
\draw[->,-latex reversed] (1,-0.5) -- (0.4,-0.2);
\draw[] (1.4,0.9) -- (1,0.5);
\draw[] (1.5,-0.75) -- (1,-0.5) -- (0.5,-0.75);
\draw[] (-0.5,0.25) -- (0,0) -- (-0.5,-0.25);
\node[] at (-1.4,0) {$p_{45}$};
\node[] at (1.65,1.05) {$p_{2}$};
\node[] at (1,-1.2) {$p_{13}$};
\node[] at (1.7,0) {$\ell-p_{13}$};
\node[] at (0.25,-0.5) {$\ell$};
\node[] at (-0.15,0.5) {$\ell+p_{45}$};
\draw[->,gray,line width=0.7] (-0.5,0) -- (-0.8,0);
\draw[->,gray,line width=0.7] (1,0.8) --++(45:0.3);
\draw[->,gray,line width=0.7] (1,-0.7) --++(-90:0.3);
\end{tikzpicture}
\end{gathered}
\label{eq:tridiagram5pt}
\end{equation}
along with all permutations of the external legs.
The amplitude in loop-momentum space is given by
\begin{equation}
    \cM^\tri =\int \frac{\rd^\D \ell}{i\pi^{\D/2}} \frac{1}{[\ell^2-i\varepsilon][(\ell-p_{13})^2-i\varepsilon] [(\ell+p_{45})^2-i\varepsilon]} \,,
\end{equation}
which evaluates to
\begin{equation}\label{eq:TRIResult}
    \cM^\tri = \frac{\Gamma\left(3- \frac{\D}{2} \right) \Gamma \left( \frac{\D}{2}-2 \right)^2}{\Gamma\left(\D-3\right)}
    \frac{(-s_{13}-i\varepsilon)^{\D/2-2}-(-s_{45}-i\varepsilon)^{\D/2-2}}{s_{45}-s_{13}}\,.
\end{equation}
This expression is valid in any physical kinematic region, and the $i \varepsilon$'s are needed only if the corresponding $s_{ij}>0$. Note that this expression is invariant under placing $\pm i\varepsilon$ in the denominator since the numerator makes corrections proportional to $\varepsilon$ at $s_{45}=s_{13}$.

Let us check the crossing equation~\eqref{eq:5ptcross1_OL} when crossing particles $2$ and $3$ in $345 \ot 12$ kinematics, i.e., starting in the region where $s_{13}<0<s_{45}$. The result of crossing is
\begin{equation}
    \left[\cM_{345\leftarrow12}^\tri\right]_{\raisebox{\depth}{\scalebox{1}[-1]{$\curvearrowright$}}s_{13}}
    =
    \frac{\Gamma\left(3- \frac{\D}{2} \right) \Gamma \left( \frac{\D}{2}-2 \right)^2}{\Gamma\left(\D-3\right)}
    \frac{(-s_{13}+i\varepsilon)^{\D/2-2}-(-s_{45}-i\varepsilon)^{\D/2-2}}{s_{45}-s_{13}+i\varepsilon} \,.
    \label{eq:MtriCrossingres}
\end{equation}
To check whether this result agrees with the right-hand side of~\eqref{eq:5ptcross1_OL}, we have to compute the conjugated amplitude and the cut in $s_{45}$ channel, according to
\begin{equation}\label{eq:exCrossingTri}
    \adjustbox{valign=c}{\begin{tikzpicture}[line width=1]
\draw[] (0,0) -- (1,0.5) -- (1,-0.5) -- (0,0);
\draw[color=Maroon] (1.4,0.5) -- (1,0.5);
\draw[color=RoyalBlue] (1.5,-0.75) -- (1,-0.5);
\draw[color=Maroon] (1,-0.5) -- (0.5,-0.75);
\draw[color=RoyalBlue] (-0.5,0.25) -- (0,0) -- (-0.5,-0.25);
\node[scale=0.75] at (1.6,0.6) {\color{Maroon}$2$};
\node[scale=0.75] at (1.7,-0.8) {\color{RoyalBlue}$1$};
\node[scale=0.75] at (0.3,-0.8) {\color{Maroon}$3$};
\node[scale=0.75] at (-0.8,0.2) {\color{RoyalBlue}$4$};
\node[scale=0.75] at (-0.8,-0.2) {\color{RoyalBlue}$5$};
    \begin{scope}[xshift=-25pt,yshift=28pt]  
    \draw[line width=1, line cap=round,yshift=-5pt] (0,-1.8) -- (-0.135,-1.8) (0,0) -- (-0.135,0) (-0.135,-1.8) -- (-0.135,0);
    \end{scope}
    \begin{scope}[xshift=51pt,yshift=28pt,xscale=-1]  
    \draw[line width=1, line cap=round,yshift=-5pt] (0,-1.8) -- (-0.135,-1.8) (0,0) -- (-0.135,0) (-0.135,-1.8) -- (-0.135,0) node[right,yshift=-50]{${}_{\raisebox{\depth}{\scalebox{1}[-1]{$\curvearrowright$}}s_{13}}$};
    \end{scope}
 \begin{scope}[yshift=-28pt,xshift=23pt]
      \draw[line width=0.4, line cap=round] (-1.6,0) -- (-1.6,-0.135) (0.9,0) -- (0.9,-0.135) (0.9,-0.135)  -- (-1.6,-0.135) node[below, midway]{$\subset \cM$};
 \end{scope}
\end{tikzpicture}}\stackrel{?}{=}\quad
\adjustbox{valign=c}{\begin{tikzpicture}[line width=1]
\draw[] (0,0) -- (1,0.5) -- (1,-0.5) -- (0,0);
\draw[color=Maroon] (1,0.5) -- (0.4,0.5);
\draw[color=RoyalBlue] (1.5,-0.75) -- (1,-0.5);
\draw[color=Maroon] (1,-0.5) -- (1.5,-0.3);
\draw[color=RoyalBlue] (-0.5,0.25) -- (0,0) -- (-0.5,-0.25);
\node[scale=0.75] at (0.2,0.5) {\color{Maroon}$\bar{2}$};
\node[scale=0.75] at (1.7,-0.8) {\color{RoyalBlue}$1$};
\node[scale=0.75] at (1.7,-0.35) {\color{Maroon}$\bar{3}$};
\node[scale=0.75] at (-0.8,0.2) {\color{RoyalBlue}$4$};
\node[scale=0.75] at (-0.8,-0.2) {\color{RoyalBlue}$5$};
\begin{scope}[xshift=-20pt,yshift=0pt]
    \draw[dashed,color=Orange!] (0.5,-0.7) -- (0.5,0.7) ;
\end{scope}
\begin{scope}[xshift=20pt,yshift=0pt]
    \draw[dashed,color=Orange!] (0.5,-0.7) -- (0.5,0.7) ;
\end{scope}
    \begin{scope}[xshift=27.5pt,yshift=-20pt]
    \begin{scope}[xshift=-1pt,yshift=0pt]
    \draw[line width=0.4, line cap=round,yshift=-7] (-1.105,0) -- (-1.105,-0.135) (0.18,0)  -- (0.18,-0.135) (-1.105,-0.135) -- (0.18,-0.135) node[below, midway,yshift=2pt]{$\subset \cM^\dag$};
    \end{scope}
    \draw[line width=0.4, line cap=round,yshift=-7] (-1.8,0) -- (-1.8,-0.135) (-1.24,0)  -- (-1.24,-0.135) (-1.8,-0.135) -- (-1.24,-0.135) node[below, midway,yshift=0pt]{$\subset S$};
    \draw[line width=0.4, line cap=round,yshift=-7] (0.28,0) -- (0.28,-0.135) (0.8,0)  -- (0.8,-0.135) (0.28,-0.135) -- (0.8,-0.135) node[below, midway,yshift=0pt]{$\subset S$};
    \end{scope}
\end{tikzpicture}}\quad+\quad
\adjustbox{valign=c}{\begin{tikzpicture}[line width=1]
\draw[] (0,0) -- (1,0.5) -- (1,-0.5) -- (0,0);
\draw[color=Maroon] (1,0.5) -- (0.4,0.5);
\draw[color=RoyalBlue] (1.5,-0.75) -- (1,-0.5);
\draw[color=Maroon] (1,-0.5)-- (1.5,-0.3);
\draw[color=RoyalBlue] (-0.5,0.25) -- (0,0) -- (-0.5,-0.25);
\begin{scope}[xshift=-3pt,yshift=0pt]
\draw[dashed,color=Orange!] (0.5,-0.7) -- (0.5,0.7) ;
\end{scope}
\node[scale=0.75] at (0.2,0.5) {\color{Maroon}$\bar{2}$};
\node[scale=0.75] at (1.7,-0.8) {\color{RoyalBlue}$1$};
\node[scale=0.75] at (1.7,-0.35) {\color{Maroon}$\bar{3}$};
\node[scale=0.75] at (-0.8,0.2) {\color{RoyalBlue}$4$};
\node[scale=0.75] at (-0.8,-0.2) {\color{RoyalBlue}$5$};
\begin{scope}[xshift=20pt,yshift=0pt]
    \draw[dashed,color=Orange!] (0.5,-0.7) -- (0.5,0.7) ;
\end{scope}
    \begin{scope}[xshift=27.5pt,yshift=-20pt]
    \begin{scope}[xshift=-1pt,yshift=0pt]
    \draw[line width=0.4, line cap=round,yshift=-7] (-0.47,0) -- (-0.47,-0.135) (0.18,0)  -- (0.18,-0.135) (-0.47,-0.135) -- (0.18,-0.135) node[below, midway,yshift=2pt]{$\subset \cM^\dag$};
    \end{scope}
    \draw[line width=0.4, line cap=round,yshift=-7] (-1.8,0) -- (-1.8,-0.135) (-0.64,0)  -- (-0.64,-0.135) (-1.8,-0.135) -- (-0.64,-0.135) node[below, midway,yshift=0pt]{$\subset S$};
    \draw[line width=0.4, line cap=round,yshift=-7] (0.28,0) -- (0.28,-0.135) (0.8,0)  -- (0.8,-0.135) (0.28,-0.135) -- (0.8,-0.135) node[below, midway,yshift=0pt,xshift=2pt]{$\subset S$};
    \end{scope}
\end{tikzpicture}}
\end{equation}
In an equation, this reads
\begin{equation}
    \left[\cM_{345\leftarrow12}^\tri\right]_{\raisebox{\depth}{\scalebox{1}[-1]{$\curvearrowright$}}s_{13}} \stackrel{?}{=} \cM^\text{tri\dag}_{245 \ot 1 3} + \cut_{s_{45}} \cM^\text{tri}_{245 \ot 1 3}\,,
\end{equation}
where the cut term on the right-hand side is defined by the rightmost picture in~\eqref{eq:exCrossingTri}. The conjugated amplitude $\cM^\text{tri\dag}_{245 \ot 1 3}$ is easily obtained from~\eqref{eq:TRIResult}, and the cut is given by
\begin{equation}
        \cut_{s_{45}} \cM^\text{tri}_{245 \ot 1 3}  = - \left(2\pi \right)^2  \int \frac{\d^\D \ell}{i\pi^{\D/2}} \frac{\delta^-[\ell^2] \delta^+\big[(\ell+p_{45})^2\big]}{(\ell-p_{13})^2+i\varepsilon}\,.
\end{equation}
The calculation of the cut term proceeds analogously to the $s_{45}$ cut of the six-point triangle diagram from Sec.~\ref{sec:triloop}, with the modifications that we now take one external leg to be massless and work in $\D$ dimensions instead of four.
We define $\ell^0_{\ast} \equiv -\frac{\sqrt{s_{45}}}{2}$ as the value of $\ell^0$ and $|\vec{\ell}|$ imposed by the delta functions. After performing the two delta functions, we can write the cut in spherical coordinates as
\begin{equation}
        \Cut_{s_{45}} \cM^\text{tri}_{2456 \ot 13}=- \frac{i \Omega_{\D-2}}{2 \pi^{\D/2-2}} \left( \frac{s_{45}}{4} \right)^{\frac{\D-4}{2}}  \int_{-1}^1  \frac{(1-\cos^2\theta)^{\frac{\D-4}{2}}\d \cos \theta}{s_{13}-2 \ell^0_\ast (p_{13}^0-\vert \vec{p}_{13}\vert \cos \theta)- i \varepsilon}\,,
\end{equation}
where $\Omega_{d} = \frac{2 \pi^{d/2}}{\Gamma(d/2)}$ is the solid angle of a sphere in $d$-dimensions. In the center-of-mass frame of $s_{13}$, we have
\begin{equation}
    p_{13}^0 = - \frac{s_{45}+s_{13}}{2 \sqrt{s_{45}}} \quad \text{and} \quad |\vec{p}_{13}| = - \frac{s_{45}-s_{13}}{2\sqrt{s_{45}}} \,,
\end{equation}
such that 
\begin{equation}
        \Cut_{s_{45}} \cM^\text{tri}_{2456 \ot 13} = 
        \frac{2\pi i s_{45}^{\D/2-2} \Gamma\left(\frac{\D}{2}-2 \right)}{(s_{45}-s_{13}+i\varepsilon)\Gamma(\D-3)} \,.
\end{equation}
Using the identity
\begin{equation}
    (-s_{45}+i\varepsilon)^{\D/2-2} - (-s_{45}-i\varepsilon)^{\D/2-2} = \frac{2 \pi i}{\Gamma\left( \frac{\D}{2}-2 \right) \Gamma\left(3-\frac{\D}{2}\right)} s_{45}^{\D/2-2} \,,
\end{equation}
where we have assumed that $s_{45}>0$, we get
\begin{equation}
    \begin{split}
    \cM^\text{tri\dag}_{245 \ot 1 3} + \cut_{s_{45}} \cM^\text{tri}_{245 \ot 1 3}
    & =
    \frac{\Gamma\left(3- \frac{\D}{2} \right) \Gamma \left( \frac{\D}{2}-2 \right)^2}{\Gamma\left(\D-3\right)(s_{45}-s_{13}+i\varepsilon)}
    \\
    & \hspace{-3cm}\times
    \left[
    (-s_{13}+i\varepsilon)^{\D/2-2}-(-s_{45}+i\varepsilon)^{\D/2-2}
    +
    \frac{2 \pi i}{\Gamma\left( \frac{\D}{2}-2 \right) \Gamma\left(3-\frac{\D}{2}\right)} s_{45}^{\D/2-2}
    \right]
    \\
    & \hspace{-3cm} = 
    \frac{\Gamma\left(3- \frac{\D}{2} \right) \Gamma \left( \frac{\D}{2}-2 \right)^2}{\Gamma\left(\D-3\right)}
    \frac{(-s_{13}+i\varepsilon)^{\D/2-2}-(-s_{45}-i\varepsilon)^{\D/2-2}}{s_{45}-s_{13}+i\varepsilon}\,,
    \end{split}
\end{equation}
where an $i \varepsilon$ can be freely added to $\cM^\text{tri\dag}_{245 \ot 1 3}$, as explained below~\eqref{eq:TRIResult}. This result agrees with the one obtained by crossing in~\eqref{eq:MtriCrossingres}, thus verifying the crossing equation for this diagram. Other kinematic channels can be checked analogously.

\subsection{Boxes (at four and five points)}
\label{sec:boxes_oneloop}

\paragraph{Warm-up: Four-point massless box}
Shifting gears, we now consider the four-point four propagators box 
\begin{equation}
\begin{gathered}
\begin{tikzpicture}[line width=1]
    \coordinate (a) at (-0.5,-0.5);
    \coordinate (b) at (-0.5,0.5);
    \coordinate (c) at (0.5,0.5);
    \coordinate (d) at (0.5,-0.5);
    \draw[solid] (a) -- (b);
    \draw[solid] (b) -- (c);
    \draw[solid] (c) -- (d);
    \draw[solid] (d) -- (a);
    \draw[] (a) -- ++(-145:0.5);
    \draw[] (b) -- ++(145:0.5);
    \draw[] (c) -- ++(45:0.5);
    \draw[] (d) -- ++(-45:0.5);
    \node[] at (1.8,0) {$s$};
    \node[] at (0,1.5) {$t$};
    \draw[->,line width=0.75] (1.5,0) -- ++(180:0.3);
    \draw[->,line width=0.75] (0,1.2) -- ++(-90:0.3);
    \node[scale=0.75] at (1.1,0.95) {$1$};
    \node[scale=0.75] at (1.1,-0.95) {$2$};
    \node[scale=0.75] at (-1.1,-0.95) {$3$};
    \node[scale=0.75] at (-1.1,0.95) {$4$};
    \node[color=white] at (1.1,-1.25) {$p_{2}$};
    \node[color=white] at (1.1,1.25) {$p_{1}$};
    \node[color=white] at (-0.8,1.25) {$p_{45}$};
    \node[color=white] at (-1.1,-1.25) {$p_{3}$};
\end{tikzpicture}
\end{gathered}
\hspace{1cm}
\begin{gathered}
\begin{tikzpicture}[line width=1]
\coordinate (a) at (-0.5,-0.5);
    \coordinate (b) at (-0.5,0.5);
    \coordinate (c) at (0.5,0.5);
    \coordinate (d) at (0.5,-0.5);
    \coordinate (e) at (0,0);
    \draw[-latex-=.65] (a) -- (b);
    \draw[-latex-=.65] (b) -- (c);
    \draw[-latex-=.65] (c) -- (d);
    \draw[-latex-=.65] (d) -- (a);
    \draw[] (a) -- ++(-135:0.5);
    \draw[] (b) -- ++(135:0.5);
    \draw[] (c) -- ++(45:0.5);
    \draw[] (d) -- ++(-45:0.5);
    \node[] at (1.1,-1.25) {$p_{2}$};
    \node[] at (1.1,1.25) {$p_{1}$};
    \node[] at (-1.1,1.25) {$p_{4}$};
    \node[] at (-1.1,-1.25) {$p_{3}$};
    \node[] at (1.4,0) {$\ell-p_{1}$};
    \node[] at (-1.4,0) {$\ell-p_{123}$};
    \node[] at (0,-0.95) {$\ell-p_{12}$};
    \node[] at (0,0.95) {$\ell$};
    \node[color=white] at (0,1.5) {$t$};
    \draw[->,gray,line width=0.7] (0.8,0.5) --++(45:0.3);
    \draw[->,gray,line width=0.7] (-0.8,0.5) --++(135:0.3);
     \draw[<-,gray,line width=0.7] (-1,-0.7) --++(45:0.3);
    \draw[->,gray,line width=0.7] (0.8,-0.5) --++(-45:0.3);
\end{tikzpicture}
\end{gathered}
\label{eq:boxdiagram}
\end{equation}
The corresponding amplitude is given by
\begin{equation}
\label{eq:box0}
    \cM_{34\leftarrow12}^{\boxx} = \int \frac{\d^\D \ell}{i\pi^{\D/2}} \frac{1}{[\ell^2-i\varepsilon][(\ell-p_{1})^2-i\varepsilon] [(\ell-p_{12})^2-i\varepsilon] [(\ell-p_{123})^2-i\varepsilon]} \,,
\end{equation}
with $p_i^2=0$, $-p_{12}^2=s$ and $-p_{23}^2=t$. We also define $-p_{13}^2=u$. The physical $s$-channel is defined using $s>0$, $t<0$, $u<0$, and the $t$ and $u$ channels are defined analogously by symmetry. Therefore, we can check the crossing relations between any two of these kinematic regions.

In the Euclidean region (corresponding to the $u$-channel where $s<0$ and $t<0$), the integral in \eqref{eq:box0} evaluates to~\cite{Bern:1993kr}
\begin{multline}
    \cM_{34\leftarrow12}^{\boxx} = \lim_{\varepsilon\to 0^+}\frac{C(\epsilon)}{\epsilon^2 s t} \Big[ 
    (-s)^{-\epsilon} \,_2F_1 \left(1,-\epsilon,1-\epsilon;z_1 \pm i\varepsilon\right)
    \\ +
    (-t)^{-\epsilon} \,_2F_1 \left(1,-\epsilon,1-\epsilon;z_2 \mp i\varepsilon \right) \Big]\,,
    \label{eq:masslessbox}
\end{multline}
in $\D=4-2\epsilon$. Here, $C(\epsilon)=\frac{2 \Gamma (1-\epsilon)^2 \Gamma(\epsilon+1)}{\Gamma (1-2 \epsilon)}$ and the variables $z_1$ and $z_2$ are given in terms of the kinematic invariants as
\begin{equation}
    z_1=1+\frac{s}{t} \quad \text{and} \quad z_2=1+\frac{t}{s}\,.
\end{equation}
 Note that in the kinematic regions where $\frac{s}{t}>0$, the $ i\varepsilon$ in the hypergeometric functions are arbitrary as long as they have opposite signs, resulting in a continuous function in the region where $s<0$ and $t<0$.
 
In the $s$- and $t$-channel kinematic regions, the amplitude is given by \eqref{eq:masslessbox} after the replacements $(-s)^{-\epsilon}\to (-s-i\varepsilon)^{-\epsilon}$ and $(-t)^{-\epsilon} \to (-t - i\varepsilon)^{-\epsilon}$, respectively. Whenever $s$ and $t$ have opposite signs, the $i\varepsilon$ in the hypergeometric functions are not needed. The $i \varepsilon$-prescriptions are summarized in Tab.~\ref{tab:masslessbox}.

\begin{table}[t]
\centering
\begin{tabular}{c|c|c|c|c} 
 Region & $s$ & $t$ & $z_1$ & $z_2$ \\ 
 \hline
 $u>0$, $s<0$, $t<0$ & $\bullet$ & $\bullet$ & $\pm i \varepsilon$ & $\mp i \varepsilon$  \\ 
 $s>0$, $t<0$, $u<0$ & $+i\varepsilon$ & $\bullet$ & $\bullet$ & $\bullet$ \\
 $t>0$, $u<0$, $s<0$ & $\bullet$ & $+i\varepsilon$ & $\bullet$ & $\bullet$ \\
\end{tabular}
\caption{Signs of the imaginary parts of $s$ and $t$ and the arguments of the hypergeometric functions for the amplitude $\mathcal{M}$ in different kinematic regions. A bullet ``$\bullet$'' denotes that no $i\varepsilon$ is needed. The conjugated amplitude $\mathcal{M}^\dagger$ is obtained by flipping the signs of all $i \varepsilon$.}
\label{tab:masslessbox}
\end{table}

The crossing relations for the massless box are straightforward to establish. For a $2\leftarrow2$ process, we are only allowed to cross two particles thanks to stability. We start by looking at crossings starting in the $u$-channel region, where $s<0$ and $t<0$. We can, for example, cross particles $2$ and $3$, which rotates $s$ to be timelike with a negative imaginary part. During the rotation, the hypergeometric functions in  \eqref{eq:masslessbox} remain on their principal sheet, and after rotation the $i\varepsilon$ for $z_1$ and $z_2$ will not play any role
\begin{equation}
\begin{gathered}
\begin{tikzpicture}
  \draw[->,thick] (-1.5, 0) -- (1.5, 0);
  \draw[->,thick,white] (0, -1.6) -- (0, 1.6);
  \draw[->,thick] (0, -1.25) -- (0, 1.25);
  \node[] at (1.1,1.05) {$s$};
  \draw[] (1.35,0.85) -- (0.8,0.85) -- (0.8,1.25);
  \draw[Maroon,fill=Maroon,thick] (-1,0) circle (0.05);
  \draw[->,Maroon,thick] (-1,0) arc (0:175:-1);
  \draw[decorate, decoration={zigzag, segment length=6, amplitude=2}, black!80] (0,0) -- (1.5,0);
  \draw[black!80,fill=black!80,thick] (0,0) circle (0.05);
\end{tikzpicture}
\hspace{2.5cm}
\begin{tikzpicture}
  \draw[->,thick,white] (0, -1.6) -- (0, 1.6);
  \draw[->,thick] (-1, 0) -- (2.2, 0);
  \draw[->,thick] (0, -1.5) -- (0, 1.5);
  \node[] at (1.8,1.3) {$z_i$};
  \draw[] (2.0,1.1) -- (1.6,1.1) -- (1.6,1.5);
  \draw[decorate, decoration={zigzag, segment length=6, amplitude=2}, black!80] (1,0) -- (2,0);
  \draw[black!80,fill=black!80,thick] (1,0) circle (0.05);
  \draw[Maroon,fill=Maroon,thick] (1.7,0.1) circle (0.05);
  \draw[->,Maroon,thick] (1.7,0.1) arc (0:178:1.2);
  \node[Maroon] at (-0.6,1.1) {$z_1$};
  \draw[Maroon,fill=Maroon,thick] (1.2,-0.1) circle (0.05);
  \draw[->,Maroon,thick] (1.2,-0.1) arc (0:-178:0.25);
  \node[Maroon] at (0.6,-0.4) {$z_2$};
\end{tikzpicture}
\end{gathered}
\end{equation}
Moreover, the prefactor $(-s)^{-\epsilon}$ will rotate to become $(-s+i\varepsilon)^{-\epsilon}$, so the $u$-channel amplitude has rotated to the conjugated $s$-channel amplitude
\begin{equation}\label{eq:box1}
\left[\cM_{24\leftarrow13}^{\boxx}\right]_{ 
\rotatedown s}
 =
 \cM_{34\leftarrow12}^{\boxx \dagger }\,.
\end{equation}
Similarly, we obtain the rotations from the $u$-channel to the $t$-channel by symmetry. We note that for purely massless $2\leftarrow2$ scattering, there is no mass scale on which a corner cut could depend on, resulting in the vanishing of the box $p_i^2$-cut for all $i$ \cite[Eq.~(4.20)]{Abreu:2017ptx}. Thus, equation \eqref{eq:box1} (as well as its $u$- and $t$-channel analogues) agrees with the crossing-equation prediction.

Next, let us check crossing when starting in the $s$-channel ($s>0$, $t<0$ and $u<0$). In this channel, the amplitude is given by \eqref{eq:masslessbox} after replacing $(-s)^{-\epsilon} \to (-s-i\varepsilon)^{-\epsilon}$. After the analytic continuation, we get $z_1\to z_1-i\varepsilon$ and $z_2 \to z_2 + i\varepsilon$,
\begin{equation}
\begin{gathered}
\begin{tikzpicture}
  \draw[->,thick] (-1.5, 0) -- (1.5, 0);
  \draw[->,thick,white] (0, -1.6) -- (0, 1.6);
  \draw[->,thick] (0, -1.25) -- (0, 1.25);
  \node[] at (1.1,1.05) {$s$};
  \draw[] (1.35,0.85) -- (0.8,0.85) -- (0.8,1.25);
  \draw[Maroon,fill=Maroon,thick] (1,0.1) circle (0.05);
  \draw[->,Maroon,thick] (1,0.1) arc (0:180:1);
  \draw[decorate, decoration={zigzag, segment length=6, amplitude=2}, black!80] (0,0) -- (1.5,0);
  \draw[black!80,fill=black!80,thick] (0,0) circle (0.05);
\end{tikzpicture}
\hspace{2.5cm}
\begin{tikzpicture}
  \draw[->,thick,white] (0, -1.6) -- (0, 1.6);
  \draw[->,thick] (-1, 0) -- (2.2, 0);
  \draw[->,thick] (0, -1.5) -- (0, 1.5);
  \node[] at (1.8,1.3) {$z_i$};
  \draw[] (2.0,1.1) -- (1.6,1.1) -- (1.6,1.5);
  \draw[decorate, decoration={zigzag, segment length=6, amplitude=2}, black!80] (1,0) -- (2,0);
  \draw[black!80,fill=black!80,thick] (1,0) circle (0.05);
  \draw[Maroon,fill=Maroon,thick] (-0.5,0.0) circle (0.05);
  \draw[->,Maroon,thick] (-0.5,0.0) arc (0:170:-1.2);
  \node[Maroon] at (1.8,-1.3) {$z_1$};
  \draw[Maroon,fill=Maroon,thick] (0.5,0.0) circle (0.05);
  \draw[->,Maroon,thick] (0.5,0.0) arc (0:-170:-0.5);
  \node[Maroon] at (1.0,0.7) {$z_2$};
\end{tikzpicture}
\end{gathered}
\end{equation}
According to Tab.~\ref{tab:masslessbox}, the resulting expression is, once again, the complex-conjugated amplitude
\begin{equation}
\left[\cM_{34\leftarrow12}^{\boxx} \right]_{\rotateup s}
 =
 \cM_{24\leftarrow13}^{\boxx\dag}\,.
\end{equation}
All other crossing relations are obtained by symmetry.

\paragraph{Five-point massless boxes}
Next, we verify the crossing relations for the box amplitude,
\begin{equation}
\begin{gathered}
\begin{tikzpicture}[line width=1]
    \coordinate (a) at (-0.5,-0.5);
    \coordinate (b) at (-0.5,0.5);
    \coordinate (c) at (0.5,0.5);
    \coordinate (d) at (0.5,-0.5);
    \draw[solid] (a) -- (b);
    \draw[solid] (b) -- (c);
    \draw[solid] (c) -- (d);
    \draw[solid] (d) -- (a);
    \draw[] (a) -- ++(-135:0.5);
    \draw[] (b) -- ++(135:0.5);
    \draw[] (b) -- ++(155:0.5);
    \draw[] (c) -- ++(45:0.5);
    \draw[] (d) -- ++(-45:0.5);
    \draw[] node at (-1.45,1.2) {$s_{45}$};
    \node[] at (1.8,0) {$s_{12}$};
    \node[] at (0,1.5) {$s_{23}$};
    \draw[->,line width=0.75] (1.5,0) -- ++(180:0.3);
    \draw[->,line width=0.75] (0,1.2) -- ++(-90:0.3);
    \node[scale=0.75] at (1.1,0.95) {$1$};
    \node[scale=0.75] at (1.1,-0.95) {$2$};
    \node[scale=0.75] at (-1.05,-0.95) {$3$};
    \node[scale=0.75] at (-0.9,1) {$5$};
    \node[scale=0.75] at (-1.1,0.8) {$4$};
    \node[color=white] at (1.1,-1.25) {$p_{2}$};
    \node[color=white] at (1.1,1.25) {$p_{1}$};
    \node[color=white] at (-0.8,1.25) {$p_{45}$};
    \node[color=white] at (-1.1,-1.25) {$p_{3}$};
\end{tikzpicture}
\end{gathered}
\hspace{1cm}
\begin{gathered}
\begin{tikzpicture}[line width=1]
\coordinate (a) at (-0.5,-0.5);
    \coordinate (b) at (-0.5,0.5);
    \coordinate (c) at (0.5,0.5);
    \coordinate (d) at (0.5,-0.5);
    \coordinate (e) at (0,0);
    \draw[-latex-=.65] (a) -- (b);
    \draw[-latex-=.65] (b) -- (c);
    \draw[-latex-=.65] (c) -- (d);
    \draw[-latex-=.65] (d) -- (a);
    \draw[] (a) -- ++(-135:0.5);
    \draw[] (b) -- ++(155:0.5);
    \draw[] (b) -- ++(135:0.5);
    \draw[] (c) -- ++(45:0.5);
    \draw[] (d) -- ++(-45:0.5);
    \node[] at (1.1,-1.25) {$p_{2}$};
    \node[] at (1.1,1.25) {$p_{1}$};
    \node[] at (-1.1,1.25) {$p_{45}$};
    \node[] at (-1.1,-1.25) {$p_{3}$};
    \node[] at (1.4,0) {$\ell-p_{1}$};
    \node[] at (-1.4,0) {$\ell-p_{123}$};
    \node[] at (0,-0.95) {$\ell-p_{12}$};
    \node[] at (0,0.95) {$\ell$};
    \node[color=white] at (0,1.5) {$s_{23}$};
    \draw[->,gray,line width=0.7] (0.8,0.5) --++(45:0.3);
    \begin{scope}[xshift=0,yshift=6]
        \draw[->,gray,line width=0.7] (-0.8,0.5) --++(147:0.3);
    \end{scope}
     \draw[<-,gray,line width=0.7] (-1,-0.7) --++(45:0.3);
    \draw[->,gray,line width=0.7] (0.8,-0.5) --++(-45:0.3);
\end{tikzpicture}
\end{gathered}
\label{eq:1mboxdiagram}
\end{equation}
It is defined by the expression
\begin{equation}
\label{eq:box1m}
    \mathcal{M}_{345 \ot 12}^{\onembox} = \int \frac{\d^\D \ell}{i\pi^{\D/2}} \frac{1}{[\ell^2-i\varepsilon][(\ell-p_{1})^2-i\varepsilon] [(\ell-p_{12})^2-i\varepsilon] [(\ell-p_{123})^2-i\varepsilon]} \,,
\end{equation}
with $p_i^2=0$, $-p_{45}^2=s_{45}$, $-p_{12}^2=s_{12}$ and $-p_{23}^2=s_{23}$.
The integral was computed exactly in $\D=4-2\epsilon$ spacetime dimensions in \cite{Bern:1993kr,Kozlov:2015kol} and can be written as a sum of hypergeometric functions
\begin{multline}
    \cM_{345\leftarrow12}^{\onembox} =  \frac{C(\epsilon)}{\epsilon^2 s_{12} s_{23}} \big[ 
    (-s_{12})^{-\epsilon} \,_2F_1^A \left(1,-\epsilon,1-\epsilon;z_1\right)
    +
    (-s_{23})^{-\epsilon} \,_2F_1^A \left(1,-\epsilon,1-\epsilon;z_2\right)
    \\-
    (-s_{45})^{-\epsilon} \,_2F_1^A \left(1,-\epsilon,1-\epsilon;z_3\right)\big] \,,
    \label{eq:onemassbox}
\end{multline}
with $C(\epsilon)$ defined below \eqref{eq:masslessbox}. Here, the variables $z_1$, $z_2$ and $z_3$ are given in terms of the kinematic invariants as
\begin{equation}
    z_1 = 1-\frac{s_{45}-s_{12}}{s_{23}} \,, \qquad 
    z_2 = 1-\frac{s_{45}-s_{23}}{s_{12}} \,, \qquad 
    z_3 = 1-\frac{(s_{45}-s_{12})(s_{45}-s_{23})}{s_{12} s_{23}} \,.
    \label{eq:zs}
\end{equation}
Note that despite the dependence on $\epsilon$, which makes the equation more concise, \eqref{eq:onemassbox} is valid in any spacetime dimension. Above, we have used the superscript $A$ to denote that the branch of the hypergeometric functions must be specified and it depends on the kinematic region, as discussed in detail below. 

\paragraph{Kinematics and branch cuts}
\begin{table}
\centering
\begin{tabular}{c|c|c|c|c} 
 Channel & Region & $z_1$ & $z_2$ & $z_3$ \\ 
 \hline
 $345\leftarrow12$ & $s_{23}<s_{45}<s_{12}$ & $\bullet$ & $\bullet$ & $\bullet$ 
 \\
 $245\leftarrow13$ & $s_{12}<s_{23}<s_{45}$ & $+i\varepsilon$ & $-i\varepsilon$ & $\bullet$
 \\
 $245\leftarrow13$ & $s_{23}<s_{12}<s_{45}$ & $-i\varepsilon$ & $+i\varepsilon$ & $\bullet$
 \\
 $235\leftarrow14$ & $s_{12},s_{45}<s_{23}$ & $- i\varepsilon $ & $\bullet$ & $-i\varepsilon$ \\
 $135\leftarrow24$  & $s_{12},s_{23}<s_{45}<0$ & $\pm i\varepsilon$ & $\mp i\varepsilon$ & $\bullet$ \\
 $135\leftarrow24$  & $s_{45}<s_{12},s_{23}<s_{12}$ & $\bullet$ & $ \bullet$ & $\bullet$ \\
 $135\leftarrow24$ & $s_{12}<s_{45}<s_{23}$ & $- i\varepsilon $ & $\bullet$ & $-i\varepsilon$ \\
 $135\leftarrow24$ & $s_{23}<s_{45}<s_{12}$ & $\bullet$ & $-i\varepsilon$ & $-i\varepsilon $ \\
 $123\leftarrow45$ & $s_{12},s_{23}<s_{45}$, $s_{45}>0$ & $\bullet$ & $\bullet$ & $\bullet$ \\
\end{tabular}
\caption{Signs of the imaginary parts of $z_i$ for the amplitude in different kinematic regions. All other physical channels are obtained by relabeling symmetry and/or time reversal. A bullet ``$\bullet$'' denotes that no $i\varepsilon$-prescription is needed. If an imaginary part is listed for a specific region, there may still be subregions in which it is superfluous since the corresponding $z_i<0$, but we still include it to avoid further subdivisions. The conjugated amplitude is obtained by flipping the signs of all $i \varepsilon$. The entries with $\pm i \eps$ mean that one can choose either, as long as one is consistent.
}
\label{tab:onemassbox}
\end{table}
To get the correct expression in different kinematic regions, we must be careful to be on the correct side of the branch cuts of the hypergeometric functions in \eqref{eq:onemassbox}. In the Euclidean region carved by $s_{12}<0$, $s_{23}<0$ and $s_{45}<0$, we can get the correct form by using the real values: $\,_2F_1^A\left(1,-\epsilon,1-\epsilon;z_i\right) = \frac{1}{2} \sum_{\pm} \,_2F_1^A\left(1,-\epsilon,1-\epsilon;z_i \pm i \varepsilon\right)$. Equivalently, we can assign $+i\varepsilon$ and $-i\varepsilon$ to the $z_i$ such that the discontinuities vanish, which requires the two hypergeometric functions with $z_i>1$ to have opposite $i \varepsilon$.
This follows from the formula for the discontinuity of the hypergeometric function:
\begin{subequations}
\begin{align}
    \disc_{z > 1} [_2F_1(1,-\epsilon,1-\epsilon,z)]
    &\equiv 
    \,_2F_1(1,-\epsilon,1-\epsilon,z+i\varepsilon)
    -
    \,_2F_1(1,-\epsilon,1-\epsilon,z-i\varepsilon)\\
    & =
    -2 \pi i \epsilon z^\epsilon \Theta(z-1)\,,
    \label{eq:disc2F1}
\end{align}
\end{subequations}
which gives that the discontinuity of $\cM_{345\leftarrow12}^{\onembox}$ in each $z_i$ is equal and given by
\begin{equation}
    \disc_{z_i > 1} \mathcal{M}_{345 \ot 12}^{\onembox}
    = \frac{2 \pi i C(\epsilon)}{\epsilon s_{12} s_{23}} \, \left(-\frac{s_{45}}{z_3}\right)^{-\epsilon} \Theta(z_i-1) \,.
\end{equation}

In the kinematic region where $s_{45}<s_{23}<0$ and $s_{45}<s_{12}<0$, we can see from \eqref{eq:zs} that $z_i<1$ for all $i$ such that the amplitude does not have any branch cuts coming from the hypergeometric functions. Therefore, we can unambiguously write, when $s_{45}<s_{12},s_{23}<0$,

\begin{multline}
    \mathcal{M}_{345 \ot 12}^{\onembox} = \frac{C(\epsilon)}{\epsilon^2 s_{12} s_{23}} \big[ 
    (-s_{12})^{-\epsilon} \,_2F_1 \left(1,-\epsilon,1-\epsilon;z_1\right)
    +
    (-s_{23})^{-\epsilon} \,_2F_1 \left(1,-\epsilon,1-\epsilon;z_2\right)
    \\-
    (-s_{45})^{-\epsilon} \,_2F_1 \left(1,-\epsilon,1-\epsilon;z_3\right)\big] \,.
    \label{eq:onemassbox2}
\end{multline}

We will use this expression to analytically continue to other kinematic regions. Note that it is not sufficient to simply take $s_{ij} \to s_{ij}+i\varepsilon$ for the invariants that are positive; the function is written in a form that requires more care to match the correct Feynman $i\varepsilon$ prescription. A list of the $i \varepsilon$ prescriptions needed for the $z_i$ in each region is given in Tab.~\ref{tab:onemassbox}. The analytic continuation of the prefactors is simply $(-s_{ij})^{-\epsilon} \to (-s_{ij}-i\varepsilon)^{-\epsilon}$ for $s_{ij}>0$.

\paragraph{Crossing prediction}
To show that the crossing equation~\eqref{eq:5ptcross1_OL} holds, we need to check it for the crossing of any pair of particles, between any allowed kinematic regions. While we present only one example in detail, we have checked, using analogous computations, that the crossing equation holds for any other crossing channels.

As an example, we start in $345 \ot 12$ kinematics, in the region $s_{23}<0<s_{45}\ll s_{12}$, and cross particles $2$ and $3$. This places us in the region $245 \ot 13$. In the first kinematic region, one can check that the amplitude is simply given by substituting $s_{12}\to s_{12}+i\varepsilon$ and $s_{45}\to s_{45} + i\varepsilon$ into \eqref{eq:onemassbox2}. The crossing results in $s_{12}$ becoming large and negative, while $s_{23}$ and $s_{45}$ stay fixed. After the analytic continuation, we end up in the region $s_{12}\ll s_{23}<0<s_{45}$ with $z_1\to z_1-i\varepsilon$ and $z_2 \to z_2 + i\varepsilon$. This is summarized below:
\begin{equation}
\begin{gathered}
\begin{tikzpicture}
  \draw[->,thick] (-2, 0) -- (2, 0);
  \draw[->,thick,white] (0, -1.6) -- (0, 1.6);
  \draw[->,thick] (0, -1.25) -- (0, 1.25);
  \node[] at (1.1,1.05) {$s_{ij}$};
  \draw[] (1.35,0.85) -- (0.8,0.85) -- (0.8,1.25);
  \draw[Maroon,fill=Maroon,thick] (1,0.1) circle (0.05);
  \draw[->,Maroon,thick] (1,0.1) arc (0:175:1);
  \node[] at (-0.5,1.3) {$\textcolor{Maroon}{s_{12}}$};
  \draw[Maroon,fill=Maroon,thick] (-0.4,0) circle (0.05);
  \node[] at (-0.3,0.3) {$\textcolor{Maroon}{s_{23}}$};
  \draw[RoyalBlue,fill=RoyalBlue,thick] (0.5,0.1) circle (0.05);
  \node[] at (0.35,0.4) {$\textcolor{RoyalBlue}{s_{45}}$};
  \draw[black!80,fill=black!80,thick] (0,0) circle (0.05);
  \draw[decorate, decoration={zigzag, segment length=6, amplitude=2}, black!80] (0,0) -- (1.5,0);
\end{tikzpicture}
\hspace{2.5cm}
\begin{tikzpicture}
  \draw[->,thick,white] (0, -1.6) -- (0, 1.6);
  \draw[->,thick] (-1, 0) -- (2.2, 0);
  \draw[->,thick] (0, -1.5) -- (0, 1.5);
  \node[] at (1.8,1.3) {$z_i$};
  \draw[] (2.0,1.1) -- (1.6,1.1) -- (1.6,1.5);
  \draw[decorate, decoration={zigzag, segment length=6, amplitude=2}, black!80] (1,0) -- (2,0);
  \draw[black!80,fill=black!80,thick] (1,0) circle (0.05);
  \draw[Maroon,fill=Maroon,thick] (-0.5,0.0) circle (0.05);
  \draw[->,Maroon,thick] (-0.5,0.0) arc (0:170:-1.2);
  \node[Maroon] at (1.8,-1.3) {$z_1$};
  \draw[Maroon,fill=Maroon,thick] (0.5,0.0) circle (0.05);
  \draw[->,Maroon,thick] (0.5,0.0) arc (0:-170:-0.5);
  \node[Maroon] at (1.0,0.7) {$z_2$};
  \draw[RoyalBlue,fill=RoyalBlue,thick] (0.25,0.0) circle (0.05);
  \node[RoyalBlue] at (0.25,0.3) {$z_3$};
\end{tikzpicture}
\end{gathered}
\end{equation}
According to Tab.~\ref{tab:onemassbox}, the expression we end up with has the same $i\varepsilon$ prescriptions for the $z_i$'s as $\mathcal{M}^{\onembox\dag}_{245 \ot 13} $. Therefore, the analytically continued amplitude differs from $\mathcal{M}^{ \onembox\dag}_{245 \ot 13} $ by a discontinuity of the prefactor $(-s_{45})^{-\epsilon}$, which is given by
\begin{equation}
    \disc_{s_{45} > 0}(-s_{45})^{-\epsilon}= \lim_{\varepsilon \to 0^+} \left[ (-s_{45}-i\varepsilon)^{-\epsilon} - (-s_{45}+i\varepsilon)^{-\epsilon} \right]
    =
    2 i \sin (\pi \epsilon) s_{45}^{-\epsilon} \,.
\end{equation}
Putting everything together, we have, after the analytic continuation in $s_{12}$,
\begin{equation}
\left[\mathcal{M}_{345\ot 12}^{\onembox}\right]_{ 
 \rotateup s_{12}}
    = \mathcal{M}^{\onembox\dag}_{245 \ot 13} - \underbracket[0.4pt]{\frac{2iC(\epsilon)\sin(\pi \epsilon)}{\epsilon^2 s_{12} s_{23}s_{45}^{\epsilon}}  \,_2F_1 \left(1,-\epsilon,1-\epsilon;z_3\right)}_{\Delta \mathcal{M}} \,.
\label{eq:crossingpred_box}
\end{equation}

Based on the crossing equation \eqref{eq:5ptcross1_OL}, the continued amplitude should be equal to the inclusive amplitude
\begin{equation}
\label{eq:cross_cuts45}
    \left[\cM_{345\ot 12}^{\onembox}\right]_{ \rotateup
    s_{12}}
    \stackrel{?}{=} \cM^{ \onembox\dag}_{245\leftarrow13} + \Cut_{s_{45}} \cM^{\onembox}_{245\leftarrow13}\,.
\end{equation}
Therefore, to establish the crossing relation, we need to check whether the expression for the inclusive amplitude obtained by analytic continuation agrees with a direct computation.  When comparing \eqref{eq:crossingpred_box} with \eqref{eq:cross_cuts45}, one sees that this check amounts to verifying that the cut in the $s_{45}$ channel equals the difference between $\cM^{\onembox}_{245\leftarrow13}$ and its complex-conjugate, namely to check if
\begin{equation}\label{eq:toCheck_box}
    \Cut_{s_{45}} \cM^{\onembox}_{245\leftarrow13} \stackrel{?}{=}\Delta \cM \,.
\end{equation}
We now proceed to compute the cut $\Cut_{s_{45}} \cM^{\onembox}_{245\leftarrow13}$ to show that this formula is indeed true.

\paragraph{Cut computation}
The cut integral is defined by
\begin{equation}
     \Cut_{s_{45}} \cM^{\onembox}_{245\leftarrow13} = - (2\pi)^2 \int \frac{\rd^\D \ell}{i\pi^{\D/2}} \frac{\delta^-(\ell^2)\delta^+[(\ell-p_{123})^2]}{[(\ell-p_{1})^2+i\varepsilon] [(\ell-p_{12})^2+i\varepsilon]}\,.
\end{equation}
To compute this integral, we use the \emph{embedding space formalism} described in \cite{Abreu:2017ptx,Abreu:2017mtm}, which allows us to rewrite the $C$-particle cut of a $\D$-dimensional $n$-gon as an $(n-|C|+1)$-gon in $(\D-|C|)$ dimensions. We review the arguments in App.~\ref{sec:app_embeddingspace}. 
In doing so, we end up with an integral with kinematics resembling a lower-dimensional triangle integral. In Schwinger-parameter space, it is given by
\begin{multline}
\Cut_{s_{45}} \cM^{\onembox}_{245\leftarrow13} = - \frac{2 i C(\epsilon) \sin(\pi \epsilon)}{s_{45}} \int_0^\infty \frac{\rd^3 \alpha}{\GL (1)} \alpha_0^{-1-2\epsilon} \\ \times \left[
- \alpha_0 \alpha_1 \left(\frac{s_{12}}{s_{45}} - 1 \right)
- \alpha_0 \alpha_2 \left(\frac{s_{23}}{s_{45}} - 1 \right)
- \alpha_1 \alpha_2 \left(- \frac{s_{12} s_{23}}{s_{45}} \right)
+ \frac{1}{s_{45}} \alpha_0^2
\right]^{\epsilon-1}\,.
\label{eq:Cutints45}
\end{multline}
Its kinematic dependence resembles a triangle integral with one internal mass:
\begin{equation}
    \begin{gathered}
\begin{tikzpicture}[line width=1]
\draw[] (0,0) -- (1,0.5) -- (1,-0.5) -- (0,0);
\draw[line width=2] (1,0.5) -- (1,-0.5);
\draw[] (1.5,0.75) -- (1,0.5) -- (0.5,0.75);
\draw[] (1.5,-0.75) -- (1,-0.5) -- (0.5,-0.75);
\draw[] (-0.5,0.25) -- (0,0) -- (-0.5,-0.25);
\node[] at (-1.5,0) {$-\frac{s_{12} s_{23}}{s_{45}}$};
\node[] at (1,1.2) {$\frac{s_{23}}{s_{45}} - 1$};
\node[] at (1,-1.2) {$\frac{s_{12}}{s_{45}} - 1$};
\node[] at (2.2,0) {$m_0^2=\frac{1}{s_{45}}$};
\node[scale=0.75] at (1.7,0.8) {$6$};
\node[scale=0.75] at (0.3,0.8) {$5$};
\node[scale=0.75] at (1.7,-0.8) {$1$};
\node[scale=0.75] at (0.3,-0.8) {$2$};
\node[scale=0.75] at (-0.8,0.2) {$3$};
\node[scale=0.75] at (-0.8,-0.2) {$4$};
\end{tikzpicture}
\end{gathered}
\end{equation}

We can directly evaluate the integral in \eqref{eq:Cutints45} by gauge fixing the $\GL(1)$ symmetry to set $\alpha_0=1$, before performing the remaining integrals over $\alpha_1$ and $\alpha_2$. We find
\begin{multline}
\Cut_{s_{45}} \cM^{\onembox}_{245\leftarrow13} =
- \frac{2iC(\epsilon)}{\epsilon s_{45}^{\epsilon}}  \Bigg[\frac{\pi}{s_{12} s_{23}} \left(-  \frac{(s_{12}+s_{23}-s_{45})s_{45}}{s_{12} s_{23}} \right)^{\epsilon}
\\ +
\frac{\sin(\pi \epsilon)}{(1+\epsilon) (s_{45}-s_{12})(s_{45}-s_{23})}
\,_2F_1 \Big(1,1;2+\epsilon;\frac{s_{12} s_{23}}{(s_{45}-s_{12})(s_{45}-s_{23})} \Big)
\Bigg] \,.
\label{eq:Cutints45_res}
\end{multline}
In the kinematic region we end up in after crossing ($s_{12}\ll s_{23}<0<s_{45}$), we can rewrite \eqref{eq:Cutints45_res} as
\begin{equation}
\Cut_{s_{45}} \cM^{\onembox}_{245\leftarrow13}= - \frac{2iC(\epsilon)\sin(\pi \epsilon)}{\epsilon^2 s_{12} s_{23}s_{45}^{\epsilon}}
\,_2F_1 \big(1,-\epsilon;1-\epsilon;z_3 \big)\,,
\label{eq:Cutints45_2}
\end{equation}
thanks to the hypergeometric identity
\begin{multline}
    _2F_1 \Big(1,1;2+\epsilon;\frac{s_{12} s_{23}}{(s_{45}-s_{12})(s_{45}-s_{23})} \Big)
    =
     \frac{(s_{45}-s_{12})(s_{45}-s_{23})}{s_{12} s_{23}}
     \frac{1+\epsilon}{\epsilon}
     \,
    \\ \times 
    \left[\, _2F_1 \Big(1,-\epsilon;1-\epsilon;z_3\Big)
    - \frac{\pi \epsilon}{\sin(\pi \epsilon)} \left(-\frac{(s_{12}+s_{23}-s_{45})s_{45}}{s_{12} s_{23}} \right)^{\epsilon}
    \right]\,.
\end{multline}
The result in \eqref{eq:Cutints45_2} matches perfectly with the crossing prediction made in \eqref{eq:toCheck_box}.
\paragraph{Other crossing channels}
For completeness, we checked the crossing relation in the other $3\leftarrow2$ crossing channels. To check other channels, we also needed to compute the different cuts. The box's $s_{12}$ and $s_{23}$-channel cuts are given by diagrams that resemble triangles with the following shifted kinematics:
\begin{equation}
\begin{gathered}
\begin{tikzpicture}[line width=1]
\draw[] (0,0) -- (1,0.5) -- (1,-0.5) -- (0,0);
\draw[line width=2] (1,0.5) -- (1,-0.5);
\draw[] (1.5,0.75) -- (1,0.5) -- (0.5,0.75);
\draw[] (1.5,-0.75) -- (1,-0.5) -- (0.5,-0.75);
\draw[] (-0.5,0.25) -- (0,0) -- (-0.5,-0.25);
\node[] at (-1.5,0) {$s_{12}$};
\node[] at (1,1.2) {$-1$};
\node[] at (1,-1.2) {$\frac{s_{45}}{s_{23}} - 1$};
\node[] at (2.2,0) {$m_0^2=\frac{1}{s_{23}}$};
\node[scale=0.75] at (1.7,0.8) {$6$};
\node[scale=0.75] at (0.3,0.8) {$5$};
\node[scale=0.75] at (1.7,-0.8) {$1$};
\node[scale=0.75] at (0.3,-0.8) {$2$};
\node[scale=0.75] at (-0.8,0.2) {$3$};
\node[scale=0.75] at (-0.8,-0.2) {$4$};
\end{tikzpicture}
\end{gathered}
\hspace{1cm}
\begin{gathered}
\begin{tikzpicture}[line width=1]
\draw[] (0,0) -- (1,0.5) -- (1,-0.5) -- (0,0);
\draw[line width=2] (1,0.5) -- (1,-0.5);
\draw[] (1.5,0.75) -- (1,0.5) -- (0.5,0.75);
\draw[] (1.5,-0.75) -- (1,-0.5) -- (0.5,-0.75);
\draw[] (-0.5,0.25) -- (0,0) -- (-0.5,-0.25);
\node[] at (-1.5,0) {$s_{23}$};
\node[] at (1,1.2) {$-1$};
\node[] at (1,-1.2) {$\frac{s_{45}}{s_{12}} - 1$};
\node[] at (2.2,0) {$m_0^2=\frac{1}{s_{12}}$};
\node[scale=0.75] at (1.7,0.8) {$6$};
\node[scale=0.75] at (0.3,0.8) {$5$};
\node[scale=0.75] at (1.7,-0.8) {$1$};
\node[scale=0.75] at (0.3,-0.8) {$2$};
\node[scale=0.75] at (-0.8,0.2) {$3$};
\node[scale=0.75] at (-0.8,-0.2) {$4$};
\end{tikzpicture}
\end{gathered}
\end{equation}
As expected, analogous computations gave label permutations of \eqref{eq:Cutints45_2}, namely
\begin{align}
\Cut_{s_{12}} \cM & = \frac{2iC(\epsilon)\sin(\pi \epsilon)}{\epsilon^2 s_{12}^{\epsilon}s_{23}s_{45}} 
\,_2F_1 \big(1,-\epsilon;1-\epsilon;z_1 \big)\,, \\
\Cut_{s_{23}} \cM & = \frac{2iC(\epsilon)\sin(\pi \epsilon)}{\epsilon^2 s_{12}s_{23}^{\epsilon}s_{45}}
\,_2F_1 \big(1,-\epsilon;1-\epsilon;z_2 \big) \,.
\label{eq:Cutints}
\end{align}

\paragraph{A midway summary} So far, we have checked the crossing equation for, essentially, all massless one-loop topologies in $\D=4-2\epsilon$ up to boxes. Each time, we followed the same strategy: we wrote down the crossing prediction, isolated the conjectured cut term, and compared it explicitly with the result of a direct calculation using methods like Schwinger parameters or the embedding space formalism. To achieve the goal of this section, it remains to perform similar checks on the five-point pentagon topology. As we will detail shortly, we will undertake these checks using the method of differential equations. This method is quite general and should be suitable to rigorously test the crossing equation for higher-loop examples or for examples that depend on more kinematic scales. At the same time, this demonstrates how the differential equation approach can be used as a more economical method to perform checks for all bubbles and boxes \emph{simultaneously}, as these are master integrals for the pentagon family.

\subsection{Pentagon (via differential equations)}\label{sec:pentagon}
For the massless pentagon amplitude
\begin{equation}\label{eq:pent_ampl}
    \mathcal{M}^{\text{pent}}=\int~ \frac{\mathrm{d}^\text{D} \ell}{i\pi^{\text{D}/2}}~ \frac{1}{\textsf{D}_1\textsf{D}_2\textsf{D}_3\textsf{D}_4\textsf{D}_5} \quad \text{with} \quad \textsf{D}_n=\Big(\ell+\sum_{i=2}^np_{j-1}\Big)^2-i\varepsilon\,,
\end{equation}
we will put crossing to the test in two ways: first, using the conventional method we just outlined (Sec.~\ref{sec:pentagon}) in the above summary, and then using a different approach (Sec.~\ref{sec:pentagonApplications}), which will allow us to make a new kind of useful predictions. In the latter case, we will explore how crossing can be used to navigate between \emph{time-ordered} amplitudes in disconnected physical channels or, equivalently, between initial conditions in different channels. The crossing equation is then tested by comparing the results with the explicit evaluation of the integral. In Sec.~\ref{sec:pentagonApplications}, we will find perfect agreement between the two.

\begin{figure}
    \centering
      \adjustbox{valign=c}{\tikzset{every picture/.style={line width=1pt}}     
\begin{tikzpicture}[x=0.75pt,y=0.75pt,yscale=-0.9,xscale=0.9]
\draw   (40,46.05) .. controls (40,36.63) and (53.36,29) .. (69.83,29) .. controls (86.31,29) and (99.67,36.63) .. (99.67,46.05) .. controls (99.67,55.46) and (86.31,63.1) .. (69.83,63.1) .. controls (53.36,63.1) and (40,55.46) .. (40,46.05) -- cycle node[above,midway,yshift=-7,xshift=20]{$s_{51}$};
\draw   (152,45.05) .. controls (152,35.63) and (165.36,28) .. (181.83,28) .. controls (198.31,28) and (211.67,35.63) .. (211.67,45.05) .. controls (211.67,54.46) and (198.31,62.1) .. (181.83,62.1) .. controls (165.36,62.1) and (152,54.46) .. (152,45.05) -- cycle node[above,midway,yshift=-7,xshift=20]{$s_{34}$};
\draw   (259,45.05) .. controls (259,35.63) and (272.36,28) .. (288.83,28) .. controls (305.31,28) and (318.67,35.63) .. (318.67,45.05) .. controls (318.67,54.46) and (305.31,62.1) .. (288.83,62.1) .. controls (272.36,62.1) and (259,54.46) .. (259,45.05) -- cycle node[above,midway,yshift=-7,xshift=20]{$s_{12}$};
\draw   (364,45.05) .. controls (364,35.63) and (377.36,28) .. (393.83,28) .. controls (410.31,28) and (423.67,35.63) .. (423.67,45.05) .. controls (423.67,54.46) and (410.31,62.1) .. (393.83,62.1) .. controls (377.36,62.1) and (364,54.46) .. (364,45.05) -- cycle node[above,midway,yshift=-7,xshift=20]{$s_{45}$};
\draw   (473,45.75) .. controls (473.1,36.34) and (486.53,28.84) .. (503,29) .. controls (519.48,29.16) and (532.76,36.93) .. (532.67,46.34) .. controls (532.57,55.76) and (519.14,63.26) .. (502.66,63.09) .. controls (486.19,62.93) and (472.91,55.17) .. (473,45.75) -- cycle node[above,midway,yshift=-7,xshift=20]{$s_{23}$};
\draw    (40,46.05) -- (20.2,36) ;
\draw    (21.2,59) -- (40,46.05) ;
\draw    (100.16,46.62) -- (119.61,57.33) ;
\draw    (119.38,34.31) -- (100.16,46.62) ;
\draw  [draw opacity=0] (532.67,46.34) .. controls (533.78,47.95) and (534.37,49.64) .. (534.37,51.37) .. controls (534.37,63.03) and (507.74,72.49) .. (474.88,72.52) .. controls (467.62,72.52) and (460.67,72.06) .. (454.23,71.23) -- (474.87,51.41) -- cycle ; \draw   (532.67,46.34) .. controls (533.78,47.95) and (534.37,49.64) .. (534.37,51.37) .. controls (534.37,63.03) and (507.74,72.49) .. (474.88,72.52) .. controls (467.62,72.52) and (460.67,72.06) .. (454.23,71.23) ;  
\draw    (152,45.05) -- (132.2,35) ;
\draw    (133.2,58) -- (152,45.05) ;
\draw    (212.16,45.62) -- (231.61,56.33) ;
\draw    (230.38,33.31) -- (211.16,45.62) ;
\draw    (152,45.05) -- (131.2,45) ;
\draw    (318.67,45.05) -- (338.12,55.76) ;
\draw    (337.89,32.74) -- (318.67,45.05) ;
\draw    (259,44.05) -- (239.2,34) ;
\draw    (240.2,57) -- (259,44.05) ;
\draw  [draw opacity=0] (318.67,45.05) .. controls (319.54,43.56) and (320,42.01) .. (320,40.42) .. controls (320.04,28.76) and (295.62,19.24) .. (265.46,19.15) .. controls (257.55,19.13) and (250.03,19.76) .. (243.23,20.9) -- (265.39,40.26) -- cycle ; \draw   (318.67,45.05) .. controls (319.54,43.56) and (320,42.01) .. (320,40.42) .. controls (320.04,28.76) and (295.62,19.24) .. (265.46,19.15) .. controls (257.55,19.13) and (250.03,19.76) .. (243.23,20.9) ;  
\draw    (364,45.05) -- (346.8,33) ;
\draw    (347.8,55) -- (364,45.05) ;
\draw    (423.67,45.05) -- (443.12,55.76) ;
\draw  [draw opacity=0] (423.67,45.05) .. controls (424.54,43.56) and (425,42.01) .. (425,40.42) .. controls (425.04,28.76) and (400.62,19.24) .. (370.46,19.15) .. controls (362.55,19.13) and (355.03,19.76) .. (348.23,20.9) -- (370.39,40.26) -- cycle ; \draw   (423.67,45.05) .. controls (424.54,43.56) and (425,42.01) .. (425,40.42) .. controls (425.04,28.76) and (400.62,19.24) .. (370.46,19.15) .. controls (362.55,19.13) and (355.03,19.76) .. (348.23,20.9) ;  
\draw  [draw opacity=0] (364,45.05) .. controls (363.13,46.54) and (362.67,48.09) .. (362.66,49.68) .. controls (362.63,61.33) and (387.04,70.86) .. (417.2,70.96) .. controls (425.11,70.99) and (432.63,70.36) .. (439.42,69.22) -- (417.27,49.86) -- cycle ; \draw   (364,45.05) .. controls (363.13,46.54) and (362.67,48.09) .. (362.66,49.68) .. controls (362.63,61.33) and (387.04,70.86) .. (417.2,70.96) .. controls (425.11,70.99) and (432.63,70.36) .. (439.42,69.22) ;  
\draw    (473,45.75) -- (455.92,33.53) ;
\draw    (456.7,55.54) -- (473,45.75) ;
\draw    (532.67,46.34) -- (550.2,39.2) ;
\draw  [draw opacity=0] (473,45.75) .. controls (472.14,44.24) and (471.7,42.68) .. (471.71,41.08) .. controls (471.84,29.42) and (496.52,20.25) .. (526.83,20.58) .. controls (534.6,20.67) and (542,21.37) .. (548.69,22.56) -- (526.59,41.68) -- cycle ; \draw   (473,45.75) .. controls (472.14,44.24) and (471.7,42.68) .. (471.71,41.08) .. controls (471.84,29.42) and (496.52,20.25) .. (526.83,20.58) .. controls (534.6,20.67) and (542,21.37) .. (548.69,22.56) ;  
\draw  [draw opacity=0] (100.16,46.62) .. controls (101.27,48.23) and (101.86,49.91) .. (101.86,51.64) .. controls (101.87,63.3) and (75.23,72.77) .. (42.37,72.79) .. controls (35.11,72.79) and (28.16,72.34) .. (21.73,71.5) -- (42.36,51.68) -- cycle ; \draw   (100.16,46.62) .. controls (101.27,48.23) and (101.86,49.91) .. (101.86,51.64) .. controls (101.87,63.3) and (75.23,72.77) .. (42.37,72.79) .. controls (35.11,72.79) and (28.16,72.34) .. (21.73,71.5) ;  
\draw   (42.98,127.69) -- (69.95,100.64) -- (97,127.62) -- (70.03,154.67) -- cycle node[above,midway,yshift=2.5,xshift=9]{$s_{34}$};
\draw   (152.98,126.69) -- (179.95,99.64) -- (207,126.62) -- (180.03,153.67) -- cycle node[above,midway,yshift=2.5,xshift=9]{$s_{23}$};
\draw   (258.98,127.69) -- (285.95,100.64) -- (313,127.62) -- (286.03,154.67) -- cycle node[above,midway,yshift=2.5,xshift=9]{$s_{12}$};
\draw   (368.98,127.69) -- (395.95,100.64) -- (423,127.62) -- (396.03,154.67) -- cycle node[above,midway,yshift=2.5,xshift=9]{$s_{51}$};
\draw   (478.98,127.69) -- (505.95,100.64) -- (533,127.62) -- (506.03,154.67) -- cycle node[above,midway,yshift=2.5,xshift=9]{$s_{45}$};
\draw    (42.98,127.69) -- (26.27,127.37) ;
\draw    (70.03,154.67) -- (53.32,154.34) ;
\draw    (69.95,100.64) -- (53.25,100.32) ;
\draw    (97,127.62) -- (116.45,138.33) ;
\draw    (116.23,115.31) -- (97,127.62) ;
\draw    (180.03,153.67) -- (163.32,153.34) ;
\draw    (196.73,153.99) -- (180.03,153.67) ;
\draw    (223.7,126.94) -- (207,126.62) ;
\draw    (152.98,126.69) -- (136.27,126.37) ;
\draw    (179.95,99.64) -- (163.25,99.32) ;
\draw    (302.73,154.99) -- (286.03,154.67) ;
\draw    (329.7,127.94) -- (313,127.62) ;
\draw    (285.95,100.64) -- (269.25,100.32) ;
\draw    (258.98,127.69) -- (239.18,117.65) ;
\draw    (240.18,140.65) -- (258.98,127.69) ;
\draw    (395.95,100.64) -- (376.15,90.6) ;
\draw    (375.27,100.37) -- (395.95,100.64) ;
\draw    (368.98,127.69) -- (352.27,127.37) ;
\draw    (412.73,154.99) -- (396.03,154.67) ;
\draw    (439.7,127.94) -- (423,127.62) ;
\draw    (522.66,100.97) -- (505.95,100.64) ;
\draw    (505.95,100.64) -- (489.25,100.32) ;
\draw    (478.98,127.69) -- (462.27,127.37) ;
\draw    (506.03,154.67) -- (489.32,154.34) ;
\draw    (549.7,127.94) -- (533,127.62) ;
\draw   (313.95,181.05) -- (313.95,228.32) -- (264.69,228.32) -- (253.32,204.68) -- (264.69,181.05) -- cycle ;
\draw    (264.69,181.05) -- (247.98,180.72) node[left]{$5$};
\draw    (253.32,204.68) -- (236.62,204.36) node[left]{$1$};
\draw    (264.69,228.32) -- (247.98,228) node[left]{$2$};
\draw    (330.66,228.65) -- (313.95,228.32) node[pos=0,right]{$3$};
\draw    (330.66,181.37) -- (313.95,181.05) node[pos=0,right]{$4$};
\end{tikzpicture}}
    \caption{The massless pentagon master topologies presented below, with the order of appearance being from left to right. The diagrams are drawn in the $215\leftarrow34$ channel.}
    \label{fig:masslessMasters}
\end{figure}
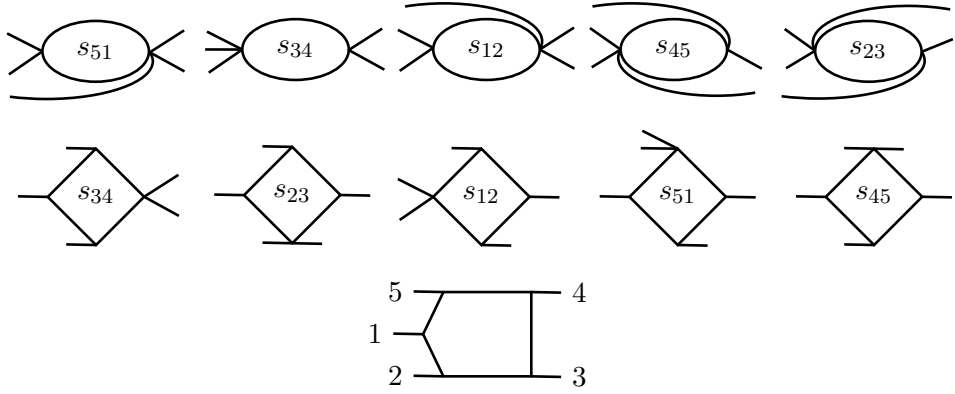
Since the known \say{closed-form} expression for \eqref{eq:pent_ampl} has obscure analytic properties (see e.g., \cite{Kozlov:2015kol}), we opted to use the method of \emph{canonical differential equations} \cite{Henn:2013pwa} in order to assess crossing for the pentagon amplitude. The associated master integrals are defined by relaxing the notation in \eqref{eq:pent_ampl} to include arbitrary integer-power inverse propagators
\begin{equation}\label{eq:penta}     \mathcal{M}^{\text{pent}}_{a_1a_2a_3a_4a_5} = \int~ \frac{\mathrm{d}^\text{D} \ell}{i\pi^{\text{D}/2}}~ \frac{1}{\textsf{D}_1^{a_1}\textsf{D}_2^{a_2}\textsf{D}_3^{a_3}\textsf{D}_4^{a_4}\textsf{D}_5^{a_5}}\, \qquad (a_i\in\mathbb{Z})\,.
\end{equation}

\paragraph{Pure masters and differential equation}  In this section, we record a basis pure master integrals and discuss its differential equation. For the integral family under consideration, there are 11 master integrals (drawn in Fig.~\ref{fig:masslessMasters}). There are five bubbles defined by:
\begin{equation}
\begin{aligned}
     I_{1}&=-\epsilon\e^{\gamma_\text{E}\epsilon} s_{51}~\mathcal{M}^{\text{pent}}_{02001},  \quad I_{2}=-\epsilon\e^{\gamma_\text{E}\epsilon} s_{34}~\mathcal{M}^{\text{pent}}_{00201},  \quad I_{3}=-\epsilon\e^{\gamma_\text{E}\epsilon}s_{12}~\mathcal{M}^{\text{pent}}_{20100},\\ \quad I_{4}&=-\epsilon\e^{\gamma_\text{E}\epsilon} s_{45}~ \mathcal{M}^{\text{pent}}_{20010}, \quad I_{5}=-\epsilon\e^{\gamma_\text{E}\epsilon} s_{23}~\mathcal{M}^{\text{pent}}_{02010}\,,
\end{aligned}
\end{equation}
five boxes defined by:
\begin{equation}
\begin{aligned}
I_{6}&=\epsilon^2\e^{\gamma_\text{E}\epsilon} s_{12}s_{51}~ \mathcal{M}^{\text{pent}}_{11101}, \quad I_{7}=\epsilon^2\e^{\gamma_\text{E}\epsilon} s_{51}s_{45}~\mathcal{M}^{\text{pent}}_{11011},\quad I_{8}=\epsilon^2\e^{\gamma_\text{E}\epsilon} s_{45}s_{34}~\mathcal{M}^{\text{pent}}_{10111},\\
     I_{9}&=\epsilon^2\e^{\gamma_\text{E}\epsilon} s_{34}s_{23}~ \mathcal{M}^{\text{pent}}_{01111},\quad I_{10}=\epsilon^2\e^{\gamma_\text{E}\epsilon} s_{23}s_{12}~\mathcal{M}^{\text{pent}}_{11110}\,.
\end{aligned}
\end{equation}
Finally, one five-propagator integral, which is taken to be proportional to the pentagon integral near \emph{six} spacetime dimensions $P^{(6-2\varepsilon)}$:\footnote{An expression for $P^{(6-2\varepsilon)}$ in terms of its $\D=4-2\epsilon$ analogue is given in \cite{Kozlov:2015kol} and the (lengthy) decomposition of the former in terms of $\mathcal{M}^{\text{pent}}_{a_1a_2a_3a_4a_5}$'s can be found in \cite{Syrrakos:2020kba}.}
\begin{equation}
    I_{11}=-\epsilon^{3}\e^{2\gamma_\text{E}\epsilon}\sqrt{\Delta}~P^{(6-2\epsilon)}\ \quad \text{with} \quad \Delta=\det(2p_i\cdot p_j)\mid_{i,j=1,2,3,4} \,.
\end{equation}

We checked with $\texttt{FIRE6}$ \cite{Smirnov:2019qkx} that the vector $\vec{I}$ of master integrals near four spacetime dimensions indeed satisfies a differential equation in the canonical form
\begin{equation}\label{eq:diffEq}
\textnormal{d}\vec{I}=\epsilon~\textnormal{d}\boldsymbol{\Omega}\cdot\vec{I}\,,
\end{equation}
where $\boldsymbol{\Omega}$ is a matrix of logarithms with rational coefficients, with arguments limited to a subset of the planar pentagon alphabet \cite{Chicherin:2017dob}, namely $\{W_{i\in [1,5] \cup [11,20] \cup [26,31]}\}$. The detailed expression can be found in App.~\ref{app:details}. Provided we have a boundary condition, we can solve \eqref{eq:diffEq}  iteratively in $\epsilon$ and get a solution in terms of iterated integrals of logarithmic kernels. Our numerical integration strategy is summarized in App.~\ref{app:numerics}.

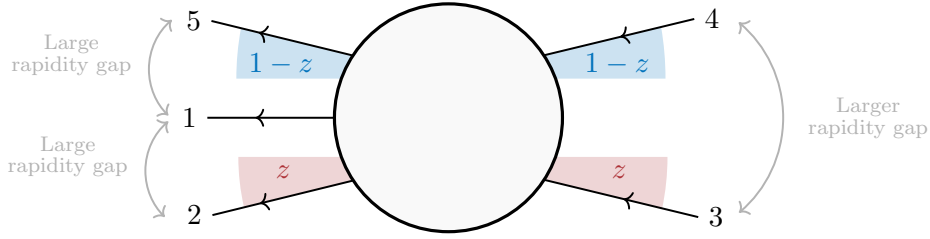
\begin{figure}
        \centering
   \adjustbox{valign=c}{\tikzset{every picture/.style={line width=0.75pt}}      
\begin{tikzpicture}[x=0.75pt,y=0.75pt,yscale=-1,xscale=1]
\tikzset{ma/.style={decoration={markings,mark=at position 0.8 with {\arrow[scale=1]{>}}},postaction={decorate}}}
\tikzset{ma2/.style={decoration={markings,mark=at position 0.3 with {\arrow[scale=1]{>}}},postaction={decorate}}}
\tikzset{mar/.style={decoration={markings,mark=at position 0.2 with {\arrowreversed[scale=1]{>}}},postaction={decorate}}}
\draw  [draw opacity=0][fill=RoyalBlue  ,fill opacity=0.2]  (428,129.11) .. controls (428,120.16) and (427.23,111.42) .. (425.77,102.99) -- (318.03,129.11) -- cycle node[pos=0.5,xshift=23,yshift=5.5,fill opacity=1,color=RoyalBlue] {$1-z$};
\draw  [draw opacity=0] [fill=RoyalBlue  ,fill opacity=0.2]  (211.55,129.11) .. controls (211.55,129.11) and (211.55,129.11) .. (211.55,129.11) .. controls (211.55,120.45) and (212.35,112) .. (213.87,103.85) -- (318.03,129.11) -- cycle node[pos=0.5,xshift=-23,yshift=5.5,fill opacity=1,color=RoyalBlue] {$1-z$};
\draw  [draw opacity=0] [fill=Maroon  ,fill opacity=0.2] (429,168.62) .. controls (429,168.62) and (429,168.62) .. (429,168.62) .. controls (429,177.56) and (428.07,186.26) .. (426.3,194.63) -- (319.03,168.62) -- cycle node[pos=0.5,xshift=23,yshift=-5.5,fill opacity=1,color=Maroon] {$z$} ;
\draw  [draw opacity=0] [fill=Maroon  ,fill opacity=0.2] (212.55,168.62) .. controls (212.55,168.62) and (212.55,168.62) .. (212.55,168.62) .. controls (212.55,177.27) and (213.52,185.68) .. (215.36,193.76) -- (319.03,168.62) -- cycle node[pos=0.5,xshift=-23,yshift=-5.5,fill opacity=1,color=Maroon] {$z$};
\draw [ma] (318.03,129.11) -- (199,100) node[pos=1, left]{$5$};
\draw  [mar]  (196.93,148.86) -- (318.53,148.87) node[pos=0, left]{$1$};
\draw  [ma2]  (444.5,199) -- (319.03,168.62) node[pos=0, right]{$3$};
\draw  [ma]  (319.03,168.62) -- (199.5,198) node[pos=1, left]{$2$};
\draw  [ma2]  (442.5,99) -- (318.03,129.11) node[pos=0, right]{$4$};
\draw [fill=gray!5  ,fill opacity=1][very thick] (261.03,148.87) .. controls (261.03,117.11) and (286.77,91.37) .. (318.53,91.37) .. controls (350.29,91.37) and (376.03,117.11) .. (376.03,148.87) .. controls (376.03,180.62) and (350.29,206.37) .. (318.53,206.37) .. controls (286.77,206.37) and (261.03,180.62) .. (261.03,148.87) -- cycle ;
\begin{scope}[xshift=-10]
\draw  [draw opacity=0] (190.59,147.7) .. controls (184.55,141.78) and (180.79,133.53) .. (180.79,124.4) .. controls (180.79,114.4) and (185.29,105.45) .. (192.38,99.47) -- (213.4,124.4) -- cycle ; \draw [<->,color=gray!60]  (190.59,147.7) .. controls (184.55,141.78) and (180.79,133.53) .. (180.79,124.4) .. controls (180.79,114.4) and (185.29,105.45) .. (192.38,99.47) node[left,midway,xshift=-5,yshift=-5]{$\substack{\text{Large}\\ \text{rapidity gap}}$} ;  
\draw [draw opacity=0] (188.81,198.93) .. controls (182.76,193.01) and (179.01,184.75) .. (179.01,175.62) .. controls (179.01,165.63) and (183.51,156.68) .. (190.59,150.7) -- (211.61,175.62) -- cycle ; \draw [<->,color=gray!60]  (188.81,198.93) .. controls (182.76,193.01) and (179.01,184.75) .. (179.01,175.62) .. controls (179.01,165.63) and (183.51,156.68) .. (190.59,150.7) node[left,midway,xshift=-5,yshift=-5]{$\substack{\text{Large}\\ \text{rapidity gap}}$};  
\end{scope}
\begin{scope}[xshift=10]
\draw  [draw opacity=0] (451.58,100.69) .. controls (465.29,112.21) and (474.01,129.48) .. (474.01,148.79) .. controls (474.01,168.04) and (465.34,185.27) .. (451.7,196.79) -- (411.22,148.79) -- cycle ; \draw [<->,color=gray!60]  (451.58,100.69) .. controls (465.29,112.21) and (474.01,129.48) .. (474.01,148.79) .. controls (474.01,168.04) and (465.34,185.27) .. (451.7,196.79) node[right,midway,xshift=10,yshift=20]{$\substack{\text{Larger}\\ \text{rapidity gap}}$};  
\end{scope}
\end{tikzpicture}}
    \caption{A pictorial representation of the multi-Regge kinematics is shown in the $s_{34}$ channel. The strong ordering of the rapidities is displayed, along with the physical interpretation of the $z$ parameter. Specifically, as $z\to0$, particles $2$ and $3$ become collinear, and as $z\to 1$, particles $4$ and $5$ become collinear. The precise ordering of external legs defines the $215\ot34$ channel multi-Regge limit. This limit is, for example, \emph{different} from the $215\ot43$ channel multi-Regge limit, which arises from a different ordering of the rapidities.}
    \label{fig:mrkCoords}
\end{figure}

\paragraph{Multi-Regge limit and analytic boundary conditions}
The multi-Regge kinematics is defined as a scattering process where the final-state particles are strongly ordered in rapidity and have comparable transverse momenta \cite{Kuraev:1976ge}. For example, setting $s_{34}$ to be the dominating variable in a five-point process, its \emph{leading order} effect is equivalent to the parameterization
\begin{align}\label{eq:mrkScaling}
s_{34}= \frac{s}{x^2},\quad
s_{23}= -\frac{s_1 s_2}{s}z \bar{z},\quad
s_{12}= \frac{s_1}{x},\quad
s_{51}= \frac{s_2}{x},\quad
s_{45}= -\frac{s_1 s_2}{s}(1-z)(1-\bar{z})\,.
\end{align}
Here, $s,s_1,s_2,z$ and $\bar{z}$ are all fixed and of $\mathcal{O}(1)$ in the limit $0<x\ll1$. The limit is shown in Fig.~\ref{fig:mrkCoords}.
The notation $z$ and $\bar{z}$ is natural since, in the physical scattering region where $\Delta\leq0$, $\bar{z}$ is the complex conjugate of $z$ \cite{Caron-Huot:2020vlo}. 
The simplicity of the one-loop pentagon alphabet is greatly influenced by the multi-Regge limit rationalizing the root $\sqrt{\Delta}$, since
\begin{equation}
    \Delta=\frac{s_1^2s_2^2}{x^4}(z-\bar{z})^2+\mathcal{O}\left(x^{-3}\right)\,,
\end{equation} 
in this kinematics. This leads to significant consequences, as the differential equation reveals that the alphabet can be factorized into three distinct and independent subalphabets \cite{Chicherin:2018yne}
\begin{equation}\label{eq:alph}
\{ x \}\, \cup  \{s,s_1,s_2, s_1\pm s_2\} \cup \{z, \bar{z}, 1-z, 1-\bar{z},z-\bar{z}, 1-z - \bar{z}\}\,.
\end{equation}
For the example considered, the multi-Regge regime is particularly advantageous for establishing a basepoint (or initial condition). Among other things, it makes the transcendental weight structure of the master integrals manifest. A convenient point in this limit is the collinear point (equivalent to $s_{23}\to0$)
\begin{equation}\label{eq:basePoint}
    X_0=\{s=1,\; s_1=1,\; s_2=1,\; z\to0^+,\; \bar{z}\to0^+\}\,.
\end{equation}
At $X_0$, we are at the boundary of the physical region ($\Delta=0$) and the integrals can be computed directly to give a simple vector of constants in terms of $\zeta$-values and powers of $\pi$. This can be observed by substituting $s=s_1=s_2=1$ into the exact expression
\begin{equation}\label{eq:bdryCTE}
\begin{split}
   \hspace{-0.4cm} \frac{\vec{I}_{ 215\leftarrow34}}{-r_\Gamma}&= \bigg( \e^{i \pi\epsilon}  s_2^{-\epsilon},\e^{i \pi  \epsilon} s^{-\epsilon},\e^{i \pi  \epsilon} s_1^{-\epsilon},\mathcal{R}, \mathcal{R},-2 \pi  \epsilon  \mathcal{R} (\cot \pi\epsilon{+}i),-2 \epsilon \mathcal{R} \big(\log (-\tfrac{s_1}{s}) {+}\Psi_1^\epsilon\big),\\
   &\quad\qquad -2 \epsilon \mathcal{R}\big(\log\tfrac{s_1}{s}{+}\Psi_2^\epsilon \big),
   -2 \epsilon \mathcal{R}\big(\log\tfrac{s_2}{s}{+}\Psi_2^\epsilon \big),-2 \epsilon \mathcal{R} \big(\log (-\tfrac{s_2}{s}) {+}\Psi_2^\epsilon\big),0 \bigg)^\top\,,
\end{split}
\end{equation}
with $r_\Gamma=\e^{\epsilon\gammaE}C(\epsilon)$, $\mathcal{R}=\big(\frac{s}{s_1 s_2}\big)^{\epsilon}$ and 
\begin{gather}
 \qquad \Psi_1^\epsilon=2 \pi  \cot
   \pi  \epsilon+\psi(\epsilon)+\gammaE \quad \text{and} \quad \Psi_2^\epsilon=\tfrac{1}{\epsilon}+ \pi  \cot
   \pi  \epsilon+\psi(\epsilon)+\gammaE\,,
\end{gather}
where $\psi(x)=\partial_x\log\Gamma(x)$ is the digamma function. Note that \eqref{eq:bdryCTE} is valid within the $215\leftarrow34$ channel region carved by $\{s>0, s_1>0, s_2>0, z\to0^+, \bar{z}\to0^+\}$. This expression was obtained by directly expanding in the multi-Regge-collinear limit the properly normalized bubble and box expressions in \cite{Kozlov:2015kol}.

\paragraph{Checking crossing} From a computation similar to the one leading to \eqref{eq:bdryCTE}, we find
\begin{equation}\label{eq:bdryCTE2}
\begin{split}
\hspace{-0.4cm}\frac{\vec{I}_{ 315\leftarrow24}^{~\dag}}{-r_\Gamma}&= \!\bigg(\! \e^{-i \pi  \epsilon}  s_2^{-\epsilon},(-s)^{-\epsilon},(-s_1)^{-\epsilon},\mathcal{R}, \mathcal{R},-2 \epsilon\pi \mathcal{R}(\cot \pi\epsilon{-}i),-2 \epsilon \mathcal{R}\big(\log\tfrac{s_1}{s}{-}i\pi {+}\Psi_1^\epsilon\big),\\
   &-2 \epsilon \mathcal{R}~\big(\log\tfrac{s_1}{s}{+}\Psi_2^\epsilon \big),
   -2 \epsilon \mathcal{R}~\big(\log(\tfrac{s_2}{-s}){+}i\pi{+}\Psi_2^\epsilon \big),-2 \epsilon \mathcal{R}~\big(\log (\tfrac{s_2}{-s}){+}\Psi_2^\epsilon\big),0 \bigg)^\top,
\end{split}
\end{equation}
within the $315\leftarrow24$ channel $\{s<0, s_1<0, s_2>0, z\to0^+, \bar{z}\to0^+\}$ multi-Regge-collinear region.

Together, the standard amplitudes in \eqref{eq:bdryCTE} and \eqref{eq:bdryCTE2} provide enough information to predict a cut using our crossing conjecture. Starting with \eqref{eq:bdryCTE2} in the $315\leftarrow24$ channel, suppose we cross particles $2$ and $3$. Under this crossing, both $s$ and $s_1$ rotate from negative to positive in the \emph{upper} half-plane (because we start with the \emph{conjugated} amplitude \eqref{eq:bdryCTE2}), while all the other Mandelstam variables remain fixed. Substituting $s\to -|s|\e^{-i\phi}$ and $s_1\to -|s_1|\e^{-i\phi}$ into \eqref{eq:bdryCTE2}, we find (at $\phi=\pi$)
\begin{align}\label{eq:bdryCTE2cont}
   \hspace{-0.4cm} {\left[\eqref{eq:bdryCTE2}\right]_{ 
 \substack{\raisebox{\depth}{\scalebox{1}[1]{$\curvearrowright$}}s_{\phantom{1}}\\ \raisebox{\depth}{\scalebox{1}[1]{$\curvearrowright$}}s_1}}}&=\left( \e^{-i \pi  \epsilon}  s_2^{-\epsilon},\e^{i \pi  \epsilon} s^{-\epsilon},\e^{i \pi  \epsilon} s_1^{-\epsilon},\mathcal{R}, \mathcal{R},-2 \pi  \epsilon  \mathcal{R} (\cot \pi\epsilon{-}i),-2 \epsilon \mathcal{R} \big(\log\tfrac{s_1}{s}{-}i\pi {+}\Psi_1^\epsilon\big),\right.\notag\\& \hspace{-0.7cm} \left.-2 \epsilon \mathcal{R}\big(\log\tfrac{s_1}{s}{+}\Psi_2^\epsilon \big),
   -2 \epsilon \mathcal{R}\big(\log\tfrac{s_2}{s}{+}2\pi i{+}\Psi_2^\epsilon \big),-2 \epsilon \mathcal{R} \big(\log (-\tfrac{s_2}{s}){+}\Psi_2^\epsilon\big),0 \right)^\top\,.
\end{align}
Verifying crossing in the multi-Regge-collinear regime $\{s>0, s_1>0, s_2>0, z\to0^+, \bar{z}\to0^+\}$ thus amounts to checking if
\begin{equation}
    \left[\vec{I}_{ 315\leftarrow24}^{~\dag}\right]_{\substack{ \substack{\raisebox{\depth}{\scalebox{1}[1]{$\curvearrowright$}}s_{\phantom{1}}\\ \raisebox{\depth}{\scalebox{1}[1]{$\curvearrowright$}}s_1}}}-\vec{I}_{ 215\leftarrow34}\stackrel{?}{=}\Cut_{s_{15}}\vec{I}_{ 215\leftarrow34}\,.\label{eq:pentagonPrediction}
\end{equation}
Note that the right-hand-side here is a non-trivial cut that cannot be deduced solely from the Cutkosky rules.

To check if \eqref{eq:pentagonPrediction} is correct, we compute the $s_{15}$-cut directly and expand the result in the multi-Regge-collinear region where \eqref{eq:bdryCTE} is defined. Using the embedding space formalism outlined earlier, we obtain (in generic kinematics):
\begin{align}\label{eq:cuts15}
&\Cut_{s_{15}}\vec{I}_{ 215\leftarrow34}=2ir_\Gamma s_{51}^{-\epsilon}\sin
   (\pi  \epsilon)\left(
        1,\vec{0}_4, 
   -\mathcal{F}\left(\tfrac{s_{34,51}}{s_{12}}\right), 
   -\mathcal{F}\left(\tfrac{s_{23,51}}{s_{45}}\right), 0,
   \mathcal{F}\left(\tfrac{s_{51,23}
   s_{51,34}}{s_{23}
   s_{34}}\right), 0,  \bullet
    \right)^\top\notag\\&
    \text{with} \quad s_{ij,kl}\equiv s_{ij}-s_{kl} \ \text{and} \ \mathcal{F}(Z)\equiv 2~{}_2F_1\left(1,-\epsilon;1-\epsilon;1-Z\right)\,.
\end{align}
The dot \say{$\bullet$} indicates that the computation of the pentagon cut is omitted, since it is sufficient to know it vanishes at $\Delta=0$ for the purpose of this calculation.

Upon comparing the multi-Regge-collinear expansion of \eqref{eq:cuts15}  with \eqref{eq:pentagonPrediction}, we observe a perfect match.

At this point, we would like to emphasize that there is no loss of generality in checking crossing in just the multi-Regge-collinear regime. Indeed, since both the master integrals and the cuts satisfy the \emph{same} differential equation, checking crossing at a single point in the kinematic space is sufficient. Checking it elsewhere (for instance, away from $\Delta=0$ where the pentagon contribution does \emph{not} vanish) is equivalent to integrating the differential equation in App.~\ref{app:numerics} with the initial condition determined by the data where the crossing is already established.

\subsection{Application: All time-ordered amplitudes from a single one}\label{sec:pentagonApplications}

An interesting application of crossing symmetry involves asking if there are paths in the kinematic space that could serve as bridges between the $ k\ell m\leftarrow ij$ \emph{ordinary} amplitude and its relabelings \cite{Giroux:2021aom}, 
instead of relating amplitudes to generalized observables or cuts as above.
In the discussion below, we will exemplify this idea for the one-loop pentagon amplitude, assuming knowledge of a differential equation and a single boundary condition as our only input.

Together, ordering the external particles for a five-point process can be done in $5!=120$ different ways. The claim is that the amplitude for each ordering can be obtained \emph{from a single one} using three elementary relabeling moves of the external particles. In practice, these moves are realized by transporting, say, the amplitude in the $k\ell m\leftarrow ij$ channel (initial condition) along paths in the kinematic space whose endpoints are associated with time-ordered amplitudes with permuted labels. These three moves are summarized as follows:
\begin{equation}\label{eq:moves}
    \adjustbox{valign=c}{\tikzset{every picture/.style={line width=1pt}}      
\begin{tikzpicture}[x=0.75pt,y=0.75pt,yscale=-1.4,xscale=1.4]
\tikzset{ma/.style={decoration={markings,mark=at position 0.5 with {\arrow[scale=0.7]{>}}},postaction={decorate}}}
\tikzset{ma2/.style={decoration={markings,mark=at position 0.3 with {\arrow[scale=0.7,red!70!black]{>}}},postaction={decorate}}}
\tikzset{ma3/.style={decoration={markings,mark=at position 0.7 with {\arrow[scale=0.7,blue!80!black]{>}}},postaction={decorate}}}
\tikzset{mar/.style={decoration={markings,mark=at position 0.5 with {\arrowreversed[scale=0.7]{>}}},postaction={decorate}}}
\tikzset{mar2/.style={decoration={markings,mark=at position 0.3 with {\arrowreversed[scale=0.7]{>}}},postaction={decorate}}}
\draw[mar2]    (326.27,136.78) -- (344.76,137.01) node[right,scale=0.75] {$i$};
\draw[mar2]    (326.24,115.18) -- (344.73,115.4) node[right,scale=0.75] {$j$};
\draw[ma]    (312.33,115.05) -- (293.84,115.23) node[left,scale=0.75] {$k$};
\draw[ma]    (312.33,136.63) -- (293.84,136.8) node[left,scale=0.75] {$m$};
\draw[ma]  (307.1,125.84) -- (294,125.76) node[left,scale=0.75] {$\ell$};
\draw[line width = 0.7, <->]  [color=gray!60]  (320,81.43) -- (320.14,107.86) node[midway, right, yshift=-1.6pt, font=\footnotesize, color=gray!60, line width = 0.7]{$\substack{\textstyle \text{permutation}\\ \textstyle \text{path}}$};
\draw[mar2]    (240.27,72.78) -- (258.76,73.01) node[right,scale=0.75] {$j$};
\draw[mar2]    (240.24,51.18) -- (258.73,51.4) node[right,scale=0.75] {$i$};
\draw[ma]    (226.33,51.05) -- (207.84,51.23) node[left,scale=0.75] {$k$};
\draw[ma]    (226.33,72.63) -- (207.84,72.8) node[left,scale=0.75] {$m$};
\draw[ma]    (221.1,61.84) -- (208,61.76) node[left,scale=0.75] {$\ell$};
\draw[mar2]    (209.27,136.78) -- (227.76,137.01) node[right,scale=0.75] {$j$};
\draw[mar2]    (209.24,115.18) -- (227.73,115.4) node[right,scale=0.75] {$i$};
\draw[ma]    (195.33,115.05) -- (176.84,115.23) node[left,scale=0.75] {$m$};
\draw[ma]    (195.33,136.63) -- (176.84,136.8) node[left,scale=0.75] {$k$};
\draw[ma]  (190.1,125.84) -- (177,125.76) node[left,scale=0.75] {$\ell$};
\draw[mar2]    (444.27,135.78) -- (462.76,136.01) node[right,scale=0.75] {$i$};
\draw[mar2]    (444.24,114.18) -- (462.73,114.4) node[right,scale=0.75] {$m$};
\draw[ma]    (430.33,114.05) -- (411.84,114.23) node[left,scale=0.75] {$k$};
\draw[ma]    (430.33,135.63) -- (411.84,135.8) node[left,scale=0.75] {$j$};
\draw[ma]   (425.1,124.84) -- (412,124.76) node[left,scale=0.75] {$\ell$};
\draw[line width = 0.7, <->]  [color=gray!60] (278.33,124.63) -- (235.84,124.8) node[midway, above, yshift=-1.6pt, font=\footnotesize, color=gray!60, line width = 0.7]{$\substack{\textstyle \text{relabeling}\\ \textstyle \text{path}}$};
\draw[mar2]    (415.27,72.78) -- (433.76,73.01) node[right,scale=0.75] {$i$};
\draw[mar2]    (415.24,51.18) -- (433.73,51.4) node[right,scale=0.75] {$j$};
\draw[ma]    (401.33,51.05) -- (382.84,51.23) node[left,scale=0.75] {$\ell$};
\draw[ma]    (401.33,72.63) -- (382.84,72.8) node[left,scale=0.75] {$m$};
\draw[ma]   (396.1,61.84) -- (383,61.76) node[left,scale=0.75] {$k$};
\draw[mar2]    (326.27,72.78) -- (344.76,73.01) node[right,scale=0.75] {$i$};
\draw[mar2]    (326.24,51.18) -- (344.73,51.4) node[right,scale=0.75] {$j$};
\draw[ma]    (313.33,51.05) -- (294.84,51.23) node[left,scale=0.75] {$k$};
\draw[ma]    (312.33,72.63) -- (293.84,72.8) node[left,scale=0.75] {$\ell$};
\draw[ma]  (307.1,61.84) -- (294,61.76) node[left,scale=0.75] {$m$};
\draw[line width = 0.7, <->]  [color=gray!60]  (396.33,123.63) -- (353.84,123.8) node[midway, above, yshift=-1.6pt, font=\footnotesize, color=gray!60, line width = 0.7]{$\substack{\textstyle \text{crossing}\\ \textstyle \text{path}}$};

\draw  [fill=gray!5  ,fill opacity=1 ][line width=1]  (306.81,61.84) .. controls (306.81,54.75) and (312.56,49) .. (319.65,49) .. controls (326.74,49) and (332.49,54.75) .. (332.49,61.84) .. controls (332.49,68.93) and (326.74,74.68) .. (319.65,74.68) .. controls (312.56,74.68) and (306.81,68.93) .. (306.81,61.84) -- cycle ;
\draw  [fill=gray!5  ,fill opacity=1 ][line width=1]  (395.81,61.84) .. controls (395.81,54.75) and (401.56,49) .. (408.65,49) .. controls (415.74,49) and (421.49,54.75) .. (421.49,61.84) .. controls (421.49,68.93) and (415.74,74.68) .. (408.65,74.68) .. controls (401.56,74.68) and (395.81,68.93) .. (395.81,61.84) -- cycle ;
\draw  [fill=gray!5  ,fill opacity=1 ][line width=1]  (424.81,124.84) .. controls (424.81,117.75) and (430.56,112) .. (437.65,112) .. controls (444.74,112) and (450.49,117.75) .. (450.49,124.84) .. controls (450.49,131.93) and (444.74,137.68) .. (437.65,137.68) .. controls (430.56,137.68) and (424.81,131.93) .. (424.81,124.84) -- cycle ;
\draw  [fill=gray!5  ,fill opacity=1 ][line width=1]  (189.81,125.84) .. controls (189.81,118.75) and (195.56,113) .. (202.65,113) .. controls (209.74,113) and (215.49,118.75) .. (215.49,125.84) .. controls (215.49,132.93) and (209.74,138.68) .. (202.65,138.68) .. controls (195.56,138.68) and (189.81,132.93) .. (189.81,125.84) -- cycle ;
\draw  [fill=gray!5,fill opacity=1 ][line width=1]  (220.81,61.84) .. controls (220.81,54.75) and (226.56,49) .. (233.65,49) .. controls (240.74,49) and (246.49,54.75) .. (246.49,61.84) .. controls (246.49,68.93) and (240.74,74.68) .. (233.65,74.68) .. controls (226.56,74.68) and (220.81,68.93) .. (220.81,61.84) -- cycle ;
\draw  [fill=gray!5  ,fill opacity=1 ][line width=1]  (306.81,125.84) .. controls (306.81,118.75) and (312.56,113) .. (319.65,113) .. controls (326.74,113) and (332.49,118.75) .. (332.49,125.84) .. controls (332.49,132.93) and (326.74,138.68) .. (319.65,138.68) .. controls (312.56,138.68) and (306.81,132.93) .. (306.81,125.84) -- cycle ;

\draw (314.49,120.94) node [anchor=north west][inner sep=0.75pt]  {$S$};
\draw (228.49,56.94) node [anchor=north west][inner sep=0.75pt]  {$S$};
\draw (197.49,120.94) node [anchor=north west][inner sep=0.75pt]    {$S$};
\draw (432.49,119.94) node [anchor=north west][inner sep=0.75pt]    {$S$};
\draw (403.49,56.94) node [anchor=north west][inner sep=0.75pt]   {$S$};
\draw (314.49,56.94) node [anchor=north west][inner sep=0.75pt]   {$S$};
\draw (267,58) node [anchor=north west][inner sep=0.75pt]   [align=left] {or};
\draw (356,58) node [anchor=north west][inner sep=0.75pt] [align=left] {or};
\end{tikzpicture}}
\end{equation}
Note that, in contrast to previous sections where the permutations were considered to have no effect within final states, or within initial states, in the above diagrams the ordering \emph{does} matter since the placement of labels indicates the choice of a rapidity ordering, see Fig.~\ref{fig:mrkCoords}. 

On one hand, the simplest move is realized through the \emph{relabeling path}. This path lies entirely within a given multi-Regge limit, meaning that it is fully contained into a single large rapidity region of the kinematic space. Simple instances are paths connecting the endpoints $(z_0,\bar{z}_0)$ to $(1-z_0,1-\bar{z}_0)$ in the $(z,\bar{z})$ space. A general example would be 
\begin{equation}\label{eq:relPath}
    \gamma_{\text{rel}}=\left\{(z(\theta),\bar{z}(\theta))=\big(z_0+\tfrac{(1-2z_0)\sin\theta}{\cos\theta+\sin\theta},\Bar{z}_0+\tfrac{(1-2\Bar{z}_0)\sin\theta}{\cos\theta+\sin\theta}\big) \;|\; 0<\theta<\pi/2\right\}\,.
\end{equation}
To better appreciate our choice of nomenclature, it is sufficient to examine the special case of \eqref{eq:relPath} where $(z_0,\bar{z}_0)=(0,0)$ in the $ k\ell m\leftarrow ij$ channel: while particles $i$ and $m$ are collinear at $\theta=0$ ($s_{im}\ll s_{jk}$), particles $j$ and $k$ become collinear at $\theta=\pi/2$ ($s_{jk} \ll s_{im}$)
\begin{equation}
\begin{gathered}
     \stackrel{\adjustbox{valign=c}{
     \tikzset{every picture/.style={line width=1pt}}
\begin{tikzpicture}[x=0.75pt,y=0.75pt,yscale=-1,xscale=1]
\draw   (180.24,116.81) -- (180.19,159.24) -- (147.69,159.2) -- (140.21,137.98) -- (147.74,116.77) -- cycle ;
\draw    (147.69,159.2) -- (130.69,159.2) node[left,scale=0.75] {$m$};
\draw    (140.21,137.98) -- (130.21,137.98) node[left,scale=0.75] {$\ell$};
\draw    (147.74,116.77) -- (130.74,113.77) node[left,scale=0.75] {$k$};
\draw   (180.24,116.81) -- (197.24,113.81) node[right,scale=0.75] {$j$};
\draw  (180.19,159.24) -- (197.19,159.24) node[right,scale=0.75] {$i$};
\end{tikzpicture}
}}{\;(\theta=0)}\qquad
     \stackrel{\adjustbox{valign=c}{
\tikzset{every picture/.style={line width=1pt}}
\begin{tikzpicture}[x=0.75pt,y=0.75pt,yscale=-1,xscale=1]
\draw   (180.24,116.81) -- (180.19,159.24) -- (147.69,159.2) -- (140.21,137.98) -- (147.74,116.77) -- cycle ;
\draw    (147.69,159.2) -- (130.69,162.2)  node[left,scale=0.75] {$m$};
\draw    (140.21,137.98) -- (130.21,137.98)  node[left,scale=0.75] {$\ell$};
\draw    (147.74,116.77) -- (130.74,116.77) node[left,scale=0.75] {$k$};
\draw  (180.24,116.81) -- (197.24,116.81) node[right,scale=0.75] {$j$};
\draw  (180.19,159.24) -- (197.19,162.24) node[right,scale=0.75] {$i$};
\end{tikzpicture}
}}{\;(\theta=\tfrac{\pi}{2})}
\end{gathered}
\end{equation}
It is evident that these two limits are equivalent through an up-down reflection of the diagrams. In fact, one can verify through direct computations that the vectors of master integrals $\vec{I}$ at $\theta=0$ and $\theta=\pi/2$ are permutations of each other (thus justifying the name!), precisely matching the reflection prediction.

On the other hand, the most complicated move is realized through the \emph{crossing path}, which will be our main focus below. While most of the technical details associated with it are postponed to the next section and example, let us just say that, starting from the $ \textcolor{RoyalBlue}{k}\textcolor{RoyalBlue}{\ell} \textcolor{Maroon}{j}\leftarrow \textcolor{RoyalBlue}{i}\textcolor{Maroon}{m}$ amplitude, this move returns the $\textcolor{RoyalBlue}{k}\textcolor{RoyalBlue}{\ell} \textcolor{Maroon}{m} \leftarrow \textcolor{RoyalBlue}{i}\textcolor{Maroon}{j}$ one.
 
 Finally, even though the \emph{permutation paths} are encountered as interim steps of the crossing path (as will be discussed below), they stand as legitimate paths in their own right: they are used to exchange either a pair of incoming or outgoing particles. Along such paths, the sign of individual invariant remains unchanged, while the overall rapidity ordering gets shuffled around.

  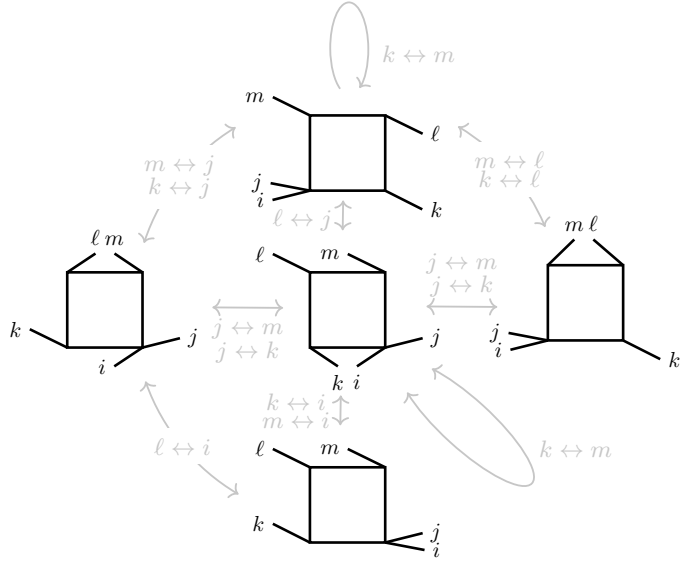
\begin{figure}
        \centering
   \adjustbox{valign=c}{\tikzset{every picture/.style={line width=1pt}}     
\begin{tikzpicture}[x=0.75pt,y=0.75pt,yscale=-1.2,xscale=1.2]
\tikzset{ma/.style={decoration={markings,mark=at position 0.5 with {\arrow[scale=0.1]{>}}},postaction={decorate}}}
\tikzset{mar/.style={decoration={markings,mark=at position 0.5 with {\arrowreversed[scale=0.1]{>}}},postaction={decorate}}}
\draw   (310.03,203.11) -- (341.67,203.11) -- (341.67,234.75) -- (310.03,234.75) -- cycle ;
\draw    (310.03,203.11) -- (294.45,195.4) node[left,scale=0.75] {$\ell$};
\draw    (310.03,234.75) -- (294.45,226.88) node[left,scale=0.75] {$k$};
\draw   (341.67,234.75) -- (358.2,237.9) node[right,scale=0.75] {$i$};
\draw  (341.67,234.75) -- (357.41,230.81) node[right,scale=0.75] {$j$};
\draw    (341.67,203.11) -- (325.93,195.4) node[left,scale=0.75] {$m$};
\draw   (310.03,121.11) -- (341.67,121.11) -- (341.67,152.75) -- (310.03,152.75) -- cycle ;
\draw    (310.03,121.11) -- (294.45,113.4) node[left,scale=0.75] {$\ell$};
\draw  (341.67,152.75) -- (329.8,160.66) node[below,scale=0.75] {$i$};
\draw    (341.67,152.75) -- (357.41,148.81)  node[right,scale=0.75] {$j$};
\draw    (341.67,121.11) -- (325.93,113.4) node[left,scale=0.75] {$m$};
\draw    (310.03,152.75) -- (321.8,160.66) node[below,scale=0.75] {$k$};
\draw[<->, color=gray!45  ,draw opacity=1, line width = 0.7]   (324,90.86) -- (323.8,103.46) node[midway, left, yshift=-1.6pt, font=\footnotesize]{$\textstyle \ell\leftrightarrow j$};
\draw   (208.03,121.11) -- (239.67,121.11) -- (239.67,152.75) -- (208.03,152.75) -- cycle ;
\draw    (208.03,121.11) -- (219.9,113.2) node[above,scale=0.75] {$\ell$};
\draw  (239.67,152.75) -- (255.41,148.81) node[right,scale=0.75] {$j$};
\draw    (208.03,152.75) -- (192.3,145.04) node[left,scale=0.75] {$k$};
\draw  (239.67,121.11) -- (227.9,113.2) node[above,scale=0.75] {$m$};
\draw  (239.67,152.75) -- (227.8,160.66) node[left,scale=0.75] {$i$};
\draw[<->, line width = 0.7] [color=gray!45  ,draw opacity=1]   (269,135.86) -- (298.8,136.06) node[midway,below, yshift=-0.2pt, font=\footnotesize]{$\substack{\textstyle j\leftrightarrow m\\ \textstyle j\leftrightarrow k}$};
\draw[<->] [color=gray!45  ,draw opacity=1, line width = 0.7]   (359,135.46) -- (388.8,135.66) node[midway,above, yshift=-0.2pt, font=\footnotesize]{$\substack{\textstyle j\leftrightarrow m\\ \textstyle j\leftrightarrow k}$};
\draw   (310.03,55.11) -- (341.67,55.11) -- (341.67,86.75) -- (310.03,86.75) -- cycle ;
\draw    (310.03,55.11) -- (294.45,47.4) node[left,scale=0.75] {$m$};
\draw   (341.67,86.75) -- (357.25,94.62) node[right,scale=0.75] {$k$};
\draw    (310.03,86.75) -- (293.51,83.6) node[left,scale=0.75] {$j$};
\draw    (310.03,86.75) -- (294.3,90.68) node[left,scale=0.75] {$i$};
\draw  (341.67,55.11) -- (357.41,62.82) node[right,scale=0.75] {$\ell$};
\draw[<->, line width = 0.7] [color=gray!45  ,draw opacity=1 ]   (323,172.86) -- (322.8,185.46) node[midway,left, yshift=-0.2pt, font=\footnotesize]{$\substack{\textstyle k\leftrightarrow i\\ \textstyle m\leftrightarrow i}$};
\draw  [draw opacity=0] (240.3,108.77) .. controls (246.97,88.54) and (260.68,71.22) .. (278.86,60.09) -- (325.85,136.93) -- cycle ; \draw[<->, line width = 0.7]  [color=gray!45  ,draw opacity=1 ] (240.3,108.77) .. controls (246.97,88.54) and (260.68,71.22) .. (278.86,60.09) node[midway, fill=white,font=\footnotesize]{$\substack{\textstyle m\leftrightarrow j\\ \textstyle k\leftrightarrow j}$} ;  
\draw  [draw opacity=0] (280.06,214.48) .. controls (262.17,203.89) and (248.07,187.23) .. (240.93,166.94) -- (325.85,136.93) -- cycle ; \draw[<->, line width = 0.7]  [color=gray!45  ,draw opacity=1 ] (280.06,214.48) .. controls (262.17,203.89) and (248.07,187.23) .. (240.93,166.94) node[midway, fill=white,font=\footnotesize]{$\textstyle \ell\leftrightarrow i$};  
\draw   (410.03,118.11) -- (441.67,118.11) -- (441.67,149.75) -- (410.03,149.75) -- cycle ;
\draw    (410.03,118.11) -- (421.14,107.29) node[above,scale=0.75] {$m$};
\draw  (441.67,149.75) -- (457.25,157.62) node[right,scale=0.75] {$k$};
\draw    (410.03,149.75) -- (393.51,146.6) node[left,scale=0.75] {$j$};
\draw    (410.03,149.75) -- (394.3,153.68) node[left,scale=0.75] {$i$};
\draw  (441.67,118.11) -- (429.14,107.29) node[above,scale=0.75] {$\ell$};
\draw  [draw opacity=0] (372.26,59.75) .. controls (388.42,69.49) and (401.42,84.2) .. (408.91,102.07) -- (325.85,136.93) -- cycle ; \draw[<->, line width = 0.7]  [color=gray!45  ,draw opacity=1 ] (372.26,59.75) .. controls (388.42,69.49) and (401.42,84.2) .. (408.91,102.07) node[midway, fill=white,font=\footnotesize]{$\substack{\textstyle m\leftrightarrow \ell\\ \textstyle k\leftrightarrow \ell}$};  
\draw  [draw opacity=0] (322.73,43.9) .. controls (320.3,40.56) and (318.63,34.28) .. (318.57,27.06) .. controls (318.49,16.33) and (322,7.61) .. (326.42,7.57) .. controls (330.84,7.54) and (334.49,16.21) .. (334.57,26.94) .. controls (334.63,33.99) and (333.13,40.18) .. (330.84,43.6) -- (326.57,27) -- cycle ; \draw[->, line width = 0.7]  [color=gray!45  ,draw opacity=1 ] (322.73,43.9) .. controls (320.3,40.56) and (318.63,34.28) .. (318.57,27.06) .. controls (318.49,16.33) and (322,7.61) .. (326.42,7.57) .. controls (330.84,7.54) and (334.49,16.21) .. (334.57,26.94) .. controls (334.63,33.99) and (333.13,40.18) .. (330.84,43.6) ;  
\draw  [draw opacity=0] (361.25,161.66) .. controls (367.5,164.8) and (375.2,170.22) .. (382.84,177.17) .. controls (398.01,191) and (407.19,205.63) .. (403.34,209.85) .. controls (399.49,214.08) and (384.07,206.3) .. (368.9,192.48) .. controls (360.9,185.19) and (354.57,177.68) .. (350.93,171.59) -- (375.87,184.83) -- cycle ; 
\draw[<->, line width = 0.7]  [color=gray!45  ,draw opacity=1 ] (361.25,161.66) .. controls (367.5,164.8) and (375.2,170.22) .. (382.84,177.17) .. controls (398.01,191) and (407.19,205.63) .. (403.34,209.85) .. controls (399.49,214.08) and (384.07,206.3) .. (368.9,192.48) .. controls (360.9,185.19) and (354.57,177.68) .. (350.93,171.59) ;  

\draw (405,190.26) node [anchor=north west][inner sep=0.75pt, font=\footnotesize]  [font=\footnotesize,color=gray!45  ,opacity=1 ]  {$k{}\leftrightarrow{}m$};
\draw (339,24.26) node [anchor=north west][inner sep=0.75pt, font=\footnotesize]  [font=\footnotesize,color=gray!45  ,opacity=1 ]  {$k{}\leftrightarrow{}m$};
\end{tikzpicture}}
    \caption{All crossing relations between permutations of the one-mass box.
     Looping arrows denote crossing moves that result in the same configuration up to a relabeling of the Mandelstam variables.}
    \label{fig:allBoxes}
\end{figure}
 
 To see this, it is helpful to circle back to momenta. For a generic five-point process $ k\ell m\leftarrow ij$, the multi-Regge kinematics is characterized by a fast propagation of the in states ($i$ and $j$) along the $\mp$ light-cone axes, and a strong ordering of the out-states ($k$, $\ell$, and $m$) based on the rapidity scale $0<x\ll 1$. In other words, while the outgoing momenta are organized as
\begin{equation}\label{eq:glc1}
    p_k^\mu=\left(\frac{p_k^+}{x},\, x p_k^-,\, p^\perp_k\right),\quad \
    p_\ell^\mu=\left(p_\ell^+,\, p_\ell^-,\, p^\perp_\ell\right), \quad \
    p_m^\mu=\left(x p_m^+,\, \frac{p_m^-}{x},\, p^\perp_m\right)\,,
\end{equation}
momentum conservation requires the incoming  momenta to be
\begin{equation}\label{eq:glc2}
p_i^\mu=-\left(0,\frac{p_m^-}{x}+p_\ell^-+x p_k^-, 0^\perp\right) \quad \text{and} \quad p_j^\mu=-\left(\frac{p_k^+}{x}+p_\ell^++x p_m^+,0,0^\perp\right)\,,
\end{equation}
where $p^\perp_k + p^\perp_\ell + p^\perp_m= 0^\perp$ in the transverse direction. The on-shell condition is $p^\perp\cdot p^\perp=p^+p^-$, and all \say{$\pm$} quantities are assumed to be non-zero. Plugging \eqref{eq:glc1} and \eqref{eq:glc2} into
\begin{equation}
s_{ab}=p_a^+p_b^- + p_b^+ p_a^- - 2p_a^\perp\cdot p_b^\perp \quad \text{for} \quad a,b\in\{i,j,k,\ell,m\}\,,
\end{equation}
it is not hard to see the dramatic effect on the kinematics following the label permutation in e.g., \ref{step:2} below: some of the original scales of order $1/x^2$ become finite after the rotation (e.g., $s_{ij}$), and vice versa (e.g., $s_{im}$ and $s_{kj}$). In other words, the original kinematic domain and the target kinematic domain are described by two \emph{parametrically disconnected} multi-Regge limits. 
 
 Therefore, to move from one domain to the other, we first need to step \emph{out} of the former and then step \emph{into} the latter, through a finite rapidity region. At the very end of this subsection, we will illustrate this procedure through an explicit example. 
 
Before moving on to the full master basis example, we find it instructive to pause and discuss how crossing relates amplitudes with different labels on a single four-propagator master integral.  For instance, one may be interested in relating different channels of the same diagram.  This is shown in Fig.~\ref{fig:allBoxes} in the case
of the five-point one-mass box $\cM^\text{box}=\cM_{0a_ja_ka_\ell a_m}^\text{pent}$ (whose functional form was given and studied earlier around \eqref{eq:onemassbox}).

Of particular interest is the fact that the central diagram connects with all the others. Thus, as soon as the central diagram and all its cuts are known, one can deduce the amplitude in all the other channels. Since the central diagram only has one cut, it can be obtained from its imaginary part from the usual Cutkosky rules. Hence, it suffices to start with the conventional time-ordered diagram to deduce all the others by crossing.

\paragraph{Crossing path between time-ordered amplitudes}
The purpose of this section is to provide the schematic instructions on how to obtain the $\textcolor{RoyalBlue}{k}\textcolor{RoyalBlue}{\ell} \textcolor{Maroon}{m} \leftarrow \textcolor{RoyalBlue}{i}\textcolor{Maroon}{j}$ time-ordered amplitude from the $ \textcolor{RoyalBlue}{k}\textcolor{RoyalBlue}{\ell} \textcolor{Maroon}{j}\leftarrow \textcolor{RoyalBlue}{i}\textcolor{Maroon}{m}$ one, by a composition of crossing moves and permuting labels. This procedure is a special case of \eqref{eq:crossing-BC-AB} and generalizes to arbitrary multiplicity \cite{Mizera:2021fap}. This section is followed by a concrete example that provides explicit equations to support what is discussed here. 

The sequence of steps required to define the crossing path between time-ordered amplitudes is given by:
\begin{equation}\label{eq:crossinPathTOA}
   \hspace{-0.5cm} \adjustbox{valign=c}{\tikzset{every picture/.style={line width=1pt}}
\begin{tikzpicture}[x=0.75pt,y=0.75pt,yscale=-0.9,xscale=0.9]
\tikzset{ma/.style={decoration={markings,mark=at position 0.5 with {\arrow{>}}},postaction={decorate}}}
\tikzset{mar/.style={decoration={markings,mark=at position 0.5 with {\arrowreversed{>}}},postaction={decorate}}}
\draw[->, line width=0.7pt]    (153.37,57.72) -- (214.89,57.45) node[midway, above, yshift=-1.6pt, font=\footnotesize]{$\substack{\textstyle \text{cross}\\ \textstyle i\leftrightarrow m}$};
\draw[->, line width=0.7pt]    (216.04,168.67) -- (156.22,168.39) node[midway, above, yshift=-1.6pt, font=\footnotesize]{$\substack{\textstyle \text{cross}\\ \textstyle i\leftrightarrow j}$};
\draw  [fill=gray!30 ,fill opacity=1 ][line width=1]  (284.64,45.44) -- (343.92,45.44) -- (343.92,66.19) -- (284.64,66.19) -- cycle node[midway, above, yshift=15pt, xshift=19pt, font=\footnotesize]{$\textstyle (B)$};
\draw  [fill=gray!30 ,fill opacity=1 ][line width=1]  (285.49,152.15) -- (344.77,152.15) -- (344.77,172.9) -- (285.49,172.9) -- cycle node[midway, above, yshift=15pt, xshift=19pt, font=\footnotesize]{$\textstyle (C)$};
\draw[ma, color=RoyalBlue]    (72.91,57.72) -- (53.44,58.01)  node[left,scale=0.75] {$\ell$};
\draw[ma, color=RoyalBlue]    (83.92,38.24) -- (53.44,38.53) node[left,scale=0.75] {$k$};
\draw[ma, color=Maroon]    (84.77,75.51) -- (54.28,75.79) node[left,scale=0.75] {$m$};
\begin{scope}[xshift=-2]
\draw[mar, color=Maroon] (105.94,38.53) -- (136.43,38.24) node[right,scale=0.75] {$j$};
\draw[mar, color=RoyalBlue]  (105.94,74.94) -- (136.43,74.66) node[right,scale=0.75] {$i$};
\end{scope}
\draw[ma, color=RoyalBlue]    (260.08,38.24) -- (229.59,38.53) node[left,scale=0.75] {$k$};
\draw[ma, color=RoyalBlue]    (260.93,75.51) -- (230.44,75.79) node[left,scale=0.75] {$\ell$};
\draw[mar, color=Maroon]    (375.26,56.06) -- (394.74,55.83) node[right,scale=0.75] {$m$};
\draw[ma, color=RoyalBlue]    (364.2,75.51) -- (312.87,92.45) node[below,scale=0.75] {$i$};
\draw[mar, color=Maroon]    (363.46,38.24) -- (393.94,38.05) node[right,scale=0.75] {$j$};
\draw[ma, color=RoyalBlue]    (75.45,166.97) -- (55.98,167.26) node[left,scale=0.75] {$\ell$};
\draw[ma, color=RoyalBlue]    (86.46,147.5) -- (55.98,147.78) node[left,scale=0.75] {$k$};
\draw[ma, color=Maroon]    (87.31,184.76) -- (56.82,185.04) node[left,scale=0.75] {$j$};
\draw[mar, color=Maroon]  (108.48,148.62) -- (138.97,148.34) node[right,scale=0.75] {$m$};
\draw[mar, color=RoyalBlue]  (108.48,184.19) -- (138.97,183.91) node[right,scale=0.75] {$i$};
\draw[ma, color=RoyalBlue]    (260.93,144.95) -- (230.44,145.24) node[left,scale=0.75] {$k$};
\draw[ma, color=RoyalBlue]    (261.77,182.22) -- (231.29,182.5) node[left,scale=0.75] {$\ell$};
\draw[mar, color=Maroon]    (376.11,162.77) -- (395.59,162.54) node[right,scale=0.75] {$j$};
\draw[ma, color=RoyalBlue]    (365.04,182.22) -- (314.56,197.46) node[below,scale=0.75] {$i$};
\draw[mar, color=Maroon]    (364.3,144.95) -- (394.79,144.76) node[right,scale=0.75] {$m$};
\draw [color=Orange  ,draw opacity=1, dashed]  (312.59,138.71) -- (312.59,201.25) ;
\draw [color=Orange  ,draw opacity=1, dashed]  (311.89,30) -- (311.74,95.54) ;
\draw  [draw opacity=0] (403.46,134.42) .. controls (406.65,127.84) and (408.91,115.55) .. (409.12,101.44) .. controls (409.31,88.12) and (407.62,76.37) .. (404.91,69.48) -- (398.69,101.29) -- cycle ; \draw[<-, line width=0.7pt]   (403.46,134.42) .. controls (406.65,127.84) and (408.91,115.55) .. (409.12,101.44) .. controls (409.31,88.12) and (407.62,76.37) .. (404.91,69.48) node[midway, yshift=-12pt, fill=white, font=\footnotesize]{$\substack{\textstyle \text{permute}\\ \textstyle j\leftrightarrow m}$};  
\draw  [fill=gray!5  ,fill opacity=1 ][line width=1]  (249.92,163.59) .. controls (249.92,151.89) and (259.4,142.41) .. (271.09,142.41) .. controls (282.78,142.41) and (292.26,151.89) .. (292.26,163.59) .. controls (292.26,175.28) and (282.78,184.76) .. (271.09,184.76) .. controls (259.4,184.76) and (249.92,175.28) .. (249.92,163.59) -- cycle ;
\draw  [fill=gray!5  ,fill opacity=1 ][line width=1]  (376.11,163.62) .. controls (376.07,175.31) and (366.57,184.76) .. (354.87,184.73) .. controls (343.18,184.7) and (333.73,175.19) .. (333.76,163.5) .. controls (333.79,151.81) and (343.3,142.35) .. (354.99,142.39) .. controls (366.69,142.42) and (376.14,151.93) .. (376.11,163.62) -- cycle ;
\draw  [fill=gray!5  ,fill opacity=1 ][line width=1]  (75.45,166.13) .. controls (75.45,154.43) and (84.93,144.95) .. (96.63,144.95) .. controls (108.32,144.95) and (117.8,154.43) .. (117.8,166.13) .. controls (117.8,177.82) and (108.32,187.3) .. (96.63,187.3) .. controls (84.93,187.3) and (75.45,177.82) .. (75.45,166.13) -- cycle node[midway, above, yshift=15pt, xshift=15pt, font=\footnotesize]{$\textstyle (D)$};
\draw  [fill=gray!5  ,fill opacity=1 ][line width=1]  (249.07,56.88) .. controls (249.07,45.18) and (258.55,35.7) .. (270.24,35.7) .. controls (281.94,35.7) and (291.42,45.18) .. (291.42,56.88) .. controls (291.42,68.57) and (281.94,78.05) .. (270.24,78.05) .. controls (258.55,78.05) and (249.07,68.57) .. (249.07,56.88) -- cycle ;
\draw  [fill=gray!5  ,fill opacity=1 ][line width=1]  (375.26,56.91) .. controls (375.23,68.6) and (365.72,78.05) .. (354.03,78.02) .. controls (342.33,77.99) and (332.88,68.48) .. (332.91,56.79) .. controls (332.95,45.1) and (342.45,35.64) .. (354.15,35.68) .. controls (365.84,35.71) and (375.29,45.21) .. (375.26,56.91) -- cycle ;
\draw  [fill=gray!5  ,fill opacity=1 ][line width=1]  (72.91,56.88) .. controls (72.91,45.18) and (82.39,35.7) .. (94.09,35.7) .. controls (105.78,35.7) and (115.26,45.18) .. (115.26,56.88) .. controls (115.26,68.57) and (105.78,78.05) .. (94.09,78.05) .. controls (82.39,78.05) and (72.91,68.57) .. (72.91,56.88) -- cycle node[midway, above,yshift=15pt, xshift=15pt, font=\footnotesize]{$\textstyle (A)$};
\draw (87.09,50.28) node [anchor=north west][inner sep=0.75pt]    {$S$};
\draw (263.24,50.28) node [anchor=north west][inner sep=0.75pt]    {$S$};
\draw (303.74,48.28) node [anchor=north west][inner sep=0.75pt, yshift=-0.2]    {\footnotesize $X$};
\draw (345.86,48.21) node [anchor=north west][inner sep=0.75pt, yshift=1]    {$S^{\dagger }$};
\draw (89.63,159.53) node [anchor=north west][inner sep=0.75pt]    {$S$};
\draw (264.09,156.99) node [anchor=north west][inner sep=0.75pt]    {$S$};
\draw (304.59,154.99) node [anchor=north west][inner sep=0.75pt, yshift=-0.2]    {\footnotesize $X$};
\draw (346.7,154.99) node [anchor=north west][inner sep=0.75pt, yshift=1]    {$S^{\dagger}$};
\draw (-155,90.99) node [anchor=north west][inner sep=0.75pt]    {$\substack{\textstyle \text{{Crossing path between}}\\ \textstyle \text{{time-ordered amplitudes:}}}$};
\end{tikzpicture}}
\end{equation}
Above, the color code mimics the one used in the previous sections: the red lines are the ones crossed between the two time-ordered amplitudes. In particular, we see that it is organized into three steps. To each of them, we associate a path along which we transport an initial condition such that:
 \begin{enumerate}[label=(\roman*)]
 \item Transporting the initial condition along the first path, we \emph{cross} particles $i$ and $m$.\label{step:1}
 \item Transporting the result of \ref{step:1} along the second path, we \emph{permute} particles $j$ and $m$.\label{step:2}
 \item Transporting the result of \ref{step:2} along the third path, we \emph{cross} particles $j$ and $i$.\label{step:3}
 \end{enumerate}
At this stage, the distinction between \say{crossing} and \say{permuting} may not be clear. The key observation is that, while \ref{step:1} and \ref{step:3} only involve flipping the signs of certain Mandelstam invariants (according to \eqref{eq:crossingpath}), \ref{step:2} leaves the sign of the invariant unchanged, but \emph{significantly} alter the rapidity dependence of certain scales.

We now illustrate how to use the crossing path between time-ordered amplitudes through an explicit example.

\paragraph{Example: the $514\leftarrow 32$ time-ordered amplitude from the $512\leftarrow 34$ one}
From now on, we use the explicit labeling $\{i,j,k,\ell,m\}=\{3,4,5,1,2\}$. Without loss of generality, we take our initial condition to be at the coplanar multi-Regge point $A$ in the $512\leftarrow 34$ channel, which is defined in terms of momenta by $p_1^\perp = p_2^\perp = p_5^\perp = 0$ and $p_1^\pm=p_2^\pm=p_5^\pm=-1$. At $A$, the integrals evaluate to
\begin{equation}\label{eq:point0}
   \vec{I}_{512\leftarrow 34}^A= \begin{blockarray}{cccccc}
   {}&{}&{}&{}&{}&\substack{\text{\textcolor{gray}{Row}}\\\text{\textcolor{gray}{mult.}}} \\
\begin{block}{(ccccc)c}
-1 & -i \pi  & \frac{7}{12}\pi ^2 & \frac{7}{3}\zeta_3+\frac{i}{4}\pi ^3 & \frac{7 i}{3}\pi  \zeta_3-\frac{73}{1440}\pi
   ^4&\textcolor{gray}{3} \\
 -1 & 0 & \frac{1}{12}\pi ^2& \frac{7}{3}\zeta_3& \frac{47}{1440}\pi ^4&\textcolor{gray}{2} \\
2 & 2 i \pi  & -\frac{5}{6} \pi ^2& -\big(\frac{14}{3}\zeta_3+\frac{i}{6}\pi ^3\big) & -\big(\frac{13}{240}\pi
   ^4+\frac{14 i}{3}\pi  \zeta_3\big)&\textcolor{gray}{2} \\
2 & 0 & -\frac{1}{2}\pi ^2 & -\frac{20}{3}\zeta_3& -\frac{43}{720}\pi ^4&\textcolor{gray}{2} \\
 2 & 2 i \pi  & -\frac{5}{6}\pi ^2 & -\big(\frac{14}{3}\zeta_3+\frac{i}{6}\pi ^3\big) & -\big(\frac{13}{240}\pi
   ^4+\frac{14 i}{3}\pi  \zeta_3\big)&\textcolor{gray}{1} \\
 0 & 0 & 0 & \substack{\frac{1}{4} \pi\mathfrak{Li}_2-\frac{i}{108}\pi ^3} & \substack{\frac{i}{18}\pi ^2 \mathfrak{Li}_2-\frac{3}{5}\pi 
  \mathfrak{Li}_3-\frac{i}{6} \mathfrak{Li}_4\\+\frac{17}{720}\pi ^4+\frac{1}{80}\pi^2 \log^23-\frac{i}{108} \pi ^3\log3} & \textcolor{gray}{1}\\
\end{block}
\end{blockarray}\,,
\end{equation}
where $\mathfrak{Li}_2=\Im\big(\text{Li}_2\big(\e^{\frac{i \pi }{3}}\big)\big)$, $\mathfrak{Li}_3= \Im\big(\text{Li}_3\big(\frac{i}{\sqrt{3}}\big)\big)$ and $\mathfrak{Li}_4=\Im\big(\text{Li}_4\big(\e^{\frac{i \pi }{3}}\big)\big)$.\footnote{The transcendental numbers appearing here are, as expected, part of the set of $\mathbb{Q}$-linearly independent constants that arise in \emph{two-loop} pentagon integrals in the multi-Regge kinematics at the exact same point (see \cite[Tab.~2]{Caron-Huot:2020vlo}).} To avoid clutters, the boundary constants are given as $(5\times11)$ matrices: the first index labels the weight/order in $\epsilon$, the second labels the master integral. Furthermore, the gray label next to each row denotes its multiplicity. For example, the notation in \eqref{eq:point0} indicates that $I_1^A$, $I_2^A$ and $I_3^A$ evaluates to the same expression.

From $\vec{I}_{512\leftarrow 34}^A$, our goal is to calculate the $514\leftarrow32$ channel time-ordered amplitude. This requires explicit prescriptions for each path in \ref{step:1}, \ref{step:2} and \ref{step:3}.

Let us first choose our paths' endpoints. We choose our target (final) endpoint $D$ in the $514\leftarrow32$ channel to be the multi-Regge point defined by $\vec{p}_4^\perp\cdot\vec{p}_5^\perp=0$ and $p_1^\pm=p_4^\pm=p_5^\pm=-1$.

To efficiently interpolate between $A$ and $D$, it will be convenient to introduce three intermediate points lying in the kinematic region where particles $2$ and $4$ are incoming. These are:
\begin{subequations}
    \begin{align}
    B&=\left\{-\tfrac{1}{x}-x,\; -1+x+x^2,\;
   -\tfrac{1}{x^2}+3-x^2,\; -1-x+x^2,\;
   \tfrac{1}{x}+x \left|~0^+\right.\right\}\,,\\
   O&=\{-x,\; -x,\; -x,\; -x,\; x ~|~ 1\}\,,
   \\
     C&=\left\{-\tfrac{1}{x}-1+x,\; -\tfrac{1}{x^2}+3-x^2,\; -1+x+x^2,\;
   -\tfrac{1}{x^2}-x^2,\; \tfrac{1}{x}+2+x\left|~0^+\right.\right\}\,,
\end{align}
\end{subequations}
where the notation $\{s_{12}(x),s_{23}(x),s_{34}(x),s_{45}(x),s_{51}(x)|x\}$ is used to denote a point in the kinematic space as a function of the rapidity variable $x$. Note that, while $B$ and $C$ are,  respectively, multi-Regge points in the $513\leftarrow 24$ and $513\leftarrow 42$ channels, $O$ is defined at \emph{finite} rapidity. 

\begin{figure}
    \centering
        \adjustbox{valign=c}{\input{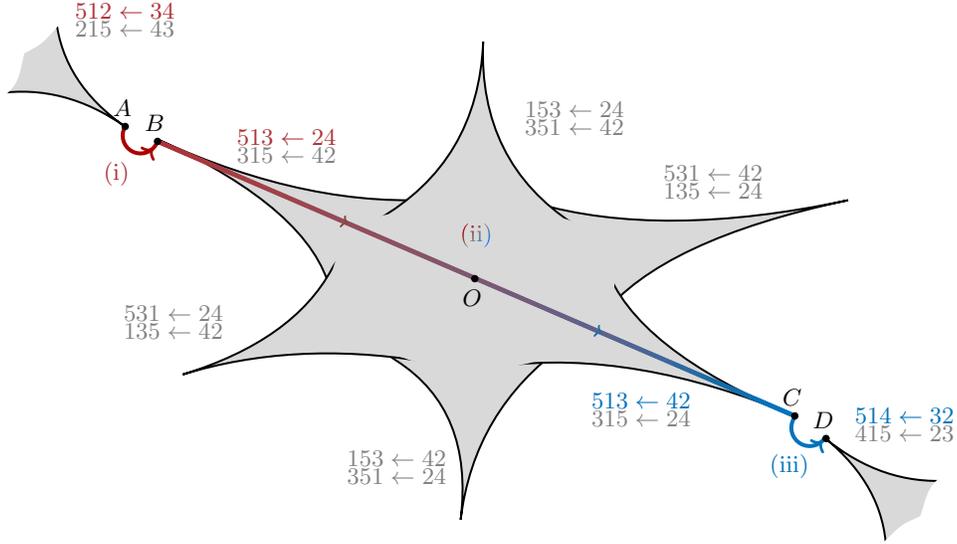}}
        \caption{
        The crossing path \eqref{eq:crossinPathTOA} between the (multi-Regge) time-ordered amplitudes is shown. The original and target channels are marked in red and blue, respectively. The three intermediate paths in \eqref{eq:crossinPathTOA} are labeled by \ref{step:1}, \ref{step:2} and \ref{step:3}. The central blob's spikes denote the twelve singular multi-Regge limits accessible within the channel with incoming particles $2$ and $4$. Each spike is tied to two Regge-collinear limits, which are related by a relabeling path.  
        An incomplete selection of spikes for blobs associated with other incoming particle pairs is also displayed. We move between disconnected blobs via two-particle crossings.
        }
        \label{fig:twoMRKS}
\end{figure}

So, in order to obtain the desired vector $\vec{I}_{514\leftarrow32}^D$, we simply transport $\vec{I}_{512\leftarrow 34}^A$ successively between the points
\begin{equation}\label{eq:listEndPoints}
    A\xrightarrow{\text{\ref{step:1}}} {B\xrightarrow{\text{\ref{step:2}}} O\xrightarrow{\text{\ref{step:2}}}C} \xrightarrow{\text{\ref{step:3}}} D\,.
\end{equation}
Here, the label above each arrow/path indicates the specific part of the crossing path \eqref{eq:crossinPathTOA} they contribute to (e.g., two-particle crossing or label permutation). The entire path is depicted in Fig.~\ref{fig:twoMRKS}.

In what follows, we provide an explicit series of integration contours that connect $A,B,O,C$ and $D$ in \eqref{eq:listEndPoints}.

As indicated in \ref{step:1}, the first step involves crossing particles $2$ and $3$. We achieve this by transporting the boundary condition $\vec{I}_{512\leftarrow 34}^A$ along large counter-clockwise arcs in the $s_{34}$ and $s_{12}$ upper half-planes, which are parametrized from \eqref{eq:mrkScaling} by
 \begin{equation}\label{eq:path1}
    \gamma_{A\to B}=\left\{
    s(\phi)=\e^{i\phi} \left| s \right|,
    s_1(\phi)=\e^{i\phi} \left| s_1 \right|
    \;\left|\; 0\leq \phi\leq\pi \right.\right\}\,.
\end{equation}  
Next, \ref{step:2} instructs us to permute particles $2$ and $4$, which is realized by transporting our initial condition from $B$ to $C$. A useful observation at this stage is that, while the processes $513\leftarrow 24$ and $513\leftarrow 42$ are different for $0<x\ll 1$, they agree at the \say{symmetric} point $O$. Heuristically, this should be enough to ensure the existence of a (polygonal) path contained in the kinematic region where $2$ and $4$ are incoming, which connects $B$ to $O$ and then $O$ to $C$. One simple candidate is the path $\gamma_{B\to C}$ connecting these points with straight lines.

Finally, as indicated in \ref{step:3}, the last step involves crossing particles $3$ and $4$. This is done by transporting the initial condition at $C$ obtained above to the point $D$ along the path 
\begin{equation}\label{eq:path34}
    \gamma_{C\to D}=\left\{
    s(\phi)=\e^{-i\phi} \left| s \right|,
    s_{1}(\phi)=\e^{-i\phi} \left| s_{1} \right|
    \;\left|\; 0\leq \phi\leq\pi \right.\right\}\,,
\end{equation}
parametrized from \eqref{eq:mrkScaling} after the permutation $\{s_{34},s_{23},s_{12},s_{51},s_{45}\}\to\{s_{23},s_{34},s_{41},s_{51},s_{25}\}$, where $s=s_1=-s_2=-1$ and $(z,\bar{z})=(\e^{i\pi/3},\e^{-i\pi/3})$.

We note that the crossing steps \ref{step:1} and \ref{step:3} (or, equivalently, the integration along $\gamma_{A\to B}$ and $\gamma_{C\to D}$) are technically very simple and similar to \eqref{eq:bdryCTE2cont}. The numerically non-trivial step is $\gamma_{B\to C}$, which connects two limits of a single kinematic region.

To summarize, the $514\leftarrow 32$ time-ordered amplitude at the point $D$ is obtained by transporting $\vec{I}_{512\leftarrow 34}^A$ along the union of contours $\gamma_{A\to D}=\gamma_{C\to D}\circ\gamma_{B\to C}\circ\gamma_{A\to B}\,$, namely
\begin{equation}
\begin{split} \vec{I}_{514\leftarrow32}^D&=\mathcal{P}\exp\left(\epsilon\int_{\gamma_{A\to D}}\d\boldsymbol{\Omega}\right)\cdot\vec{I}_{512\leftarrow 34}^A\\&=
        \begin{blockarray}{cccccc}
    {}&{}&{}&{}&{}&\substack{\text{\textcolor{gray}{Row}}\\\text{\textcolor{gray}{mult.}}} \\
\begin{block}{(ccccc)c}
 -1 & -i \pi  & \frac{7}{12} \pi ^2 & \frac{7}{3}\zeta_3+\frac{i}{4}\pi ^3 & \frac{7 i}{3}\pi  \zeta_3-\frac{73}{1440}\pi ^4&\textcolor{gray}{1} \\
 -1 & 0 & \frac{1}{12}\pi ^2 & \frac{7}{3}\zeta_3& \frac{47}{1440}\pi ^4&\textcolor{gray}{2} \\
-1 & -i \pi   & \frac{7}{12}\pi ^2 & \big(\frac{7}{3}\zeta_3+\frac{i}{4}\pi ^3\big) & \big(\frac{7 i}{3}\pi  \zeta_3-\frac{73}{1440}\pi ^4\big)&\textcolor{gray}{2} \\
2 & 2 i \pi  & -\frac{7}{6}\pi ^2 & -\big(\frac{14}{3}\zeta_3+\frac{i}{2}\pi ^3\big) & \big(\frac{73}{720}\pi ^4-\frac{14 i}{3}\pi  \zeta_3\big) &\textcolor{gray}{2} \\
 2 & 2 i \pi  & -\frac{3}{2}\pi ^2 & -\big(\frac{20}{3}\zeta_3+\frac{i}{6}\pi ^3\big) & -\big(\frac{43}{720}\pi ^4+\frac{14 i}{3}\pi  \zeta_3\big)&\textcolor{gray}{1} \\
 2 & 0 & -\frac{1}{2}\pi ^2 & -\frac{20}{3}\zeta_3 & -\frac{43}{720}\pi ^4& \textcolor{gray}{1}\\
 2 & 0 & -\frac{1}{6}\pi ^2 & -\frac{14}{3}\zeta_3 & -\frac{47}{720}\pi ^4& \textcolor{gray}{1}\\
 0 & 0 & 0 & \frac{1}{2}\pi  \mathfrak{Li}_2-\frac{i}{54}\pi ^3 & \substack{\frac{9}{65}
   \mathfrak{Li}_2^2+\frac{131 i}{9}\pi ^2 \mathfrak{Li}_2+\frac{97}{520}\pi
   \mathfrak{Li}_3\\-\frac{11i}{9}  \mathfrak{Li}_4+\frac{7 i}{9}\pi  \zeta_3+\frac{23}{520}\pi ^4\\-\frac{69}{520}\pi ^2 \log
   ^23-\frac{13i}{3}\pi ^3\log 3} & \textcolor{gray}{1}\\
\end{block}
\end{blockarray}.
\end{split}
\end{equation}
The above result for the bubble and box integrals were cross-verified by directly evaluating the closed-form formulae in \cite{Kozlov:2015kol} at endpoint $D$ in the multi-Regge limit. The agreement between these quantities is yet another successful stress test for the crossing equation.
\def\Tri{\mathcal{M}_{\mathrm{tri}}}
\def\ellnorm{|\ell|}

\section{Analytic obstruction to crossing massive particles in gapless theories}
\label{sec:anomalous}

Our conjecture for crossing described in Sec.~\ref{sec:conjectureI} and \ref{sec:conjectureII} is based on two key ingredients:
analyticity of the Green's function in off-shell kinematics, and the existence of an extension of this domain to scattering amplitudes on the mass shell.
In this section, we show that the second ingredient can, in principle, fail. We present a simple example where the time-ordered amplitude cannot be analytically continued to an inclusive observable without leaving the mass shell. This feature is closely tied to the existence of anomalous thresholds and is not unexpected from the axiomatic point of view. Indeed, it is known that beyond $4$-point scattering, it might no longer be possible to represent S-matrix elements in terms of a \emph{single} analytic function \cite{Bros:1972jh}. What happens in practice is that two thresholds (here, normal and anomalous) need to be approached from opposite directions in the complex plane, giving rise to an analytic obstruction or a gridlock, even at high energies. We will explain this phenomenon from several different points of view.

The key features of this example will be the presence of massive external \emph{and} massless internal lines.
Physically, this creates the possibility that internal excitations travel faster than external ones,
even in the Regge limit where they are both nearly luminal.

\subsection{Example: Triangle and anomalous thresholds}
\label{sec:exTridiagram}

The example we will study in detail is
the following triangle diagram for the time-ordered amplitude $1'2'3 \ot 12$:
\begin{equation}
	\label{eq:Tri diagram}
\begin{gathered}
\begin{tikzpicture}[line width=1]
\draw[photon] (0,0) -- (1,0.5);
\draw[] (1,0.5) -- node[right,scale=0.9] {$m_2$} (1,-0.5);
\draw[dashed] node[below,xshift=10,yshift=-5,scale=0.9] {$m_1$} (1,-0.5) -- (0,0);
\draw[Maroon] (1.5,0.75) -- (1,0.5);
\draw[dashed,RoyalBlue] (1.5,-0.75) -- (1,-0.5);
\draw[Maroon] (1,-0.5) -- (0.5,-0.75);
\draw[dashed,RoyalBlue] (-0.5,0.25) -- (0,0);
\draw[photon,RoyalBlue] (0,0) -- (-0.5,-0.25);
\node[] at (-1.5,0) {$s_{1}$};
\node[] at (1,1.1) {$s_{2}$};
\node[] at (1,-1.2) {$s$};
\node[scale=0.75,Maroon] at (1.7,0.8) {$2'$};
\node[scale=0.75,RoyalBlue] at (1.7,-0.8) {$1$};
\node[scale=0.75,Maroon] at (0.3,-0.8) {$2$};
\node[scale=0.75,RoyalBlue] at (-0.8,0.25) {$1'$};
\node[scale=0.75,RoyalBlue] at (-0.8,-0.25) {$3$};
\end{tikzpicture}
\end{gathered}
\hspace{1.5cm}
\begin{gathered}
\begin{tikzpicture}[line width=1]
\draw[photon] (0,0) -- (1,0.5);
\draw[] (1,0.5) -- (1,-0.05) node[left,color=gray]{$\circlearrowleft$};
\draw[dashed] (1,-0.5) -- (0.4,-0.2);
\draw[photon] (0,0) -- (1,0.5);
\draw[] (1,0.5) -- (1,-0.5);
\draw[dashed] (1,-0.5) -- (0,0);
\draw[] (1.5,0.75) -- (1,0.5);
\draw[dashed] (1.5,-0.75) -- (1,-0.5);
\draw[] (1,-0.5) -- (0.5,-0.75);
\draw[dashed] (-0.5,0.25) -- (0,0);
\draw[photon] (0,0) -- (-0.5,-0.25);
\draw[->,gray,line width=0.7] (-0.5,0) -- (-0.8,0);
\draw[->,gray,line width=0.7] (1,0.8) --++(30:0.3);
\draw[->,gray,line width=0.7] (1,-0.7) --++(-90:0.3);
\node[] at (-1.3,0) {$p_{1}$};
\node[] at (1,1.2) {$p_{2}$};
\node[] at (1,-1.2) {$p$};
\node[] at (1.7,0) {$\ell+p_{2}$};
\node[] at (0.1,-0.5) {$\ell-p_{1}$};
\node[] at (0.2,0.5) {$\ell$};
\end{tikzpicture}
\end{gathered}
\end{equation}
The dashed and solid lines represent particles with mass $m_1$ and $m_2$, while the wiggly line is a massless particle representing a graviton. This diagram depends only on three Mandelstam invariants, which we call $s$, $s_1$, and $s_2$, in addition to the two masses. For later convenience, at the start of the analytic continuation we will consider the complex-conjugated amplitude:
\be\label{eq:triDef}
\Tri^\dag = -\int \frac{\text{d}^\D\ell}{i \pi^{\D/2}}
  \frac{1}{[(\ell {-} p_1)^2+m_1^2+i\eps][(\ell{+}p_2)^2+m_2^2+i\eps][\ell^2+i\eps]}\,,
\ee
which amounts to reversing the signs of $i$ and the Feynman $i\eps$ (we have omitted coupling constants for simplicity). The prediction of the crossing equation is that exchanging particles $2 \leftrightarrow 2'$ leads to
\begin{align}
 \begin{gathered}
		\begin{tikzpicture}[baseline= {($(current bounding box.base)+(10pt,10pt)$)},line width=1, scale=1]
			\coordinate (a) at (0,0) ;
			\coordinate (b) at (-1,0.5) ;
			\coordinate (c) at (-1,-0.5);
			\coordinate (d) at (-1,0);
			\draw[dashed] (a) -- (b);
			\draw[photon] (b) -- (c);
			\draw[] (c) -- (a);
			\draw[dashed,RoyalBlue] (b)-- ++(170:0.5) node[left,scale=0.75] {$1'$};
			\draw[photon,RoyalBlue] (b)-- ++(140:0.5) node[left,scale=0.75,yshift=5] {$3$};
			\draw[Maroon] (c)-- ++(-15:0.5) node[right,scale=0.75,xshift=0] {$2'$};
			\draw[Maroon] (a)-- ++(180:0.5) node[left,scale=0.75,xshift=0] {$2$};
			\draw[dashed,RoyalBlue] (a)-- ++(-30:0.5) node[right,scale=0.75,xshift=-3] {$1$};
                 \begin{scope}[xshift=20pt,yshift=-25pt]  
                \draw[line width=1, line cap=round] (0,2) -- (0.135,2) (0,0) -- (0.135,0) (0.135,2) -- (0.135,0) node[pos=0,xshift=-10,yshift=-3]{$\dagger$} node[pos=1, right,yshift=5]{${\small\curvearrowright}s$};
                \end{scope}
                \begin{scope}[xshift=-50pt,yshift=31.9pt]  
                \draw[line width=1, line cap=round] (0,-2) -- (-0.135,-2) (0,0) -- (-0.135,0) (-0.135,-2) -- (-0.135,0);
                \end{scope}
                   \begin{scope}[xshift=0pt,yshift=-24pt]
                   \draw[line width=0.4, line cap=round] (-1.7,0) -- (-1.7,-0.135) (0.65,0)  -- (0.65,-0.135) (-1.7,-0.135) -- (0.65,-0.135) node[below, midway,yshift=2pt]{$\subset \cM^\dag$};
                \end{scope}
		\end{tikzpicture}
	\end{gathered}
	\quad  \stackrel{?}{=}  \quad 
	\begin{gathered}
		\begin{tikzpicture}[baseline= {($(current bounding box.base)+(10pt,10pt)$)},line width=1, scale=1]
			\coordinate (a) at (0,0) ;
			\coordinate (b) at (-1,0.5) ;
			\coordinate (c) at (-1,-0.5);
			\coordinate (d) at (-1,0);
			\draw[dashed] (a) -- (b);
			\draw[photon] (b) -- (c);
			\draw[] (c) -- (a);
			\draw[dashed,RoyalBlue] (b)-- ++(170:0.5) node[left,scale=0.75] {$1'$};
			\draw[photon,RoyalBlue] (b)-- ++(140:0.5) node[left,scale=0.75,yshift=5] {$3$};
			\draw[Maroon] (c)-- ++(-170:0.5) node[left,scale=0.75,xshift=-3] {$\bar{2}'$};
			\draw[Maroon] (a)-- ++(30:0.5) node[right,scale=0.75,xshift=-3] {$\bar{2}$};
			\draw[dashed,RoyalBlue] (a)-- ++(-30:0.5) node[right,scale=0.75,xshift=-3] {$1$};
			\draw[dashed,Orange] ($(d)+(180:0.25)+(0,1)$) -- ($(d)+(180:0.25)+(0,-1)$);
			\draw[dashed,Orange] ($(a)+(0:0.25)+(0,1)$) -- ($(a)+(0:0.25)+(0,-1)$);
                \begin{scope}[xshift=0pt,yshift=-25pt]
                    \draw[line width=0.4, line cap=round] (1,0) -- (1,-0.135) (0.3,0) -- (0.3,-0.135) (0.3,-0.135) -- (1,-0.135) node[below, midway]{$\subset S^\dag$};
                    \draw[line width=0.4, line cap=round] (-1.2,0) -- (-1.2,-0.135) (0.2,0) -- (0.2,-0.135) (-1.2,-0.135) -- (0.2,-0.135) node[below, midway,yshift=-2pt]{$\subset \cM$};
                    \draw[line width=0.4, line cap=round] (-2,0) -- (-2,-0.135) (-1.3,0) -- (-1.3,-0.135) (-2,-0.135) -- (-1.3,-0.135) node[below, midway]{$\subset S^\dag$};
                \end{scope}
		\end{tikzpicture}
	\end{gathered}
	\quad
	+
	\;
	\begin{gathered}
		\begin{tikzpicture}[baseline= {($(current bounding box.base)+(10pt,10pt)$)},line width=1, scale=1]
			\coordinate (a) at (0,0) ;
			\coordinate (b) at (-2,0.5) ;
			\coordinate (c) at (-1,-0.5);
			\coordinate (d) at (-1,0);
			\draw[dashed] (a) -- (b);
			\draw[photon] (b) -- (c);
			\draw[] (c) -- (a);
			\draw[dashed,RoyalBlue] (b)-- ++(170:0.5) node[left,scale=0.75] {$1'$};
			\draw[photon,RoyalBlue] (b)-- ++(140:0.5) node[left,scale=0.75,yshift=5] {$3$};
			\draw[Maroon] (c)-- ++(-170:0.5) node[left,scale=0.75,xshift=-3] {$\bar{2}'$};
			\draw[Maroon] (a)-- ++(30:0.5) node[right,scale=0.75,xshift=-3] {$\bar{2}$};
			\draw[dashed,RoyalBlue] (a)-- ++(-30:0.5) node[right,scale=0.75,xshift=-3] {$1$};
			\draw[dashed,Orange] ($(d)+(180:0.25)+(0,1)$) -- ($(d)+(180:0.25)+(0,-1)$);
			\draw[dashed,Orange] ($(a)+(0:0.25)+(0,1)$) -- ($(a)+(0:0.25)+(0,-1)$);
            \begin{scope}[xshift=0pt,yshift=-25pt]
                    \draw[line width=0.4, line cap=round] (1,0) -- (1,-0.135) (0.3,0) -- (0.3,-0.135) (0.3,-0.135) -- (1,-0.135) node[below, midway]{$\subset S^\dag$};
                    \draw[line width=0.4, line cap=round] (-1.2,0) -- (-1.2,-0.135) (0.2,0) -- (0.2,-0.135) (-1.2,-0.135) -- (0.2,-0.135) node[below, midway, yshift=-2pt]{$\subset \cM$};
                    \draw[line width=0.4, line cap=round] (-3,0) -- (-3,-0.135) (-1.3,0) -- (-1.3,-0.135) (-3,-0.135) -- (-1.3,-0.135) node[below, midway]{$\subset S^\dag$};
                \end{scope}
		\end{tikzpicture}\label{eq:atCrossing}
	\end{gathered}
\end{align}
Here, $s = -p^2$ rotates from negative to positive in the upper half-plane, with all the other invariants fixed. The reason why the rotation is clockwise is because we started with complex-conjugated amplitude.
In equations, we have
\be\label{Tri test with cut}
[\Tri^\dag]_{\raisebox{\depth}{\scalebox{1}[1]{$\curvearrowright$}}s} \; \stackrel{?}{=}\; \Tri + \Cut_{s_1} \Tri \,\equiv\, \Exp_2\,,
\ee
where $\Cut_{s_1} \Tri$ denotes the contribution from the second diagram on the right-hand side of \eqref{eq:atCrossing}. 
The quantity on the right-hand side is the in-in expectation value of an observable with momentum $2'$, $\Exp_{2}$.

The goal of this section is to explain why this equation breaks down when particle $2$ is  massive and on shell with $s_2 = m_2^2$. To illustrate this point, we will start with the off-shell configuration, say $s_2<m_2^2$, and show that the crossing path is cut off as we tune $s_2 \to m_2^2$. Physically, this phenomenon will have an explanation originating from the triangle anomalous threshold and that we cannot analytically continue around it when $s_2$ is on-shell. We also take $s_1 > m_1^2$ in order to make the problem non-trivial:
if $s_1$ were spacelike, the triangle could be viewed as part of a $2\to 2$ scattering process for which crossing is already well established.

Our interest in studying this specific configuration comes from inquiring into limitations of crossing symmetry when crossing a massive particle in a gapless theory. This question becomes particularly important in applications to computing gravitational waveforms, in which potential-scattering diagrams of the form \eqref{eq:Tri diagram} appear. The intuition developed in this section will help us decide about the path of analytic continuation between time-ordered amplitudes and expectation values of gravitational radiation. In particular, we will be able to understand why the issues described here go away when crossing a graviton instead of a massive particle.

\subsection{Analytic continuation in Schwinger parameters}
\label{sec:analytic-Schwinger}

While we could illustrate this idea with explicit expressions, they would not be entirely illuminating. For example, in $\D=4$, the result of $\Tri^\dag$ is written in terms of dilogarithms with square roots depending on the scales $s,s_1,s_2,m_1$ and $m_2$. Instead, we are going to manipulate $\Tri^\dag$ entirely at the level of integrals to work out under which conditions \eqref{Tri test with cut} holds. It will also illustrate how the unitarity cuts that arise can be represented in their Schwinger-parametric form.

Denoting Schwinger parameters of the three propagators by $\alpha_1, \alpha_2$ and $\alpha_3$, corresponding to the internal edges of masses $m_1$, $m_2$ and $0$, respectively, the parametric form of the integral reads
\be\label{Tri Schwinger}
\Tri^\dag = \int_0^\infty \frac{\text{d}\alpha_1 \text{d}\alpha_3}{(\alpha_1 + 1 + \alpha_3)\left[A(\alpha_1)+ \alpha_3 B(\alpha_1)+i\eps\right]}\,,
\ee
with
\be
A \equiv (\alpha_1+1)(\alpha_1 m_1^2+m_2^2)- \alpha_1 s \quad \text{and} \quad B \equiv \alpha_1(m_1^2-s_1)+m_2^2-s_2\,.
\ee
Recall that the $\alpha_i$'s are determined only up to an overall rescaling, and here we set $\alpha_2 = 1$, with the other two integrated from $0$ to $\infty$.
The Feynman $i\varepsilon$ prescription can be alternatively implemented by giving $s$ a small \emph{negative} imaginary part, i.e., by approaching the real axis from the lower half-plane.
However, the crossing equation instructs us to continue in the upper half-plane.

It turns out that the integral representation \eqref{Tri Schwinger} gives a branch cut along the whole real $s$-axis. (Recall that branch points are properties of the function itself, but placements of branch cuts are specific to a given representation.) It is the so-called \emph{external mass singularity}, see \cite{Hannesdottir:2022bmo}. The way to understand why it arises is to think of assigning $\pm i \varepsilon$ to $s_1$ instead of including it as a separate factor, which is entirely equivalent. Since $s_1 > m_1^2$ is above the threshold, the two choices $s_1 \pm i \eps$ clearly give different answers: they differ by a unitarity cut in the $s_1$-channel. But mathematically, this procedure is analogous to setting $s \pm i \eps$ for any real $s$.
This is why the above representation exhibits a discontinuity across the whole $s$-axis. This is not a problem by itself. We just need to analytically continue the representation through the branch cut at $s<0$ to get access to the upper half-plane, instead of blindly inserting $s+i\eps$. 

As promised, we will perform this analytic continuation at the level of the integral. Since $s<0$ and $m_1,m_2>0$ at the beginning of the continuation, we know that $A > 0$ across the entire integration domain. On the other hand, $B$ can change sign depending on the value of $\alpha_1$. In particular, we have $B<0$ for
\be\label{eq:alpha1-ast}
\alpha_1 > \alpha_{1*} = - \frac{s_2 - m_2^2}{s_1 - m_1^2} > 0\, .
\ee
For any fixed $\alpha_1$ satisfying this inequality, the discontinuity we are looking for is given by 
\be\label{eq:pvEq}
\frac{1}{A + \alpha_3 B + i\eps} = \frac{1}{A + \alpha_3 B - i\eps} - 2\pi i\, \delta(A + \alpha_3 B)\,.
\ee
As such, we can rewrite the conjugated amplitude as
\be 
\Tri^\dag = 
\Tri - 2 \pi i  \int_0^\infty \frac{\text{d}\alpha_1 \text{d}\alpha_3}{(\alpha_1 + 1 + \alpha_3)}\, \delta \left[A(\alpha_1)+ \alpha_3 B(\alpha_1)\right]\,.
\ee 
The first term on the right-hand side is simply the amplitude $\Tri$ without the complex conjugate, for any real $s$ when approaching from the upper half-plane. The second one gives a new term, which allows us to localize $\alpha_3$ around $\alpha_{3*} = -\tfrac{A}{B} > 0$ using the $\delta$-function with $\delta(A + \alpha_3 B) = -\frac{1}{B}\delta(\alpha_3 - \alpha_{3*})$.
This procedure is, of course, equivalent to deforming the $\alpha_3$-contour and picking up an extra residue. In the end, we find that
\be\label{eq:crossingPredictionCut}
[\Tri^\dag]_{\raisebox{\depth}{\scalebox{1}[1]{$\curvearrowright$}}s} = \Exp_2 + \Delta\,,
\ee
where we have factored out the expected answer \eqref{Tri test with cut}
and the ``error'' $\Delta$ is the difference between the continued residue and a cut:
\be\label{eq:C2}
\Delta = \underbracket[0.4pt]{2\pi i \int_{\alpha_{1*}}^{\infty} \frac{\d \alpha_1}{(\alpha_1 + 1)B(\alpha_1) - A(\alpha_1)}}_{C(s)} \,-\, \Cut_{s_1} \Tri \, .
\ee
The integral does not have any singularities in the upper half-plane of $s$ when the remaining kinematics is real. This is guaranteed by the fact that $\Im A < 0$ for any $\Im s >0$ and hence its denominator always stays non-zero.
The remaining challenge is to show that,
after taking its limit to the real
axis at large positive $s$, it agrees precisely with the cut term, so that
\be\label{eq:goal} 
\Delta \stackrel{?}{=} 0\, .
\ee
In the above, $\Cut_{s_1} \Tri$ is defined for real positive $s$ by the on-shell cut integral in \eqref{eq:atCrossing}; we will also discuss below its continuation into the upper half-plane of $s$.

At this stage, it is instructive to analyze the singularities that can potentially arise from the integrals $\Tri$ and $\Cut_{s_1} \Tri$ when approached from the upper half-plane. There are three relevant singularities: the $s$-channel normal threshold, the leading second-type singularity, and the triangle anomalous thresholds. The first one arises when 
\begin{equation}
(\alpha_1 : \alpha_2 : \alpha_3) = \left(\pm\tfrac{m_2}{m_1} : 1 : 0 \right) \quad \text{and} \quad  s = (m_1 \pm m_2)^2\,.  
\end{equation}
The second one happens for
\begin{equation}
    (\alpha_1 : \alpha_2 : \alpha_3) = \left(\pm \tfrac{\sqrt{s_2}}{\sqrt{s_1}} : 1 : -1 \mp \tfrac{\sqrt{s_2}}{\sqrt{s_1}} \right) \quad \text{and} \quad s=(\sqrt{s_1} \pm \sqrt{s_2})^2\,.
\end{equation}
In both cases, only the first choice of signs gives a singularity of $\Tri$ on the relevant sheet.
Finally, the triangle anomalous threshold occurs for
\be\label{eq:triangle-alphas}
(\alpha_1 : \alpha_2 : \alpha_3 ) = (\alpha_{1*} : 1 : \alpha_{3*}(\alpha_{1*})) \quad\text{and}\quad
s = s_* \equiv m_1^2 \left( 1 - \frac{s_2 - m_2^2}{s_1-m_1^2} \right) + (1\leftrightarrow 2)\,.
\ee
Here, crucially, $\alpha_{3*}(\alpha    _{1*}) = -\lim_{\alpha_1\to\alpha_{1\ast}}\tfrac{A(\alpha_{1})}{B(\alpha_{1})}$ is evaluated at the specific value $\alpha_{1}=\alpha_{1*}$ (defined in \eqref{eq:alpha1-ast}) and does \emph{not} have to be positive. Explicitly, it reads
\be
\alpha_{3*}(\alpha_{1*}) = \frac{m_2^2}{s_2 - m_2^2} - m_1^2 \frac{s_2 - m_2^2}{(s_1 - m_1^2)^2}\, .
\label{eq:alpha3-ast}
\ee
Since we are interested in the kinematics with $s_1 > m_1^2$, as $s_2 < m_2^2$ approaches the threshold $s_2 \to m_2^2$, the corresponding $\alpha_{3*} \to -\infty$. In particular, the fact that $\alpha_{3*}$ is negative means that the singularity does not occur on the sheet of $\Tri$ in the upper half-plane we are exploring now. This argument does not apply to the correction term $C$ in \eqref{eq:C2} nor to $\Cut_{s_1} \Tri$, because $\alpha$-positivity only holds for singularities of time-ordered amplitudes on the first sheet.
We will see that, indeed, the anomalous threshold \emph{does} appear for these. 

The anomalous threshold at $s=s_\ast$ manifests itself as an endpoint singularity of $C$ at $\alpha_1 = \alpha_{1*}$, where the integrand denominator for $C$ becomes
\be
(\alpha_{1*} + 1)B(\alpha_{1*}) - A(\alpha_{1*}) = \alpha_{1*} (s - s_\ast)\,,
\ee
which shows that $C$ develops a logarithmic singularity at $s = s_\ast$. The sign of $\alpha_{3*}$ is irrelevant since it is no longer an integration variable. This can be verified explicitly by evaluating its integral representation in \eqref{eq:C2}. We indeed find
\begin{align}
C(s) &= \frac{2\pi}{\sqrt{-\lambda}}
 \left(\log\frac{i\sqrt{-\lambda}+(s-s_1-s_2-2s_1 \alpha_{1\ast})}{i\sqrt{-\lambda}-(s-s_1-s_2-2s_1 \alpha_{1\ast})} - i\pi \right),
\end{align}
where $\lambda\equiv s^2 + s_1^2 + s_2^2 - 2s s_1 - 2s s_2 - 2s_1s_2\,$ is the \emph{K\"all\'en/triangle function}. One of its zeros is at $s = (\sqrt{s_1} + \sqrt{s_2})^2$, which is the square-root second-type singularity. This threshold is also the lower limit of the physical region, $s > (\sqrt{s_1} + \sqrt{s_2})^2$ for the $1'23 \ot 12'$ kinematics.
The branches of the square roots were carefully chosen so that the above formula is valid in the upper half-plane of $s$.

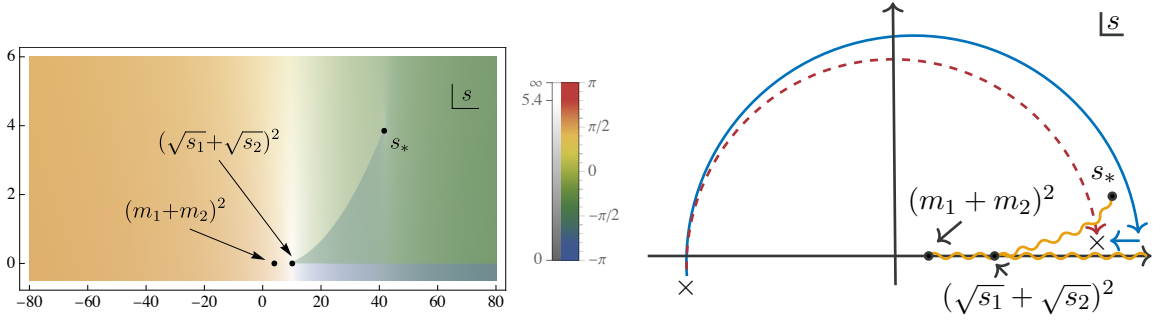
\begin{figure}[t]    \centering
\includegraphics[valign=b,width=0.53\textwidth]{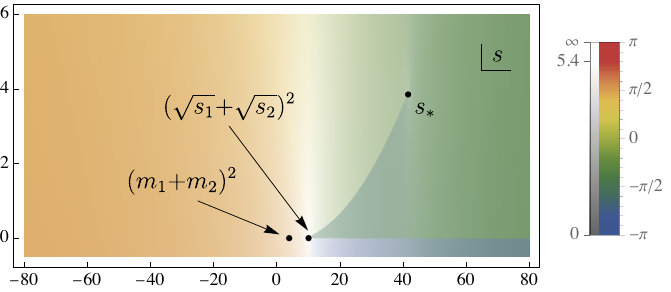}
\adjustbox{valign=b,width=0.45\textwidth}{
\tikzset{every picture/.style={line width=1}}  
\begin{tikzpicture}[yscale=-0.03,xscale=0.03,draw=charcoal] 
\draw[->]  [color=RoyalBlue,draw opacity=1 ] (64.84,173.89) .. controls (61.55,120.05) and (101.96,73.24) .. (155.97,68.88) .. controls (209.08,64.58) and (255.81,102.91) .. (262.48,155.23) ;  
\draw[->,dashed]  [color=Maroon ,draw opacity=1 ] (64.74,171.01) .. controls (63.77,123.61) and (99.92,83.07) .. (147.8,79.23) .. controls (195.42,75.41) and (237.37,109.25) .. (244.33,155.8);
\draw[->]    (47.7,165.07) -- (201,165.07) -- (268.24,165.07) ;
\draw [Orange, draw opacity=1 ]    (170.64,165.07) .. controls (172.31,163.4) and (173.97,163.4) .. (175.64,165.07) .. controls (177.31,166.74) and (178.97,166.74) .. (180.64,165.07) .. controls (182.31,163.4) and (183.97,163.4) .. (185.64,165.07) .. controls (187.31,166.74) and (188.97,166.74) .. (190.64,165.07) .. controls (192.31,163.4) and (193.97,163.4) .. (195.64,165.07) .. controls (197.31,166.74) and (198.97,166.74) .. (200.64,165.07) -- (201,165.07) -- (201,165.07) ;
\draw [color=Orange  ,draw opacity=1 ]   (199.5,166.57) .. controls (200.59,164.37) and (202.18,163.78) .. (204.26,164.8) .. controls (206.33,165.85) and (208,165.29) .. (209.26,163.12) .. controls (210.35,161.02) and (211.83,160.55) .. (213.7,161.71) .. controls (215.79,162.8) and (217.39,162.31) .. (218.5,160.22) .. controls (219.59,158.13) and (221.18,157.63) .. (223.26,158.72) .. controls (225.56,159.72) and (227.2,159.17) .. (228.19,157.07) .. controls (229.09,154.97) and (230.65,154.39) .. (232.86,155.33) .. controls (235.09,156.2) and (236.62,155.54) .. (237.44,153.33) .. controls (238.03,151.16) and (239.45,150.38) .. (241.69,151.01) .. controls (244.14,151.34) and (245.47,150.31) .. (245.7,147.92) .. controls (245.46,145.67) and (246.42,144.36) .. (248.57,143.98) .. controls (250.73,142.67) and (251.04,141.05) .. (249.51,139.13) -- (249.5,138.9) ;
\draw [color=Orange  ,draw opacity=1 ]    (201,165.07) .. controls (202.67,163.4) and (204.33,163.4) .. (206,165.07) .. controls (207.67,166.74) and (209.33,166.74) .. (211,165.07) .. controls (212.67,163.4) and (214.33,163.4) .. (216,165.07) .. controls (217.67,166.74) and (219.33,166.74) .. (221,165.07) .. controls (222.67,163.4) and (224.33,163.4) .. (226,165.07) .. controls (227.67,166.74) and (229.33,166.74) .. (231,165.07) .. controls (232.67,163.4) and (234.33,163.4) .. (236,165.07) .. controls (237.67,166.74) and (239.33,166.74) .. (241,165.07) .. controls (242.67,163.4) and (244.33,163.4) .. (246,165.07) .. controls (247.67,166.74) and (249.33,166.74) .. (251,165.07) .. controls (252.67,163.4) and (254.33,163.4) .. (256,165.07) .. controls (257.67,166.74) and (259.33,166.74) .. (261,165.07) .. controls (262.67,163.4) and (264.33,163.4) .. (266,165.07) -- (266.24,165.07) -- (266.24,165.07) ;
\draw[<-]    (155,54) -- (156.14,177.86) ;
\draw[yshift=-550]   (256,87.58) -- (246.04,87.56) -- (246.06,77.61) ;
\draw[<-] [color=RoyalBlue  ,draw opacity=1 ]   (250,158.4) -- (263,158.4) ;
\draw[->]    (185,150) -- (174.33,159) ;
\draw[<-]    (200,168.57) -- (205,173.57) ;
\draw  [fill=black,fill opacity=1 ] (252.5,138.9) .. controls (252.5,138.07) and (251.83,137.4) .. (251,137.4) .. controls (250.17,137.4) and (249.5,138.07) .. (249.5,138.9) .. controls (249.5,139.73) and (250.17,140.4) .. (251,140.4) .. controls (251.83,140.4) and (252.5,139.73) .. (252.5,138.9) -- cycle ;
\draw  [fill=black, fill opacity=1 ] (201,165.07) .. controls (201,164.24) and (200.33,163.57) .. (199.5,163.57) .. controls (198.67,163.57) and (198,164.24) .. (198,165.07) .. controls (198,165.9) and (198.67,166.57) .. (199.5,166.57) .. controls (200.33,166.57) and (201,165.9) .. (201,165.07) -- cycle ;
\draw  [fill=black,fill opacity=1 ] (172.14,165.07) .. controls (172.14,164.24) and (171.47,163.57) .. (170.64,163.57) .. controls (169.81,163.57) and (169.14,164.24) .. (169.14,165.07) .. controls (169.14,165.9) and (169.81,166.57) .. (170.64,166.57) .. controls (171.47,166.57) and (172.14,165.9) .. (172.14,165.07) -- cycle ;
\draw (247,58.8) node [anchor=north west][inner sep=0.75pt]   {$s$};
\draw (58,173.4) node [anchor=north west][inner sep=0.75pt]  {$\times$};
\draw (238,153.4) node [anchor=north west][inner sep=0.75pt] {$\times$};
\draw (175,174.4) node [anchor=north west][inner sep=0.75pt]  [scale=1]  {$(\sqrt{s_1} + \sqrt{s_2})^2$};
\draw (240,124.4) node [anchor=north west][inner sep=0.75pt]  [scale=1]  {$s_{\tiny *}$};
\draw (160,132.4) node [anchor=north west][inner sep=0.75pt]  [scale=1]  {$( m_{1} +m_{2})^{2}$};
\end{tikzpicture}}
\caption{\label{fig:offshell triangle}
\textbf{Left:}
The analytic structure of $\Exp_2$ 
in the complex $s$-plane for $m_1 = m_2 = 1$, $s_1 = 5$, and $s_2 = 0.9 +   
0.01 i$. We observe a branch cut starting at the normal threshold at $s = (m_1 + m_2)^2 = 4$ up to infinity, another extending to the triangle anomalous threshold at $s = s_\ast \approx 41.6 + 4 i$ from the leading second-type singularity at $s = (\sqrt{s_1} + \sqrt{s_2})^2 \approx 10.1 + 0.03 i$. The physical region in the $s$ channel starts at the second-type singularity $s = (\sqrt{s_1} + \sqrt{s_2})^2$, and extends to $s\to \infty$.
In the on-shell limit, $s_2 \to m_2^2$, the anomalous threshold behaves as $s_\ast \to - m_2^2 \frac{s_1 - m_1^2}{s_2 - m_2^2}$ and hence moves to infinity, thus cutting off 
the inclusive observable from the analytic continuation of $\Tri^\dag$. The branch cut along the real axis is the external mass singularity.
\textbf{Right:} A schematic figure of the analytic structure of $\Exp_2$ in the complex $s$-plane. The branch cuts are marked in yellow. We observe two possible paths of analytic continuation in the upper half-plane, marked in red and blue, which differ by how the anomalous threshold at $s^\ast$ is traversed. The red dashed path is the one that would be reached without deforming $s^\ast$ to be complex, and does not correspond to $\text{Exp}_{2}$. The solid blue path, on the other hand, is the one that reaches the real axis where the inclusive observable $\text{Exp}_{2}$ is defined.
}
\end{figure}

The analyticity of ${\rm Exp}_{2}$ in the $s$-plane is summarized in Fig.~\ref{fig:offshell triangle}, where we added a small positive imaginary part to $s_2$ to more clearly illustrate the sheet structure.

\subsection{Comparison with unitarity cuts}\label{ssec:comparision}

Let us now compare $C$ with the cut $\Cut_{s_1} \Tri$ predicted by \eqref{Tri test with cut} and check if \eqref{eq:goal} is satisfied. The latter is given by
\be\label{def C tri}
\Cut_{s_1} \Tri \equiv -(2\pi i)^2
    \int \frac{\text{d}^4\ell}{i\pi^{2}}
    \frac{\delta^+(\ell^2)\, \delta^-[(\ell{-}p_1)^2+m_1^2]}{(\ell {+} p_2)^2+m_2^2 - i\eps}\, .
\ee
Our goal is to massage this expression into a form similar to $C$. In the $s_1$ rest frame, the momentum vectors can be written as
\begin{subequations}\label{eq:restFrame1}
\begin{align}
p_1 &= \big(\sqrt{s_1},\, 0,\, 0,\, 0\big)\,,\\
p_2 &= \big(\tfrac{s {-} s_1 {-} s_2}{2\sqrt{s_1}},\, 0,\, 0,\, \tfrac{\sqrt{\lambda}}{2\sqrt{s_1}}\big)\,, \label{eq:p2} \\
p &= \big(\tfrac{{-}s {-} s_1 {+} s_2}{2\sqrt{s_1}},\, 0,\, 0,\, {-}\tfrac{\sqrt{\lambda}}{2\sqrt{s_1}}\big)\,,\\
\ell &= \big(\ell^0,\, |\vec{\ell}| \cos \varphi \sin \theta,\, |\vec{\ell}| \sin \varphi \sin \theta,\, |\vec{\ell}| \cos \theta\big)\, . 
\end{align}
\end{subequations}
In this frame, the two $\delta$-functions impose
\be
-(\ell^0)^2 + |\vec{\ell}|^2 = 0 \quad \text{and} \quad -(\ell^0 - \sqrt{s_1})^2 + |\vec{\ell}|^2 + m_1^2 = 0\, .
\ee
Together, they fix $\ell^0_* = |\vec{\ell}|_* = \frac{s_1 - m_1^2}{2\sqrt{s_1}}$. Imposing positivity of the energies flowing through the cut ($\ell^0 > 0$ and $\ell^0 - p_1^0 < 0$) amounts to $s_1 > m_1^2$, meaning we are, as needed, above the $s_1$-threshold. On the support of these constraints, the leftover propagator in \eqref{def C tri} is
\be
(\ell {+} p_2)^2 + m_2^2 = -s_2 - 2 \ell^0_* (p_2^0 - p_2^3 \cos \theta) + m_2^2\, ,
\ee
where the components $p_2^0$ and $p_2^3$ are given in \eqref{eq:p2}. Hence, after going to spherical coordinates
and localizing along $\ell^0 = \ell^0_*$ and $|\vec{\ell}| = |\vec{\ell}|_*$ (with a Jacobian of $\tfrac{1}{4\sqrt{s_1} \ell^0_*}$ coming from the $\delta$-functions) we arrive at
\begin{align}\label{def C tri 2}
\Cut_{s_1} \Tri &= - \frac{2\pi i \ell^0_*}{\sqrt{s_1}} \int_{-1}^1
\frac{\text{d}\cos \theta}
{-s_2 - 2 \ell^0_* \big(p_2^0 -p_2^3 \cos \theta \big) + m_2^2 - i\eps}\, .
\end{align}
The only kinematic assumption here is that $s_1>m_1^2$, ensuring that the cut has support.  This representation can be used for any real $p_2^\mu$ and thus real $s$.
Its continuation to complex $s$ will turn out to be different depending on whether $s$ is above or below the triangle anomalous threshold at $s=s_\ast$.

Let us first check whether $\Delta=0$ in real $s$-channel kinematics (i.e., for $s>(\sqrt{s_1} + \sqrt{s_2})^2$), by massaging the cut contribution \eqref{def C tri 2} into a form that resembles the $\alpha_1$ integral from \eqref{eq:C2}. To this end, we start by combining the regions with $\pm \cos\theta$ and 
change variable to $x = \frac{1}{\cos \theta}$ to get
\be 
    \Cut_{s_1} \Tri = 4 \pi i \int_1^\infty \rd x \frac{-s+s_1+s_2+2 s_1 \alpha_{1\ast}}{\lambda-x^2 (-s+s_1+s_2+2 s_1 \alpha_{1\ast})^2 - i \varepsilon} \,,
\ee 
where we have assumed that $s>s_1+s_2+2 s_1 \alpha_{1 \ast}$ for the change of variables, which always holds when $s_2 \to m_2^2$.
Next, we shift and rescale $x$ by defining $\alpha_1 = \tfrac{-s+s_1+s_2+2 s_1 \alpha_{1\ast}}{2 s_1} x + \frac{s-s_1-s_2}{2 s_1}$ and end up with
\be
    \Cut_{s_1} \Tri =  2\pi i \int^{-\infty}_{\alpha_{1*}} \frac{\d \alpha_1}{(\alpha_1 + 1)B(\alpha_1) - A(\alpha_1) - i \varepsilon} \,.
    \label{eq:cuteps}
\ee
Here, we have $-\infty$ and not $+\infty$ in the integration domain because $s$ is taken to be large enough such that $s>s_1+s_2+2s_1 \alpha_{1\ast}$. Moreover, the coefficient of $i\varepsilon$ after these changes of variables is 
$(s_1+s_2-s+2s_1 \alpha_1)^2$, which we drop because it is always positive in physical kinematics.

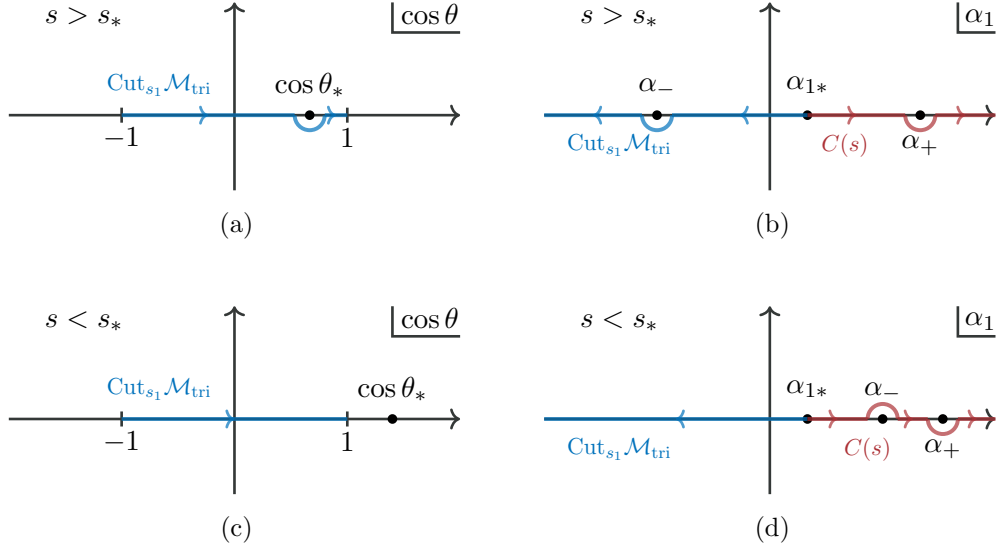
\begin{figure}[t]
    \centering
    \tikzset{every picture/.style={line width=1}}  
    \begin{subfigure}[t]{0.45\textwidth}
      \centering
    \begin{tikzpicture}[draw=charcoal,decoration={markings, mark=at position 0.5 with {\arrow[scale=0.7]{>}}}]
    \coordinate (aast) at (1,0);
    \coordinate (ay1) at (0,-1);
    \coordinate (ay2) at (0,1.5);
    \coordinate (ax1) at (-3,0);
    \coordinate (ax2) at (3,0);
    \coordinate (b1) at (-1.5,0);
    \coordinate (b2) at (1.5,0);
  \draw[] ($(b1)-(0,0.1)$) -- ($(b1)+(0,0.1)$);
  \draw[] ($(b2)-(0,0.1)$) -- ($(b2)+(0,0.1)$);
  \node[] at ($(b1)-(0,0.3)$) {$-1$};
  \node[] at ($(b2)-(0,0.3)$) {$1$};
  \draw[->] (ax1) -- (ax2);
  \draw[->] (ay1) -- (ay2);
  \node[] at ($(ax2)+(ay2)-(0.4,0.15)$) {$\cos \theta$};
  \draw[] ($(ax2)+(ay2)-(0.9,0)$) -- ($(ax2)+(ay2)-(0.9,0.4)$) -- ($(ax2)+(ay2)-(0,0.4)$);
  \node[black] at ($(aast)+(0,0.4)$) {$\cos \theta_{\ast}$};
  \draw[black,fill=black,thick] (aast) circle (0.05);
  \node[] at ($(ay2)+(-2,-0.2)$) {$s>s_\ast$};
  \draw[RoyalBlue,line width=1.5,opacity=0.7,postaction=decorate] (b1) -- ($(aast)-(0.2,0)$);
  \draw[RoyalBlue,line width=1.5,opacity=0.7,postaction=decorate] ($(aast)+(0.2,0)$) -- ($(b2)$);
  \centerarc[RoyalBlue,line width=1.5,opacity=0.7](aast)(0:180:-0.2);
  \node[RoyalBlue,scale=0.8] at ($(b1)+(0.5,0.4)$) {$\Cut_{s_1} \Tri$};
    \end{tikzpicture}
    \caption{}
    \end{subfigure}
    \begin{subfigure}[t]{0.45\textwidth}
      \centering
    \begin{tikzpicture}[draw=charcoal,decoration={markings, mark=at position 0.5 with {\arrow[scale=0.7]{>}}}]
    \coordinate (aast) at (0.5,0);
    \coordinate (r1) at (-1.5,0);
    \coordinate (r2) at (2,0);
    \coordinate (ay1) at (0,-1);
    \coordinate (ay2) at (0,1.5);
    \coordinate (ax1) at (-3,0);
    \coordinate (ax2) at (3,0);
  \draw[->] (ax1) -- (ax2);
  \draw[->] (ay1) -- (ay2);
  \node[] at ($(ax2)+(ay2)-(0.2,0.2)$) {$\alpha_1$};
  \draw[] ($(ax2)+(ay2)-(0.5,0)$) -- ($(ax2)+(ay2)-(0.5,0.4)$) -- ($(ax2)+(ay2)-(0,0.4)$);
  \draw[black,fill=black,thick] (r1) circle (0.05);
  \node[black] at ($(r1)+(0,0.36)$) {$\alpha_-$};
  \draw[black,fill=black,thick] (r2) circle (0.05);
  \node[black] at ($(r2)-(0,0.4)$) {$\alpha_+$};
  \node[black] at ($(aast)+(0,0.4)$) {$\alpha_{1 \ast}$};
  \draw[black,fill=black,thick] (aast) circle (0.05);
  \draw[Maroon,line width=1.5,opacity=0.7,postaction=decorate] (aast) -- ($(r2)-(0.2,0)$);
  \draw[Maroon,line width=1.5,opacity=0.7,postaction=decorate] ($(r2)+(0.2,0)$) -- (ax2);
  \draw[RoyalBlue,line width=1.5,opacity=0.7,postaction=decorate] (aast) -- ($(r1)+(0.2,0)$);
  \draw[RoyalBlue,line width=1.5,opacity=0.7,postaction=decorate] ($(r1)-(0.2,0)$) -- (ax1);
  \centerarc[RoyalBlue,line width=1.5,opacity=0.7](r1)(0:-180:0.2);
  \centerarc[Maroon,line width=1.5,opacity=0.7](r2)(0:-180:0.2);
  \node[] at ($(ay2)+(-2,-0.2)$) {$s>s_\ast$};
  \node[RoyalBlue,scale=0.8] at ($(aast)+(-2.5,-0.4)$) {$\Cut_{s_1} \Tri$};
  \node[Maroon,scale=0.8] at ($(r2)+(-1,-0.4)$) {$C(s)$};
    \end{tikzpicture}
    \caption{}
    \end{subfigure}
    \\[20pt]
    \begin{subfigure}[t]{0.45\textwidth}
        \centering
    \begin{tikzpicture}[draw=charcoal,decoration={markings, mark=at position 0.5 with {\arrow[scale=0.7]{>}}}]
    \coordinate (aast) at (2.1,0);
    \coordinate (ay1) at (0,-1);
    \coordinate (ay2) at (0,1.5);
    \coordinate (ax1) at (-3,0);
    \coordinate (ax2) at (3,0);
    \coordinate (b1) at (-1.5,0);
    \coordinate (b2) at (1.5,0);
  \draw[] ($(b1)-(0,0.1)$) -- ($(b1)+(0,0.1)$);
  \draw[] ($(b2)-(0,0.1)$) -- ($(b2)+(0,0.1)$);
  \node[] at ($(b1)-(0,0.3)$) {$-1$};
  \node[] at ($(b2)-(0,0.3)$) {$1$};
  \draw[->] (ax1) -- (ax2);
  \draw[->] (ay1) -- (ay2);
  \node[] at ($(ax2)+(ay2)-(0.4,0.15)$) {$\cos \theta$};
  \draw[] ($(ax2)+(ay2)-(0.9,0)$) -- ($(ax2)+(ay2)-(0.9,0.4)$) -- ($(ax2)+(ay2)-(0,0.4)$);
  \draw[RoyalBlue,line width=1.5,opacity=0.7,postaction=decorate] (b1) -- (b2);
  \node[black] at ($(aast)+(0,0.4)$) {$\cos \theta_{\ast}$};
  \draw[black,fill=black,thick] (aast) circle (0.05);
  \node[] at ($(ay2)+(-2,-0.2)$) {$s<s_\ast$};
  \node[RoyalBlue,scale=0.8] at ($(b1)+(0.5,0.4)$) {$\Cut_{s_1} \Tri$};
    \end{tikzpicture}
    \caption{}
    \end{subfigure}
    \begin{subfigure}[t]{0.45\textwidth}
        \centering
    \begin{tikzpicture}[draw=charcoal,decoration={markings, mark=at position 0.5 with {\arrow[scale=0.7]{>}}}]
    \coordinate (aast) at (0.5,0);
    \coordinate (r1) at (1.5,0);
    \coordinate (r2) at (2.3,0);
    \coordinate (ay1) at (0,-1);
    \coordinate (ay2) at (0,1.5);
    \coordinate (ax1) at (-3,0);
    \coordinate (ax2) at (3,0);
  \draw[->] (ax1) -- (ax2);
  \draw[->] (ay1) -- (ay2);
  \node[] at ($(ax2)+(ay2)-(0.2,0.2)$) {$\alpha_1$};
  \draw[] ($(ax2)+(ay2)-(0.5,0)$) -- ($(ax2)+(ay2)-(0.5,0.4)$) -- ($(ax2)+(ay2)-(0,0.4)$);
  \draw[black,fill=black,thick] (r1) circle (0.05);
  \node[black] at ($(r1)+(0,0.36)$) {$\alpha_-$};
  \draw[black,fill=black,thick] (r2) circle (0.05);
  \node[black] at ($(r2)-(0,0.4)$) {$\alpha_+$};
  \node[black] at ($(aast)+(0,0.4)$) {$\alpha_{1 \ast}$};
  \draw[black,fill=black,thick] (aast) circle (0.05);
  \draw[RoyalBlue,line width=1.5,opacity=0.7,postaction=decorate] (aast) -- (ax1);
  \draw[Maroon,line width=1.5,opacity=0.7,postaction=decorate] (aast) -- ($(r1)-(0.2,0)$);
  \draw[Maroon,line width=1.5,opacity=0.7,postaction=decorate] ($(r1)+(0.2,0)$) -- ($(r2)-(0.2,0)$);
  \draw[Maroon,line width=1.5,opacity=0.7,postaction=decorate] ($(r2)+(0.2,0)$) -- (ax2);
  \centerarc[Maroon,line width=1.5,opacity=0.7](r1)(0:180:0.2);
  \centerarc[Maroon,line width=1.5,opacity=0.7](r2)(0:-180:0.2);
  \node[] at ($(ay2)+(-2,-0.2)$) {$s<s_\ast$};
  \node[RoyalBlue,scale=0.8] at ($(aast)+(-2.5,-0.4)$) {$\Cut_{s_1} \Tri$};
  \node[Maroon,scale=0.8] at ($(r2)+(-1,-0.4)$) {$C(s)$};
    \end{tikzpicture}
    \caption{}
    \end{subfigure}
    \caption{
    \textbf{Left:} The integration contour in $\cos \theta$ for $\Cut_{s_1} \Tri$. The integral has a real part when $s>s_\ast$ (a), but is purely imaginary when $s<s_\ast$ (c).
    \textbf{Right:} The two contributions to the integration contour for $\Delta$, from $\Cut_{s_1} \Tri$  (blue) and $C(s)$ (red). The points $\alpha_-$ and $\alpha_+$ correspond to the two roots of the denominator that appear in the integrand of $\Delta$. When $s>s_\ast$ (b), the red and blue contours can be deformed into each other, resulting in $\Delta=0$. When $s<s_\ast$ (d), the difference between the red and blue contours can be deformed into one that encircles the root $\alpha_-$ counter-clockwise. 
    }
    \label{fig:rootsalpha}
\end{figure}

Comparing~\eqref{eq:C2} and~\eqref{eq:cuteps} and dropping the $i\varepsilon$'s, we see that the \emph{integrands} for $\Cut_{s_1} \Tri$ and $C(s)$ are the same. Their integration contours are different, however, so to find the contribution to $\Delta$, we must first analyze which way the contour traverses around the singularities of the integrand. The singularities are at the two roots of the denominators: 
\be 
    \alpha_{\pm} = \frac{s-s_1-s_2 \pm \sqrt{\lambda}}{2 s_1} \,.
    \label{eq:aroots}
\ee 
One can see from this expression that in the physical region for $\lambda>0$ we have
\be 
    \alpha_{1 \ast}<\alpha_{\pm} \quad \text{for }  s<s_\ast \quad \text{or} \quad
    \alpha_- < \alpha_{1 \ast} < \alpha_{+} \quad \text{for }  s>s_\ast\,.
\ee
The contour for $C(s)$ runs from $\alpha_{1 \ast}$ to $\infty$, and approaches the roots with $s\to s+ i\varepsilon$, which translates into the integration contour for $C(s)$ being deformed into the upper half-plane around $\alpha_-$ and into the lower half-plane around $\alpha_+$, as shown in Fig.~\ref{fig:rootsalpha}. The contour for $-\Cut_{s_1} \Tri$ runs from $-\infty$ to $\alpha_{1 \ast}$, and when the root $\alpha_-$ is on the contour for $s>s_\ast$, it must be approached by deforming the contour into the lower half-plane, as dictated by the $i \varepsilon$ in \eqref{eq:cuteps}. This is possible since the integrand in \eqref{eq:cuteps} has no residue at infinity.

We can now read the integration contour for $\Delta$ from Fig~\ref{fig:rootsalpha}, as the sum of the contours for $C(s)$ and $-\Cut_{s_1} \Tri$ in the $\alpha$ plane. On one hand, when $s>s_\ast$ (see Fig~\ref{fig:rootsalpha} (b)), the contributions from $C(s)$ and $-\Cut_{s_1} \Tri$ add up to give a contour that can be deformed to zero, and hence $\Delta=0$ when $s>s_\ast$. On the other hand, when $s<s_\ast$ (see Fig~\ref{fig:rootsalpha} (d)), the two contours can be deformed around a small clockwise circle around $\alpha_-$ (from \eqref{eq:aroots}) in the $\alpha_1$-plane:
\be 
    \Delta \vert_{s<s_\ast} =  2\pi i\ointclockwise_{\alpha_-} \frac{\d \alpha_1}{(\alpha_1 + 1)B(\alpha_1) - A(\alpha_1)} = \frac{4 \pi^2}{\sqrt{\lambda}} \,,
\ee 
where we have evaluated the integral using the residue theorem, noting the overall minus sign because of the orientation.
To summarize,
\be\label{eq:residue}
    \Delta = \begin{cases}
        0 & \qquad \text{if } s>s_\ast\,, \\
    \frac{4 \pi^2}{\sqrt{\lambda}} & \qquad \text{if } (\sqrt{s_1} + \sqrt{s_2})^2 < s<s_\ast \,.
    \end{cases}
\ee

Finally, we can ask whether it is possible to analytically continue $C(s)$ such that it agrees with $\Cut_{s_1} \Tri$.
Inspection of the lower-right panel of Fig.~\ref{fig:rootsalpha} indicates that we need to continue in such a way that the root $\alpha_-$ goes clockwise around $\alpha_{1\ast}$.  Note that the collision $\alpha_-=\alpha_{1\ast}$ occurs precisely at the anomalous threshold $s=s_\ast$.
In the $s$-plane, this can thus be achieved by taking $s$ clockwise below $s_\ast$.
In other words, starting from large real $s>s_\ast$, we lower $s$ passing at $s=s_\ast-i\eps$ \emph{below} the threshold, as indicated by the arrows
in Fig.~\ref{fig:offshell triangle}.

This construction is not possible in the on-shell limit when $s_2 \to m_2^2$, since the branch point at $s_\ast$ moves to infinity:
a crossing path exists only off-shell.

The fact that the anomalous threshold must be avoided with $s-i\eps$ can also be seen directly from the endpoint singularities in \eqref{def C tri 2}.  Since $\ell_\ast^0>0$, the contribution near the upper endpoint at $\cos\theta=1$ is analytic for ${\rm Im}\,p_2^+>0$, while that from the lower endpoint $\cos\theta=-1$ is analytic for ${\rm Im}\, p_2^->0$.  Since $p_2^+\propto s$ while $p_2^-\propto \frac{1}{s}$, these singularities lie on opposite sides of the real $s$ axis
(this is the ``gridlock'' mentioned at the top of this section).
The $s=s_\ast$ branch point is caused by the lower endpoint.

To conclude this subsection, we briefly consider the consequences of analytically continuing $\Tri^\dag$ from~\eqref{eq:triDef} in the loop-momentum space representation of $\Tri$ instead of in Schwinger-parameter space. Rather than computing the discontinuity across the branch cut in $s$ for $s<0$ as the term $C(s)$ from~\eqref{eq:C2}, we could instead, by unitarity, evaluate $\Cut_{s_1} \Tri$ from~\eqref{def C tri 2}. For $s<0$, we can, therefore, write
\begin{equation}
    \Tri^\dag = \Tri + \Cut_{s_1} \Tri \,.
    \label{eq:Tridagsneg}
\end{equation}
Note that although this expression looks like the crossing equation, we have not done any analytic continuation in $s$, so we cannot conclude anything about crossing yet. To do so, we have to analytically continue \emph{both} the left-hand side and right-hand side in the upper half-plane of $s$, and check whether
\begin{equation}
    \left[ \Tri^\dag \right]_{\raisebox{\depth}{\scalebox{1}[1]{$\curvearrowright$}}s} \stackrel{?}{=} \Tri + \Cut_{s_1} \Tri \,,
    \label{eq:TriContLoop}
\end{equation}
where the right-hand side should now be evaluated at $s>0$.

\begin{figure}[]    \centering
    \centering
        \begin{minipage}[c]{0.9\textwidth}
        \adjustbox{valign=c,scale={1}{1},rotate=0}{\tikzset{every picture/.style={line width=1pt}} 

\begin{tikzpicture}[x=0.75pt,y=0.75pt,yscale=-1.3,xscale=-1.3]
\tikzset{ma/.style={decoration={markings,mark=at position 0.5 with {\arrow[scale=0.7]{>}}},postaction={decorate}}}
\tikzset{ma2/.style={decoration={markings,mark=at position 0.4 with {\arrow[scale=0.7]{>}}},postaction={decorate}}}
\tikzset{mar/.style={decoration={markings,mark=at position 0.5 with {\arrowreversed[scale=0.7]{>}}},postaction={decorate}}}
\tikzset{mar1/.style={decoration={markings,mark=at position 0.1 with {\arrowreversed[scale=0.7]{>}}},postaction={decorate}}}
\tikzset{mar2/.style={decoration={markings,mark=at position 0.85 with {\arrowreversed[scale=0.7]{>}}},postaction={decorate}}}
\tikzset{mar3/.style={decoration={markings,mark=at position 0.2 with {\arrowreversed[scale=0.7]{>}}},postaction={decorate}}}

\begin{scope}
    \draw[color=white]    (0,160.58) -- (100,170.58) ;
\end{scope}
\begin{scope}[xshift=-100,yshift=0]
\begin{scope}[xshift=-200,yshift=0]
\draw[<-]    (177.67,160.58) -- (429.98,160.58) ;
\draw[->]    (299.4,180.4) -- (299.4,110.4) node[above]{
};
\begin{scope}[xshift=0]
\draw  [color=Maroon  ,draw opacity=1,ma] (190.56,160.51) .. controls (190.45,158.57) and (190.86,156.72) .. (191.78,154.94) .. controls (192.73,153.12) and (194.09,151.63) .. (195.87,150.47) .. controls (197.7,149.29) and (199.71,148.62) .. (201.9,148.48) .. controls (204.15,148.33) and (206.3,148.74) .. (208.32,149.73) .. controls (210.4,150.74) and (212.06,152.21) .. (213.33,154.13) .. controls (214.63,156.11) and (215.32,158.28) .. (215.41,160.67) .. controls (215.49,163.12) and (214.95,165.46) .. (213.74,167.67) .. controls (212.52,169.94) and (210.79,171.77) .. (208.55,173.17) .. controls (206.26,174.6) and (203.76,175.39) .. (201.04,175.54) .. controls (198.26,175.68) and (195.64,175.13) .. (193.17,173.89) .. controls (190.64,172.62) and (188.63,170.8) .. (187.12,168.44) .. controls (185.57,166.02) and (184.76,163.37) .. (184.69,160.47) .. controls (184.62,157.51) and (185.32,154.7) .. (186.79,152.05) .. controls (188.29,149.35) and (190.4,147.17) .. (193.1,145.51) .. controls (195.84,143.82) and (198.85,142.91) .. (202.09,142.77) .. controls (205.38,142.63) and (208.5,143.3) .. (211.41,144.81) .. controls (214.38,146.33) and (216.74,148.5) .. (218.5,151.31) .. controls (218.97,152.06) and (219.38,152.83) .. (219.73,153.62) ;
\draw  [draw opacity=0] (190.84,160.9) .. controls (190.27,155.68) and (192.53,150.32) .. (197.18,147.12) .. controls (203.9,142.5) and (213.08,144.19) .. (217.67,150.88) .. controls (218.33,151.84) and (218.86,152.84) .. (219.26,153.88) -- (205.5,159.24) -- cycle ; \draw[color=RoyalBlue,dash pattern=on 3pt off 2pt,ma]   (190.84,160.9) .. controls (190.27,155.68) and (192.53,150.32) .. (197.18,147.12) .. controls (203.9,142.5) and (213.08,144.19) .. (217.67,150.88) .. controls (218.33,151.84) and (218.86,152.84) .. (219.26,153.88) ; 
\draw  [fill={rgb, 255:red, 0; green, 0; blue, 0 }  ,fill opacity=1 ] (189.68,160.9) .. controls (189.68,160.26) and (190.2,159.74) .. (190.84,159.74) .. controls (191.48,159.74) and (192,160.26) .. (192,160.9) .. controls (192,161.55) and (191.48,162.07) .. (190.84,162.07) .. controls (190.2,162.07) and (189.68,161.55) .. (189.68,160.9) -- cycle ;
\draw  [fill={rgb, 255:red, 0; green, 0; blue, 0 }  ,fill opacity=1 ] (218.1,153.88) .. controls (218.1,153.24) and (218.62,152.72) .. (219.26,152.72) .. controls (219.9,152.72) and (220.42,153.24) .. (220.42,153.88) .. controls (220.42,154.52) and (219.9,155.04) .. (219.26,155.04) .. controls (218.62,155.04) and (218.1,154.52) .. (218.1,153.88) -- cycle ;
\end{scope}

\draw [color=Orange!  ,draw opacity=1 ][line width=1.5]    (201.52,160.58) -- (406.13,160.58) node[below]{$-1$};
\draw [shift={(406.13,160.58)}, rotate = 0] [color=Orange!  ,draw opacity=0 ][fill=Orange!  ,fill opacity=1 ][line width=1.5]      (0, 0) circle [x radius= 1.74, y radius= 1.74]   ;
\draw [shift={(201.52,160.58)}, rotate = 0] [color=Orange!  ,draw opacity=0 ][fill=Orange!  ,fill opacity=1 ][line width=1.5]      (0, 0) circle [x radius= 1.74, y radius= 1.74]   node[below]{$1$};

\begin{scope}[xshift=-140]
\draw  (410,114.52) -- (410,127) -- (380.59,127) ;
\begin{scope}[xshift=5,yshift=3]
\draw (376.59,110.92) node [anchor=north east][inner sep=0.75pt]  [font=\normalsize]  {$\cos \theta $};
    
\end{scope}
\end{scope}

\end{scope}

\end{scope}

\end{tikzpicture}}
    \end{minipage}
    \begin{minipage}[b]{0.32\textwidth}
        \includegraphics[width=\linewidth]{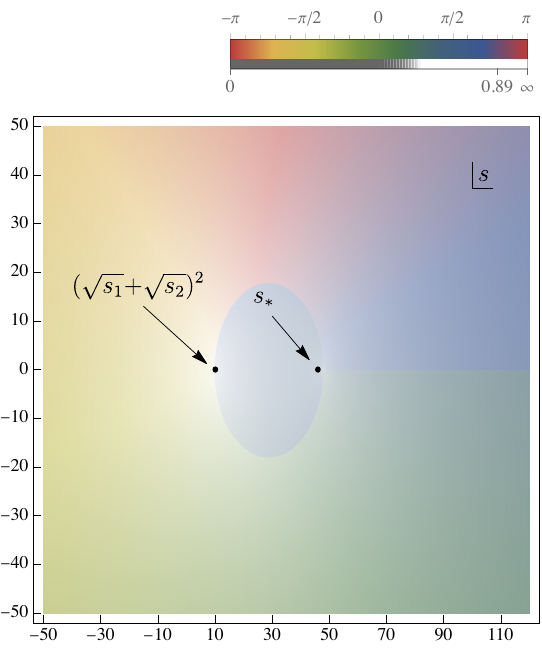}
    \end{minipage}
    \hfill
    \begin{minipage}[b]{0.32\textwidth}
        \includegraphics[width=\linewidth]{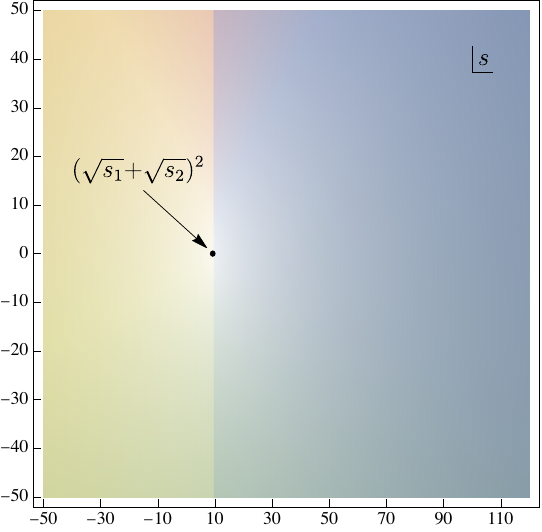}
    \end{minipage}
    \hfill
    \begin{minipage}[b]{0.32\textwidth}
        \includegraphics[width=\linewidth]{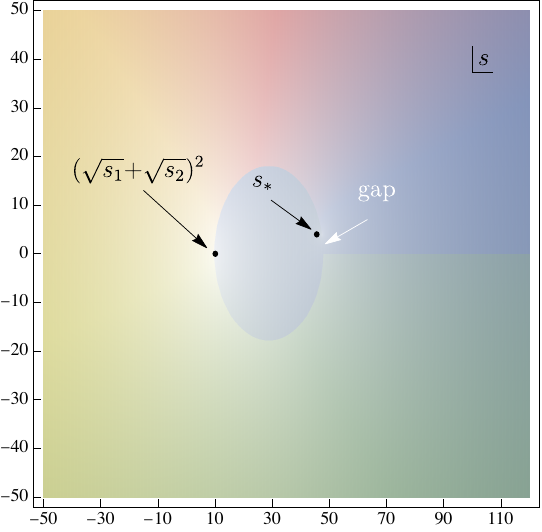}
    \end{minipage}
\caption{\label{fig:cutCos}
\textbf{Top:} Trajectory of the pole in the integrand of $\Cut_{s_1} \Tri$ in \eqref{def C tri 2} as we analytically continue along our constant-$\vert s\vert$ crossing path. For $\vert s\vert < s_\ast$ (red spiral), the pole loops around the integration domain (yellow thick line) and ends up on a different sheet from where the (cut of the) inclusive observable is defined. For $\vert s\vert > s_\ast$ (dashed blue arc), it does not cross the integration domain and therefore remains on the correct sheet. \textbf{Bottom left:} Analytic structure of \eqref{def C tri 2} in the $s$-plane for \emph{real} $s_2<m_2^2$ and $s_1>m_1^2$.
 For the plot we used $m_1^2=m_2^2=1$, $s_1=5$, and $s_2=0.9$. In particular, we observe that \eqref{def C tri 2} is analytic in the upper half-plane of $s$ when $\vert s\vert>s_\ast$; a branch cut runs along an arc from $(\sqrt{s_1}+\sqrt{s_2})^2$ to $s_\ast$. 
\textbf{Bottom middle:} In the on-shell limit where $s_2\to m_2^2$, the anomalous threshold $s_\ast$ approaches infinity, causing an analytic obstruction between $s\lessgtr(\sqrt{s_1}+\sqrt{s_2})^2$. As a result, \eqref{def C tri 2} is represented by at least \emph{two} analytic functions (one on the left and one on the right of the vertical cut). \textbf{Bottom right:} The correct integral \eqref{def C tri 2} in the region $(\sqrt{s_1}+\sqrt{s_2})^2<s<s_\ast$ with $s$ slightly above the axis can be reached from the large-$s$ region by analytically continuing below $s=s_*$. In the plot we show this connection by adding a small positive imaginary component $s_2=0.9+0.01i$ to open a small gap above the real axis.  The analytic structure of the cut \eqref{def C tri 2} is identical to that of $\Exp_2$ in Fig.~\ref{fig:offshell triangle} in the upper half-plane;
its behavior in the lower half-plane is not relevant in that context.
}
\end{figure}
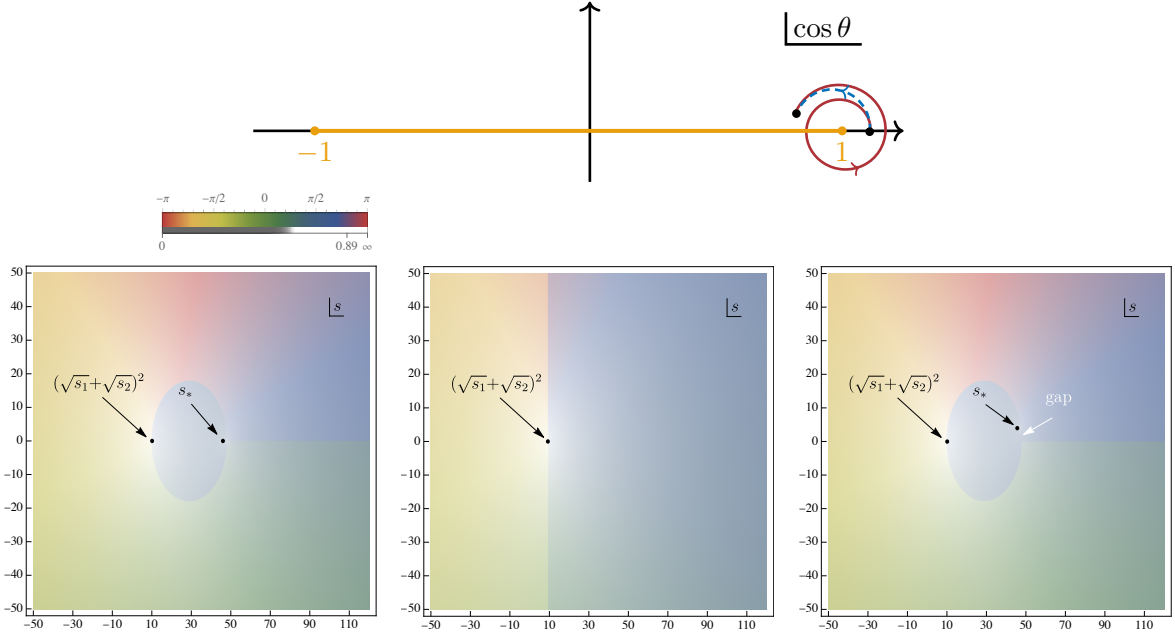

We recall that $\Tri$ is analytic in the upper half-plane of $s$, so
to answer the question of whether the crossing equation~\eqref{eq:TriContLoop} holds, it is sufficient to ask whether $\Cut_{s_1} \Tri$ from~\eqref{def C tri 2} evaluates to the same function for $s>0$ and $s<0$. Let us take this opportunity to stress the important point that even though we write the same expression~\eqref{def C tri 2}, for both $s<0$ and $s>0$, it does not immediately follow that it evaluates to the same analytic function in both kinematic channels. In fact, as we found previously, the function for $s>0$ is the same as the one for $s<0$ if and only if $s>s_{\ast}$. We summarize the analytic properties of~\eqref{def C tri 2} in Fig.~\ref{fig:cutCos}, both in the $s$-plane and the $\cos\theta$-plane. We see from the figure that~\eqref{def C tri 2} is the same analytic function for $s>0$ as for $s<0$ only if $s>s^\ast$ (Fig.~\ref{fig:cutCos} bottom left). As we found previously using the Schwinger-parametric form of $C(s)$, we can access the relevant sheet of $\Exp_2$ by adding a small imaginary part to $s_2$ (Fig.~\ref{fig:cutCos} bottom right). However, when $s_2$ is on shell (Fig.~\ref{fig:cutCos} bottom middle), it is not possible to reach the sheet of $\Exp_2$ starting from $\cM^\dag$.

\subsection{Spacetime interpretation}\label{sec:spacetime1}

The presence of the triangle anomalous threshold for the ${\rm Exp}_{2}$ observable could have been anticipated by thinking about the position-space support of the inclusive amplitude.
In this subsection, we show this by considering the associated trajectories of particles using a time-folded version of the Coleman--Norton interpretation of anomalous thresholds for exclusive scattering amplitudes \cite{Coleman:1965xm}. This will also allow us to understand why the anomalous threshold branch cut must be approached with an opposing kinematic $s \pm i \varepsilon$ prescription compared to the normal-threshold cut, causing a clash in the on-shell limit.

Recall that anomalous thresholds correspond to kinematic configurations where all the on-shell internal momenta $q_i^\mu$ of the diagram follow classical trajectories of particles interacting at the vertices located at $x_j^\mu$. The spacetime displacement between two vertices $(x_j, x_k)$ connected by the momentum $q_i$ is $x_{k} - x_{j} = \alpha_{i_\ast} q_{i_\ast}$, where $\alpha_{i_\ast}$ and $q_{i_\ast}$ are solutions of Landau equations, see, e.g., \cite{Hannesdottir:2022bmo}. Time-ordered amplitudes additionally require $\alpha_i \geq 0$ for all $i$, but cuts do not have the same restriction. Since we found in~\eqref{eq:alpha3-ast} that $\alpha_{3\ast} <0$ when $s_2 \to m_2^2$, the triangle threshold is accessible only through the cut contribution in \eqref{Tri test with cut}. Let us adopt the following labeling of vertices for the cut diagram:
\begin{align}\label{eq:triangle-cut}
\begin{gathered}
		\begin{tikzpicture}[baseline= {($(current bounding box.base)+(10pt,10pt)$)},line width=1.2, scale=1,yscale=-1]
			\coordinate (a) at (0,0) ;
			\coordinate (b) at (-2,0.5) ;
			\coordinate (c) at (-1,-0.5);
			\coordinate (d) at (-1,0);
			\draw[dashed] (a) -- (b);
            \draw[->,photon] (c) -- (-1.5,0);
			\draw[photon] (c) -- (b);
			\draw[] (c) -- (a);
			\draw[dashed,RoyalBlue] (b)-- ++(170:0.5) node[left,scale=0.75] {$1'$};
			\draw[photon,RoyalBlue] (b)-- ++(140:0.5) node[left,scale=0.75,yshift=0] {$3$};
			\draw[Maroon] (c)-- ++(-170:0.5) node[left,scale=0.75,xshift=-3] {$\bar{2}'$};
			\draw[Maroon] (a)-- ++(30:0.5) node[right,scale=0.75,xshift=-3] {$\bar{2}$};
			\draw[dashed,RoyalBlue] (a)-- ++(-30:0.5) node[right,scale=0.75,xshift=-3] {$1$};
			\draw[dashed,Orange] ($(d)+(-0.1,1)$) -- ($(d)+(-0.1,-1)$);
            \node[] at (-1.8,-0.2) {$\ell_\ast$};
            \filldraw[black] (a) circle (1pt) node[anchor=south]{$x$};
            \filldraw[black] (b) circle (1pt) node[anchor=south east]{$x_1$};
            \filldraw[black] (c) circle (1pt) node[anchor=south west]{$x_2$};
                \begin{scope}[xshift=4.4pt,yshift=25pt,yscale=-1]
                    \draw[line width=0.4, line cap=round, color=gray] (-1.2,0) -- (-1.2,-0.135) (0.8,0) -- (0.8,-0.135) (-1.2,-0.135) -- (0.8,-0.135) node[below, midway]{\footnotesize I};
                    \draw[line width=0.4, line cap=round, color=gray] (-3,0) -- (-3,-0.135) (-1.3,0) -- (-1.3,-0.135) (-3,-0.135) -- (-1.3,-0.135) node[below, midway]{\footnotesize II};
                \end{scope}
		\end{tikzpicture}
	\end{gathered}
\end{align}
Here, we labeled the vertices $x,x_1,x_2$, and added Schwinger--Keldysh labels I and II
to the fields on the two sides of the cut, which will be useful below.
The Landau solution corresponding to the anomalous threshold then corresponds to
\be\label{eq:spacetime1}
x-x_1 = \alpha_{1*} (\ell_\ast - p_1)\,, \qquad x_2-x = \alpha_{2*} (\ell_\ast + p_2)\,, \qquad x_1 - x_2 = \alpha_{3*} \ell_\ast\, ,
\ee
where
\begin{equation}
    \alpha_{1 \ast} = \sigma \frac{m_2^2-s_2}{s_1 - m_1^2} \,, \quad \alpha_{2 \ast} = \sigma \,, \quad \alpha_{3 \ast} = \sigma \left( \frac{m_2^2}{s_2 - m_2^2} - m_1^2 \frac{s_2 - m_2^2}{(s_1 - m_1^2)^2} \right) \,,
\end{equation}
as previously given in~\eqref{eq:alpha1-ast} and~\eqref{eq:alpha3-ast} and repeated here for convenience. Compared to the previous solution, we have included a rescaling parameter $\sigma>0$.
To ensure that these constraints are satisfied simultaneously, the loop momentum is fixed to $\ell_\ast = \frac{\alpha_{1*} p_1 - \alpha_{2*} p_2}{\alpha_{1*} + \alpha_{2*} + \alpha_{3*}}$.
The positions of vertices are only determined up to translations and an overall rescaling by a positive constant, which allows us to fix two of them to be 
\begin{equation}\label{eq:x1x2}
    x_1 = (0,0,0,0) \,, \quad \text{and} \quad x_2 = \rho (1,0,0,1)\,,
\end{equation}
without loss of generality. Here, $\rho>0$ is a dilation parameter representing the invariance under rescaling the diagram. (Note that \eqref{eq:x1x2} is consistent with the choice of rest frame made in \eqref{eq:restFrame1}, since both the position and the momentum four-vectors are confined to the $(x^0,x^3)$-plane.) These two vertices are null-separated, which corresponds to the propagation of the on-shell graviton between them.

Plugging in the solution from \eqref{eq:triangle-alphas}, the loop momentum pinches at
\be
\ell_\ast = \frac{s_1-m_1^2}{2\sqrt{s_1}} (1, 0, 0, 1)\,.
\label{eq:looppinch}
\ee
This allows us to determine the shape of the triangle anomalous threshold, which has its third corner at
\be\label{eq:x}
x = \sigma \frac{m_2^2-s_2}{2\sqrt{s_1}} \left(-\tfrac{s_1 + m_1^2}{s_1-m_1^2},0,0,1 \right)\,,
\ee
with
\be 
    \frac{\sigma}{\rho} = \frac{2  \sqrt{s_1} \left(s_1-m_1^2\right) \left(m_2^2-s_2\right)}{(m_2(m_1^2-s_1))^2-(m_1(m_2^2-s_2))^2} \,.
\ee 

Consequently, the spacetime geometry (with time flowing to the left) of the anomalous threshold looks schematically as follows:
\tikzstyle{midarrow}=[decoration={markings,mark=at position 0.5 with {\arrow{>}; \draw[] (0,0) -- (0.1,0);}}, postaction={decorate}, dashed]
\begin{align}\label{eq:spacetime2}
	\begin{gathered}
		\begin{tikzpicture}[baseline= {($(current bounding box.base)+(10pt,10pt)$)},line width=1.0, scale=1, yscale=-1]
\begin{scope}[xshift=-130]
    \coordinate (a) at (0.5,0) ;
	\coordinate (b) at (-0.5,0) ;
	\coordinate (c) at (-2,-1.5);
	\coordinate (d) at (-2,0);
   \draw[->,gray] (a) -- ++ (-90:1) node[above] {$x^3$};
   \draw[->,gray] (a) -- ++ (-180:1) node[left] {$x^0$};
   \draw[gray] (a) -- ++ (90:0.2);
   \draw[gray] (a) -- ++ (0:0.2);
\end{scope}
			\coordinate (a) at (0.3,-0.2);
			\coordinate (b) at (-0.5,0);
			\coordinate (c) at (-2,-1.5);
			\coordinate (d) at (-2,0);
			\draw[dashed] (a) to[out=145,in=145,distance=84pt] (b);
            \draw[dashed,midarrow] (a) to[out=145,in=145,distance=84pt] (b);
			\draw[->,photon] (c) -- (-1.25,-0.75);
			\draw[photon] (c) -- (b);
			\draw[] (a) -- (c);
            \draw[->-=0.5] (a) -- (c);
            \draw[fill=white,draw=none,opacity=0.6] (0,0.3) circle (0.3);
			\draw[dashed,RoyalBlue] (b)-- ++(20:1) node[right,scale=0.75] {$1'$};
			\draw[photon,RoyalBlue] (b)-- ++(40:1) node[below,scale=0.75,xshift=1] {$3$};
			\draw[Maroon] (c)-- ++(-170:0.5) node[left,scale=0.75,xshift=-3] {$\bar{2}'$};
			\draw[Maroon] (a)-- ++(30:0.5) node[right,scale=0.75,xshift=-3] {$\bar{2}$};
			\draw[dashed,RoyalBlue] (a)-- ++(-30:0.5) node[right,scale=0.75,xshift=-3] {$1$};
			\draw[dashed,Orange] ($(d)+(0.1,1.5)$) -- ($(d)+(0.1,-2)$);
			\filldraw[black] (a) circle (1pt) node[anchor=south]{$x$};
			\filldraw[black] (b) circle (1pt) node[anchor=east]{$x_1\;$};
			\filldraw[black] (c) circle (1pt) node[anchor=south east]{$x_2$};
            \node[gray, xshift=20, yshift=20,scale=0.75] at ($(a)!0.5!(c)$) {$x_2{-}x$};
            \draw[->, gray] ($($(a)!0.4!(c)$) + (10pt, -10pt)$) -- ($($(a)!0.6!(c)$) + (10pt, -10pt)$);
            \node[gray, xshift=-18, yshift=-15,scale=0.75] at ($(b)!0.5!(c)$) {$x_2{-}x_1$};
            \draw[->, gray] ($($(b)!0.4!(c)$) + (-7pt, 7pt)$) -- ($($(b)!0.6!(c)$) + (-7pt, 7pt)$);
            \node[gray, xshift=-15, yshift=-35,scale=0.75] at ($(b)!0.5!(a)$) {$x_1{-}x$};
            \draw[->, gray] ($($(a)!0.2!(b)$) + (-15pt, 20pt)$) -- ($($(a)!0.8!(b)$) + (-15pt, 25pt)$);
		\end{tikzpicture}
	\end{gathered}
\end{align}
It is the same as \eqref{eq:triangle-cut} folded along the cut and with the vertices $x, x_1, x_2$ placed according to \eqref{eq:x1x2} and \eqref{eq:x}. 
This fold has a simple physical interpretation.   As discussed in \cite[Sec.~4]{Caron-Huot:2023vxl}, the in-in observable $\Exp_2$ can be computed using a Schwinger--Keldysh timefold with two branches I and II.
More precisely, it is the Fourier transform of a retarded function that vanishes unless $x$ and $x_1$ are both in the past lightcone of the observation point $x_2$. 
The vertical dashed line can be visualized as the fold in the timefold, which can be put on any spacelike surface going through $x_2$.  The dashed propagator from $x$ to $x_1$ represents a massive particle propagating forward and backward in time along the fold.

The black arrows in the above picture denote the flow of energy, as dictated by the solution in~\eqref{eq:looppinch}. The gray arrows point in the directions of the spacetime displacements between the vertices. If the arrows for energy flow and the spacetime distance between two vertices point in the same direction, the corresponding Schwinger parameter is positive, but negative if they point in opposite directions. This is a generalization of the Coleman--Norton picture to solutions with positive and negative Schwinger parameters. Note that $\alpha_{3*} < 0$ corresponds to the fact that energy flows through the graviton propagator in the direction opposite to the spacetime displacement from $x_1$ to $x_2$.
This is precisely what is physically expected for the in-in observable because positive-energy particles on the second timefold propagate backward in time.

Below, in Sec.~\ref{sec:worldsheet-geomtry}, we will illustrate the counterpart of this discussion in string scattering, where the trajectories of strings can be plotted explicitly.

What happens here is that the graviton propagates at the speed of light from $x_1$ to $x_2$. In contrast, the massive particle travels from $x$ to $x_2$ at near-luminal speed. The fact that the two meet at $x_2$ allows the anomalous threshold to be physically allowed.

Let us now understand local analyticity near this threshold. As reviewed in App.~\ref{sec:local-analyticity}, the position space support of a diagram allows us to determine from which side to approach the branch cuts in momentum space. The relevant quantity to study is $\sum_{j=1}^{3} \Im p_j \cdot x_j$, where $p_j$ is the momentum flowing out of the vertex $x_j$. Imposing that this quantity is negative gives an inequality on the kinematic invariants that ensures convergence of the Fourier transform. For example, normal thresholds always require $\Im s > 0$, see App.~\ref{sec:local-analyticity}.

We can now apply these ideas to the triangle threshold. According to the crossing path from~\eqref{eq:crossingpath}, we fix $p_1$ to be real and set $p_2^\mu=(z p_2^+,\frac{1}{z} p_2^-,p_{2}^\perp)$ where $z$ can be complex. This fixes the kinematic invariants $s_1$ and $s_2$ to be real, but $s$ can be complex.
(We should technically add a small $s_1+i\eps$ but this will not play a role here.)  By momentum conservation, $p=-p_1-p_2$ is also generally complex. This prescription fixes $s_1$ and $s_2$ to be real, but $s$ becomes complex if $z$ does. We impose
\be\label{eq:dualcone}
\Im (p_1 \cdot x_1 + p_2 \cdot x_2 + p \cdot x)<0 \implies (\Im p_2) \cdot (x_2-x)<0 \,,
\ee
where we used momentum conservation and the fact that $x$, $x_1$ and $x_2$ are all real. To analyze what conditions $p_2$ needs to satisfy, we can use the values of $x$, $x_1$ and $x_2$ at the anomalous thresholds from~\eqref{eq:x1x2} and~\eqref{eq:x}.
We can translate this condition into one on the lightcone components of $p_2^\mu$,
\be 
    \Im (z p_2^+) (x_2^--x^-) + \Im \Big(\frac{p_2^-}{z} \Big) (x_2^+-x^+)>0 \,.\label{eq:constraintIM0}
\ee 
If $x_2^--x^- \neq 0$ (in which case it is necessarily positive due to the properties of retarded products), we can take $|z|$ to be sufficiently large along the crossing path so that the term proportional to $\frac{p_2^-}{z}$ can be dropped.
This shows that for large $s$, the real axis must be approached from ${\rm Im}\,(zp_2^+)>0$, or equivalently ${\rm Im}\,s>0$, in agreement with our discussion below \eqref{eq:residue} and Fig.~\ref{fig:offshell triangle}.

Let us call $z_\ast$ the value of $z$ at the anomalous threshold, i.e., $s(z_\ast) = s_\ast$.
To understand what happens near this value,
we can rewrite the convergence condition \eqref{eq:constraintIM0} as
\be \label{eq:dualcone 1}
 \Big({\rm Im}\, z\Big) \left[\frac{\partial p_2(z)}{\partial z}{\cdot}(x_2-x)\right]_{z = z_\ast} < 0\,,
\ee
where we assumed that $z$ is close to the real axis.
On $z = z_\ast$, we have $x_2^- - x^- = 0$, so the term in the square brackets equals $\frac{p_2^-}{z_\ast^2} (x_2^+ - x^+)$, which is positive. This implies ${\rm Im}\, z<0$.  Near $s=s_*$,  \eqref{eq:constraintIM0} is thus primarily a constraint on ${\rm Im}\, \frac{p_2^-}{z}$ rather than on the $\Im (z p_2^+)$ term.  This shows that, when computing the in-in observable $\Exp_2$, the anomalous threshold singularity must be
avoided with $s-i\eps$, again in agreement with the discussion below \eqref{eq:residue} and Fig.~\ref{fig:offshell triangle}. 

This calculation gives additional insight into the origin of the anomalous threshold for the $\Exp_2$ in-in observable. Physically, $\frac{\partial p_2^\mu(z)}{\partial z}$ is a vector orthogonal to the velocity $p_2^\mu$ of the outgoing particle.
The condition that the square bracket is positive is equivalent to saying that the internal state $(x_2-x)^\mu$ propagates \emph{faster} than the external state $p_2^\mu$.
This comparison of velocity also explains why, on the mass shell $p_2^2+m_2^2=0$, increasing the energy does not help: even though the external particle $p_2^\mu$ moves closer and closer to the speed of light, the internal particle of the same mass always moves slightly faster, because it has a larger energy due to $\alpha_{3\ast}<0$ (see \eqref{eq:spacetime2}). While this mechanism could never create a singularity in $\Tri$ near real momenta, it does affect $\Exp_2$.

It is instructive to repeat the above calculation for a massless external particle, where we can choose $p_2^\mu=z(1,0,0,1)$.
Since all the separations $(x_2-x_i)^\mu$ contributing to the Fourier integral are necessarily future timelike (by the support property of retarded products, cf., the argument around \eqref{eq:R22}), the condition \eqref{eq:dualcone 1} reduces to ${\rm Im}\,z>0$.
We conclude that when computing the in-in expectation value $\Exp_2$ of a \emph{massless} particle $p_2$, the obstruction to crossing found in this section disappears.

\subsection{Analytic obstruction to multi-particle crossing}
\label{sec:analytic-obstruction}

In the multi-particle crossing conjecture, the obstruction to analyticity is not restricted to crossing of massive particles. As an example, let us look at the following triangle diagram where all particles are massless, and $1$, $2$ and $6$ are incoming,
\begin{equation}
\begin{gathered}
\begin{tikzpicture}[line width=1]
\draw[] (0,0) -- (1,0.5) -- (1,-0.5) -- (0,0);
\draw[Maroon] (1.4,0.85) -- (1,0.5);
\draw[Maroon] (1,0.5)--++(20:0.5);
\draw[RoyalBlue] (1.5,-0.75) -- (1,-0.5);
\draw[Maroon] (1,-0.5) -- (0.5,-0.75);
\draw[RoyalBlue] (-0.5,0.25) -- (0,0) -- (-0.5,-0.25);
\node[] at (-1.5,0) {$s_{45}$};
\node[] at (2.3,1.) {$s_{26}$};
\node[] at (1,-1.2) {$s_{13}$};
\node[scale=0.75,Maroon] at (1.5,1.1) {$6$};
\node[scale=0.75,Maroon] at (1.65,0.8) {$2$};
\node[scale=0.75,RoyalBlue] at (1.7,-0.8) {$1$};
\node[scale=0.75,Maroon] at (0.3,-0.8) {$3$};
\node[scale=0.75,RoyalBlue] at (-0.8,0.2) {$4$};
\node[scale=0.75,RoyalBlue] at (-0.8,-0.2) {$5$};
\end{tikzpicture}
\end{gathered}
\end{equation}
We aim to cross the cluster $B=\{2,6\}$ with $C=\{3\}$. Therefore, we start with the amplitude in the region $R^{(26)(45)}$, before rotating $s_{13}$ in the lower half-plane. This corresponds to $\zb$ starting with a negative imaginary part and $z$ starting with a positive imaginary part, and they both rotate in the clockwise direction: 
\begin{equation}
\begin{gathered}
\begin{tikzpicture}
  \draw[->,thick] (-1.5, 0) -- (1.5, 0);
  \draw[->,thick] (0, -1.25) -- (0, 1.25);
  \node[] at (1.1,1.05) {$s_{ij}$};
  \draw[] (1.35,0.85) -- (0.8,0.85) -- (0.8,1.25);
  \draw[RoyalBlue,fill=RoyalBlue,thick] (0.23,0.1) circle (0.05);
  \node[] at (0.23,0.3) {\color{RoyalBlue}$s_{45}$};
  \draw[RoyalBlue,fill=RoyalBlue,thick] (0.4,0.1) circle (0.05);
  \node[] at (0.6,0.5) {\color{RoyalBlue}$s_{26}$};
  \draw[Maroon,fill=Maroon,thick] (-1,0) circle (0.05);
  \draw[->,Maroon,thick] (-1,0) arc (0:175:-1);
  \node[] at (-0.8,-1) {\color{Maroon} $s_{13}$};
  \draw[black!80,fill=black!80,thick] (0,0) circle (0.05);
  \draw[decorate, decoration={zigzag, segment length=6, amplitude=2}, black!80] (0,0) -- (1.45,0);
\end{tikzpicture}
\end{gathered}
\hspace{1.5cm}
\begin{gathered}
\begin{tikzpicture}
  \draw[->,thick] (-1, 0) -- (2.2, 0);
  \draw[->,thick] (0, -1) -- (0, 1.5);
  \node[] at (1.8,1.3) {$z$};
  \draw[] (2.0,1.1) -- (1.6,1.1) -- (1.6,1.5);
  \draw[decorate, decoration={zigzag, segment length=6, amplitude=2}, black!80] (1,0) -- (2.14,0);
  \draw[black!80,fill=black!80,thick] (1,0) circle (0.05);
  \draw[decorate, decoration={zigzag, segment length=6, amplitude=2}, black!80] (0,0) -- (-1,0);
  \draw[black!80,fill=black!80, thick] (0,0) circle (0.05);
  \draw[Maroon,fill=Maroon,thick] (1.2,0.1) circle (0.05);
  \draw[->,Maroon,thick] (1.2,0.1) arc (0:-160:0.35);
\end{tikzpicture}
\end{gathered}
\hspace{1.5cm}
\begin{gathered}
\begin{tikzpicture}
  \draw[->,thick] (-1, 0) -- (2.2, 0);
  \draw[->,thick] (0, -1) -- (0, 1.5);
  \node[] at (1.8,1.3) {$\zb$};
  \draw[] (2.0,1.1) -- (1.6,1.1) -- (1.6,1.5);
  \draw[decorate, decoration={zigzag, segment length=6, amplitude=2}, black!80] (1,0) -- (2.14,0);
  \draw[black!80,fill=black!80,thick] (1,0) circle (0.05);
  \draw[decorate, decoration={zigzag, segment length=6, amplitude=2}, black!80] (0,0) -- (-1,0);
  \draw[black!80,fill=black!80,thick] (0,0) circle (0.05);
  \draw[Maroon,fill=Maroon,thick] (-0.2,-0.1) circle (0.05);
  \draw[->,Maroon,thick] (-0.2,-0.1) arc (0:-160:-0.35);
\end{tikzpicture}
\end{gathered}
\end{equation}
The result of the analytic continuation is the conjugated amplitude, along with a (single) discontinuity
from simultaneously crossing the branch cuts at $\zb<0$ and $z>1$:
\begin{equation}
 \left[\cM_{345\ot 126}\right]_{ 
 \raisebox{\depth}{\scalebox{1}[-1]{$\curvearrowright$}}s_{13}}
 =
 \cM_{2456\ot 13}^{\dag} + \text{Disc}_{\substack{z>1\\ \zb<0}} \cM_{345\ot 126}\,,
 \label{eq:rotDisc2}
\end{equation}
where the discontinuity can be worked out from \eqref{eq:Mtri},
\begin{equation}
    \text{Disc}_{\substack{z>1\\ \zb<0}} \cM_{345\ot 126}  = \frac{-2 \pi i}{s_{13} (z-\zb)} \left[\log  \frac{z(1-\zb)}{\zb(1-z)}+2\pi i\right]\,.
    \label{eq:discz1zb0}
\end{equation}

The crossing conjecture now predicts that the following diagrams, which fit the blob pattern in \eqref{eq:crossing_speculation}, should contribute to the analytically continued amplitude,
\begin{align}
    \begin{gathered}
    \begin{tikzpicture}[baseline= {($(current bounding box.base)+(10pt,10pt)$)},line width=1, scale=1]
    \coordinate (a) at (0,0) ;
    \coordinate (b) at (-1,0.5) ;
    \coordinate (c) at (-1,-0.5);
    \coordinate (d) at (-1,0);
    \draw[] (a) -- (b) -- (c) -- (a);
    \draw[Maroon] (b)-- ++(170:0.5) node[left,scale=0.75] {$\bar{6}$};
    \draw[Maroon] (b)-- ++(140:0.5) node[left,scale=0.75,yshift=5] {$\bar{2}$};
    \draw[RoyalBlue] (c)-- ++(-170:0.5) node[left,scale=0.75,xshift=-3] {$4$};
    \draw[RoyalBlue] (c)-- ++(-140:0.5) node[left,scale=0.75,xshift=0,yshift=-4] {$5$};
    \draw[RoyalBlue] (a)-- ++(30:0.5) node[right,scale=0.75,xshift=-3] {$1$};
    \draw[Maroon] (a)-- ++(-30:0.5) node[right,scale=0.75,xshift=-3] {$\bar{3}$};
    \draw[dashed,Orange] ($(d)+(180:0.25)+(0,1)$) -- ($(d)+(180:0.25)+(0,-1)$);
    \draw[dashed,Orange] ($(a)+(0:0.25)+(0,1)$) -- ($(a)+(0:0.25)+(0,-1)$);
    \end{tikzpicture}
    \end{gathered}
    \qquad
    \begin{gathered}
    \begin{tikzpicture}[baseline= {($(current bounding box.base)+(10pt,10pt)$)},line width=1, scale=1]
    \coordinate (a) at (0,0) ;
    \coordinate (b) at (-2,0.5) ;
    \coordinate (c) at (-1,-0.5);
    \coordinate (d) at (-1,0);
    \draw[] (a) -- (b) -- (c) -- (a);
    \draw[Maroon] (b)-- ++(170:0.5) node[left,scale=0.75] {$\bar{6}$};
    \draw[Maroon] (b)-- ++(140:0.5) node[left,scale=0.75,yshift=5] {$\bar{2}$};
    \draw[RoyalBlue] (c)-- ++(-170:0.5) node[left,scale=0.75,xshift=-3] {$4$};
    \draw[RoyalBlue] (c)-- ++(-140:0.5) node[left,scale=0.75,xshift=0,yshift=-4] {$5$};
    \draw[RoyalBlue] (a)-- ++(30:0.5) node[right,scale=0.75,xshift=-3] {$1$};
    \draw[Maroon] (a)-- ++(-30:0.5) node[right,scale=0.75,xshift=-3] {$\bar{3}$};
    \draw[dashed,Orange] ($(d)+(180:0.25)+(0,1)$) -- ($(d)+(180:0.25)+(0,-1)$);
    \draw[dashed,Orange] ($(a)+(0:0.25)+(0,1)$) -- ($(a)+(0:0.25)+(0,-1)$);
    \end{tikzpicture}
    \end{gathered} 
    \qquad
    \begin{gathered}
    \begin{tikzpicture}[baseline= {($(current bounding box.base)+(10pt,10pt)$)},line width=1, scale=1]
    \coordinate (a) at (0,0) ;
    \coordinate (b) at (-1,0.5) ;
    \coordinate (c) at (-2,-0.5);
    \coordinate (d) at (-1,0);
    \draw[] (a) -- (b) -- (c) -- (a);
    \draw[Maroon] (b)-- ++(170:0.5) node[left,scale=0.75] {$\bar{6}$};
    \draw[Maroon] (b)-- ++(140:0.5) node[left,scale=0.75,yshift=5] {$\bar{2}$};
    \draw[RoyalBlue] (c)-- ++(-170:0.5) node[left,scale=0.75,xshift=-3] {$4$};
    \draw[RoyalBlue] (c)-- ++(-140:0.5) node[left,scale=0.75,xshift=0,yshift=-4] {$5$};
    \draw[RoyalBlue] (a)-- ++(30:0.5) node[right,scale=0.75,xshift=-3] {$1$};
    \draw[Maroon] (a)-- ++(-30:0.5) node[right,scale=0.75,xshift=-3] {$\bar{3}$};
    \draw[dashed,Orange] ($(d)+(180:0.25)+(0,1)$) -- ($(d)+(180:0.25)+(0,-1)$);
    \draw[dashed,Orange] ($(a)+(0:0.25)+(0,1)$) -- ($(a)+(0:0.25)+(0,-1)$);
    \end{tikzpicture}
    \end{gathered}
\end{align}
where we have left implicit that the middle part between the orange lines should be conjugated.
To compare \eqref{eq:rotDisc2} with the result of the crossing conjecture, we therefore have to compute the unitarity cuts in the $s_{26}$ and $s_{45}$ channels. In the kinematic region $0<\zb<z<1$ we land on after crossing, we get,
\begin{align} 
    \Cut_{s_{26}} \cM^\text{tri}_{2456 \ot 13}
    & = \frac{-2 \pi i}{s_{13} (z-\zb)} \log \frac{1-\zb}{1-z}\,, \\
    \Cut_{s_{45}} \cM^\text{tri}_{2456 \ot 13}
    & = \frac{-2 \pi i}{s_{13} (z-\zb)} \log \frac{z}{\zb}\,,
\end{align}
from computations similar to the one that resulted in \eqref{eq:cut2Particles}. Comparing these results with~\eqref{eq:discz1zb0} gives
\begin{equation}
\left[\cM_{345\ot 126}\right]_{ 
 \raisebox{\depth}{\scalebox{1}[-1]{$\curvearrowright$}}s_{13}}
 =
 \underbracket[0.4pt]{\cM_{2456\ot 13}^{\dag} +\Cut_{s_{26}} \cM^\text{tri}_{2456 \ot 13}+\Cut_{s_{45}} \cM^\text{tri}_{2456 \ot 13}}_{\text{result of multi-particle crossing conjecture}} + \frac{4 \pi^2}{s_{13} (z-\zb)}\,,
\end{equation}
which does not agree with the multi-particle crossing conjecture in \eqref{eq:crossing_speculation}.

As in the previous subsections, the difference between the analytically continued amplitude and the result of the crossing conjecture looks like a discontinuity across a triangle anomalous threshold. Indeed, the maximal cut of the triangle diagram where all particles are put on shell (which is not allowed in this kinematic region) is exactly $\frac{4 \pi^2}{s_{13}(z-\zb)}$. Thus, this example seems to suggest that the factorization of the blobs of~\eqref{eq:crossing_speculation} is broken when massless particles run in the loop. We leave a more thorough study of such analytic obstructions for future work. Since the invariants $s_{ij}$ can be varied independently, it is still possible that a fully on-shell crossing path exists in this example.

\subsection{Inclusive massless observables are safe}\label{ssec:massless}

\tikzstyle{every node}=[font=\small,
dot/.style = {circle, fill, minimum size=4pt,
              inner sep=0pt, outer sep=0pt}]
\tikzset{photon/.style={decorate, decoration={snake, amplitude=1pt, segment length=6pt}}}

In the previous subsections, we explained in two complementary ways how the branch cuts coming from anomalous thresholds can potentially obstruct the analytic continuation between time-ordered amplitudes and inclusive \emph{massive} observables. In contrast, the purpose of this subsection is to illustrate how crossings leading to inclusive \emph{massless} observables are free from such anomalous threshold problems.

\paragraph{Massless limit of~\eqref{eq:atCrossing}}
We start by exploring the massless limit of the diagram we studied previously in this section. Looking back at the location of the anomalous threshold $s_\ast$ from~\eqref{eq:triangle-alphas}, which has a term proportional to $m_2^2\frac{s_1-m_1^2}{s_2^2-m_2^2}$, it is not immediately clear what the limit of $s_\ast$ is when $s_2\to 0$ and $m_2^2\to 0$, since the result depends on the order in which the limits are taken. Instead, we must directly solve the Landau equations for the case where $s_2=0$ and $m_2^2=0$. When doing so, we do find singularities at $s_1=m_1^2$ and $s=m_1^2$, $s_1=0$ and $s=0$, as well as the second-type singularity at $s=s_1$. However, the anomalous threshold is no longer present.

Let us trace what happens to the derivation in Sec.~\ref{ssec:comparision}, which previously, when $s_2<m_2^2$, led us to a condition on how large we needed to take $s$ for the crossing path to exist. When $s_2 = 0$ and $m_2^2=0$, the integrands for $\Cut_{s_1} \Tri$ and $C$ have singularities at the two roots $\alpha_{-}=0$ and $\alpha_+ = \frac{s-s_1}{s_1}$, and the integration region now has an endpoint at $\alpha_{1 \ast}=0$. The vanishing of the triangle singularity is reflected in the fact that the integrals for $C$ and $\Cut_{s_1} \Tri$ are now infrared divergent in $\D=4$ spacetime dimensions, and we must regulate them to obtain sensible expressions. In dimensional regularization, the triangle integral from~\eqref{Tri Schwinger} is given by
\be
\Tri^\dag = \Gamma(1+\epsilon) \int_0^\infty \frac{\text{d}\alpha_1 \text{d}\alpha_3}{(\alpha_1 + 1 + \alpha_3)^{1-2\epsilon} \left[A(\alpha_1)+ \alpha_3 B(\alpha_1)+i\eps\right]^{1+\epsilon}}\,,
\ee
with $A(\alpha_1) = (\alpha_1+1)\alpha_1 m_1^2-\alpha_1 s$ and $B(\alpha_1) = \alpha_1 (m_1^2-s_1)$.

For the kinematics with $\text{Im}(s_1)=-i\varepsilon$, $s<0$ and $|s|>s_1$ the conjugated amplitude can be computed using Feynman parameters and is given by
\begin{equation}
   \Tri^\dag= -\frac{\Gamma(1+\epsilon) (m_1^2)^{1-\epsilon}}{\epsilon^2 (s{-}s_1)} \left(\frac{\mathcal{F}\big(\frac{s_1}{s_1{-}m_1^2}+i\varepsilon\big)}{s_1{-}m_1^2}-\frac{\mathcal{F}\big(\frac{s}{s{-}m_1^2}\big)}{s{-}m_1^2}\right)\,,
   \label{eq:Mtrimassless}
\end{equation}
where $\mathcal{F}(z)\equiv {}_2F_1\big(1,1{-}2
   \epsilon,1{-}\epsilon,z\big)$. The last expression for $\Tri^\dag$ is analytic in the upper half-plane of $s$, and unlike in the case where $m_2 \neq 0$, there is no branch cut on the negative $s$-axis. We can therefore easily analytically continue~\eqref{eq:Mtrimassless} in the upper half-plane of $s$ to get
\be 
    \big[\Tri^\dag\big]_{\raisebox{\depth}{\scalebox{1}[1]{$\curvearrowright$}}s}
    =
    -\frac{\Gamma(1+\epsilon) (m_1^2)^{1-\epsilon}}{\epsilon^2 (s{-}s_1)} \left(\frac{\mathcal{F}\big(\frac{s_1}{s_1{-}m_1^2}+i\varepsilon\big)}{s_1{-}m_1^2}-\frac{\mathcal{F}\big(\frac{s}{s{-}m_1^2}-i\varepsilon\big)}{s{-}m_1^2}\right)\,.
    \label{eq:Mtricontinued}
\ee 
In the new kinematic channel where $s>0$, the second term on the right-hand side has the right $i \varepsilon$ prescription for $\Tri$, but the first term does not. 
Subtracting $\Tri$ from~\eqref{eq:Mtricontinued} gives
\begin{equation}\label{eq:discTilde}
        \big[\Tri^\dag\big]_{\raisebox{\depth}{\scalebox{1}[1]{$\curvearrowright$}}s}  = \Tri + \underbracket[0.4pt]{\frac{2 i \pi ^2\csc (\pi  \epsilon)  s_1^{\epsilon}(s_1-m_1^2)^{-2 \epsilon}}{\Gamma (1-2 \epsilon) \Gamma(1+\epsilon) (s_1-s)}}_{\tilde{C}(s)}\,.
\end{equation}
The result for $\Tilde{C}(s)$ is clearly analytic in the $s$-plane (recall that $|s|\neq s_1$).

The last thing we need to do to check that the crossing equation holds is to compare $\Tilde{C}(s)$ against the direct evaluation of the cut in the $s_1$-channel. Using a strategy similar to that leading to \eqref{eq:cuteps}, we find (in generic kinematics)
\begin{equation}\label{eq:cutTilde}
    \begin{split}
        \Cut_{s_1}\Tri&=-(2\pi i)^2\int \frac{\text{d}^\D \ell}{i \pi^{\D/2}}\frac{\delta^{+}(\ell^2)\delta^{-}((\ell-p_1)^2+m_1^2)}{(\ell+p_2)^2+m_2^2-i\varepsilon}\\&=\frac{2\pi i \cos(\pi \epsilon)(\ell^0_\ast)^{-2\epsilon}}{\Gamma(1-\epsilon)}\int_{-\infty}^{\alpha_{1\ast}}\text{d}\alpha_1\frac{(1-X^2(\alpha_1))^{-\epsilon}}{(\alpha_1 + 1)B(\alpha_1) - A(\alpha_1)}\,,
    \end{split}
\end{equation}
where $\ell^0_\ast$, $A$ and $B$ are as before and $X(\alpha_1)$ satisfies $\alpha_1 = \tfrac{-s+s_1+s_2+2 s_1 \alpha_{1\ast}}{2 s_1} X(\alpha_1) + \frac{s-s_1-s_2}{2 s_1}$. Setting $s_2=m_2=0$ in \eqref{eq:cutTilde} and integrating leads to 
\begin{equation}
    \begin{split}
        \Cut_{s_1}\Tri&=\frac{2\pi i \cos(\pi \epsilon)}{\Gamma(1-\epsilon)}\frac{1}{s_1^{1+\epsilon}}\left(\frac{s-s_1}{s_1-m_1^2}\right)^{2\epsilon}\int_{-\infty}^{0}\frac{\text{d}\alpha_1}{\big[\alpha_1~\big(\alpha_1-\frac{s-s_1}{s_1}\big)\big]^{1+\epsilon}} \qquad (\epsilon<0)\\&= 
        \tilde{C}(s) \,.
    \end{split}
\end{equation}

\paragraph{Crossing for gravitational waveforms}
Another massless limit we can look at is one obtained when changing the diagram to still have four massive legs, and analytically continue from a conjugated amplitude to an inclusive measurement of a massless particle. As a physically relevant example where such continuations may be useful, let us consider the inclusive expectation value of a graviton in the background of black-hole scattering. It is computed by the expression
\begin{equation} \label{KMOC}
{\rm Exp}_3\equiv
    {}_{\rm in}\< 2'1'| S^\dag a_3 S |12\>_{\rm in} = \< 0| a_{2'} a_{1'}\, b_3\, \adag_{2}\adag_{1}|0\>\,, 
\end{equation}
and is related to the gravitational waveform according to the KMOC formalism~\cite{Kosower:2018adc}. This inclusive observable is precisely the one discussed earlier in Sec.~\ref{sec:Smatrix}, and is related by analytic continuation to a conventional scattering amplitude:
\begin{equation}
    \adjustbox{valign=b}{\begin{tikzpicture}[line width=1]
\draw[] (0,-1.3) node {\small$\vac{a_{2'} a_{1'} a_{1} b_{3}^\dag b_{2}^\dag}$};
\draw[RoyalBlue] (1.2,-0.3) -- (0,-0.3) node[right,pos=0] {\small$2$};
\draw[Maroon] (0,0.45) -- (-1.2,0.45) node[left] {\small$1$};
\draw[RoyalBlue] (0,0) node[right] {} -- (-1.2,0) node[left] {\small$1'$};
\draw[photon,Maroon] (1.2,0.3) -- (0,0.3) node[pos=0,right] {\small$3$} node[left] {};
\draw[RoyalBlue] (0,-0.45) node[right] {} -- (-1.2,-0.45) node[left] {\small$2'$};
\filldraw[fill=gray!5](0,0) circle (0.6) node {$S^\dag$};
\draw[-latex] (2,0) -- (4,0);
\node[align=center] at (3,0.5) {cross ${\color{Maroon} 1} \leftrightarrow {\color{Maroon} 3}$};
\end{tikzpicture}}
\quad
\adjustbox{valign=b}{\begin{tikzpicture}[line width=1]
\draw[] (1,-1.1) node {\small$\vac{a_{2'} a_{1'} b_3 \adag_{2}\adag_1}$};
\begin{scope}[yshift=6]
    \draw[RoyalBlue] (0,0.3) -- (-1.2,0.3) node[left] {\small$1'$};
\draw[RoyalBlue] (0,-0.3) -- (-1.2,-0.3) node[left] {\small$2'$};
\draw[Maroon] (2,0.3) -- (3.2,0.3) node[right] {\small$\bar{1}$};
\draw[RoyalBlue] (2,-0.3) -- (3.2,-0.3) node[right] {\small$2$};
\draw[photon,Maroon] (2,0) -- (1,0.7) node[left] {\small$\bar{3}$};
\filldraw[fill=gray!30](0,-0.3) rectangle (2,0.3);
\draw[] (1,0) node {$X$};
\filldraw[fill=gray!5](0,0) circle (0.6) node {$S^\dag$};
\filldraw[fill=gray!5](2,0) circle (0.6) node {$S$};
\draw[dashed,orange] (1,0.7) -- (1,-0.6);
\end{scope}
\end{tikzpicture}}
\label{eq:5ptcross1}
\end{equation}
The observable in~\eqref{KMOC} was recently computed at one-loop in the classical limit by various authors (see~\cite{Brandhuber:2023hhy,Herderschee:2023fxh,Elkhidir:2023dco,Georgoudis:2023lgf}) in the \emph{eikonal} or \emph{heavy-mass effective field theory} (HEFT) expansion. In a recent work~\cite{Caron-Huot:2023vxl}, we showed how to efficiently compute the master integrals for inclusive observables such as~\eqref{KMOC} using differential equations and imposing boundary conditions where the observable is expressed as a sum of time-ordered amplitudes and cuts. In this section, we will illustrate on some of the relevant master integrals how to alternatively compute~\eqref{KMOC} using crossing symmetry. Unlike in the previous subsections, we find that the observable computed here corresponds to an inclusive measurement of a \emph{massless} particle. As such, crossing works as expected from~\eqref{eq:5ptcross1}, with no issues arising from anomalous thresholds.

A feature of the analytic continuation in~\eqref{eq:5ptcross1} is that it passes through a kinematic region in which the external graviton is hard. This kinematic region is far outside the validity of the HEFT expansion, i.e., it is incompatible with the eikonal approximation for the heavy objects labeled by $1$, $1'$ and $2$, $2'$ in~\eqref{eq:5ptcross1}. For crossing to work, we have to keep at least one of the heavy objects to be non-eikonal in the intermediate steps and take the eikonal limit only \emph{after} the analytic continuation. The fact that crossing and the HEFT limit do not commute is not surprising. In general, such non-commutativity of limits and analytic continuations is an expected feature of analytic functions. In this section, we therefore do not expand any propagators in the eikonal limit. Despite having to work beyond the eikonal expansion, our hope is that this continuation will prove valuable in computing higher-order corrections to these kinds of observables in the future.

To illustrate why there is no anomalous threshold obstruction to crossing for massless inclusive variables, we focus, for simplicity, on triangle Feynman integrals. These contribute as master integrals to the computation of the observable $\Exp_3$. For most triangle topologies, we can easily check that crossing is satisfied. An example of such check is given by the following topology,
\begin{equation}\label{eq:crossing5}
    \left[\scalebox{1}[-1]{\adjustbox{valign=c,rotate=180}{\tikzset{every picture/.style={line width=1pt}}
\begin{tikzpicture}[x=0.75pt,y=0.75pt,yscale=-0.75,xscale=0.75]
\begin{scope}[xshift=2,yshift=-0.7]
\begin{scope}[xshift=2,yshift=40]
\draw  [color=Maroon]  (137.08,119.81) .. controls (135.39,121.46) and (133.73,121.44) .. (132.08,119.75) .. controls (130.43,118.06) and (128.77,118.04) .. (127.08,119.69) .. controls (125.39,121.33) and (123.73,121.31) .. (122.08,119.62) .. controls (120.43,117.93) and (118.77,117.91) .. (117.08,119.56) .. controls (115.39,121.21) and (113.73,121.19) .. (112.08,119.5) .. controls (110.44,117.81) and (108.78,117.79) .. (107.09,119.43) -- (106.71,119.43) -- (106.71,119.43) node[pos=0,right,rotate=180]{\scalebox{0.75}[-0.75]{$3$}};
\end{scope}
\draw    (137.08,119.81) .. controls (138.79,121.44) and (138.82,123.11) .. (137.19,124.81) .. controls (135.56,126.51) and (135.6,128.18) .. (137.3,129.81) .. controls (139,131.44) and (139.04,133.11) .. (137.41,134.81) .. controls (135.78,136.51) and (135.82,138.18) .. (137.52,139.81) .. controls (139.22,141.44) and (139.26,143.11) .. (137.63,144.81) .. controls (136,146.51) and (136.04,148.17) .. (137.74,149.8) .. controls (139.44,151.43) and (139.48,153.1) .. (137.85,154.8) .. controls (136.22,156.5) and (136.26,158.17) .. (137.96,159.8) .. controls (139.66,161.43) and (139.7,163.1) .. (138.07,164.8) .. controls (136.44,166.5) and (136.48,168.17) .. (138.18,169.8) .. controls (139.88,171.43) and (139.92,173.1) .. (138.29,174.8) -- (138.33,175) -- (138.33,175) ;
\end{scope}
\coordinate (A) at (209.18,188.11);
\coordinate (B) at (127.28,106.22);
\pgfmathsetmacro{\slope}{(188.11 - 106.22) / (209.18 - 127.28)}
\pgfmathsetmacro{\angle}{atan(\slope)}
\pgfmathsetmacro{\d}{17}
\pgfmathsetmacro{\newX}{127.28 + \d*cos(\angle)}
\pgfmathsetmacro{\newY}{106.22 + \d*sin(\angle)}
\coordinate (B') at (\newX, \newY);
\draw[rounded corners=1pt,color=RoyalBlue,line cap=round] (A) -- (B') node[pos=0,left,rotate=180]{\scalebox{0.75}[-0.75]{$2$}};
\draw[color=RoyalBlue,line cap=round] (209.18,106.22) -- (127.28,188.11) node[pos=1,right,rotate=180]{\scalebox{0.75}[-0.75]{$1'$}} node[pos=0,left,rotate=180]{\scalebox{0.75}[-0.75]{$1$}}; 
\draw[line cap=round] (168.23, 147.165) -- (143,172.45) ;
\draw[line cap=round] (168.23, 147.165) -- (140.525, 119.46);
\draw[color=Maroon,line cap=round] (B') -- (180,100) node[pos=1,left,rotate=180,color=Maroon]{\scalebox{0.75}[-0.75]{$2'$}} node[right,xshift=-50,yshift=5,color=black]{$\dagger$};

\end{tikzpicture}}}\right]_{\raisebox{\depth}{\scalebox{1}[-1]{$\curvearrowleft$}}s_1} \stackrel{?}{=} \adjustbox{valign=c,rotate=180}{\tikzset{every picture/.style={line width=1pt}}
\begin{tikzpicture}[x=0.75pt,y=0.75pt,yscale=-0.75,xscale=0.75]
\begin{scope}[xshift=2,yshift=-0.7]
\begin{scope}[xshift=23,yshift=0]
\draw  [color=Maroon]  (137.08,119.81) .. controls (135.39,121.46) and (133.73,121.44) .. (132.08,119.75) .. controls (130.43,118.06) and (128.77,118.04) .. (127.08,119.69) .. controls (125.39,121.33) and (123.73,121.31) .. (122.08,119.62) .. controls (120.43,117.93) and (118.77,117.91) .. (117.08,119.56) .. controls (115.39,121.21) and (113.73,121.19) .. (112.08,119.5) .. controls (110.44,117.81) and (108.78,117.79) .. (107.09,119.43) -- (106.71,119.43) -- (106.71,119.43) node[pos=0,right,rotate=180,xshift=-30,yshift=-5]{\scalebox{0.75}[0.75]{$\bar{3}$}};
\end{scope}
\draw   (137.08,119.81) .. controls (138.79,121.44) and (138.82,123.11) .. (137.19,124.81) .. controls (135.56,126.51) and (135.6,128.18) .. (137.3,129.81) .. controls (139,131.44) and (139.04,133.11) .. (137.41,134.81) .. controls (135.78,136.51) and (135.82,138.18) .. (137.52,139.81) .. controls (139.22,141.44) and (139.26,143.11) .. (137.63,144.81) .. controls (136,146.51) and (136.04,148.17) .. (137.74,149.8) .. controls (139.44,151.43) and (139.48,153.1) .. (137.85,154.8) .. controls (136.22,156.5) and (136.26,158.17) .. (137.96,159.8) .. controls (139.66,161.43) and (139.7,163.1) .. (138.07,164.8) .. controls (136.44,166.5) and (136.48,168.17) .. (138.18,169.8) .. controls (139.88,171.43) and (139.92,173.1) .. (138.29,174.8) -- (138.33,175) -- (138.33,175) ;
\end{scope}

\draw[color=RoyalBlue,line cap=round] (209.18,106.22) -- (143,172.45) node[pos=0,left,rotate=180]{\scalebox{0.75}[0.75]{$2$}}; 
\draw[color=RoyalBlue,line cap=round] (209.18,188.11) -- (127.28,106.22)
node[pos=1,right,rotate=180]{\scalebox{0.75}[0.75]{$1'$}} node[pos=0,left,rotate=180]{\scalebox{0.75}[0.75]{$1$}}; 
\draw[line cap = round] (168.23, 147.165) -- (143,172.45) ;
\draw[line cap = round] (168.23, 147.165) -- (140.525, 119.46) ;
\draw[color=Maroon, line cap = round]  (143,172.45)--(127.28,188.11) node[pos=1,right,rotate=180,color=Maroon]{\scalebox{0.75}[0.75]{$\bar{2}'$}};

\end{tikzpicture}}+\adjustbox{valign=c,rotate=180}{\tikzset{every picture/.style={line width=1pt}}
\begin{tikzpicture}[x=0.75pt,y=0.75pt,yscale=-0.75,xscale=0.75]
\begin{scope}[xshift=2,yshift=-0.7]
\begin{scope}[xshift=23,yshift=0]
\draw  [color=Maroon]  (137.08,119.81) .. controls (135.39,121.46) and (133.73,121.44) .. (132.08,119.75) .. controls (130.43,118.06) and (128.77,118.04) .. (127.08,119.69) .. controls (125.39,121.33) and (123.73,121.31) .. (122.08,119.62) .. controls (120.43,117.93) and (118.77,117.91) .. (117.08,119.56) .. controls (115.39,121.21) and (113.73,121.19) .. (112.08,119.5) .. controls (110.44,117.81) and (108.78,117.79) .. (107.09,119.43) -- (106.71,119.43) -- (106.71,119.43) node[pos=0,right,rotate=180,xshift=-30,yshift=-5]{\scalebox{0.75}[0.75]{$\bar{3}$}};
\end{scope}
\draw   (137.08,119.81) .. controls (138.79,121.44) and (138.82,123.11) .. (137.19,124.81) .. controls (135.56,126.51) and (135.6,128.18) .. (137.3,129.81) .. controls (139,131.44) and (139.04,133.11) .. (137.41,134.81) .. controls (135.78,136.51) and (135.82,138.18) .. (137.52,139.81) .. controls (139.22,141.44) and (139.26,143.11) .. (137.63,144.81) .. controls (136,146.51) and (136.04,148.17) .. (137.74,149.8) .. controls (139.44,151.43) and (139.48,153.1) .. (137.85,154.8) .. controls (136.22,156.5) and (136.26,158.17) .. (137.96,159.8) .. controls (139.66,161.43) and (139.7,163.1) .. (138.07,164.8) .. controls (136.44,166.5) and (136.48,168.17) .. (138.18,169.8) .. controls (139.88,171.43) and (139.92,173.1) .. (138.29,174.8) -- (138.33,175) -- (138.33,175) ;
\end{scope}

\draw[color=RoyalBlue,line cap=round] (209.18,106.22) -- (143,172.45) node[pos=0,left,rotate=180]{\scalebox{0.75}[0.75]{$2$}}; 
\draw[color=RoyalBlue,line cap=round] (209.18,188.11) -- (127.28,106.22)
node[pos=1,right,rotate=180]{\scalebox{0.75}[0.75]{$1'$}} node[pos=0,left,rotate=180]{\scalebox{0.75}[0.75]{$1$}}; 
\draw[line cap=round] (168.23, 147.165) -- (143,172.45) ;
\draw[line cap=round] (168.23, 147.165) -- (140.525, 119.46);
\draw[color=Maroon, line cap = round]  (143,172.45)--(127.28,188.11) node[pos=1,right,rotate=180,color=Maroon]{\scalebox{0.75}[0.75]{$\bar{2}'$}};

\begin{scope}[xshift=-145,yshift=5]
\draw [color=Orange  ,draw opacity=1, dashed ][very thick]    (347,97) -- (347,180) ;
\end{scope}
\end{tikzpicture}}
\end{equation}
We use analogous momentum labelings as in the previous subsections, so that $s=-(p_1+p_2)^2$, $s_1=-(p_{1'}+p_3)^2$ and $s_2=-p_{2'}^2$.
In equations, we write \eqref{eq:crossing5} such that
\be\label{Tri test 2}
[\Tri^\dag]_{\raisebox{\depth}{\scalebox{1}[-1]{$\curvearrowleft$}}s_1} \; \stackrel{?}{=}\; \Tri + \Cut_{s} \Tri \,\equiv\, \Exp_3\,.
\ee
Clearly, the diagram on the right-hand side of~\eqref{eq:crossing5} admits a $s$-channel cut after crossing. However, even though it may seem that this topology could have a potentially problematic anomalous threshold branch cut obstructing the  path of analytic continuation, we find that the continuation remains safe, even when particle $2$ is on-shell ($s_2 = m_2^2$). This is what we show now.

Analogously to the previous subsections, we begin by taking $s_2$ off-shell with $s_2<m_2^2$, and show that no issues arise as we approach $s_2 \to m_2^2$. Additionally, we assume that $s\geq(m_1+m_2)^2$ such that the $s$-channel cut is allowed, making the problem is non-trivial. The crossing path from~\eqref{eq:crossingpath0} and~\eqref{eq:crossingpath3} specifies that $s>0$ and $s_2>0$ remain fixed during the analytic continuation, while $s_1$ rotates from positive to negative in the lower half-plane. As before, we work with the Schwinger-parametrized form of this diagram, which was given in~\eqref{Tri Schwinger}, and repeated here for convenience,
\be\label{Tri Schwinger 2}
\Tri^\dag = \int_0^\infty \frac{\text{d}\alpha_1 \text{d}\alpha_3}{(\alpha_1 + 1 + \alpha_3)\left[A(\alpha_1)+ \alpha_3 B(\alpha_1)+i\eps\right]}\,,
\ee
with
\be
A \equiv (\alpha_1+1)(\alpha_1 m_1^2+m_2^2)- \alpha_1 s \quad \text{and} \quad B \equiv \alpha_1(m_1^2-s_1)+m_2^2-s_2\,.
\ee
Notice that the analytic continuation of this integral in the lower half-plane of $s_1$ is straightforward: the required $+i\varepsilon$ in the denominator is obtained directly by taking $s_1$ to be in the lower half-plane. So, after continuing from positive to negative values of $s_1$,~\eqref{Tri Schwinger 2} still computes $\Tri^\dag$ in the new kinematic channel.  Whether $s_2$ is on-shell or not, the crossing equation in~\eqref{eq:crossing5} is straightforward to verify by unitarity: this is because $\Tri$ and $\Tri^\dag$ for $s_1<0$ differ only by the cut in the $s$-channel, which is the only one allowed in these kinematics.

We can contrast this with the conclusion from Sec.~\ref{sec:exTridiagram} where we found that to match with the Feynman $i \varepsilon$, we had to take $s - i\varepsilon$, but the crossing equation instructed us to continue in the upper half-plane of $s$. Thus,~\eqref{Tri Schwinger 2} was not well-suited for analytic continuation in $s$, so we rewrote it using the discontinuity across the branch cut on the real $s$ channel, given by $C(s)$, before continuing in the upper half-plane.

As a last check, we look at a different triangle topology contributing to $\Exp_3$ from~\eqref{KMOC}, 
\begin{equation}
    \left[\scalebox{1}[1]{\adjustbox{valign=c,rotate=0}{\tikzset{every picture/.style={line width=1pt}}
\begin{tikzpicture}[x=0.75pt,y=0.75pt,yscale=-1,xscale=1]
\draw  [color=Maroon] (227.39,100.01) .. controls (227.99,97.73) and (229.43,96.89) .. (231.71,97.49) .. controls (233.99,98.09) and (235.43,97.25) .. (236.02,94.97) .. controls (236.62,92.69) and (238.06,91.85) .. (240.34,92.45) .. controls (242.62,93.05) and (244.06,92.21) .. (244.66,89.93) .. controls (245.26,87.65) and (246.7,86.81) .. (248.98,87.41) .. controls (251.26,88.01) and (252.7,87.17) .. (253.3,84.89) .. controls (253.89,82.61) and (255.33,81.77) .. (257.61,82.36) -- (257.67,82.33) -- (257.67,82.33) node[right,scale=0.75]{$3$} ;
\draw [line cap=round,color=RoyalBlue]    (181.25,99.93) -- (220.48,76.89) node[right,scale=0.75]{$1$} node[right,xshift=30,yshift=10,color=black]{$\dagger$};
\draw [line cap=round,color=Maroon]    (140.67,78.33) -- (181.25,99.93) node[pos=0,left,scale=0.75]{$2$};
\draw [line cap=round,color=RoyalBlue]    (190.67,127.33) -- (227.39,100.01) node[pos=0,left,scale=0.75]{$1'$};
\draw    (145.42,99.9) .. controls (145.42,99.9) and (146.95,99.75) .. (147.78,101.95) .. controls (148.9,104.26) and (150.51,104.95) .. (152.6,104.02) .. controls (154.62,103.03) and (156.15,103.63) .. (157.19,105.81) .. controls (158.26,107.96) and (159.86,108.52) .. (161.99,107.47) .. controls (164.01,106.34) and (165.55,106.78) .. (166.62,108.8) .. controls (168.03,110.85) and (169.66,111.22) .. (171.51,109.89) .. controls (173.48,108.53) and (175.08,108.77) .. (176.31,110.61) .. controls (177.92,112.42) and (179.64,112.54) .. (181.48,110.96) .. controls (183.19,109.3) and (184.81,109.28) .. (186.34,110.9) .. controls (188.26,112.43) and (189.92,112.29) .. (191.32,110.46) .. controls (192.89,108.55) and (194.61,108.28) .. (196.48,109.65) .. controls (198.48,110.95) and (200.14,110.59) .. (201.47,108.58) .. controls (202.49,106.61) and (204.09,106.18) .. (206.28,107.31) .. controls (208.27,108.46) and (209.81,108) .. (210.88,105.93) .. controls (211.95,103.83) and (213.56,103.29) .. (215.72,104.32) .. controls (217.97,105.3) and (219.5,104.76) .. (220.32,102.69) .. controls (221.5,100.47) and (223.11,99.87) .. (225.15,100.88) -- (227.39,100.01) ;
\draw [line cap=round]    (135.11,99.85) -- (181.25,99.93) ;
\draw [line cap=round]    (181.25,99.93) -- (227.39,100.01) ;
\draw [line cap=round,color=RoyalBlue]     (135.11,99.85) -- (146.17,99.85) node[pos=0,left,scale=0.75]{$2'$};
\end{tikzpicture}}}\right]_{\substack{\raisebox{\depth}{\scalebox{1}[1]{$\curvearrowright$}}s_{\phantom{1}}\\ {\scalebox{1}[1]{$\curvearrowright$}}s_1}} \stackrel{?}{=} \adjustbox{valign=c,rotate=0}{\tikzset{every picture/.style={line width=1pt}} 
\begin{tikzpicture}[x=0.75pt,y=0.75pt,yscale=-1,xscale=1]
\draw  [color=Maroon]  (406.39,180.01) .. controls (406.98,182.3) and (406.13,183.73) .. (403.84,184.32) .. controls (401.56,184.91) and (400.71,186.34) .. (401.3,188.62) .. controls (401.89,190.91) and (401.04,192.34) .. (398.75,192.92) .. controls (396.46,193.5) and (395.61,194.93) .. (396.2,197.22) .. controls (396.79,199.5) and (395.94,200.94) .. (393.66,201.53) .. controls (391.37,202.11) and (390.52,203.54) .. (391.11,205.83) .. controls (391.7,208.12) and (390.85,209.55) .. (388.56,210.13) -- (386.67,213.33) -- (386.67,213.33) node[left,scale=0.75]{$\bar{3}$} ;
\draw [line cap=round,line cap=round,color=RoyalBlue]    (360.25,179.93) -- (399.48,156.89) node[right,scale=0.75]{$1$};
\draw [line cap=round,line cap=round,color=Maroon]    (389.49,145.13) -- (360.25,179.93) node[pos=0,right,scale=0.75]{$\bar{2}$};
\draw [line cap=round,color=RoyalBlue]    (369.67,207.33) -- (406.39,180.01) node[pos=0,left,scale=0.75]{$1'$};
\draw    (324.42,180.05) .. controls (324.42,180.05) and (325.95,179.75) .. (326.78,181.95) .. controls (327.9,184.26) and (329.51,184.95) .. (331.6,184.02) .. controls (333.62,183.03) and (335.15,183.63) .. (336.19,185.81) .. controls (337.26,187.96) and (338.86,188.52) .. (340.99,187.47) .. controls (343.01,186.34) and (344.55,186.78) .. (345.62,188.8) .. controls (347.03,190.85) and (348.66,191.22) .. (350.51,189.89) .. controls (352.48,188.53) and (354.08,188.77) .. (355.31,190.61) .. controls (356.92,192.42) and (358.64,192.54) .. (360.48,190.96) .. controls (362.19,189.3) and (363.81,189.28) .. (365.34,190.9) .. controls (367.26,192.43) and (368.92,192.29) .. (370.32,190.46) .. controls (371.89,188.55) and (373.61,188.28) .. (375.48,189.65) .. controls (377.48,190.95) and (379.14,190.59) .. (380.47,188.58) .. controls (381.49,186.61) and (383.09,186.18) .. (385.28,187.31) .. controls (387.27,188.46) and (388.81,188) .. (389.88,185.93) .. controls (390.95,183.83) and (392.56,183.29) .. (394.72,184.32) .. controls (396.97,185.3) and (398.5,184.76) .. (399.32,182.69) .. controls (400.5,180.47) and (402.11,179.87) .. (404.15,180.88) -- (406.39,180.01) ;
\draw [line cap=round]    (314.11,179.85) -- (360.25,179.93) ;
\draw [line cap=round]    (360.25,179.93) -- (406.39,180.01) ;
\draw [line cap=round,color=RoyalBlue]    (314.11,179.85) -- (325.17,179.85) node[pos=0,left,scale=0.75]{$2'$};
\end{tikzpicture}}
    \label{eq:crossing6}
\end{equation}
Here, we have used the labelings $s=-(p_1+p_2)^2$, $s_1=-(p_{1'}+p_3)^2$ and $s_2=-p_{2'}^2$. The crossing path given in~\eqref{eq:crossingpath0} and~\eqref{eq:crossingpath3} now instructs us to start with $s<0$ and $s_1<0$, followed by rotating both $s$ and $s_1$ in the upper half-plane. We have also used that for this topology, the crossing equation does not predict any cuts on the right-hand side of~\eqref{eq:crossing6} (stability forbids a cut in the $s_2$-channel). Thus, we simply want to check that
\be\label{Tri test 3}
[\Tri^\dag]_{\substack{\raisebox{\depth}{\scalebox{1}[1]{$\curvearrowright$}}s_{\phantom{1}}\\ {\scalebox{1}[1]{$\curvearrowright$}}s_1}} \; \stackrel{?}{=}\; \Tri \equiv\, \Exp_3\,.
\ee
This equation is easy to check using the Schwinger-parametrized form from~\eqref{Tri Schwinger 2}. The correct $i \varepsilon$ prescription for $\Tri^\dag$ is obtained by taking either $s \to s + i\varepsilon$ or  $s_1 \to s_1 + i\varepsilon$. After the analytic continuation, both $s$ and $s_1$ are taken to have a positive imaginary part, which is precisely the prescription that agrees with the $i\varepsilon$ for $\Tri$.

The examples above support the argument provided below \eqref{eq:dualcone 1}, stating that there is no analytic obstruction to obtain the quantity $\Exp_3$ by continuing a standard on-shell time-ordered amplitude when particle 3 is a massless particle.
\section{Crossing between observables in string theory}\label{sec:strings}

In this section, we provide another stress-test of the crossing conjecture \eqref{eq:crossing23}. We will consider scattering in open string theory at tree level. This analysis will also allow us to compute the inclusive measurement $\Exp_k$ in string theory and illustrate what happens to the
string geometry for this processes.

\subsection{\label{sec:spirals}Spirals and anti-spirals}

As a warm-up, let us consider $4$-point scattering, which will illustrate the general strategy for analytic continuation. For example, we can take the Veneziano amplitude
\be\label{eq:Veneziano}
\M_{34 \ot 12}^{\mathrm{string}} = \int_{0}^{1} z^{- s-1} (1-z)^{- t-1}\, \d z\, ,
\ee
where we work in the units $\alpha' =1$.
Different choices of external states would only multiply the integrand by a Laurent polynomial in $z$, which is not going to affect the analysis. We can therefore stick with \eqref{eq:Veneziano} without loss of generality. The result is always proportional to the Euler beta function $B(- s, - t)$.

In the $s$-channel, $s>0$ and $t<0$, which means that the integral \eqref{eq:Veneziano} is not convergent as $z \to 0$. At tree-level, string amplitudes are meromorphic functions of kinematic invariants, just like field-theory amplitudes, so one can in principle evaluate the integral in the Euclidean region $s,t < 0$ and then analytically continue back to the $s$-channel. However, the purpose of the discussion is to understand more precisely what happens to the contour as we work directly in the physical kinematics and then analytically continue in $s$ and $t$ between different channels. We refer readers to \cite{Witten:2013pra,Eberhardt:2022zay,Eberhardt:2023xck} for more detailed discussions of Lorentzian integration contour and unitarity cuts of string amplitudes.

Since the divergence comes from $z \to 0$, it is going to be useful to parameterize the integral according to $z = \e^{-\tau}$. Here, $\tau$ can be thought of as the Schwinger proper time along the neck of the worldsheet responsible for the $s$-channel Feynman diagram degeneration. The divergence comes from $\tau \to \infty$, so we can expand the integrand as
\be\label{eq:Veneziano-expanded}
\M_{34 \ot 12}^{\mathrm{string}} = \int_{0}^{\infty} \e^{ s \tau} \left[ 1 + (1 + t) \e^{-\tau} + \frac{1}{2}(1 + t )(2 + t ) \e^{-2\tau}  + \mathcal{O}(\e^{-3\tau}) \right] \d \tau\, .
\ee
Each term is a Schwinger-parametric version of a propagator.
The leading term formally integrates to a pole in $s=0$, the subleading to a pole in $s = 1$, and so on. The integration is only formal because it actually does not converge as $\tau \to \infty$ where the integrand has an essential singularity.

The above problem can be traced back to the fact that the Riemann surface (in this case a disk with four punctures) is an Euclidean manifold, while the target space is Lorentzian. The appropriate thing to do would therefore be to integrate over Lorentzian worldsheets close to the dangerous parts of the moduli space. Since we already identified the relevant Euclidean Schwinger parameter in \eqref{eq:Veneziano-expanded}, we simply switch to its Lorentzian version where it is needed: 
\be\label{eq:Wick-rotation}
\int_{0}^{\infty} \to \int_{0}^{\tau_\ast} + \int_{\tau_\ast}^{\tau_\ast + i\infty}\, .
\ee
Here, $\tau_\ast \gg 1$ is an arbitrary cutoff after which the Wick rotation is done. To understand why running the contour in the upper half-plane was the correct choice, it is enough to look at the behavior of the integrand. Approaching the $s$-channel from the upper half-plane $s + i\varepsilon$, it goes as $\e^{(s + i\varepsilon)(\tau_\ast + i\infty)}$, so at infinity it gives the exponential suppression $\sim \e^{-\varepsilon\infty}$.

\begin{figure}
	\centering
	\begin{tikzpicture}
	\begin{scope}
		\draw[->, gray] (-0.5,0) -- (5,0);
		\draw[->, gray] (0,-2) -- (0,2);
		\draw[thick] (4.5,2) -- (4.5,1.5) -- (5,1.5);
		\node at (4.75,1.75) {$\tau$};
		\draw[very thick, Maroon] (0,0.05) -- (3,0.05);
		\draw[->, very thick, Maroon] (0,0.05) -- (2.0,0.05);
		\draw[very thick, Maroon] (3,0) -- (3,2);
		\draw[->, very thick, Maroon] (3,0) -- (3,1);
        \draw[very thick, RoyalBlue] (0,-0.05) -- (3,-0.05);
		\draw[->, very thick, RoyalBlue] (0,-0.05) -- (1.5,-0.05);
		\draw[very thick, RoyalBlue] (3,0) -- (3,-2);
		\draw[->, very thick, RoyalBlue] (3,0) -- (3,-1);
        \draw[->, very thick, orange] (3.2,-2) -- (3.2,-0.9);
        \draw[->, very thick, orange] (3.2,-2) -- (3.2,1);
        \draw[very thick, orange] (3.2,-2) -- (3.2,2);
		\fill (3,0) circle (0.09) node[above left] {$\tau_\ast$};
		\node at (1.5,0.8) {\footnotesize\textcolor{Maroon}{spiral}};
        \node at (1.5,-0.8) {\footnotesize\textcolor{RoyalBlue}{anti-spiral}};
        \node at (4,0.8) {\footnotesize\textcolor{orange}{cut}};
	\end{scope}
	\begin{scope}[shift={(8,0)}]
		\draw[->, gray] (-1.5,0) -- (5,0);
		\draw[->, gray] (0,-2) -- (0,2);
		\draw[thick] (4.5,2) -- (4.5,1.5) -- (5,1.5);
		\node at (4.75,1.75) {$z$};
        \draw[dotted,very thick, Maroon, variable=\t, domain=2.8*360:2.88*360, samples=200, smooth] plot ({-\t}:{1-.0002*\t});
		\draw[->,very thick, Maroon, variable=\t, domain=0:2.8*360, samples=200, smooth] plot ({-\t}:{1-.0002*\t});
        \draw[dotted,very thick, RoyalBlue, variable=\t, domain=2.8*360:2.85*360, samples=200, smooth] plot ({\t}:{1+.0002*\t});
        \draw[->,very thick, RoyalBlue, variable=\t, domain=0:2.8*360, samples=200, smooth] plot ({\t}:{1+.0002*\t});
		\draw[very thick, Maroon] (1,-0.05) -- (4,-0.05);
        \draw[very thick, RoyalBlue] (1,0.03) -- (4,0.03);
        \draw[dotted,very thick, RoyalBlue] (4,0.03) -- (4.5,0.03);
        \draw[dotted,very thick, Maroon] (4,-0.03) -- (4.5,-0.03);
		\draw[->, very thick, RoyalBlue] (4,0.03) -- (2.5,0.03);
        \draw[->, very thick, Maroon] (2.5,-0.05) -- (2.0,-0.05);
		\fill (1,0) circle (0.09) node[below left] {$e^{-\tau_\ast}\;$};
	\end{scope}
	\end{tikzpicture}
\caption{\label{fig:spirals}Building blocks for contours of integration in string theory. \textbf{Left:} In the $\tau$-plane we plotted the spiral (red), anti-spiral (blue), and cut (orange) contours. \textbf{Right:} The shape of the contours in the $z$-plane after using $z = \e^{-\tau}$.}
\end{figure}
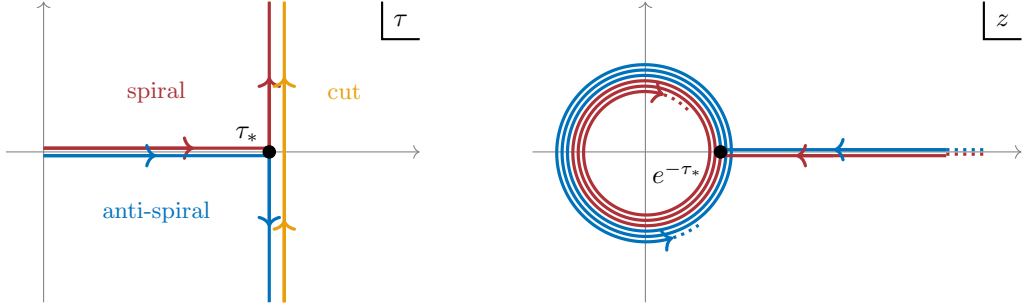

The situation flips, quite literally, if we chose to approach the $s$-channel from the lower half-plane with $s - i\varepsilon$, where the final piece of the contour in \eqref{eq:Wick-rotation} should go from $\tau_\ast$ to $\tau_\ast - i\infty$. Finally, this also allows us to define the discontinuity across the $s$-channel, which is the difference between the two choices of $\pm i\varepsilon$. In that case, the contour simply goes from $\tau_\ast - i\infty$ to $\tau_\ast + i\infty$. The discontinuity computes the unitarity cut in the $s$-channel.\footnote{In contrast with Cutkosky cuts, the part of the amplitude to the right of the cut is not complex-conjugated. Positivity of the energy of the particle going across the cut is fixed by the external kinematics.} Back in the original $z$ variable, the three types of contours take the form of a spiral, an anti-spiral, and their difference, as illustrated in Fig.~\ref{fig:spirals}. The identity
\be\label{eq:spiral-identity}
(\mathrm{spiral}) - (\text{anti-spiral}) = (\mathrm{cut})\,,
\ee
is a direct analogue of the distributional identity
\be
\frac{1}{s + i\varepsilon} - \frac{1}{s - i\varepsilon} = -2\pi i \delta(s)\,,
\ee
at the level of worldsheets. Note that we could have also added spirals on the other endpoint $z \to 1$. However, in this case they would not have any effect because $t<0$ and the two contours can be deformed into each other. Equivalently, there is no $t$-cut allowed in the $s$-channel.

We are now ready to discuss the analytic continuation in the Mandelstam invariants. On one hand, the $u$-channel ($s,t<0$) is boring because neither $s$- nor $t$-channel poles can happen there. On the other hand, the $t$-channel ($s<0$, $t>0$) has $t$-poles but not $s$. Therefore, let us perform an analytic continuation from $s$- to $t$-channel, which is equivalent to crossing particles $1 \leftrightarrow 3$. It corresponds to the simultaneous rotation of $s$ and $t$ as follows:
\begin{equation}
\begin{gathered}
\begin{tikzpicture}
  \draw[->,thick] (-1.5, 0) -- (1.5, 0);
  \draw[->,thick,white] (0, -1.6) -- (0, 1.6);
  \draw[->,thick] (0, -1.25) -- (0, 1.25);
  \node[] at (1.1,1.05) {$s$};
  \draw[] (1.35,0.85) -- (0.8,0.85) -- (0.8,1.25);
  \draw[Maroon,fill=Maroon,thick] (0.9,0.1) circle (0.05);
  \draw[->,Maroon,thick] (0.9,0.1) arc (0.1:180:1);
  \draw[black,fill=black,thick] (0,0) circle (0.05);
  \draw[black,fill=black,thick] (0.2,0) circle (0.05);
  \draw[black,fill=black,thick] (0.4,0) circle (0.05);
  \draw[black,fill=black,thick] (0.6,0) circle (0.05);
  \draw[black,fill=black,thick] (0.8,0) circle (0.05);
  \draw[black,fill=black,thick] (1.0,0) circle (0.05);
  \draw[black,fill=black,thick] (1.2,0) circle (0.05);
\end{tikzpicture}
\hspace{2.5cm}
\begin{tikzpicture}
  \draw[->,thick] (-1.5, 0) -- (1.5, 0);
  \draw[->,thick,white] (0, -1.6) -- (0, 1.6);
  \draw[->,thick] (0, -1.25) -- (0, 1.25);
  \node[] at (1.1,1.05) {$t$};
  \draw[] (1.35,0.85) -- (0.8,0.85) -- (0.8,1.25);
  \draw[Maroon,fill=Maroon,thick] (-1,0) circle (0.05);
  \draw[->,Maroon,thick] (-1,0) arc (0:175:-1);
  \draw[black,fill=black,thick] (0,0) circle (0.05);
  \draw[black,fill=black,thick] (0.2,0) circle (0.05);
  \draw[black,fill=black,thick] (0.4,0) circle (0.05);
  \draw[black,fill=black,thick] (0.6,0) circle (0.05);
  \draw[black,fill=black,thick] (0.8,0) circle (0.05);
  \draw[black,fill=black,thick] (1.0,0) circle (0.05);
  \draw[black,fill=black,thick] (1.2,0) circle (0.05);
\end{tikzpicture}
\end{gathered}
\end{equation}
The dots denote string resonances at $s, t \in \mathbb{Z}_{\geq 0}$. The only thing that matters is that analytic continuation is performed in the upper half-plane in $s$ and lower half-plane in $t$. This selects a spiral in the $s$-channel degeneration ($z \to 0$) and anti-spiral in the $t$-channel degeneration ($z \to 1$). In other words, we land on the $t$-channel amplitude complex-conjugated. This fact is consistent with the crossing equation, which in this case says
\be
[\M_{34 \ot 12}]_{\substack{\rotatedown t\\ \rotateup s}} = \M_{14 \ot 23}^\dagger\, .
\ee
One can repeat similar arguments for other color orderings without any trouble.

Similarly, one can treat the closed-string case, where divergent parts of the integral look like $I = \int_{|z| < \epsilon} |z|^{-2s-2} \d^2 z$ instead of $\int_0^{\epsilon} z^{-s-1} \d z$. Changing variables to $(z, \bar{z}) = (r \e^{i\theta}, r \e^{-i\theta})$, the angular coordinate can be integrated out giving $I = 4\pi \int_{0}^{\epsilon} r^{-2s-1} \d r$. At this stage, we can study the Lorentzian contour in $r$ using the same technology as described above for open strings. In other words, when a worldsheet develops a long neck, only its length (not the angular direction) needs to be Wick rotated.

\subsection{Example of inclusive measurement}

Let us illustrate how crossing looks like at higher multiplicity. One of the simplest case which contains a new measurement is the is the $5$-point amplitude $\M_{245 \ot 13}$. We fix the color ordering to be $(12345)$ in which case the amplitude only depends on the Mandelstam invariants $(s_{12}, s_{23}, s_{34}, s_{45}, s_{51})$. After crossing particles $2 \leftrightarrow 3$, only the $s_{12}$ and $s_{34}$ are rotated in the lower half-plane and we should arrive at
\begin{subequations}
\begin{align}\label{eq:crossing-5pt-string-theory}
[\M_{245 \ot 13}]_{\substack{\rotatedown s_{12}\\ \rotatedown s_{34}}} &= \Exp_3\\
&= \M_{345 \ot 12}^\dagger + \Cut_{s_{45}} \M_{345 \ot 12}\, .
\end{align}
\end{subequations}
The goal is to reproduce this identity from the moduli space geometry.

Fixing the positions of three vertex operators to $(z_1, z_4, z_5) = (0,1,\infty)$, we have
\be
\M_{245 \ot 13}^{\mathrm{string}} = -\int_{0 < z_2 < z_3 < 1} \hspace{-3em} \d z_2 \, \d z_3\; z_2^{-s_{12}-1} z_3^{-s_{13}} (1-z_2)^{-s_{24}} (1-z_3)^{-s_{34}-1} (z_3 - z_2)^{-s_{23}-1}\, .
\ee
The result can be expressed in terms of a ${}_3 F_2$ hypergeometric function \cite{Bialas:1969jz}.
Originally, the only positive planar Mandelstam invariant is $s_{45}$. In the above parametrization of the moduli space, the corresponding degeneration comes from the corner where both $z_2$ and $z_3$ approach $z_1 = 0$ both at once. To see this more clearly, it is necessary to blow-up the space. In practice, it amounts to the change of variables $(z_2, z_3) = (xy, x)$, where $x,y \in [0,1]$. The result is
\be
\M_{245 \ot 13}^{\mathrm{string}} = -\int_{0 < x,y < 1} \hspace{-2em} \d x \, \d y\; x^{-s_{45}-1} y^{-s_{12} - 1} (1-x)^{-s_{34}-1} (1-y)^{-s_{23}-1} (1-xy)^{-s_{24}}\, .
\ee
The $s_{45}$ poles come from the degeneration $x \to 0$. The integration contour in the neighborhood of this boundary should therefore look like a product of a spiral in the $x$-plane times an interval $y \in [0,1]$, with the remainder of the contour unchanged.

\begin{figure}
\centering
\begin{tikzpicture}
    \fill[black!10!white] (0,0) rectangle (1,1);
    \draw[snake=coil,segment length=3pt] [color=Maroon][line width=1] (0,0) -- ++ (0,1);
    \draw[snake=coil,segment length=3pt] [color=RoyalBlue][line width=1] (0,0) -- ++ (1,0);
    \draw[snake=coil,segment length=3pt] [color=RoyalBlue][line width=1] (1,0) -- ++ (0,1);
    \draw[->,thick] (-0.5, 0) -- (2, 0);
    \draw[->,thick] (0, -0.5) -- (0, 2);
    \draw[thick] (-0.5, 1) -- (2, 1);
    \draw[thick] (1, -0.5) -- (1, 2);
    \draw[thick] plot[variable=\x,domain=0.5:2,samples=73,smooth] (\x,{1/\x});
    \node[] at (2.3,0) {$x$};
    \node[] at (0,2.3) {$y$};
    \node[] at (-0.25,0.2) {\footnotesize$0$};
    \node[] at (-0.25,1.2) {\footnotesize$1$};
    \node[] at (0.2,-0.25) {\footnotesize$0$};
    \node[] at (1.2,-0.25) {\footnotesize$1$};

    \node[] at (3,0.7) {$=$};

    \begin{scope}[xshift=120pt]
    \fill[black!10!white] (0,0) rectangle (1,1);
    \draw[snake=coil,segment length=3pt] [color=RoyalBlue][line width=1] (0,0) -- ++ (0,1);
    \draw[snake=coil,segment length=3pt] [color=RoyalBlue][line width=1] (0,0) -- ++ (1,0);
    \draw[snake=coil,segment length=3pt] [color=RoyalBlue][line width=1] (1,0) -- ++ (0,1);
    \draw[->,thick] (-0.5, 0) -- (2, 0);
    \draw[->,thick] (0, -0.5) -- (0, 2);
    \draw[thick] (-0.5, 1) -- (2, 1);
    \draw[thick] (1, -0.5) -- (1, 2);
    \draw[thick] plot[variable=\x,domain=0.5:2,samples=73,smooth] (\x,{1/\x});
    \node[] at (2.3,0) {$x$};
    \node[] at (0,2.3) {$y$};
    \node[] at (-0.25,0.2) {\footnotesize$0$};
    \node[] at (-0.25,1.2) {\footnotesize$1$};
    \node[] at (0.2,-0.25) {\footnotesize$0$};
    \node[] at (1.2,-0.25) {\footnotesize$1$};
    \end{scope}

    \node[] at (7.15,0.7) {$+$};

    \begin{scope}[xshift=240pt]
    \fill[black!10!white] (0,0) rectangle (1,1);
    \draw[snake=coil,segment length=3pt] [color=orange][line width=1] (0,0) -- ++ (0,1);
    \draw[snake=coil,segment length=3pt] [color=RoyalBlue][line width=1] (0,0) -- ++ (1,0);
    \draw[snake=coil,segment length=3pt] [color=RoyalBlue][line width=1] (1,0) -- ++ (0,1);
    \draw[->,thick] (-0.5, 0) -- (2, 0);
    \draw[->,thick] (0, -0.5) -- (0, 2);
    \draw[thick] (-0.5, 1) -- (2, 1);
    \draw[thick] (1, -0.5) -- (1, 2);
    \draw[thick] plot[variable=\x,domain=0.5:2,samples=73,smooth] (\x,{1/\x});
    \node[] at (2.3,0) {$x$};
    \node[] at (0,2.3) {$y$};
    \node[] at (-0.25,0.2) {\footnotesize$0$};
    \node[] at (-0.25,1.2) {\footnotesize$1$};
    \node[] at (0.2,-0.25) {\footnotesize$0$};
    \node[] at (1.2,-0.25) {\footnotesize$1$};
    \end{scope}
    
\end{tikzpicture}
\caption{\label{fig:moduli-space}Schematic illustration of the integration contours for $5$-point string observables, encoding the crossing identity \eqref{eq:crossing-5pt-string-theory}. The original integration domain after crossing (gray) is modified by attaching respectively a spiral (red), anti-spiral (blue), or cut (orange) on the $x=0$ boundary; the other Lorentzian boundaries are both anti-spirals.
}
\end{figure}
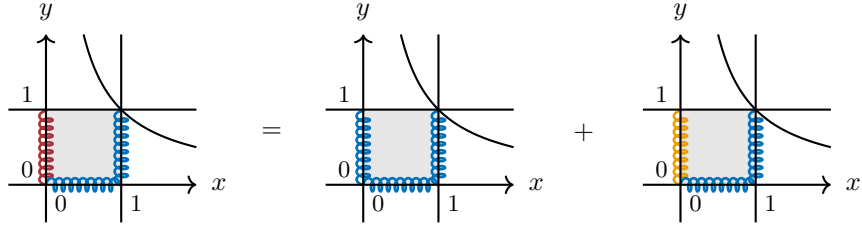

Upon crossing $2 \leftrightarrow 3$, the two Mandelstam invariants we have to analytically continue are $s_{12}$ and $s_{34}$, both in the lower half-plane:
\begin{equation}
\begin{gathered}
\begin{tikzpicture}
  \draw[->,thick] (-1.5, 0) -- (1.5, 0);
  \draw[->,thick,white] (0, -1.6) -- (0, 1.6);
  \draw[->,thick] (0, -1.25) -- (0, 1.25);
  \node[] at (1.1,1.05) {$s_{12}$};
  \draw[] (1.35,0.85) -- (0.8,0.85) -- (0.8,1.25);
  \draw[Maroon,fill=Maroon,thick] (-1,0) circle (0.05);
  \draw[->,Maroon,thick] (-1,0) arc (0:175:-1);
  \draw[black,fill=black,thick] (0,0) circle (0.05);
  \draw[black,fill=black,thick] (0.2,0) circle (0.05);
  \draw[black,fill=black,thick] (0.4,0) circle (0.05);
  \draw[black,fill=black,thick] (0.6,0) circle (0.05);
  \draw[black,fill=black,thick] (0.8,0) circle (0.05);
  \draw[black,fill=black,thick] (1.0,0) circle (0.05);
  \draw[black,fill=black,thick] (1.2,0) circle (0.05);
\end{tikzpicture}
\hspace{2.5cm}
\begin{tikzpicture}
  \draw[->,thick] (-1.5, 0) -- (1.5, 0);
  \draw[->,thick,white] (0, -1.6) -- (0, 1.6);
  \draw[->,thick] (0, -1.25) -- (0, 1.25);
  \node[] at (1.1,1.05) {$s_{34}$};
  \draw[] (1.35,0.85) -- (0.8,0.85) -- (0.8,1.25);
  \draw[Maroon,fill=Maroon,thick] (-1,0) circle (0.05);
  \draw[->,Maroon,thick] (-1,0) arc (0:175:-1);
  \draw[black,fill=black,thick] (0,0) circle (0.05);
  \draw[black,fill=black,thick] (0.2,0) circle (0.05);
  \draw[black,fill=black,thick] (0.4,0) circle (0.05);
  \draw[black,fill=black,thick] (0.6,0) circle (0.05);
  \draw[black,fill=black,thick] (0.8,0) circle (0.05);
  \draw[black,fill=black,thick] (1.0,0) circle (0.05);
  \draw[black,fill=black,thick] (1.2,0) circle (0.05);
\end{tikzpicture}
\end{gathered}
\end{equation}
The resulting contour therefore has to have an anti-spiral around the boundary at $y \to 0$ and another at $x\to 1$, responsible for poles in $s_{12}$ and $s_{34}$ respectively (on top of the spiral at $x\to 0$ that stays there). The resulting contour is schematically depicted in Fig.~\ref{fig:moduli-space}.

After analytic continuation, the contour in Fig.~\ref{fig:moduli-space} is the precise analogue of the crossing formula \eqref{eq:crossing-5pt-string-theory}. To see this, we expressed the spiral at $x=0$ as an anti-spiral plus a cut according to the identity \eqref{eq:spiral-identity}. Recall that close to boundaries, the worldsheets limit to worldlines resembling Feynman diagrams. For example, close to $(x,y) = (0,0)$, the have poles in $s_{12}$ and $s_{45}$. The identity then implies
\be
\frac{1}{(s_{12} {-} n_{12} {-} i\varepsilon)(s_{45} {-} n_{45} {+} i\varepsilon)} = \frac{1}{s_{12} {-} n_{12} {-} i\varepsilon} \left( \frac{1}{s_{45} {-} n_{45} {-} i\varepsilon} - 2\pi i \delta(s_{45} {-} n_{45}) \right),
\ee
where $n_{12}, n_{45} \in \mathbb{Z}_{\geq 0}$ are masses of the exchanged states.

Interestingly, taking the integration contour from the last panel in Fig.~\ref{fig:moduli-space} allows us to compute the inclusive measurement 
$\Exp_3$
in string theory directly. According to \eqref{eq:crossing-5pt-string-theory}, there are two contributions: conjugated amplitude and the cut. The latter is given by
\begin{equation}
    \begin{split}
        \Cut_{s_{45}} \M_{345 \ot 12}^{\mathrm{string}} &= -\oint_{\mathrm{cut}} \!\! \d x \int_{0}^{1} \d y\\&\times\; x^{-s_{45}-1} y^{-s_{12} - 1} (1-x)^{-s_{34}-1} (1-y)^{-s_{23}-1} (1-xy)^{-s_{24}}\, ,
    \end{split}
\end{equation}
which can be performed by expanding the integrand in $x$ and using the identity $\oint_{\mathrm{cut}} \d x\, x^{-s-1} = 2\pi i \delta(s)$ for every term. Explicitly, the first couple of contributions read
\begin{align}
\Cut_{s_{45}} \M_{345 \ot 12}^{\mathrm{string}} = - 2\pi i \bigg[ \delta(s_{45}) + \delta(s_{45} {-} 1) \left(1 + s_{34} + \frac{s_{12} s_{24}}{s_{12} + s_{23}} \right) + \ldots \bigg] \\
\times \frac{\Gamma(-s_{12}+i\varepsilon) \Gamma(- s_{23})}{\Gamma(- s_{12} {-} s_{23})}\, .\nonumber
\end{align}
Analogously, the full observable 
$\Exp_3$
can be computed in terms of ${}_3 F_2$ and gives
\begin{align}
\Exp_3= -& B(- s_{12}+i\varepsilon, - s_{23})
B(- s_{34}+i\varepsilon, - s_{45} - i\varepsilon)\\
&\times {}_3 F_2(- s_{12} + i\varepsilon,  s_{24}, -s_{45} - i\varepsilon; \, - s_{12}- s_{23}, - s_{34} - s_{45};\, 1)\, ,\nonumber
\end{align}
where $B$ is the Euler beta function.

Generalizations to other planar orderings, higher multiplicity, and closed strings are straightforward and essentially use the same combinatorics as for tree-level Feynman diagrams. Blow-ups of the boundaries responsible for each degeneration can be conveniently studied using dihedral coordinates \cite{Brown:2009qja}, leading to explicit formulae for inclusive measurements in string theory at tree level.

\subsection{\label{sec:worldsheet-geomtry}Spacetime geometry}

In order to get more intuition for the objects discussed above, let us visualize the spacetime geometry of the worldsheet that they correspond to. Recall that the classical solution of the worldsheet field $X^\mu$ is obtained by taking the $\alpha' \to \infty$ limit. For tree-level open-string scattering, the answer is 
\be\label{eq:X-mu}
X^\mu(z) = i \sum_{j=1}^{n} p_j^\mu \log |z - z_j^\ast|\, ,
\ee
for every $z$ in the upper half-plane, where $z_j^\ast$ are the positions of the vertex operators on the relevant $\alpha' \to \infty$ saddle point.
Note that this solution is purely imaginary. This fact is precisely the consequence of the worldsheet being Euclidean, which naively leads to strings propagating in the imaginary spacetime, interpreted as quantum tunnelling.

The solution \eqref{eq:X-mu} can be used to understand string geometries even if they are not evaluated on the saddle. We will use it to visualize the type of worldsheets close to the $(x,y) = (0,0)$ corner of the moduli space illustrated in Fig.~\ref{fig:moduli-space}, after a certain number of windings. As explained in previous sections, windings give rise to Lorentzian worldsheet evolution.

As the first step in analytic continuation, we promote \eqref{eq:X-mu} to a holomorphic function of $z$ by replacing $\log |z - z_j^\ast| \to \log (z - z_j^\ast)$. The branch of the logarithm does not matter because it drops out from $X^\mu$ by momentum conservation. We now focus on the real part of the solution, $\mathrm{Re}\, X^\mu(z)$. The Euclidean part corresponds to the interior of a polygonal Wilson loop constructed out of the momenta $p_j^\mu$. Let us first consider what happens if we simply treat asymptotic states as Lorentzian. This corresponds to taking a semi-circle $|z - z_j^\ast| \leq \varepsilon$ for every puncture $z_j^\ast$ and rotating it a large number of times $w$ clockwise. Locally, we can make the change of variables
\be
z = z_j^\ast + \varepsilon r \e^{i\theta + 2\pi i w (1-r)}\,,
\ee
with $r \in [0,1]$ and $\theta \in [0,\pi]$. The solution \eqref{eq:X-mu} looks like
\be
X^\mu(r,\theta) = i p_j^\mu \Big[ \log(\varepsilon r) +  i\theta + 2\pi i w (1{-}r) \Big] + \ldots
\ee
The real part therefore looks like $\mathrm{Re}\, X^\mu \sim -p_j^\mu [\theta + 2\pi w(1-r)]$ for small $\varepsilon$. In other words, the worldsheet turns asymptotically into a collection of worldlines aligned along the momenta of the external particles. The minus sign difference between $\mathrm{Re}\, X^\mu$ and $p_j^\mu$ arises because, e.g., the momentum vector of an incoming particle with $p_j^0 > 0$ points to the future, while the corresponding part of the string worldsheet stretches towards the past.

\begin{figure}
\centering
\begin{tikzpicture}
\begin{scope}
    \node[] at (0,0) {
    \includegraphics[valign=c,width=0.3\textwidth]{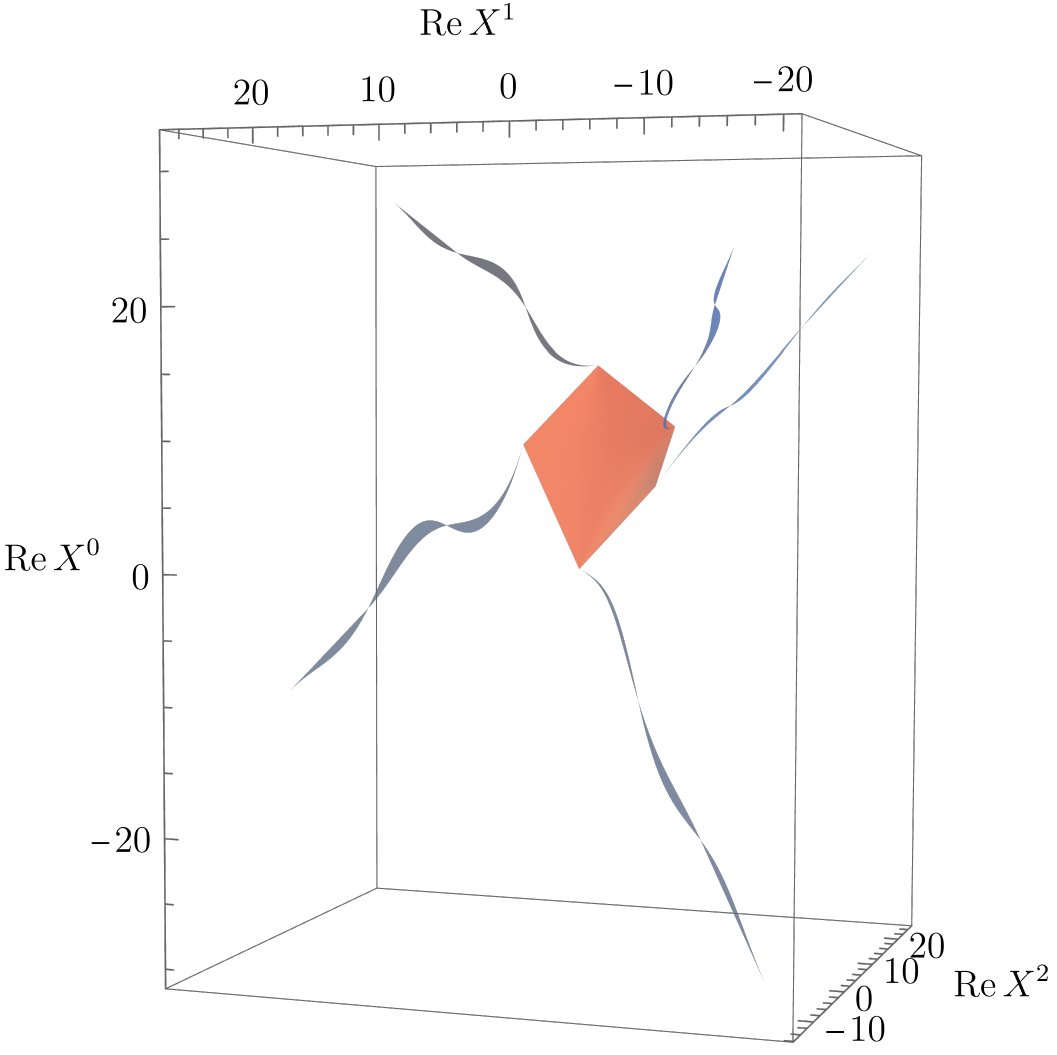}
    };
    \node[] at (-1.2,-0.8) {\tiny $1$};
    \node[] at (1,-2.1) {\tiny $2$};
    \node[] at (-0.7,1.5) {\tiny $3$};
    \node[] at (0.9,1.5) {\tiny $4$};
    \node[] at (1.6,1.2) {\tiny $5$};
\end{scope}
\begin{scope}[shift={(5,0)}]
    \node[] at (0,0) {
    \includegraphics[valign=c,width=0.3\textwidth]{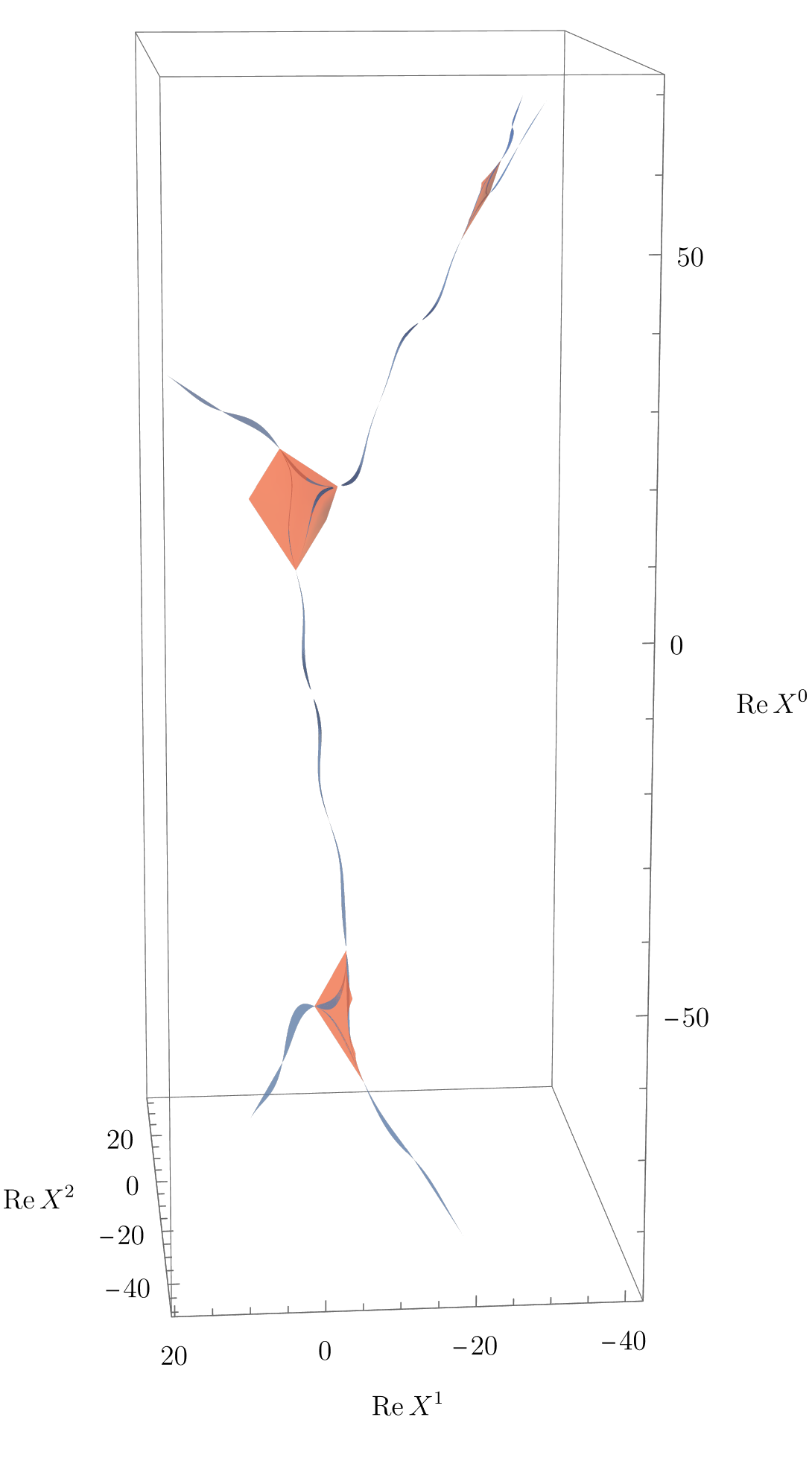} };
    \node[] at (-1.1,-2.5) {\tiny $1$};
    \node[] at (0.5,-3) {\tiny $2$};
    \node[] at (-1.5,2.1) {\tiny $3$};
    \node[] at (0.6,3.6) {\tiny $4$};
    \node[] at (0.9,3.9) {\tiny $5$};
\end{scope}
\begin{scope}[shift={(10,0)}]
    \node[] at (0,0) {
    \includegraphics[valign=c,width=0.3\textwidth]{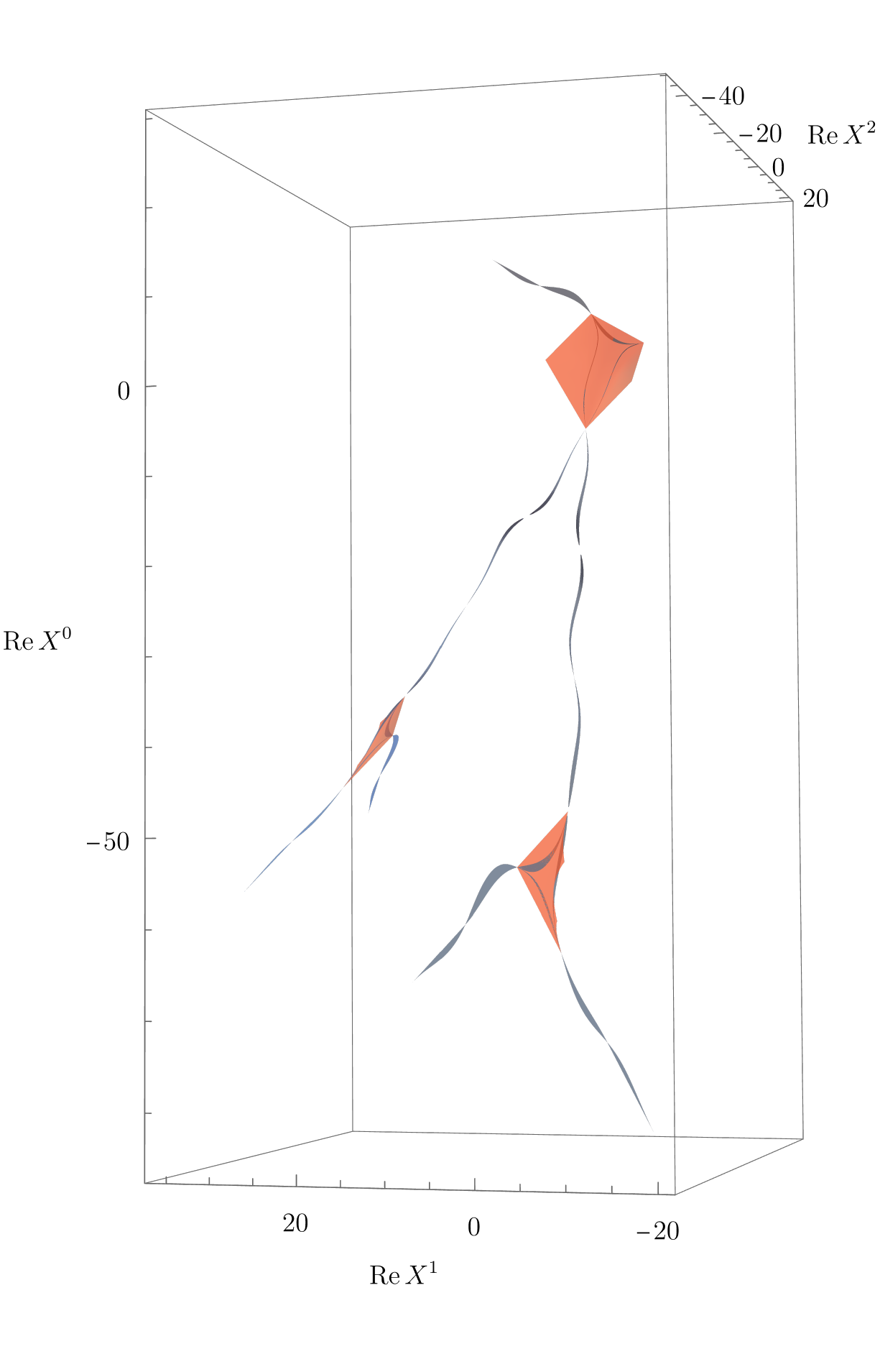}
    };
    \node[] at (-0.2,-1.7) {\tiny $1$};
    \node[] at (1.3,-2.5) {\tiny $2$};
    \node[] at (0,2.4) {\tiny $3$};
    \node[] at (-1.1,-1.2) {\tiny $4$};
    \node[] at (-0.3,-0.8) {\tiny $5$};
\end{scope}
\end{tikzpicture}
\caption{\label{fig:twisted-worldsheets}Worldsheet geometries before and after applying the crossing relation to $\M_{13 \to 245}$. The plots are 3-dimensional slices through the real Minkowski space with the time component going up. The blue ribbons are \textcolor{RoyalBlue}{Lorentzian} and the red polyhedra are \textcolor{Maroon}{Euclidean}. For illustration purposes, we used $w=2$ windings for all Lorentzian segments. \textbf{Left:} $\M_{345 \ot 12}$ at a generic point in the moduli space. \textbf{Middle:} $\M_{345 \ot 12}$ close to the $s_{12}$ and $s_{45}$ resonances. \textbf{Right:} 
$\Exp_3$
close to the same resonances.}
\end{figure}
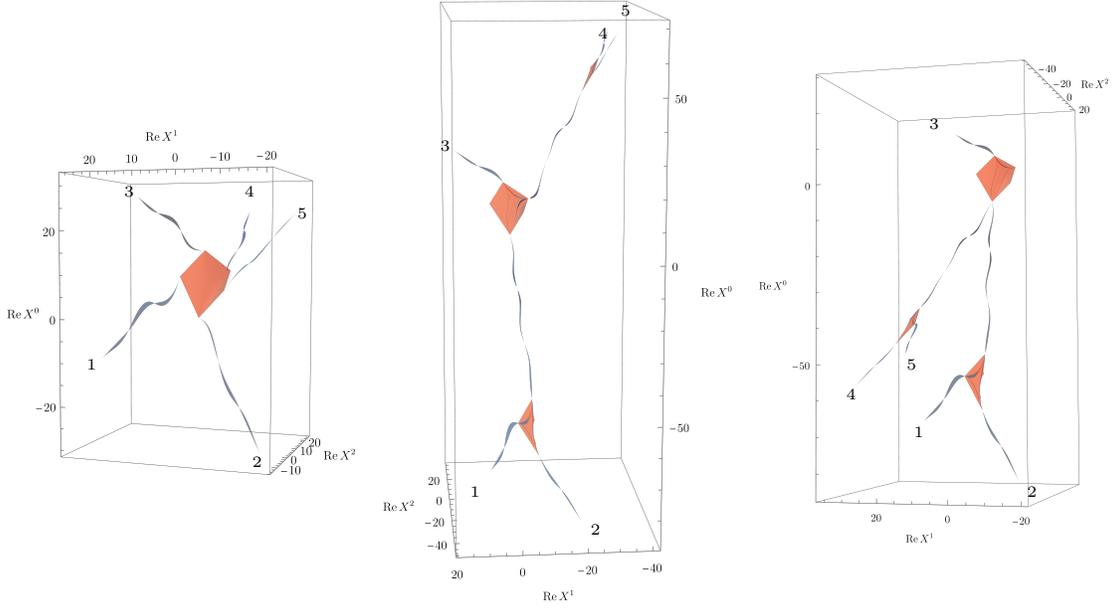

Let us now get a better approximation to the worldline picture by expanding to subleading orders in $\varepsilon$. Taking into account the remaining vertex operators, we have
\be
\mathrm{Re}\, X^{\mu}(r,\theta) = -p_j^\mu  \big[\theta + 2\pi w(1-r)\big] + \mathrm{Re} \left[ i \sum_{k \neq j} p_k^\mu \log \left(z_j^\ast - z_k^\ast + \varepsilon r \e^{i\theta + 2\pi i w (1-r)} \right) \right]\, .
\ee
Recalling that all $z_k^\ast$ are real, the leading term in the sum does not contribute since it does not have any real part. The most leading correction is therefore
\be
\mathrm{Re}\, X^{\mu}(r,\theta) = - p_j^\mu  \big[\theta + 2\pi w(1-r) \big] - \varepsilon\sin (\theta - 2\pi w r) \sum_{k \neq j}  \frac{p_k^\mu}{z_j^\ast - z_k^\ast}  + \mathcal{O}(\varepsilon^2)\, .
\ee
It means that the worldsheet oscillates $w$ times around the worldline. The size and direction of this oscillation is set by those $p_k^\mu$ whose punctures $z_k^\ast$ are the closest to $z_j^\ast$.

The corresponding worldsheet geometry is illustrated in Fig.~\ref{fig:twisted-worldsheets} (left), where five ribbon-like strands attach to the inside of a pentagon, corresponding to the mostly-Lorentzian and mostly-Euclidean worldsheet evolution respectively.

\begin{figure}[t]
	\centering
\hspace*{-3.5cm}
\adjustbox{valign=c}{\input{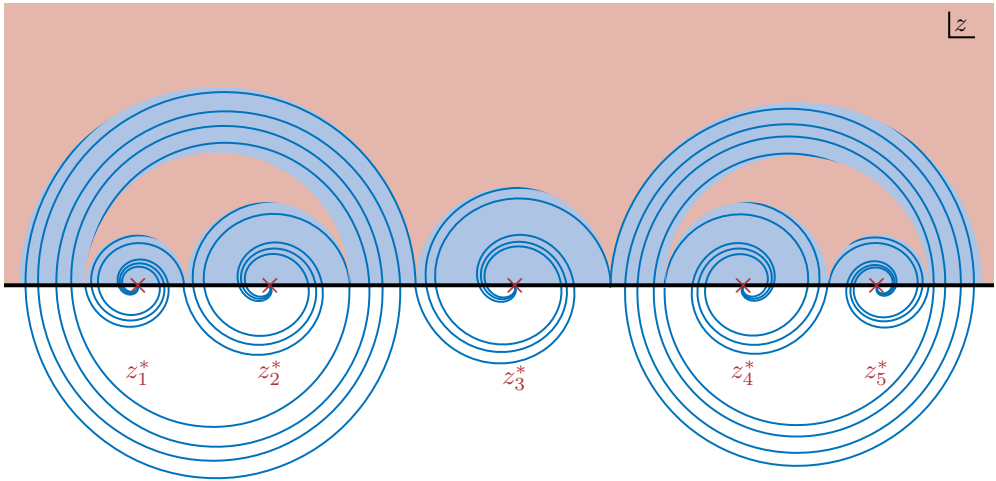}}       
\caption{\label{fig:worldsheet-regions}The decomposition of the worldsheet for a fixed configuration of vertex operators points $(z_1^\ast, z_2^\ast, \ldots, z_5^\ast)$ in the upper half-plane $z$, responsible for 
$\Exp_3$
close to the resonances $s_{12}$ and $s_{45}$. The blue regions are \textcolor{RoyalBlue}{Lorentzian} and the red are \textcolor{Maroon}{Euclidean}. The way the Lorentzian regions are sheared and glued to the Euclidean ones is illustrated by the blue curves. For readability we used only $w=2$ windings in all cases.}
\end{figure}

In the next step, let us illustrate how the spirals and anti-spirals from Sec.~\ref{sec:spirals} affect the geometry. The situation can be modelled by introducing arc-like regions illustrated in Fig.~\ref{fig:worldsheet-regions}. For concreteness, let us focus on the one surrounding the first two punctures. We pick the radii to be $r_1 < r_2$ and assume that the two punctures $z_1^\ast$ and $z_2^\ast$ are contained inside the smaller radius. The corresponding worldsheet parametrization is
\be
z = \frac{z_1^\ast + z_2^\ast}{2} + r \e^{i\theta + 2\pi i w (r - r_1)/(r_2 - r_1)}\,,
\ee
with $r \in [r_1, r_2]$ and $\theta \in [0,\pi]$. We can repeat an analysis similar to the one above. The bottom line is that this region corresponds to a Lorentzian evolution along the direction of $p_{12}^\mu$ with $w$ windings. After attaching another copy of the region around punctures $z_4^\ast$ and $z_5^\ast$ we obtain the a typical worldsheet contributing corresponding to the $(x,y) \approx (0,0)$ corner of the moduli space pictured in Fig.~\ref{fig:moduli-space}.

The corresponding worldsheet trajectory for $\M_{345 \ot 12}$ close to the resonance $s_{12}$ and $s_{45}$ is plotted in Fig.~\ref{fig:twisted-worldsheets} (middle). Here, the additional blue ribbons stretching between the pairs of Euclidean regions are consequences of adding the Lorentzian evolution described above. In Fig.~\ref{fig:twisted-worldsheets} (right), we illustrate the same configuration but for
$\Exp_3$. The difference is reversing the orientation of the Lorentzian segments around the punctures $4$ and $5$. This plot illustrates a generic worldsheet trajectory for this in-in observable. It agrees with the idea that the target space of this string worldsheet is the same Schwinger--Keldysh timefold as that which defines this observable in field theory~\cite[Sec.~4]{Caron-Huot:2023vxl}, where all sources $1,2,4,5$ are in the past of the observation point $3$.

\section{Conclusion}\label{sec:conclusions}

In this work, we found that crossing symmetry is much richer than previously believed: it relates together not only scattering amplitudes, but also a larger family of asymptotic observables. The focus of this paper was to motivate, explain, and test crossing rules summarized in Sec.~\ref{sec:crossing}. These developments open up a number of new directions, some of which are outlined below.

A natural direction is to do further explicit checks of the crossing equation on multi-loop Feynman integrals and understand to what extent analytic obstructions affect it. Alternatively, assuming the crossing relations, one could deduce and test relations between amplitudes and discontinuities in various kinematic channels, as done in Sec.~\ref{sec:individual-cuts} and \ref{sec:pentagonApplications}. 

More ambitiously, one could attempt to extend the tree-level proof given in Sec.~\ref{sec:treelevelProof} to multi-loop processes. The fact that the path of analytic continuation is expressed in lightcone coordinates suggests that looking into lightcone-ordered perturbation theory \cite{Chang:1968bh} might be worthwhile. We do not expect such a proof to be a walk in the park, as it would have to address issues with local analyticity and anomalous thresholds already encountered in Sec.~\ref{sec:anomalous}.

In the process of understanding the crossing equation better, more theoretical data is needed. As with the time-ordered amplitudes, a natural place to start with is the $\mathcal{N}=4$ super Yang--Mills theory. It would be interesting to see if the asymptotic observables can be ``bootstrapped'' in the same way as the amplitudes (as reviewed in \cite{Arkani-Hamed:2022rwr}).
We expect that they would also inherit many analytic properties of time-ordered amplitudes such as cluster adjacency.
The relation between our crossing moves and those based on the analytic (dispersive) representation of \cite{Bartels:2008ce} would also be worth exploring.

As reviewed in Sec.~\ref{sec:webs}, the specific crossing moves we proposed lead to disjoint families of observables. However, there could of course also exist other ways of deforming the energies and momenta of external particles that lead to other forms of crossing symmetry reconnecting the families. More work is required in this direction. For example, for massless particles, one natural idea is to continue through the $(2,2)$-signature kinematics as an intermediate step, which sounds particularly appealing in the view of the importance of the split signature in the on-shell recursion relations. However, to date, there is no concrete understanding for how to do such an analytic continuation systematically in practice (see, e.g., \cite{Srednyak:2013ylj} for partial progress for off-shell Green's functions). Our work highlights the importance of revisiting this problem.

Beyond field theory, it might prove useful to study crossing symmetry in string perturbation theory, which provides a yet another starting point distinct from Feynman integrals and the LSZ reduction formulae. Progress in proving it in $2\to 2$ scattering includes \cite{deLacroix:2018tml}, which using string field theory to show that string amplitudes are analytic in the primitive domain (see \eqref{eq:primitive-domain}) provided massless states are ignored. Using worldsheet methods, the immediate challenge is that analytic properties are not yet fully understood beyond genus-one four-point functions \cite{DHoker:1994gnm,Eberhardt:2023xck}.

On the more conceptual side, it is crucial to further understand the interplay between crossing and local analyticity. Recall that axiomatic field theory points to the fact that S-matrix elements may in general not be expressible in terms of a single analytic function \cite{Bros:1972jh}. Concrete examples are given in Sec.~\ref{sec:anomalous} and \cite{Hannesdottir:2022bmo}. We have seen that whenever such a local analyticity clash happens, the crossing equation is no longer valid. Nevertheless, as reviewed in App.~\ref{sec:local-analyticity}, there still exists a (non-canonical) decomposition of S-matrix elements into a sum of analytic functions. Can we characterise physically when such problems with analyticity occur, and how to then think of crossing symmetry? When is the S-matrix analytic?
We leave these foundational questions to future work.

\acknowledgments
S.C.H. and M.G.’s work is supported in parts by the National Science and Engineering Council of Canada (NSERC) and the Canada Research
Chair program, reference number CRC-2022-00421.
S.C.H.'s work is additionally supported by a Simons Fellowships in Theoretical Physics
and by the Simons Collaboration on the Nonperturbative Bootstrap. The work of H.S.H. and S.M. is supported by the U.S. Department of Energy, Office of Science, Office of High Energy Physics under Award Number DE-SC0009988. Additional funding for H.S.H. is provided by the William D. Loughlin Membership, an endowed fund of the Institute for Advanced Study (IAS). The work of S.M. is additionally supported by the Sivian Fund and the Roger
Dashen Member Fund at the IAS. 
S.C.H. thanks the IAS for hospitality during a sabbatical semester where this work got started.

\appendix

\def\deltaD{{\bf \delta^D}}

\section{Why are scattering amplitudes analytic? \texorpdfstring{\\}{}
Review of the axiomatic approach to crossing
}
\label{app:axiomatic}

As emphasized in the main text, there are two separate steps that go into establishing crossing relations for a general scattering process. The first one involves proving analyticity in certain regions of the on-shell kinematics within which analytic continuation can be performed without encountering any singularities or crossing branch cuts. The second step consists in identifying the result of this analytic continuation as another observable. In this paper, we explored the rich structure of this second problem and found that crossing can relate scattering amplitudes to more general asymptotic measurements in quantum field theories. However, we assumed that analyticity along the crossing path holds.

Our formal manipulations seem easy to make rigorous if one were allowed to go off-shell, and the main subtle question is whether our crossing paths continue to exist when restricting to the mass shell. The purpose of this appendix is to review the axiomatic arguments for analyticity that make this assumption plausible.

\subsection{Proof of crossing for \texorpdfstring{$2\to 2$}{2-to-2} scattering}

Crossing symmetry for $2\to 2$ scattering in mass-gapped theories was established by Bros, Epstein, and Glaser (BEG) in \cite{Bros:1964iho,Bros:1965kbd}. Here we offer a streamlined version of their proof that we hope to be more accessible to physicists.

\subsubsection{Overview}

The starting point is reduction formulas of the LSZ type for amputated Green's $n$-point functions (off-shell versions of scattering amplitudes):
\be\label{eq:A1}
\mathcal{G}_n(p) = \int \prod_{j=1}^{n} \d^\D x_j \, \e^{-i p_j \cdot x_j} \langle 0| \mathcal{T}\left\{ j(x_1) j(x_2) \cdots j(x_n) \right\} | 0 \rangle\, ,
\ee
where $j(x_i) = (-\partial_{x_i}^2 + m^2) \phi(x_i)$ are the currents.  As in the main text, we focus here on a real scalar theory with a single field $\phi$ of mass $m$.  The generalization to scattering amplitudes with unequal masses would complicate some formulas but does not pose any real challenge.

The immediate difficulty is that \eqref{eq:A1} does not define an analytic function of the momenta $p_j$: for the factor $\e^{-i p_j \cdot x_j}$ not to diverge as $x\to\infty$, one needs $\mathrm{Im}\, p_j \cdot x_j \leq 0$ for \emph{all} $x_j$. However, $x_j$ can be timelike or spacelike and hence a fixed complex Lorentz vector $p_j$ cannot satisfy the above inequality.

There are two natural solutions to this problem.  One is to show that the position-space correlator $\mathcal{G}_n(x) \equiv \langle 0| \mathcal{T}\left\{ j(x_1) j(x_2) \cdots j(x_n) \right\} | 0 \rangle$ itself decays exponentially in some directions,
such that a certain amount of exponential growth can be tolerated in $\e^{-i p_j \cdot x_j}$.
This leads to the concept of \emph{essential support} pursued in Sec.~\ref{sec:local-analyticity}.
The other solution, which will suffice for $2\to 2$ scattering, is to express $\mathcal{G}_n(x)$ in terms of retarded (as opposed to time-ordered) products. After an analysis of exponentials similar to the one sketched above, one arrives at the so-called \emph{primitive domain of analyticity} \cite{Steinmann1960a,Steinmann1960b,ruelle1961connection,doi:10.1063/1.1703695,araki1960properties}:
\be\label{eq:primitive-domain}
\bigcap_{S} \left\{ \mathrm{Im}\, p_S \text{ timelike} \} \cup \{ \mathrm{Im}\, p_S = 0 \text{ and } -p_S^2 < \mathfrak{m}_{S}^2  \right\}\,,
\ee
for every proper subset of external labels $S$, where $p_S = \sum_{i\in S} p_i$ and $\mathfrak{m}_S$ is the mass of the lightest threshold allowed in each channel. To be technically precise, we should say that correlators are analytic in some open neighborhood of \eqref{eq:primitive-domain} (a domain is always an open set).
Similar levels of analyticity can be proven more easily in perturbation theory, see for example \cite{deLacroix:2018tml} and \cite[Note 32]{Mizera:2021ujs}.

The domain \eqref{eq:primitive-domain} implies an off-shell version of crossing.  To see this explicitly, let us introduce coordinates which we will also use throughout this appendix.
We fix the momentum transfer and take it to be transverse and real:
\be
p_2^\mu+p_3^\mu= \left(0,\, 0,\, 2q_\perp\right)\, ,
\ee
in lightcone notation $p_i^\mu = (p^+_i,p^-_i,p^\perp_i)$ with $-p_i^2 = p_i^+ p_i^- - (p_i^\perp)^2$. Recall that $p_1^\pm, p_2^\pm < 0$ are incoming and $p_3^\pm, p_4^\pm > 0$ are outgoing.
We can then parametrize the two projectile momenta as (recall that we take $m_2=m_3=m$):
\begin{align} \label{p23 appendix}
   p_2^\mu=\left({-}p^+,\;
   {-}\frac{m^2+q_\perp^2+\xi}{p^+},\;
   q_\perp\right),
   \qquad
   p_3^\mu=\left(p^+,\;
   \frac{m^2+q_\perp^2+\xi}{p^+},\;
   q_\perp\right)\,.
\end{align}
On the one hand, we will keep $p_1$ and $p_4$ fixed and on-shell,
and we take $p^+$ to be large and positive (ultra-relativistic kinematics)
such that $s = -(p_3 + p_4)^2 \approx p^+ p_4^-$. On the other hand, $t = -(p_2 + p_3)^2 < 0$ remains constant.
To describe crossing, we must take $p^+$ along a large arc in the upper half-plane. The essential variables will be the energy $p^+$ and off-shellness parameter 
$\xi$. The variable $p^+$ plays the same role as $z$
in Sec.~\ref{eq:crossingpath} and $\xi = 0$ is the on-shell point.

The off-shell crossing path in the primitive domain \eqref{eq:primitive-domain} can be described as follows.
Since the imaginary part of $p_3$
is non-zero, the domain requires it to be timelike, i.e., ${\rm Im}\, p_3 \in V^+$. Thus, ${\rm Im}\,p_3^+$ and ${\rm Im}\,p_3^-$ must have the same sign. This can be satisfied for ${\rm Im}\, p^+ > 0$ and constant real $\xi$ if it is sufficiently negative:
\be \label{D1 criterion}
{\rm Im}\, \frac{m^2+q_\perp^2+\xi}{p^+}>0 \quad\mbox{for}\quad 0<{\rm arg}(p^+)<\pi
\quad\Rightarrow\quad \xi<-m^2-q_\perp^2\equiv \xi_0\,.
\ee
All the other conditions in \eqref{eq:primitive-domain} are then satisfied: the imaginary parts of $p_2^\mu$, $p_3^\mu$, $p_1^\mu+p_2^\mu$ and $p_1^\mu+p_3^\mu$ channels are timelike by the same condition, while $p_1^\mu$, $p_4^\mu$ and $p_1^\mu+p_4^\mu$ are real and below threshold.  It is important here that we consider correlators of currents, e.g. amputated correlation functions, so that the mass shell is strictly below the first singularity at
$-p_i^2=\mathfrak{m}_i^2$: $m_i^2<\mathfrak{m}_i^2$.
The primitive domain \eqref{eq:primitive-domain} thus contains a neighborhood of the following subset, which enables off-shell crossing:
\be \label{domain D1 app}
  D_1 = \{ (p^+, \xi) \,|\, {\rm Im}\,p^+>0 \,\,\mbox{and}\,\, \xi<\xi_0<0\}\,.
\ee
This is not yet what we want.
The crux to obtain on-shell crossing will be to show that the
correlator depends on the small lightcone momentum $p_3^-$ sufficiently mildly
that we can relax the constraint
${\rm Im}\,p_3^->0$ in \eqref{D1 criterion}.

\begin{figure}
    \centering
    \includegraphics[scale=1]{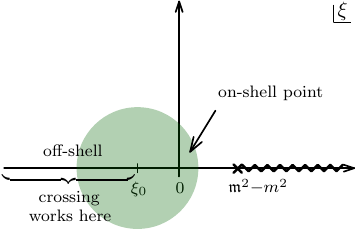}
    \caption{The origin in the $\xi$-plane represents the on-shell point.
       The primitive domain of analyticity $D_1$ ensures the existence of a complex-$s$ crossing path when
       ${\rm Re}\,\xi<\xi_0$.  The BEG argument leverages the fact that that for real $s$, $|s|>s_0$, the nearest $\xi$-singularity is strictly to the right of the origin, to extend crossing to a circular region (green) that contains the on-shell point at the origin.
    }
    \label{fig:xi plane}
\end{figure}

Physically, since $p_3^-\propto 1/p^+$ is small, most invariants are largely insensitive to its precise value except for the single-momentum invariants $p_2^2$ and $p_3^2$. Naively, one may thus expect analytic properties in $\xi$ to be controlled by the thresholds in these channels, which is at $\xi=\mathfrak{m}^2-m^2$.
Below, we will show that this is indeed the case and
review why the correlator is analytic in a neighborhood of:
\be \label{domain D2 app}
 D_2 = \{ (p^+, \xi) \,|\, {\rm Re}\,\xi<\mathfrak{m}^2-m^2 \,\,\mbox{and}\,\, |p^+|>p_0,\,\,p^+\,{\rm real}\}\,,
\ee
for some sufficiently large constant $p_0$.
Important loci in the $\xi$-plane are summarized in Fig.~\ref{fig:xi plane}.

The domains $D_1$ and $D_2$ control respectively the dependence of correlators on complex $p^+$ and $\xi$, with the other variable kept fixed and real. At face value, the union $D_1\cup D_2$ allows one to start from an on-shell $s$-channel configuration, go off-shell, cross to the $u$-channel, and then return on-shell.
It turns out that the magic of complex analysis implies a stronger result: analyticity in a neighborhood of $D_1\cup D_2$ automatically implies analyticity in a larger domain in which crossing paths can remain on-shell at all times.

In Sec.~\ref{ssec:D1D2} we elaborate on the proof of analyticity in $D_1$ and $D_2$, while the extension of $D_1\cup D_2$
is reviewed in Sec.~\ref{ssec:domain extension}.
Compared with the original proof of crossing in \cite{Bros:1965kbd} (expanded upon in \cite[Sec.~5.3]{Sommer:1970mr} and \cite[Sec.~16.3]{Bogolyubov:1990kw}), our presentation is simplified in two ways.  First, our off-shell deformation $-p_i^2 \to m_i^2+\xi$ only affects the two particles $i=2,3$, whereas in the original proof it affected all particles democratically.
Second, we focus on high-energy (Regge) kinematics.  Together, these two choices
greatly simplify the derivation of the $D_2$ domain since our $\xi$ is trivially related to a lightcone coordinate $p_3^-$.  In addition, the application of the tube theorem will be simplified since the complicated almond-shaped regions of \cite{Bros:1965kbd} reduce to circles in the Regge limit, see also \cite{EpsteinIHES} for a similar approach.  Despite these simplifications, we believe that our abridged presentation remains rigorous and at the same time more accessible to physicists.

\subsubsection{Global and local domains \texorpdfstring{$D_1$}{D1} and \texorpdfstring{$D_2$}{D2}}\label{ssec:D1D2}

Instead of relying on \eqref{eq:primitive-domain}, we find it instructive to establish analyticity in $D_1$ and $D_2$ directly by exploiting the support properties of various retarded products.
The product relevant for $D_1$ was discussed in \eqref{R amplitude} in the main text. For $2\to 2$ scattering it simplifies to
\begin{equation} \label{R amplitude app}
    \mathcal{R}_{34\ot 12}(p_3,p_2) = \int \text{d}^\D x_3~\text{d}^\D x_2\, \e^{-ip_2{\cdot}x_2-ip_3{\cdot}x_3}\, {}_{\rm out}\<4|\ [j(x_3),j(x_2)]_R\ |1\>_{\rm in}\,.
\end{equation}
Importantly, the integral only has support when $x_3$ is in the future lightcone of $x_2$.  If we perform the change of variable $x_3 \mapsto x_2+y$, we obtain 
\be\label{eq:R22}
\mathcal{R}_{34\ot 12}(p_3,p_2) = 
\int_{y\in \bar{V}^+} \!\!\! \d^\D y\, \e^{-ip_3{\cdot} y}
\int \text{d}^\D x_2\, \e^{-i(p_2+p_3){\cdot}x_2}
{}_{\rm out}\<4|\ [j(x_2 + y),j(x_2)]_R\ |1\>_{\rm in}\, .
\ee
The integration range is $y$ in the future lightcone $\bar{V}^+$, since otherwise the retarded commutator vanishes. This implies that \eqref{eq:R22} is analytic for ${\rm Im}\,p_3\in V^+$ when $p_2+p_3$ is kept real (meaning that equal and opposite imaginary parts are added to $p_2$ and $p_3$).
This is precisely the condition on $p_3$  discussed in \eqref{D1 criterion}, and hence we conclude that $\mathcal{R}_{34\ot 12}$ with $p_i(p^+,\xi)$, now viewed as a function of $p^+$ and $\xi$ only, is analytic in $D_1$ in \eqref{domain D1 app}.

To establish the domain $D_2$ in the off-shell parameter $\xi$, we use
a different reduction formula. We now fix the $s$-channel
momentum $p_3 + p_4$ to be real and study analyticity in
$p_3$ of the following two products:
\be \label{iM s-channel commutators}
\bm{\delta}_4 \times i\cM_{34 \ot 12}
 = \<0|\, [j_3,j_4]_R\, |21\>_{\rm in} 
 = \<0|\, [j_4,j_3]_R\, |21\>_{\rm in}\,.
\ee
That both correlators compute the same amplitude can be verified using reduction formulas in the same way as \eqref{R reduction}, see \cite{Caron-Huot:2023vxl}.
The first representation is analytic for ${\rm Im}\,p_3\in V^+$, while the second is analytic in the opposite cone ${\rm Im}\,p_3\in -V^+$.
They agree for real momenta below the single-current threshold $-p_i^2<\mathfrak{m}^2$ by spectral considerations, since the difference is a commutator $[j_3, j_4]$ without the step function.
In this situation, the celebrated \emph{edge-of-the-wedge theorem} (see, e.g., \cite{vladimirov2007methods})
states that $i\cM_{34 \ot 12}$ is actually analytic in some open neighborhood of the real axis below the cut, which includes ${\rm Im}\,p_3^\mu$ that can be spacelike as long as it is small enough.

This is not quite sufficient for our purposes since we will need some uniformity in that the \emph{size} of the open neighborhood does not shrink at large $p^+$. To analyze the situation at large $p^+$ (or equivalently large $s$), the following parametrization in lightcone coordinates will be useful (again we set $m_1=m_2=m$ for simplicity):
\be
 p_3^\mu = \left(\sigma,\;\frac{m^2+\xi'}{\sigma},\;
 0_\perp\right),\quad
 p_4^\mu = \left(\frac{m^2}{\sigma},\;
 \sigma-\frac{\xi'}
 {\sigma},\;
 0_\perp\right),
\ee
where we fix $\sigma$ (which grows as $\sigma \approx \sqrt{s}$ at high energy) and vary the single complex variable $\xi'$.
The two forms in \eqref{iM s-channel commutators} are then respectively analytic for ${\rm Im}\,\xi'>0$ and ${\rm Im}\,\xi'<0$ and they agree for ${\rm Re}\,\xi'<\mathfrak{m}^2-m^2$.  Thus, by single-variable complex analysis, $i\cM_{34\ot 12}$ is analytic for $\xi'$ in the cut plane $\mathbb{C}\setminus [\mathfrak{m}^2-m^2,\infty)$ similar to that in Fig.~\ref{fig:xi plane}.

To complete the proof, it remains to show that the $\xi'$ and $\xi$ planes are essentially equivalent.  More precisely,
at large $s$, the parameter $\xi'$ is effectively just the off-shellness in $p_3^2$, since that of $p_4^2$ is suppressed by $\sigma^{-2}\approx s^{-1}$:
\be
  -p_3^2=m^2+\xi',\qquad -p_4^2 = m^2 - \frac{m^2}{\sigma^2}\xi'\,.
\ee
Neglecting the change in $p_4^2$ at large $s$, we have effectively established analyticity in $p_3^2$ at fixed $s$ and $p_4^2$.  More rigorously, for sufficiently large $s$, the change in $p_4^2$ could be cancelled by a small deformation of $p_i^+$ within the neighborhood guaranteed by the edge-of-the-wedge theorem (while not explicitly computed by the theorem, it is clear from various of its proofs \cite{Bros:1971ghu,Bogolyubov:1990kw} that the size of the neighborhood can only depend on the distance to singularities in the considered four momenta, i.e., on $\mathfrak{m}^2-m^2$ but not on $s$).
To obtain the desired region $D_2$ in \eqref{domain D2 app}, we can apply an argument similar to a simultaneous shift of all four momenta in such a way that $p_1+p_2$ \emph{and} $p_2+p_3$ remain constant, resulting in $-p_2^2= -p_3^2=m^2+\xi$ acquiring the same off-shellness $\xi$. 

We note that the original BEG argument \cite[Sec.~4]{Bros:1965kbd} did not assume large $s$. Consequently, they had to control spacelike shifts in ${\rm Im}\,p_i$. This was achieved using the Jost--Lehmann--Dyson representation, which is a different way of leveraging the vanishing of commutators at spacelike separations.
Their geometry was somewhat more complicated since they shifted all four masses by the same amount.
Nonetheless, at large $s$, their resulting domain \cite[Fig.~6]{Bros:1965kbd} simplifies and contains the large disk that we will use below.

\subsubsection{Tube theorem and domain extension} \label{ssec:domain extension}

In order to state crossing as an intrinsic property of the on-shell amplitude $i\cM(s,t)$, we need to extend the domain $D_1\cup D_2$ in such a way that we can find a crossing path that remains on-shell at all times.  While the derivation so far relied on physical principles such as microcausality and stability, the final step of ``interpolating'' between the two regions will be purely mathematical.

Functions in several complex variables have a peculiar property that if they are analytic in a certain domain $D$, they can be extended to a possibly larger region called its \emph{envelope of holomorphy} $\mathcal{H}(D)$. The envelope does not depend on the function itself, but rather only on the geometry of $D$. This aspect does not have an analogue in one complex dimension, where always $D = \mathcal{H}(D)$. For example, if we only knew a function $f(z)$ is analytic in a unit disk $|z| < 1$, we cannot say anything about its analyticity properties in $|z| \geq 1$. We will see that these changes in two and higher complex dimensions.

\paragraph{Bochner's tube theorem}
While the computation of $\mathcal{H}(D)$ is in general a complicated problem \cite{fuks1963theory,kaup2011holomorphic,lebl2019tasty}, in some situations it can be solved exactly. Among the simplest examples are tubes, which are generalizations of strips to several complex variables. A tube in $m$ dimensions is a domain of the form
\be
T_B = B + i \mathbb{R}^m\, ,
\ee
where the base $B$ is connected and the imaginary parts remain unconstrained. \emph{Bochner's tube theorem} asserts that its envelope of holomorphy is
\be
\mathcal{H}(T_B) = T_{\mathrm{ch}(B)}\, ,
\ee
where $\mathrm{ch}(B)$ is the convex hull of $B$.

A simple way to understand this theorem is to consider functions $\hat{f}(q)$
that originate as the Laplace transform of some $f(x)$. 
Suppose $\hat{f}(q)$ is analytic for $q\in T_B$.
Then Bochner's theorem is simply the statement that
set of ${\Re}\,q$ for which the Laplace transform converges is convex. In more details, for the purposes of this explanation let us additionally assume that the function $\hat{f}(q)$ does not grow too fast at large imaginary $q$, such that the inverse transform makes sense (this property can be verified directly in the applications below):
\be
 f(x) = \int_{b+i\mathbb{R}^m} \frac{\d^m q}{(2\pi i)^m} \e^{q{\cdot}x} \hat{f}(q)\,.
\ee
Convergence in a neighborhood of some $b\in B$ means that $|f(x)/\e^{b{\cdot}x}|$ does not grow exponentially as $|x|\to\infty$ in any direction. Now, since the contour can be deformed to any $b\in B$, we
conclude that the direct transform converges at large $|x|$ for any point $q$ in the interior of $T_B$:
\be 
 \hat{f}(q) = \int \d^m x\, \e^{-q{\cdot}x} f(x) \quad\mbox{for}\quad {\rm Re}\,q\in B\,.
\ee
One may say that the Laplace transform and its inverse manifest the assumed analyticity of $\hat{f}(q)$.  Now, 
if $q_1$ and $q_2$ are any two points in the interior of $T_B$, then the transform will also converge anywhere on the line connecting them, because the exponential function is convex:
\be
 \left|\e^{-[\lambda q_1+(1-\lambda)q_2]{\cdot}x}\right|
 \leq
 \lambda \left|\e^{-q_1{\cdot}x}\right|+
 (1-\lambda) \left|\e^{-q_2{\cdot}x}\right| \quad\mbox{for}\quad 0\leq \lambda\leq 1\,.
\ee
This line of thought is made into a rigorous mathematical proof in \cite[Sec.~2]{Bros:1971ghu}, where the authors then proceed to generalize it to a local version.

Let us mention that the same arguments also apply to so-called \emph{flattened} tubes, which are lower-dimensional tubes, i.e., not open sets in $\mathbb{C}^m$ \cite{Epstein:1966yea}. Technically speaking, this is the version of the tube theorem used in the final step of the proof of crossing below.

\paragraph{Implication: analyticity at large energy}

We are now well-equipped to complete the proof of crossing symmetry for $2\to 2$ scattering with a mass gap (we will use that $\mathfrak{m}>m$ so that the mass shell is isolated from other cuts).
While the full details of the last step become incredibly technical \cite{Bros:1964iho,Bros:1965kbd}, their essence can be explained on a simplified toy model we consider now. Below we will argue that the model is actually justified for sufficiently large $|s|$.  We closely follow \cite[App.~A.2]{Mizera:2022dko}.

Since tube theorems become useful only starting in two complex dimensions, the simplest setup in which we can get some mileage is a two-dimensional subspace of the off-shell kinematic space.
The $(\xi,p^+)$-plane in \eqref{p23 appendix} is precisely what we need.
Since $p^+\propto s$ at large $s$ (we hold $p_4^-$ fixed), we will refer to it in this section as the $(\xi,s)$-plane for better readability.
We wish to show analyticity in the neighborhood of the on-shell kinematic point $\xi = 0$ and a subset of the upper half-plane in $s$ connecting the physical region in the $s$-channel (say, located in $s>s_2$) to that of the $u$-channel from the unphysical side (located in $s< s_1<s_2$).

\begin{figure}
\centering
\includegraphics[width=\textwidth]{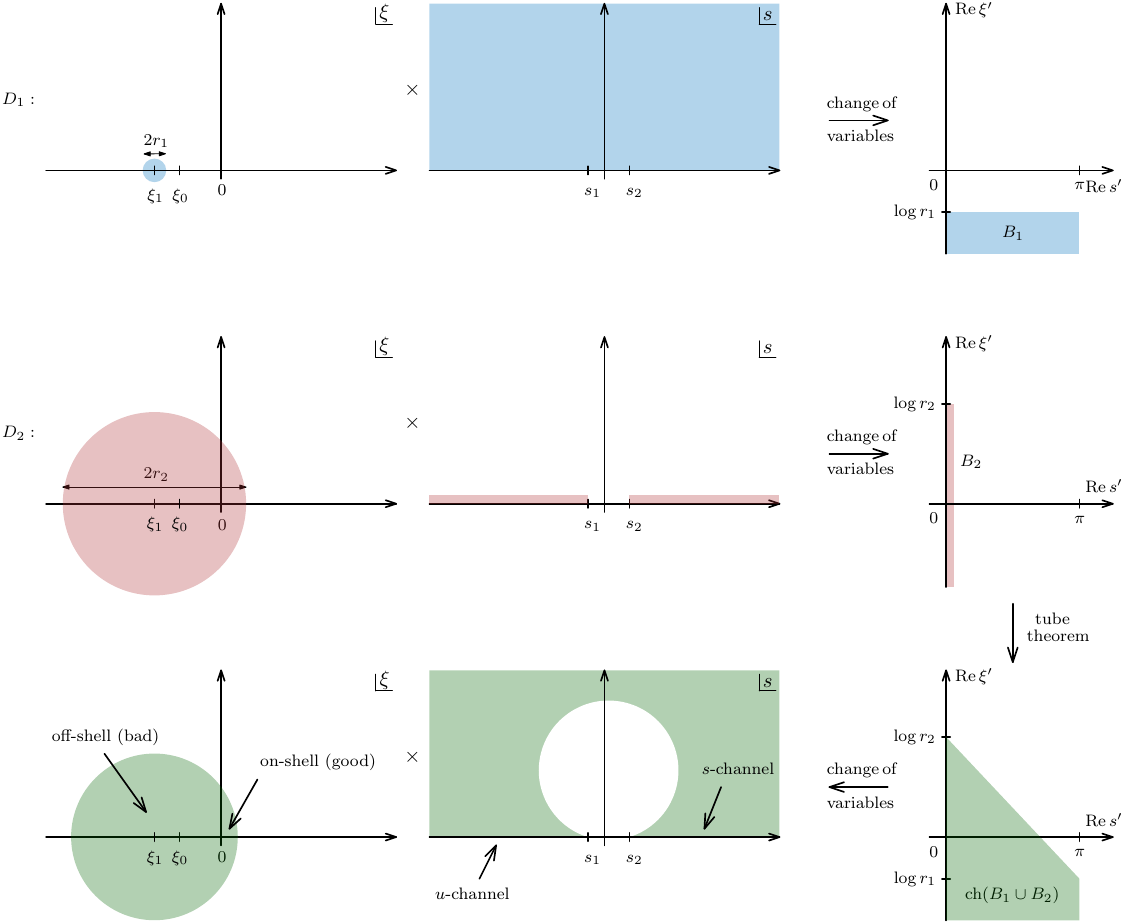}
\caption{\label{fig:tube-theorem}Application of the tube theorem to proving analyticity of gapped on-shell $2\to2$ scattering amplitudes needed for crossing symmetry. \textbf{Top:} The global but off-shell region is a topological product of a disk in $\xi$ and the upper half-plane in $s$ (blue). \textbf{Middle:} The on-shell but local region is a disk times an infinitesimal neighborhood of the physical regions (red). The change of variables to $(\xi', s')$ converts both of them into tubes. \textbf{Bottom:} The tube theorem allows us to interpolate between the two regions to one that can be simultaneously on-shell and analytic in the asymptotic region of $s$ in the upper half-plane connecting the $s$-channel to the $u$-channel from the wrong side.}
\end{figure}

The toy model starts by assuming that the domain $D_1$ in \eqref{domain D1 app} contains a neighborhood of size $r_1$ around some off-shell point $\xi = \xi_1 < \xi_0$, times the whole upper half-plane in $s$.
It is illustrated in Fig.~\ref{fig:tube-theorem} (top).
We also use the domain $D_2$, which contains the product of a disk of radius $r_2$ around $\xi_1$, times a neighborhood of the real $s$ axis for sufficiently large $s$.
This is shown in the middle row of Fig.~\ref{fig:tube-theorem}.
The disk of radius $r_2$ only needs to be sufficiently large that it contains the origin.
Note that the first assumption goes beyond what we have shown so far:  it does not follow from the above discussion that $D_1$ contains a circle of \emph{uniform} radius $r_1$.  This is the only assumption that we will need to address below.

The above regions are not yet tubes. However, we can convert them into ones using the following change of variables
\be \label{xi s change}
\xi' = \log (\xi - \xi_1) \quad \text{and} \quad s' = i \log \left(\frac{s - s_1}{s - s_2} \right).
\ee
They are chosen carefully such that lines of constant $\Re \xi'$ are circles around the point $\xi_1$ within the $\xi$-plane. Likewise, lines of constant $\Re s'$ are arcs connecting $s_1$ and $s_2$ in the upper half-plane of $s$. For both $D_1$ and $D_2$, they remain unconstrained. Therefore, in these variables both regions are tubes of the form $D_j = B_j + i \mathbb{R}^2$ with the bases illustrated in Fig.~\ref{fig:tube-theorem} (right). More concretely, they are given by (the neighborhoods of)
\begin{subequations}
\begin{align}
B_1 &= \{ \Re \xi' \,|\, \Re \xi' < \log r_1 \} \times \{ \Re s' \,|\, 0 < \Re s' < \pi \}\, , \\
B_2 &= \{ \Re \xi' \,|\, \Re \xi' < \log r_2 \} \times \{ \Re s' \,|\, \Re s' = 0 \}\, .
\end{align}
\end{subequations}
Since the base $B_2$ is one-dimensional, $D_2$ is an example of a flattened tube.

We can now apply the (flattened) tube theorem, which amounts to taking the convex hull of the two bases. That is,
\be
\mathrm{ch}(B_1 \cup B_2) = \bigcup_{0 \leq \lambda \leq 1}  \{ \Re \xi' \,|\, \Re \xi' < \log (r_1^{\lambda} r_2^{1-\lambda}) \} \times \{ \Re s' \,|\, 0 < \Re s' < \lambda\pi \}\, . \label{final region}
\ee
The result is illustrated in green in Fig.~\ref{fig:tube-theorem} (bottom right). Finally, mapping the result back to the original variables $(\xi, s)$, we obtain the green regions illustrated in Fig.~\ref{fig:tube-theorem} (bottom). They effectively provide an interpolation between the blue and red regions across both variables simultaneously, thereby enabling us to attain our final goal: connecting the $s$- and $u$-channel kinematics.
To summarize,
we found the domain \eqref{final region} which contains
a neighborhood of the on-shell point $\xi'=\log |\xi_1|$
times the region $0<\arg \frac{s-s_2}{s-s_1}< \lambda_*=\frac{\log r_2/|\xi_1|}{\log r_2/r_1}$, which links the $s$ and $u$ channels.
If $\lambda_*$ is small, the highest point of the ``bridge'' is at $s\approx \frac{s_1+s_2}{2} + i \frac{s_2-s_1}{\lambda_*}$.

It is instructive to estimate the size of the $s$-region obtained by this method.  At large $|t|\gg m^2$, we have $\xi_1 \sim \frac{t}{4}$ large, and
$r_2\approx |\xi_1|+\mathfrak{m}^2-m^2$ nearby.  On the other hand, $r_1$ does not have to shrink and we can take $r_1\sim m^2$.  Thus $\lambda_* \sim
\frac{\mathfrak{m}^2-m^2}{|t| \log(-t/m^2)}\ll 1$.  At the same time, we expect
$s_2-s_1\sim |t|$ since the separation between the physical $s$- and $u$-channel regions is linear in $t$. Thus, the highest point of the bridge is asymptotically at
\be
{\rm Im}\,s\sim \frac{t^2 \log(-t/m^2)}{\mathfrak{m}^2-m^2}\, .
\ee
(Ref. \cite{Sommer:1970mr} states without a proof the bound $\sim |t|^{3+\eps}$ for any $\eps>0$ which appears less optimal than ours.)
Note that this bound diverges as the gap between single-particle poles and multi-particle cuts shrinks, $\mathfrak{m}^2\to m^2$, which is as expected since the method leverages this gap. This is reminiscent of the analytic obstructions encountered in Sec.~\ref{sec:anomalous} when internal massless particles are involved. On the other hand, it remains unclear whether for $2\to2$ scattering there really can be $s$-plane singularities (anomalous thresholds) in this region or if the method is simply being too conservative.

\paragraph{Justifying a technical assumption}
Let us finally address the technical question of uniformity of the $\xi$ neighborhood inside $D_1$, which we assumed above to contain a disk of fixed radius $r_1$ around $\xi=\xi_1 < \xi_0$.
Away from real $s$ there is nothing to do, since the primitive domain of analyticity \eqref{D1 criterion} already contains such a region which is somewhat larger than $D_1$:
\begin{equation}
D_1' = \{(p^+, \xi) \;|\; 0{<}\arg p^+{<}\pi \quad\mbox{and}\quad
{-}\pi{+}\arg p^+ {<} \arg(\xi_0{-}\xi) {<} \arg p^+ \}\, .
\end{equation}
Thus, if we stay away from the real $p^+$ (or $s$)
axis by a finite angle, $\epsilon<\arg s<\pi{-}\epsilon$, then for large enough $s$, $D_1'$ uniformly contains a wedge $|\arg(\xi_0-\xi)|<\eps$ to the left of $\xi=\xi_0$, which certainly contains finite-radius balls centered around some point $\xi_1<\xi_0$. What we need to prove is that this ball does not shrink as $\arg s\to 0$
for any ${\rm Re}\,s>s_2$.  This is plausible because when $\arg s=0$, the domain $D_2$ tells us that the correlator is analytic in a full $\xi$ half-plane (${\rm Re}\,\xi < \mathfrak{m}^2-m^2$): it would be surprising for a $\xi$-plane singularity to appear as $\arg s\to 0$, and yet to remain invisible in some open neighborhood of a two-variable region containing the real axis ${\rm Im}\, s=0$. Mathematically, we believe that this follows simply by combining $D_1'$ and $D_2$ using the local version of the tube theorem proved by Bros and Iagolnitzer \cite{Bros:1971ghu}.

Historically, this theorem was not available at the time of the original Bros--Epstein--Glaser proof \cite{Bros:1965kbd}. Unfortunately, a simplified account of the proof incorporating these later ideas does not seem to have been published. The historical alternative was instead to combine the global tube theorem with a clever change of variables of the sort leading to \cite[Lem.~4]{Bros:1965kbd},
\begin{equation}
 \xi'' = 2\sin^{-1}\!\left(\frac{\mathfrak{m}^2-m^2-\xi}{\mathfrak{m}^2-m^2-\xi_0}\right) \quad\mbox{and}\quad s''=-i \log (s-s_2)\,.
\end{equation}
In these variables, $\xi=\xi_0$ maps to $\xi''=\pi$.
The domain $D_1'\cup D_2$ then contains a neighborhood of the (flattened) tubes
\be
\left(\{ {\rm Re}\,\xi''=\pi\}\times \{0<{\rm Re}\,s''<\pi\}\right)
\cup \left(\{0< {\rm Re}\,\xi''<\pi\}\times \{{\rm Re}\,s''=0\}\right),
\ee
whose convex hull contains the so-far missing wedge
$0<\arg(s-s_2)<\epsilon$, times the region $\epsilon<{\rm Re}\,\xi''<\pi$. The latter differs by $\epsilon$ from the original $D_2$ half-plane ${\rm Re}\,\xi < \mathfrak{m}^2-m^2$ and thus it certainly contains constant-radius balls near $\xi_0$. A similar argument with $s''=i\log(s_1-s)$ takes care of the $\epsilon$-wedge near the negative $s$ axis, thus concluding this technical aside.

\subsection{Local analyticity near the mass shell for \texorpdfstring{$m \to n$}{m-to-n} scattering}
\label{sec:local-analyticity}

The generalization of the above global arguments is not known for higher-multiplicity scattering (with the exception of the $2\to 3$ case \cite{Bros:1985gy}), however a lot is known about
local analyticity near the mass shell, which has been studied in \cite{Bros:1972jh} exploiting Bros and Iagolnitzer's powerful ``local Fourier transform'' \cite{Bros:1971ghu}. This provides the starting point of crossing paths, and, at the same time, a rigorous foundation for the LSZ reduction formula.
Here we review the basic constructions of \cite{Bros:1972jh}; other useful reference are the proceedings \cite{Pham:1975mda} and the book \cite{Iagolnitzer:1994xv}.
The basic question is: why are amplitudes analytic functions of on-shell momenta?  The answer, it turns out, is that amplitudes are often not analytic!  

\paragraph{Essential support}

The manipulations in Sec.~\ref{ssec:D1D2} use retarded functions to define functions with good global analyticity properties.
Here we will exploit that, around some \emph{given} real kinematic point $\textsf{p} = (p_1, p_2, \ldots, p_n)$, many such representations agree.  The analysis will be best done in position space by studying convergence of the Fourier integral \eqref{eq:A1}.

To simplify the discussion, let us introduce the notation
\be
\mathcal{T}(S) = \mathcal{T} \left\{ j(x_{S_1}) j(x_{S_2}) \cdots \right\}\, ,
\ee
for a multi-particle label $S$, such that the original LSZ formula in \eqref{eq:A1} contains $\mathcal{T}(12\cdots n)$. The trick, then, is to study more general chains of time-ordered products of the form
\be
\langle 0 | \mathcal{T}(I) \mathcal{T}(J) \cdots \mathcal{T}(K) | 0 \rangle\, ,
\ee
where the labels $\{1,2,\ldots,n\}$ have been partitioned into disjoint sets $I, J, \ldots, K$. To illustrate why these objects are useful, let us consider the simplest example, namely $\mathcal{T}(12\cdots n{-}1)\mathcal{T}(n)$. Inserting a complete basis of states between the two operators $\mathcal{T}$, we have
\be\label{eq:T123-T4}
\langle 0 | \mathcal{T}(12\cdots n{-}1)\mathcal{T}(n) | 0 \rangle = \sumint_X \langle 0 | \mathcal{T}(12\cdots n{-}1) |X\rangle\, \langle X| j(x_n) | 0 \rangle\,.
\ee
After integrating over $x_n$, the second expectation value vanishes unless $-p_n^2 > \mathfrak{m}_n^2$. Hence, if the $n$-th particle is stable, \eqref{eq:T123-T4} always integrates to zero.

The key idea is to subtract such quantities, which vanish in the kinematics of interest, from the integrand of \eqref{eq:A1} in order to improve its convergence. For example, we can write
\be \label{eq:A1 subtracted}
\mathcal{G}_n(p) = \int \prod_{j=1}^{n} \d^\D x_j \, \e^{-i p_j \cdot x_j} \langle 0| \mathcal{T}(123\cdots n)  - \mathcal{T}(12\cdots n{-}1)\mathcal{T}(n) | 0 \rangle\, ,
\ee
which computes the same function as \eqref{eq:A1} in a neighborhood of the mass shell.
Written this way, the integrand has a \emph{smaller support} in position space: it vanishes if $x_n$ is in the past of, or is spacelike from, each $x_1, x_2, \ldots, x_{n-1}$. This implies \emph{more analyticity} in the momentum (Fourier dual) space.

Repeating the same logic for different choices of sprinkling particles among time-ordered products results in a collection of \emph{support regions} $R_1, R_2, \ldots, R_q$. Just as in the above example, we have to commit to specific kinematic conditions of the type $-p_S^2 < \mathfrak{m}_S^2$ in various channels and hence the collection of regions $R_a$ that can be used to compute the same amplitude depends on the kinematic point $\textsf{p}$. The \emph{essential support} is then the intersection of all these regions, namely
\be
\mathrm{ES}_{\textsf{p}} = \bigcap_{a=1}^{q} R_a\, .
\ee
Another way to understand $\mathrm{ES}_{\textsf{p}}$ is as describing the coordinate-space region in which the correlator does \emph{not} decay exponentially.
More precisely, for $x\notin \mathrm{ES}_{\textsf{p}}$, the correlator either decays exponentially at large $|x|$ \emph{or} oscillates at frequencies that are distinct from $\textsf{p}$ (and therefore decay exponentially after applying a suitable smearing function).

The following theorem \cite{Bros:1972jh} lies at the heart of establishing local analyticity using this concept.
The theorem requires that we first reorganize the essential support as a union of convex pointed\footnote{A cone is pointed if it does not contain any complete line. That is, it is not possible to find two points that define an infinite line which lies entirely within the pointed cone. The \emph{future} light-cone $V^+$ defined earlier in the text is an example of such cone.  The dual of a pointed cone is an open cone.} cones $C_1, C_2, \ldots, C_r$:
\be \label{ES union}
\mathrm{ES}_\textsf{p} = \bigcup_{b=1}^{r} C_b\, . 
\ee
The theorem then states that the Fourier transform $\mathcal{G}_n$ can be written as the sum
\be \label{ES G sum}
\mathcal{G}_n(\textsf{p}) = \sum_{b=1}^{r} G_b(\textsf{p})\, ,
\ee
where each $G_b$ is an analytic function in a complex neighborhood of $\textsf{p}$ intersected with the dual cone
\be
\check{C}_b = \bigg\{ \textsf{p} \,\bigg|\, {-}\sum_{j=1}^{n} \mathrm{Im}\, p_j \cdot x_j > 0 \text{ for all } x \in C_b \bigg\}\, .
\ee
In effect, the theorem states that $\mathcal{G}_n$ enjoys the same analyticity properties as the Fourier transform of a function which decays exponentially for $x$ outside $\mathrm{ES}_\textsf{p}$.
The ``neighborhood'' part means that the theorem is about the \emph{directions} in ${\rm Im}\, p_j$ along which small imaginary parts can be added; the overall magnitude of allowed imaginary parts is not computed (although the proof gives some insight, as discussed below).

When $r=1$ in \eqref{ES G sum} we say that the correlator is the boundary value of an analytic function of momenta near $\textsf{p}$. This is the familiar situation where for example
the $4$-point scattering amplitude in $s$-channel kinematics
is analytic upon adding the correct infinitesimal shift:
$\cM_{34\ot 12}\equiv \lim_{\varepsilon \to 0^+ }\cM_{34\ot 12}(s+i\varepsilon,t)$.

Situations that require $r>1$ mean that the correlator is a sum of boundary values of analytic functions $G_b$'s whose
$i\eps$ directions are \emph{mutually incompatible}.
A familiar example is the discontinuity ${\rm Disc}_{s>s_2}\,\cM_{34\ot 12}(s,t)$, which (again in real $s$-channel kinematics, $s>s_2$) is the difference between a function analytic with $s+i\varepsilon$ and one with $s-i\varepsilon$.
While each piece can be separately analytically continued, in general their sum cannot: for example,
knowing ${\rm Disc}_{s>s_2}\,\cM_{34\ot 12}(s,t)$ for pion scattering below the four-particle threshold
does not determine the discontinuity above that threshold.\footnote{
In practice, it is only when $\textsf{p}$ is directly on a singularity, like $s=16m_\pi^2$ here, that we expect the essential support \eqref{ES union} to require more than one component.
}
A surprising fact is that, at higher multiplicity, this phenomenon can affect even the conventional time-ordered amplitude $\cM$!  Explicit examples have been found in perturbation theory \cite{Chandler:1969bd,PhysRev.174.1749,Chandler:1969nd,doi:10.1063/1.1703936,10.1143/PTP.51.912,Hannesdottir:2022bmo}.

\paragraph{A simple example}

To illustrate this construction more clearly, let us focus on the simplest case of $34 \ot 12$ (or $s$-channel) scattering, in which $p_3$ and $p_4$ have positive energies, while $p_1$ and $p_2$ have negative energies, all of them being close to the mass shell. We wish to determine the essential support. 

We already saw that $\mathcal{T}(1234)-\mathcal{T}(123)\mathcal{T}(4)$ is supported in the region
\be
R_1 = \{ (x_1, x_2, x_3) \le x_4 \}\, , \label{R1sup}
\ee
where $\{x\le y\}$ means that $x$ is in the \emph{past} lightcone of $y$,
and the parentheses are \emph{or} statements: $R_1$ is the region where $x_1$, $x_2$ \emph{or} $x_3$ is in the last lightcone of $x_4$.  There is a similar region from $\mathcal{T}(4)\mathcal{T}(123)$, which reverses the time ordering leading to
\be
R_2 = \{ x_4 \le (x_1, x_2, x_3) \}\, . \label{R2sup}
\ee
Permuting the labels, we obtain $2\times \binom{4}{1}=8$ similar regions if all external particles are stable.
In addition, we could subtract two-particle products like $\mathcal{T}(12)\mathcal{T}(34)$, whose integral vanishes provided that $p_3+p_4$ is below all the particle-production thresholds. Since both $p_3$ and $p_4$ absorb particles, this is certainly the case and gives the next support region
\be
R_9 = \{ (x_1, x_2) \le (x_3, x_4) \}\, . 
\ee
We can permute the labels in all $\binom{4}{2} = 6$ ways. The only possibility we have to exclude is $\mathcal{T}(34) \mathcal{T}(12)$, which does not give rise to a vanishing contribution. This is because $p_1 + p_2$ can be above a production threshold, by definition of the $s$-channel.

All in all, we found $13$ regions, whose intersection $\mathrm{ES}_{34 \ot 12} = \bigcap_{a=1}^{13} R_a$ upper-bounds the essential support at any kinematic point in the $s$-channel. The intersection turns out to be quite simple
\be
\mathrm{ES}_{34 \ot 12}\,=\, \{ x_1 \asymp x_2 \le x_3 \asymp x_4\}\,\equiv\, R \,=\,
\vcenter{\hbox{
\begin{tikzpicture}[thick]
\draw (1,0)--(-1,0); \draw[->] (1,0)--(0,0);
\filldraw (-1,0) circle (2pt) (1,0) circle (2pt);
\draw (-1.3,0.3)--(-1,0)--(-1.3,-0.3);
\draw (1.3,0.3)--(1,0)--(1.3,-0.3);
\node[left] at (-1.3,0.3) {$\small 4$};
\node[left] at (-1.3,-0.3) {$\small 3$};
\node[right] at (1.3,0.3) {$\small 1$};
\node[right] at (1.3,-0.3) {$\small 2$};
\end{tikzpicture}}}~.
\ee
Here, the relation $x_1 \asymp x_2$ means that the two points must coincide in spacetime and similarly for $x_3 \asymp x_4$; the first pair has to be in the past of the second.
The set $R$ is simply the support of an (amputated) tree-level $s$-channel exchange diagram where the exchanged particle could propagate along any future timelike direction.  Adding identities from more general products of three time-ordered products (a simple algorithm to produce combinations with restricted support is described in \cite{Epstein:1973pq,Iagolnitzer:1994xv}) does not lead to a stronger result:  the set $R$ is the smallest support that can be derived using generic causality arguments that apply uniformly throughout the $s$-channel.

We stress that the above calculation, being uniform throughout the $s$-channel, only provides an upper bound for ${\rm ES}_\textsf{p}$ at a particular momentum $\textsf{p}$. Physically, we expect from the Landau--Coleman--Norton picture of singularities that vertices can only separate along the (positive) line $(x_3-x_2)^\mu\propto (p_3+p_4)^\mu$ rather than along an arbitrary future-timelike direction. Furthermore, this should only be possible exactly on a Landau singularity: for a generic kinematical point $\textsf{p}$ we expect ${\rm ES}_{\textsf{p}}=\{x_1 \asymp x_2 \asymp x_3 \asymp x_4\}$ to be a simple contact diagram, synonymous with the amplitude being analytic in a complex
neighborhood of $\textsf{p}$.

The cone $R$ is pointed and therefore we only need a single term in the decomposition \eqref{ES union}.  The dual cone of $R$ can be computed by imposing
\begin{subequations}
\begin{align}
{\textstyle\sum}_{j=1}^{4} &\mathrm{Im}\, p_j {\cdot} x_j<0\quad \forall\,x\in R
\\\Leftrightarrow\quad &
\mathrm{Im} (p_1 {+} p_2){\cdot} x_1 + \mathrm{Im} (p_3 {+} p_4){\cdot} x_3 <0
\quad\forall\,\, x_1\le x_3
\\
\Leftrightarrow\quad &
\mathrm{Im} (p_3 {+} p_4){\cdot} (x_3{-}x_1)<0\quad\forall\,\,0\le(x_3{-}x_1)
\\
\Leftrightarrow\quad &
\mathrm{Im} (p_3^\mu {+} p_4^\mu) \in V^+\,.
\label{eq:dualConeCondition1}
\end{align}
\end{subequations}
In the third line we used momentum conservation.
This shows that the amputated correlator is analytic in a neighborhood of the mass shell for arbitrary (but sufficiently small) ${\rm Im}\,p_i^\mu$,
as long as ${\rm Im}(p_3^\mu+p_4^\mu)>0$.
Note that this condition implies 
\be
\mathrm{Im}\, s = - 2\, \Re (p_3 + p_4) \cdot \Im (p_3 + p_4) > 0\, .
\ee
Hence, the off-shell Green's function $\mathcal{G}_4$ is analytic in the $s$-channel approached from the $s+i\eps$ direction, thus defining the scattering amplitude in this channel.  However there are no constraints on the imaginary part of $t$ (as long as it is sufficiently small).

We stress that the domain \eqref{eq:dualConeCondition1} is much larger than the primitive domain \eqref{eq:primitive-domain} obtained from generic considerations of retarded products.
In particular, the method directly explains why the Fourier transform \eqref{eq:A1} makes sense even when spacelike imaginary parts are added to individual external momenta $p_i^\mu$, as is required to realize complex on-shell momenta.  Instead of repeatedly applying the edge-of-the-wedge theorem as done in the early proofs of analyticity near the mass shell \cite{Bros:1964iho} (exemplified below \eqref{iM s-channel commutators}),
one now has to calculate intersections of physically-transparent causal sets.

\paragraph{Higher points and outlook} Following a similar logic, some explicit results on the shape of the essential support of some $n$-point correlation functions have been obtained in \cite{Bros:1972jh}. For example,
for $2\to 3$ scattering one finds $r=1$ in the decomposition \eqref{ES union} at sufficiently high energies, however in nonrelativistic kinematics the calculation leads to $r=3$.  See also \cite{Iagolnitzer:1994xv} for a review.

This phenomenon should be related to the examples \cite{Chandler:1969bd,PhysRev.174.1749,Chandler:1969nd,doi:10.1063/1.1703936,10.1143/PTP.51.912,Hannesdottir:2022bmo}, but the precise connection is not entirely clear to us.
More generally, calculations based on generic causality considerations can only provide an upper bound on the essential support and it would be useful to be able to 
calculate the actual (minimal) essential support around a given point $\textsf{p}$, equivalent to its ``wave front set'' \cite{Hormander1971,kashiwara1973microfunctions}.
Adding some dynamical information (like the notion that scattering slow particles in a gapped theory cannot produce intermediate states that propagate ultra-relativistically over macroscopic distances) could conceivably lead to stronger rigorous results, perhaps of the sort conjectured in \cite{Chandler:1969bd}. 

Let us conclude with two comments on the decomposition \eqref{ES G sum}.  First, it is generally not unique when $r>1$. The ambiguities correspond roughly to the choice of a partition of unity when integrating over the union of regions \eqref{ES union} \cite{Bros:1972jh}.  In fact, the theorem could be proved rather straightforwardly by inserting a partition of unity in the Fourier transform, \emph{if} the real neighborhood of $\textsf{p}$ on which the different representations (Eqs.~\eqref{eq:A1}, \eqref{eq:A1 subtracted} etc.)
agree were replaced by the full $\mathbb{R}^n$. The non-trivial feature of the theorem is its local nature near $\textsf{p}$.

Second, its proof uses Bros and Iagolnitzer's local Fourier transform \cite{Bros:1971ghu} mentioned above, which is constructive and leads to concrete estimates on the size of the complex neighborhood in which \eqref{ES G sum} holds.  If the different representations all agree in a real ball $|p-p_0|<r$ around $p_0$, then (as far as we understand) the resulting
complex domain is guaranteed to contain the points $p=p_0+iq$ with $|p-p_0| + |q|<r$.
The local Fourier transform suppresses ``wrong-frequency'' components simply by convolving the correlator ${\cal G}_n(x)$ against a Gaussian with
\emph{position-dependent} width $\Delta x \sim \sqrt{|x|}$. The transform remains largely unexplored and it is interesting to ask whether the same idea could lead to stronger results for the regions
$-p_S^2<\mathfrak{m}_S^2$ that appear in physics.
\section{Cuts of one-loop integrals in embedding space}
\label{sec:app_embeddingspace}
In this section, we review the embedding space formalism and how it can be used to compute cuts \cite{Abreu:2017ptx}, as was done in Sec.~\ref{sec:boxes_oneloop}. We warn the reader that the cuts computed in this way do not necessarily land on the right sides of the branch cuts, as explained in \cite{Abreu:2017ptx}, but this was not a concern for us since the branch of the box cut in Sec.~\ref{sec:boxes_oneloop} was unambiguous.
For a given Feynman integral, the parameterization to embedding space is achieved by first associating to each inverse propagator (momenta labelled with the subscript ${}_{\textnormal{E}}$ are Wick-rotated),
\begin{equation}
    \textsf{D}_i=(\ell_{\textnormal{E}}+p_{\textnormal{E},I_i})^2+m_{I_i}^2\,,
\end{equation}
a point in $\mathbb{CP}^{\D+1}$ defined by
\begin{equation}
    X_i=\begin{pmatrix}
    p_{\textnormal{E},i}^\mu\\
    p_{\textnormal{E},I_i}^2+m_{I_i}^2\\1
    \end{pmatrix}\,.
\end{equation}
Given the two extra lightlike coordinates 
\begin{gather}
    Y=\begin{pmatrix}
    \ell_{\textnormal{E}}^\mu\\
    \ell_{\textnormal{E}}^2\\1
    \end{pmatrix} \quad \textnormal{and} \quad X_\infty=\begin{pmatrix}
    0^\mu\\
    1\\0
    \end{pmatrix},
\end{gather}
together with the bilinear form on $\mathbb{CP}^{\D+1}$
\begin{equation}
    (ZW) = - z^\mu w_\mu-\frac{1}{2}Z^+W^- - \frac{1}{2}Z^-W^+,
\end{equation}
where we have used the notation
\begin{equation}
    Z =\begin{pmatrix}
    z^\mu\\
    Z^+\\Z^-
    \end{pmatrix}\,,
\end{equation}
a one-loop integral in our conventions takes the following form
\begin{equation}\label{eq:uncutApp}
    \mathcal{M}^{n\text{-gon}} = \frac{1}{\pi^{\D/2}}\int_\gamma  \frac{\text{d}^{\D+2}Y\,\delta[(YY)]}{\text{GL}(1)}\frac{[-2(X_{\infty}Y)]^{n-\D}}{[-2(X_1Y)]\cdots [-2(X_{n}Y)]}\,.
\end{equation}
Here, the integration contour ${\gamma}$ runs over the real quadric defined by $(YY)=0$ in $\mathbb{CP}^{\D+1}$ as indicated with the delta function, and $n$ is the number of internal edges in the diagram.
One can demonstrate that all one-loop integrals, whether cut or uncut, can be expressed using the same type of functions $\mathcal{Q}_{n}^D$, defined according to
\begin{equation}
\label{eq:Q_n_def}
\mathcal{Q}_{n}^\D (X_1,\ldots,X_n,X_{0}) = \frac{1}{\pi^{\D/2}} \int_{\gamma}\frac{\text{d}^{\D+2}Y\,\delta[(YY)]}{\text{GL}(1)}\,\frac{[-2(X_{0}Y)]^{n-\D}}{[-2(X_1 Y)]\ldots [-2(X_n Y)]}\,.
\end{equation}
Note that the normalization of this integral differs from the one in~\cite{Abreu:2017ptx} to match our conventions.
On one hand, if $X_0 = X_\infty$, then \eqref{eq:Q_n_def} simplifies to a standard one-loop integral, namely
\begin{equation}\label{eq:IComp}
\mathcal{M}^{n\text{-gon}} =  \mathcal{Q}_{n}^\D(X_1,\ldots,X_n,X_{\infty})\,.
\end{equation}
On the other hand, the cut integral $\text{Cut}_C \mathcal{M}^{n\text{-gon}}$, where $C$ comprises a subset of the propagators in
\eqref{eq:uncutApp}, takes the form
\begin{equation}\label{eq:cutEmb}
\text{Cut}_C \mathcal{M}^{n\text{-gon}} = \frac{(-\pi i)^c}{\pi^{c/2}\sqrt{Y_C}}\,\mathcal{Q}_{n-c}^{\D-c}(X'_{C,c+1},\ldots,X'_{C,n},X'_{C,\infty})\,,
\end{equation}
where $c=|C|$ and where $Y_C$ is a (modified) Cayley determinant, which has a simple expression in terms of the $X_i$'s
\begin{equation}\label{eq:compact_Gram}
    Y_C=(-1)^c\det(X_iX_j)_{i,j\in C} \qquad (\text{assuming} \ \infty \notin C)\,.
\end{equation}
In practice, it is often useful to write the scalar products $(X'_{C,i}X'_{C,j})$ in terms of the $X_i$:
\begin{equation}\label{eq:proj_scalar_product}
(X'_{C,i}X'_{C,j}) = \frac{1}{(-1)^c\, Y_C}\,\det\left(\begin{array}{cccc}
(X_1X_1) & \ldots &(X_cX_1)& (X_iX_1)\\
\vdots && \vdots&\vdots \\
(X_1X_c) & \ldots &(X_cX_c)& (X_iX_c)\\
(X_1X_j) & \ldots &(X_cX_j)& (X_iX_j)
\end{array}\right)\,.
\end{equation}
Furthermore, the integral $\mathcal{Q}_{n}^\D$ is easily written as a parametric integral in Feynman parameter space \cite{Simmons-Duffin:2012juh}. The result is
\begin{equation}
\mathcal{Q}_{n}^\D(X_1,\ldots,X_n,X_0) = \frac{\Gamma(\D/2)}{\Gamma(\D-n)}\int \frac{\d a_0 \cdots \d a_n}{\text{GL}(1)}\, a_0^{\D-n-1}\,[- (\xi\xi)]^{-\D/2}\,,
\end{equation}
where $\xi=\sum_{i=0}^n a_i X_i$. Plugging this result in \eqref{eq:cutEmb} gives a Schwinger-parameter representation for the cut integral.
\section{Massless pentagon differential equation}
\label{app:details}
In this appendix, we provide additional details regarding the differential equation that we derived for the massless pentagon. The (transpose of) the differential equation is given by
\begin{equation}\label{eq:OmegaT}
\text{d}\boldsymbol{\Omega}^\top=\begin{pmatrix}\boldsymbol{A} & \boldsymbol{B}\\
\boldsymbol{0}_{6\times5}&\boldsymbol{C}\end{pmatrix},
\end{equation}
where
\begin{equation*}
    \boldsymbol{A}=-\text{diag}\left([W_5],[W_3],[W_1],[W_4],[W_2]\right),
\end{equation*}
\begin{equation*}
    \boldsymbol{B}=-2\begin{pmatrix}
        \left[\tfrac{W_{15}}{W_{20}}\right]&\left[\tfrac{W_{12}}{W_{19}}\right]&0&\left[\tfrac{W_5W_{17}}{W_{12}W_{15}}\right]&0&-\tfrac{1}{32}\left[\tfrac{W_{28}
W_{29}}{W_{26}}\right]\\
\left[\tfrac{W_3W_{20}}{W_{13}W_{15}}\right]&0&\left[\tfrac{W_{13}}{W_{18}}\right]&\left[\tfrac{W_{15}}{W_{17}}\right]&0&-\tfrac{1}{32}\left[\tfrac{W_{26}
W_{27}}{W_{29}}\right]\\
\left[\tfrac{W_{13}}{W_{20}}\right]&0&\left[\tfrac{W_1W_{18}}{W_{11}W_{13}}\right]&0&\left[\tfrac{W_{11}}{W_{16}}\right]&-\tfrac{1}{32}\left[\tfrac{W_{29}
W_{30}}{W_{27}}\right]\\
0&\left[\tfrac{W_{14}}{W_{19}}\right]&\left[\tfrac{W_{11}}{W_{18}}\right]&0&\left[\tfrac{W_4W_{16}}{W_{11}W_{14}}\right]&-\tfrac{1}{32}\left[\tfrac{W_{27}
W_{28}}{W_{30}}\right]\\
0&\left[\tfrac{W_2W_{19}}{W_{12}W_{14}}\right]&0&\left[\tfrac{W_{12}}{W_{17}}\right]&\left[\tfrac{W_{14}}{W_{16}}\right]&-\tfrac{1}{32}\left[\tfrac{W_{26}
W_{30}}{W_{28}}\right]
    \end{pmatrix},
\end{equation*}
\begin{equation*}
    \boldsymbol{C}=\begin{pmatrix}
        \left[\tfrac{W_{20}}{W_1W_{5}}\right]&0&0&0&0&\tfrac{1}{32}[W_{29}]\\
0&\left[\tfrac{W_{19}}{W_4W_{5}}\right]&0&0&0&\tfrac{1}{32}[W_{28}]\\
0&0&\left[\tfrac{W_{18}}{W_3W_{4}}\right]&0&0&\tfrac{1}{32}[W_{27}]\\
0&0&0&\left[\tfrac{W_{17}}{W_2W_{3}}\right]&0&\tfrac{1}{32}[W_{26}]\\
0&0&0&0&\left[\tfrac{W_{16}}{W_1W_{2}}\right]&\tfrac{1}{32}[W_{30}]\\
0&0&0&0&0&2[W_{31}]-[W_1W_2W_3W_4W_{5}]
    \end{pmatrix},
\end{equation*}
and where the symbol $[X]$ is a shorthand for $\text{d}\log(X)$. The differential equation alphabet consists of a proper subset of the planar pentagon alphabet denoted as $\mathbb{A}_{\text{P}}=\{W_1,\ldots, W_{31}\}$  \cite{Chicherin:2017dob}. Setting $\{s_{12},s_{23},s_{34},s_{45},s_{51}\}=\{v_1,v_2,v_3,v_4,v_5\}$, these are explicitly given by
\begin{align}\label{fullpentagonalphabet}
W_{1} =& \; v_{1} \,,  \notag\\
W_{6} =& \; v_{3} + v_{4} \,,\notag  \ \\
W_{11} =& \; v_{1} - v_{4} \,,  \notag\\ 
W_{16} =& \; v_{4} - v_{1}- v_{2} \,, \\
W_{21} =& \; v_3 + v_4 - v_1 - v_2 \,,  \notag\\
W_{26} =& \; \frac{v_{1} v_{2} -v_{2} v_{3} +v_{3} v_{4} -v_{1} v_{5} -v_{4} v_{5} - \sqrt{\Delta} }{v_1 v_2 -v_2 v_3 +v_3 v_4 -v_1 v_5 -v_4 v_5 + \sqrt{\Delta}} \,,  \notag\\
W_{31} =& \; \sqrt{\Delta} \,,\notag
\end{align}
with $W_{1+i}, W_{6+i}, W_{11+i}, W_{16+i}, W_{21+i}, W_{26+i}$, with $i=1\ldots 5$, defined by cyclic symmetry.
It is worth noting that the symbols $W_i$, where $i$ ranges from $26$ to $30$, are of parity-odd nature. Specifically, these symbols transform to their inverse under the transformation $\Delta \to -\Delta$, while all other symbols are of parity-even nature under the same transformation.
\section{Numerical method for solving differential equations}
\label{app:numerics}
In this section, we present details on a simple numerical method for solving canonical differential equation satisfied by a set of (polylogarithmic) master integrals. While the approach is discussed with more details in \cite{Caron-Huot:2020vlo} (see also \cite{Abreu:2020jxa,Hidding:2020ytt}), this section will focus on a specific example to illustrate the practical implementation of the method.

We will focus on the massless pentagon family, which is introduced in Sec.~\ref{sec:pentagon}. The differential equation satisfied by the pure master integrals is provided in \eqref{eq:OmegaT} for  generic kinematics. To keep things simple, we will limit our discussion to the multi-Regge regime, although this is not necessary for the method outlined below to work (for example, we used it to move between disconnected multi-Regge limits in Sec.~\ref{sec:pentagon}). 

We will use \eqref{eq:bdryCTE} as the boundary condition and further set $s=s_1=s_2=1$. We will be referring to it as $\vec{I}_0$. Our goal is to transport $\vec{I}_0$ (defined at the singular point $z=\bar{z}=0$) to  $z=\bar{z}=1$ by solving
\begin{equation}
    \text{d}\vec{I}=\epsilon~\text{d}\boldsymbol{\Omega}(z,\bar{z})\cdot \vec{I}\,,
\end{equation}
along the straight line $z=\bar{z}$. 

As a first step, we set a sequence of paths in the $(z,\bar{z})$ plane, each starting at a singularity of the differential equation and ending at a regular point that is approximately equidistant to the nearest singularity (this choice ensures a better numerical convergence of the algorithm). On the line $z=\bar{z}$, the singularities are located at $(0,0), (1/2,1/2),$ and $(1,1)$. A natural set of paths is 
\begin{gather}
    \gamma_1(t)=\left\{(z,\bar{z})=(t,t) \;|\; 0\leq t\leq1/3\right\},\notag\\
    \label{eq:exContour}
    \gamma_2(t)=\left\{(z,\bar{z})=(1/2-t,1/2-t) \;|\; 0\leq t\leq1/6\right\},\\
    \gamma_3(t)=\left\{(z,\bar{z})=(1/2+t,1/2+t) \;|\; 0\leq t\leq1/6\right\},\notag\\
    \gamma_4(t)=\left\{(z,\bar{z})=(1-t,1-t) \;|\; 0\leq t\leq1/3\right\}.\notag
\end{gather}
Transporting along these paths, the solution is formally given by
\begin{equation}\label{eq:solPOE}
\begin{split}
\vec{I}_1&=\mathcal{P}\exp\left(\epsilon\int_{\gamma_4^{-1}}\text{d}\boldsymbol{\Omega}(t)\right)\cdot\mathcal{P}\exp\left(\epsilon\int_{\gamma_3}\text{d}\boldsymbol{\Omega}(t)\right)\\& \qquad \cdot\mathcal{P}\exp\left(\epsilon\int_{\gamma_2^{-1}}\text{d}\boldsymbol{\Omega}(t)\right)\cdot\mathcal{P}\exp\left(\epsilon\int_{\gamma_1}\text{d}\boldsymbol{\Omega}(t)\right)\cdot\vec{I}_0\,.
\end{split}
\end{equation}
Subsequently, we evaluate the path-ordered exponential using a series expansion. The method relies on the fact that the differential equation has, at worst, simple poles. In particular, the pullback of $\text{d}\boldsymbol{\Omega}(t)$ along the path $\gamma_i(t)$ takes the form
\begin{equation}\label{eq:expDE}
    \text{d}\boldsymbol{\Omega}(t)=\left(\frac{\boldsymbol{A}_0^{(i)}}{t}+\sum_{k\ge 0}t^k \boldsymbol{A}_{k+1}^{(i)}\right)\text{d}t\,,
\end{equation}
where the matrices $\boldsymbol{A}_\bullet^{(i)}$ are purely numerical. Note that if square roots were present, \eqref{eq:expDE} would contain half-integer powers of $t$. Moreover, if the expansion is done near a regular point, it is understood that $\boldsymbol{A}_0^{(i)}=0$.

Next, the gauge transformation $\vec{I}=\boldsymbol{T}^{(i)}\cdot\vec{I}^{(i)}$ isolating the pole at $t=0$
\begin{equation}\label{eq:newDE}
    \left(\boldsymbol{T}^{(i)}\right)^{-1}\cdot\left(\epsilon \partial_t\boldsymbol{\Omega}(t)\cdot \boldsymbol{T}^{(i)}-\partial_t\boldsymbol{T}^{(i)}\right)=\epsilon\frac{\boldsymbol{A}_0^{(i)}}{t}\,,
\end{equation}
can be recursively constructed as the double series
\begin{equation}\label{eq:Tmatrix}
    \boldsymbol{T}^{(i)}(t,\epsilon)= \mathbbm{1}+ \sum_{k= 1}^\infty\sum_{m=1}^\infty t^k \epsilon^m \boldsymbol{T}_{k,m}^{(i)}\,,
\end{equation}
where 
\begin{equation}\label{eq:recRel}
\begin{aligned}
& \boldsymbol{T}_{k,1}^{(i)} = \frac{1}{k} \boldsymbol{A}_k^{(i)}, \, \\
& \boldsymbol{T}_{k,m}^{(i)} = \frac{1}{k} \left( \left[\boldsymbol{A}_0^{(i)},\boldsymbol{T}_{k,m-1}^{(i)}\right]+\sum_{j=1}^{k-1} \boldsymbol{A}_{k-j}^{(i)}\cdot \boldsymbol{T}_{j,m-1}^{(i)} \right)  \,,\;\;\; \forall m > 1\,.
\end{aligned}
\end{equation}
The recurrence relations are derived at the very end of this section. The equation can therefore be solved systematically in terms of $t$ and $\epsilon$ order by order (we will in general truncate the series in $\epsilon$ to $\mathcal{O}(\epsilon^4)$), yielding an explicit expression for $\boldsymbol{T}^{(i)}$. The gauge transformed differential equation in \eqref{eq:newDE} is easily solved as a power-law in $t_{\text{e.p.}}^{(i)}$, where $t_{\text{e.p.}}^{(i)}$ labels the endpoint of the $i^\text{th}$-contour. Inverting the gauge transformation gives us the solution to the original set of master integrals. In particular,
\begin{equation}\label{eq:POE}
\mathcal{P}\exp\left(\epsilon\int_{\gamma_i}\text{d}\boldsymbol{\Omega}(t)\right)=\boldsymbol{T}^{(i)}(t_{\text{e.p.}}^{(i)},\epsilon)\cdot \exp\left(\epsilon \log(t_{\text{e.p.}}^{(i)})A_0^{(i)}\right)\,,
\end{equation}
and similarly 
\begin{equation}\label{eq:POEinv}
\mathcal{P}\exp\left(\epsilon\int_{\gamma_i^{-1}}\text{d}\boldsymbol{\Omega}(t)\right)=\exp\left(-\epsilon \log(t_{\text{e.p.}}^{(i)})A_0^{(i)}\right)\cdot(\boldsymbol{T}^{(i)}(t_{\text{e.p.}}^{(i)},\epsilon))^{-1}\,.
\end{equation}
Note that the matrix $\boldsymbol{T}^{(i)}$ is typically defined as an \emph{infinite} series. However, since the (regular) endpoint of each integration contour $\gamma_i$ in \eqref{eq:exContour} is roughly equidistant to its nearest singularity (which is the starting point of $\gamma_{i+1}$), the series $\boldsymbol{T}^{(i)}$ converges quickly at a finite order in $t$, providing an accurate approximation of the initial condition on which either $\boldsymbol{T}^{(i+1)}$ acts.

By substituting \eqref{eq:POE} and \eqref{eq:POEinv} into \eqref{eq:solPOE}, we can achieve an approximation of the solution at $z=\bar{z}=1$ with a desired level of numerical accuracy. In this specific calculation, we limited each series in $t$ to $\mathcal{O}(t^{1500})$, which provides confidence in the result up to roughly $\sim 265$ digits.

To conclude this section, we obtain the recursive relationships stated in \eqref{eq:recRel} by substituting the ansatz
\begin{equation}
    \boldsymbol{T}^{(i)}=\mathbbm{1}+\sum_{k\ge1} t^k \boldsymbol{T}_k^{(i)}\,,
\end{equation}
and equation \eqref{eq:expDE} into \eqref{eq:newDE} and organizing the resulting terms according to their powers of $t$. It can be shown through a straightforward computation that
\begin{gather}
\epsilon \partial_t\boldsymbol{\Omega}(t)\cdot \boldsymbol{T}^{(i)}-\partial_t\boldsymbol{T}^{(i)}-\epsilon\boldsymbol{T}^{(i)}\cdot\frac{\boldsymbol{A}_0^{(i)}}{t}=0\,,\\
\iff \notag\\
\sum_{k\ge1}t^{k-1}\left(\epsilon\left[\boldsymbol{A}_0^{(i)},\boldsymbol{T}_k^{(i)}\right]+\epsilon\boldsymbol{A}_{k}^{(i)}+\epsilon\sum_{j\ge 1}t^{j}\boldsymbol{A}_{k}^{(i)}\cdot\boldsymbol{T}_j^{(i)}- k\boldsymbol{T}_k^{(i)}\right)=0\,.
\end{gather}
To factor the $t$-dependence completely from the third term, we use 
\begin{subequations}
\begin{align}
     \sum_{\substack{k\ge1\\j\ge 1}}t^{k-1+j}\boldsymbol{A}_{k}^{(i)}\cdot\boldsymbol{T}_j^{(i)}&=\left(\sum_{k=1+j}^\infty t^{k-1-j}\boldsymbol{A}_{k-j}^{(i)}\right)\cdot\left(\sum_{k=1}^\infty t^{j}\boldsymbol{T}_{j}^{(i)}\right)
     \\&=\sum_{k=1}^\infty\sum_{j=1}^\infty t^{k-1}\boldsymbol{A}_{k-j}^{(i)}\cdot\boldsymbol{T}_{j}^{(i)}-\sum_{j=1}^\infty\sum_{k=1}^j t^{k-1}\boldsymbol{A}_{k-j}^{(i)}\cdot\boldsymbol{T}_{j}^{(i)}\\&=\sum_{k=1}^\infty\sum_{j=1}^{k-1} t^{k-1}\boldsymbol{A}_{k-j}^{(i)}\cdot\boldsymbol{T}_{j}^{(i)}\,.
\end{align}
\end{subequations}
To derive the final term from the third term, we made use of the fact that $\boldsymbol{A}_{k-j}^{(i)}$ is non-zero only for $k\ge j$, and that the double series initially begins at order $t^1$ to exclude the case where $k=j$. By combining these observations, we arrive at the contiguous relation
\begin{equation}\label{eq:contigous}
\boldsymbol{T}_k^{(i)}=\frac{\epsilon}{k}\left(\boldsymbol{A}_{k}^{(i)}+\left[\boldsymbol{A}_0^{(i)},\boldsymbol{T}_k^{(i)}\right]+\sum_{j=1}^{k-1}\boldsymbol{A}_{k-j}^{(i)}\cdot\boldsymbol{T}_j^{(i)}\right)\,.
\end{equation}
In particular, we see that $\boldsymbol{T}_k^{(i)}=\mathcal{O}(\epsilon)$. This instructs us to expand $\boldsymbol{T}_k^{(i)}$ further in $\epsilon$
\begin{equation}\label{eq:epsExp}
    \boldsymbol{T}_k^{(i)}=\sum_{m\ge 1}\epsilon^m\boldsymbol{T}_{k,m}^{(i)}\,.
\end{equation}
Plugging \eqref{eq:epsExp} into \eqref{eq:contigous} and collecting powers of $\epsilon$, we recover \eqref{eq:recRel}.

\bibliographystyle{JHEP}
\bibliography{refs}

\end{document}